\documentclass[apj]{emulateapj}

\usepackage{amsmath}
\usepackage{rotating}
\usepackage{graphicx}

\newcommand{\be}{\begin{equation}}
\newcommand{\ee}{\end{equation}}

\newcommand{\gtsima}{$\; \buildrel > \over \sim \;$}
\newcommand{\ltsima}{$\; \buildrel < \over \sim \;$}
\newcommand{\prosima}{$\; \buildrel \propto \over \sim \;$}
\newcommand{\gsim}{\lower.5ex\hbox{\gtsima}}
\newcommand{\lsim}{\lower.5ex\hbox{\ltsima}}
\newcommand{\simgt}{\lower.5ex\hbox{\gtsima}}
\newcommand{\simlt}{\lower.5ex\hbox{\ltsima}}
\newcommand{\simpr}{\lower.5ex\hbox{\prosima}}

\newcommand{\cxo}{\textit{Chandra}}

\newcommand{\spi}{\textit{Spitzer}}

\shorttitle{AMUSE-VIRGO. III: MIR imaging}
\shortauthors{Leipski et al.}

\begin{document}


\title{AMUSE-VIRGO. III: mid-infrared photometry of early-type galaxies and limits on obscured nuclear emission}


\author{Christian Leipski\altaffilmark{1}}
\author{Elena Gallo\altaffilmark{2}}
\author{Tommaso Treu\altaffilmark{3,4}}
\author{Jong-Hak Woo\altaffilmark{5}}
\author{Brendan P. Miller\altaffilmark{2}}
\author{Robert Antonucci\altaffilmark{3}}

\altaffiltext{1}{Max-Planck Institut f\"ur Astronomie (MPIA), K\"onigstuhl 17, D-69117 Heidelberg, Germany; email: {\tt leipski@mpia-hd.mpg.de}}
\altaffiltext{2}{Department of Astronomy, University of Michigan, 500 Church St., Ann Arbor, MI 48109}
\altaffiltext{3}{Department of Physics, University of California, Santa Barbara, CA 93106}
\altaffiltext{4}{Packard Fellow}
\altaffiltext{5}{Department of Physics \& Astronomy, Seoul National University, Seoul, 151-742, Republic of Korea}

\begin{abstract}

We complete our census of low-level nuclear activity in Virgo Cluster
early-type galaxies by searching for obscured emission using {\it
Spitzer} Space Telescope mid-infrared (MIR) imaging at 24\,$\mu$m. Of
a total sample of 95 early-type galaxies, 53 objects are detected,
including 16 showing kiloparsec-scale dust in optical images. One
dimensional and two dimensional surface photometry of the 37
detections without extended dust features reveals that the MIR light
is more centrally concentrated than the optical light as traced by
{\it Hubble} Space Telescope F850LP-band images.  No such modeling was 
performed for the sources with dust detected in the optical images. We 
explore several possible sources of the MIR excess emission, including obscured
nuclear emission. We find that radial metallicity gradients in the
stellar population appear to be a natural and most likely explanation
for the observed behavior in a majority of the sources. Alternatively,
if the concentrated MIR emission were due to nuclear activity, it
would imply a MIR-to-X luminosity ratio $\sim$\,$5-10$ for the low
luminosity AGN detected in X-rays by our survey. This ratio is an order
of magnitude larger than that of typical low-luminosity AGN and would
imply an unusual spectral energy distribution. We conclude that the
black holes found by our survey in quiescent early-type galaxies in Virgo
have low bolometric Eddington ratios arising from low accretion rates
and/or highly radiatively inefficient accretion.

\end{abstract}


\section{Introduction}

Perhaps the most traditional technique to identify accretion-powered
activity in galactic nuclei relies on identifying sources with an
ultraviolet excess, i.e. sources whose spectral energy distribution
(SED) does not steeply decline on the blue side of the stellar
spectrum peak \citep[e.g.][]{sch83}.
Albeit successful, this method fails to detect both obscured and
high-redshift AGN (Active Galactic Nuclei), which disappear from the
UV due to absorption from local dust and the Lyman forests,
respectively.  This shortcoming is not negligible, since the presence
of a large population of heavily obscured AGN seems required by the
hard spectrum of the X-ray background \citep{gil07}.

More generally, optical, UV, and X-ray observations, which are
traditionally employed to detect unobscured AGN activity, are still
known to miss a large fraction of the obscured AGN population, and
nearly all of the Compton-thick AGN which are thought to dominate AGN
number counts at high redshifts \citep{dad07}.

The obscuring dust that hides AGN from ultraviolet, optical, and soft
X-ray surveys should be a strong, largely isotropic emitter in the
mid-to-far-infrared. Inactive galaxies and AGN have different SEDs at
short infrared wavelengths: while the composite blackbody spectra of the
stellar population of normal galaxies produce an SED that peaks at
approximately 1.6$\mu$m (in F$_{\nu}$), AGN have roughly power-law shaped
SEDs. Searching for AGN at these wavelengths suffers very modestly
from extinction by dust or gas, as demonstrated for example by studies
based on the Two Micron All-Sky Survey~\citep[e.g.][]{gli04}.

More recently, observations taken with the Multiband Imaging
Photometer (MIPS; \citealt{rie04}) and Infrared Array Camera (IRAC;
\citealt{faz04}) aboard {\it Spitzer} have been successfully
employed to select AGN candidates independently of their optical
and/or X-ray properties. In addition, they have proven capable of
identifying heavily obscured AGN that are missed even in the deepest
X-ray fields, making IR selection a viable alternative to traditional
AGN selection methods \citep[e.g.][]{lac04,hat05,pol06,fio09}.

Analyzing a 24 $\mu$m-selected sample in GOODS-S,  \citet{don08} 
found that the fraction of MIR sources dominated by an AGN decreases
with decreasing flux density, but only down to a 24 $\mu$m flux
density of about 300 $\mu$Jy.  {Below this limit, the AGN fraction
levels out at ~10\%}, indicating that a substantial fraction of faint
24 $\mu$m sources are primarily powered by mechanisms other than star
formation.  Furthermore, the majority of AGN with low 24 $\mu$m flux
densities are missed by X-ray surveys, suggesting that X-ray emission
alone cannot be used to unambiguously pinpoint AGN activity.

Unfortunately, direct measurements of MIR Eddington ratios are mostly
limited to samples of known AGN -- here loosely defined as emitting
above a few per cent of their Eddington luminosity. At fainter levels
of nuclear luminosity, disentangling accreation powered and star
formation powered MIR emission becomes a challenge. While
contamination can be reduced by focusing on quiescent (in terms of
their star formation) early-type galaxies, pushing the threshold to
the lowest accretion rates/luminosities -- that is, below a few
thousandths of the Eddington limit in terms of nuclear,
accretion-powered activity -- necessarily means facing additional
sources of contamination. Those include the stellar population itself,
mainly made of photospheric emission in the Rayleigh-Jeans tail, and
emission from dust produced in the atmospheres and outflows of evolved
stars \citep[e.g.][]{bre98}.

The AMUSE-Virgo (AGN Multiwavelength Survey of Early-Type Galaxies in
the Virgo Cluster; see URL \url{http://tartufo.physics.ucsb.edu/$\sim$amuse/})
project set out to effectively bridge the gap between formally
inactive galactic nuclei and 'traditional' AGN. In order to detect and
characterize super-massive black holes accreting at
extremely sub-Eddington rates over a broad range in host stellar
masses, we were awarded \cxo\ and \spi\ observations (SPITZER
PID-30958, PI T.~Treu) of the 100 early-type galaxies which compose the
Virgo Cluster Survey \citep[VCS,][]{cot04}.  The VCS sample is
selected from the 163 Virgo spheroids brighter than B$_{\rm T}<16$
(including 44\% of all Virgo spheroids to this limit).  It is complete
to B$_{\rm T}=12$ (i.e. M$_{\rm B}\sim-18$) and it is a random
subsample for fainter magnitudes, unbiased with respect to the nuclear
properties. The sample covers over 4 orders of magnitude in black hole
mass, as estimated from the the M$_{\rm BH}$-$\sigma$ relation and
M$_{\rm BH}$-$L_{\rm B}$ relation.

Archival F475W and F850LP band (hereafter $g$ and $z$) {\it Hubble}
ACS (Advanced Camera for Surveys) images are available for each target
(see \citealt{fer06} for a detailed isophotal analysis), while
\cxo\ observations of the same sample have been presented in Paper
I. and II. of this series \citep{gal08,gal10}.  In
this paper, we present the results of the MIPS \spi\ 24 $\mu$m
observations.

\begin{figure}[t!]
\centering
\includegraphics[angle=0,scale=.5]{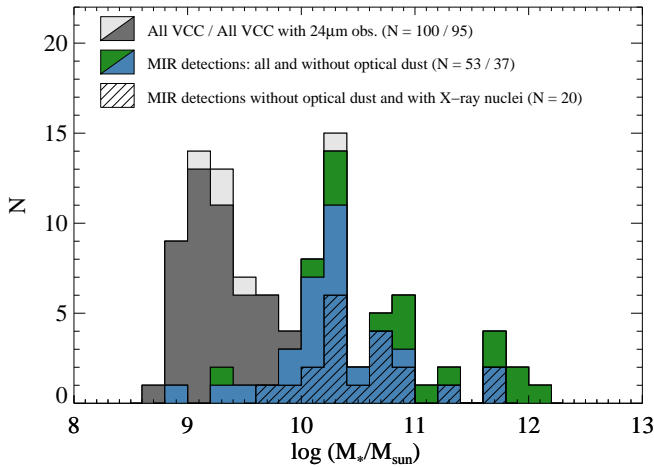}
\caption{The VCC sources as a function of their stellar mass. 
\label{histogram} }
\end{figure}

\begin{figure*}[t!]
\centering
\includegraphics[angle=0,scale=.28]{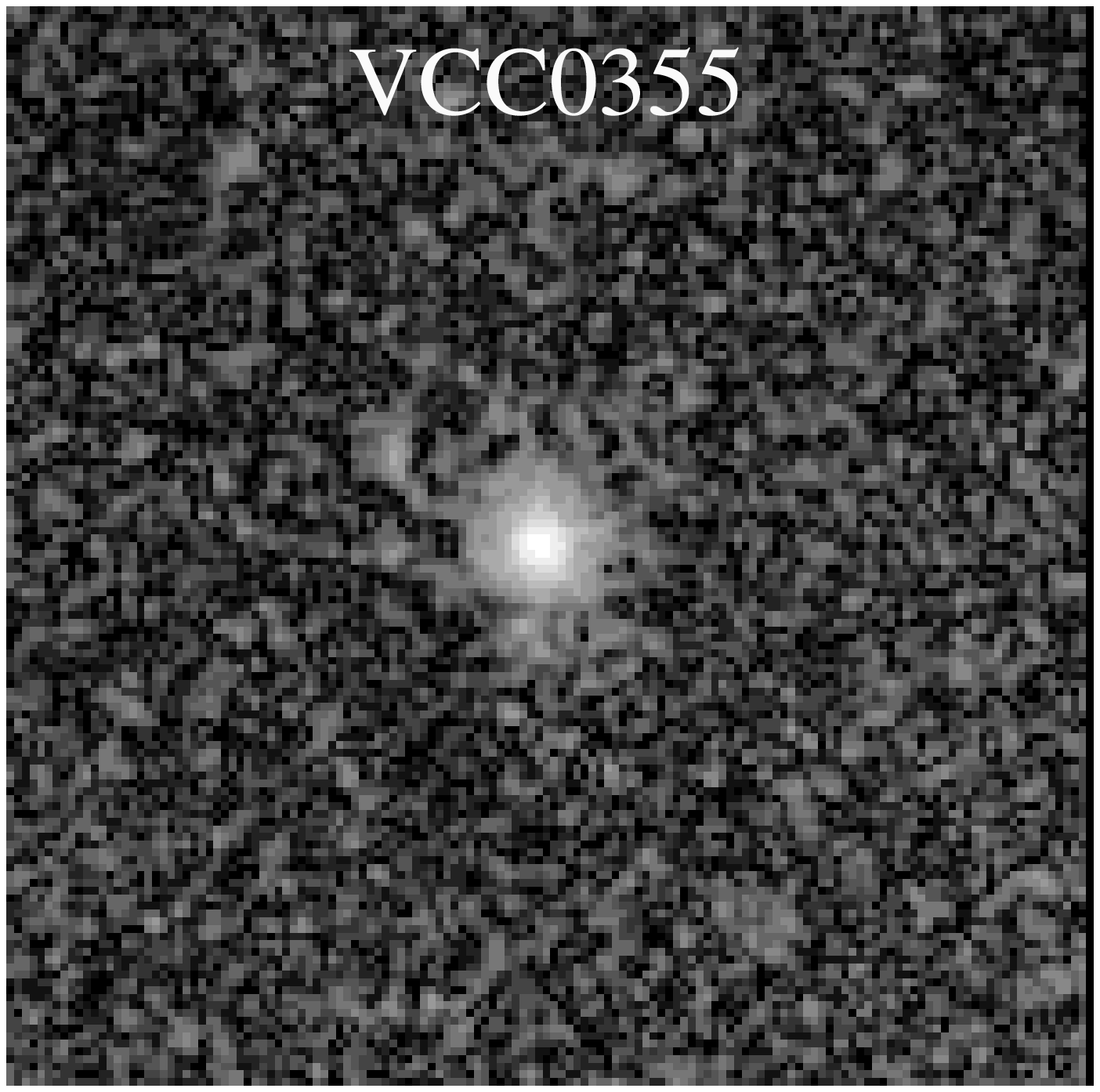}
\includegraphics[angle=0,scale=.28]{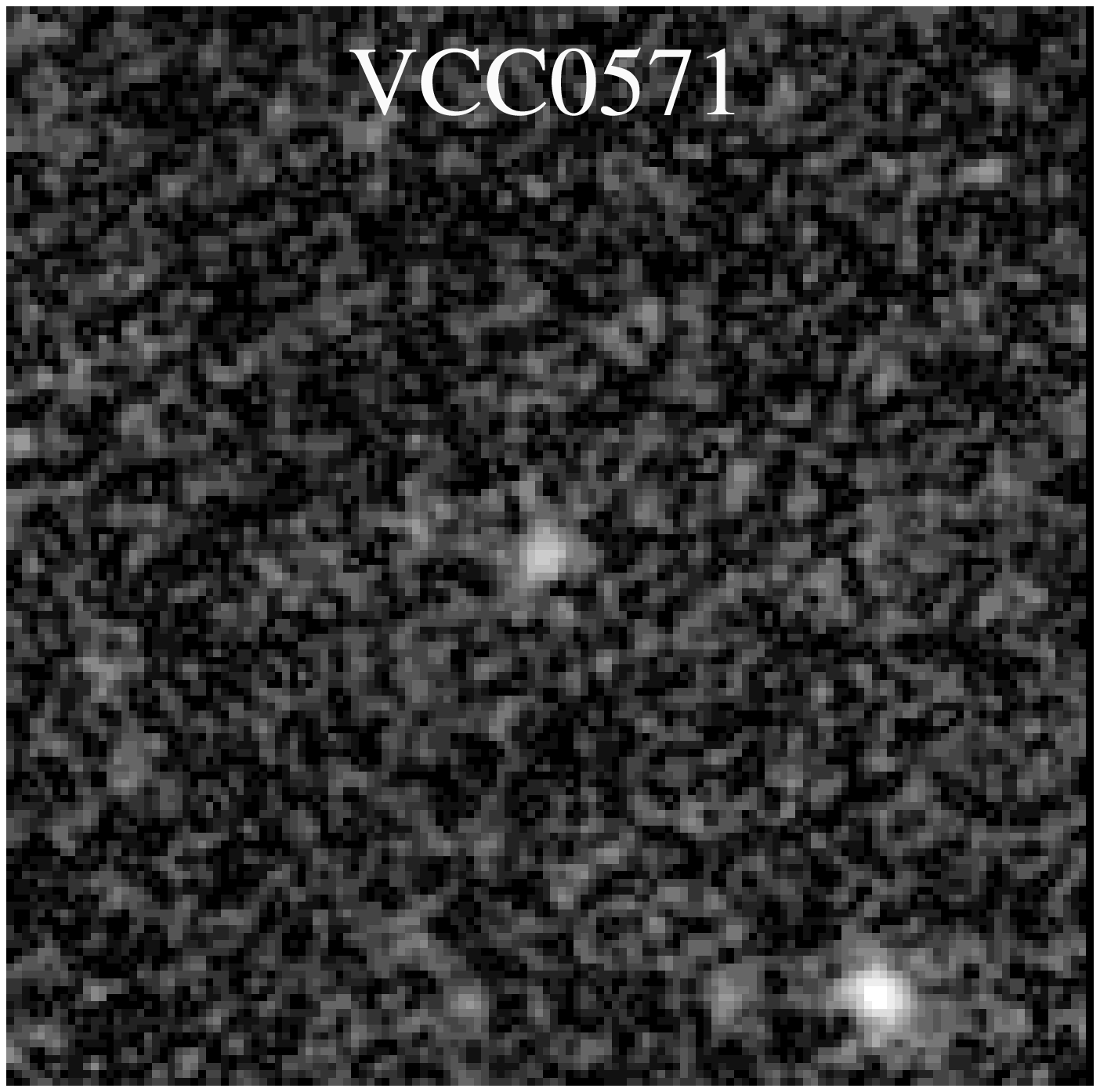}
\includegraphics[angle=0,scale=.28]{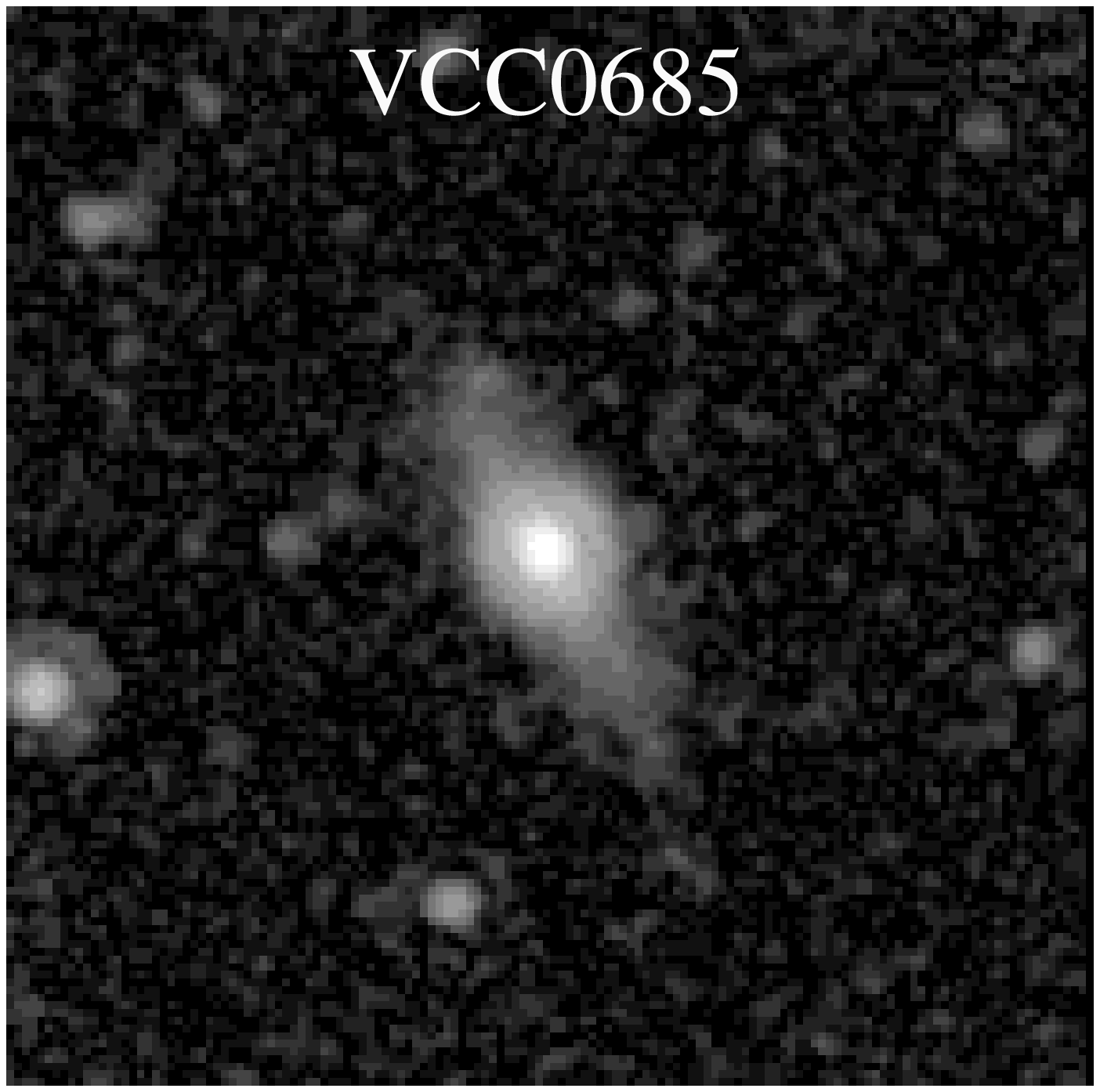}
\includegraphics[angle=0,scale=.28]{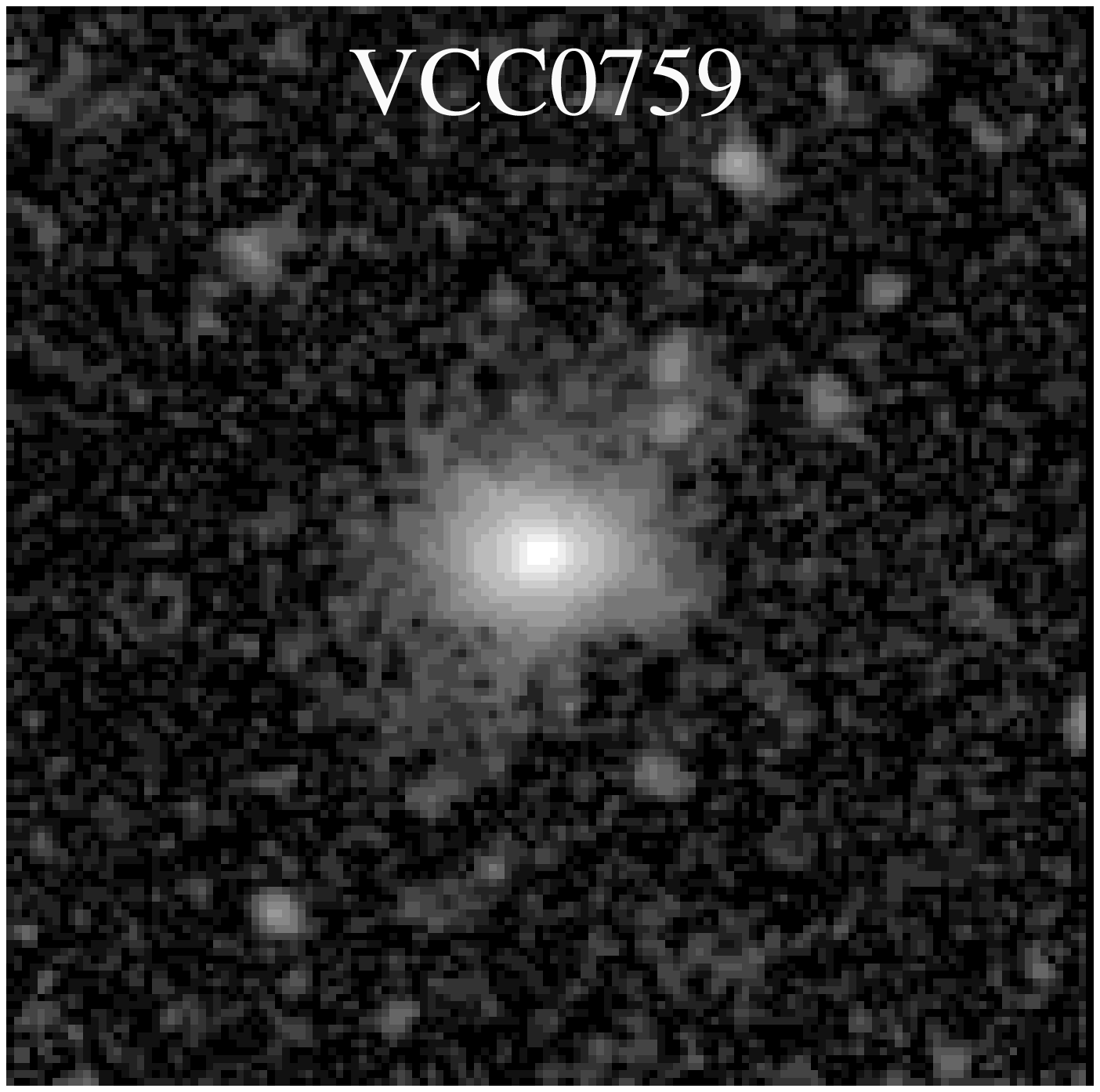}\vspace{0.025cm}
\includegraphics[angle=0,scale=.28]{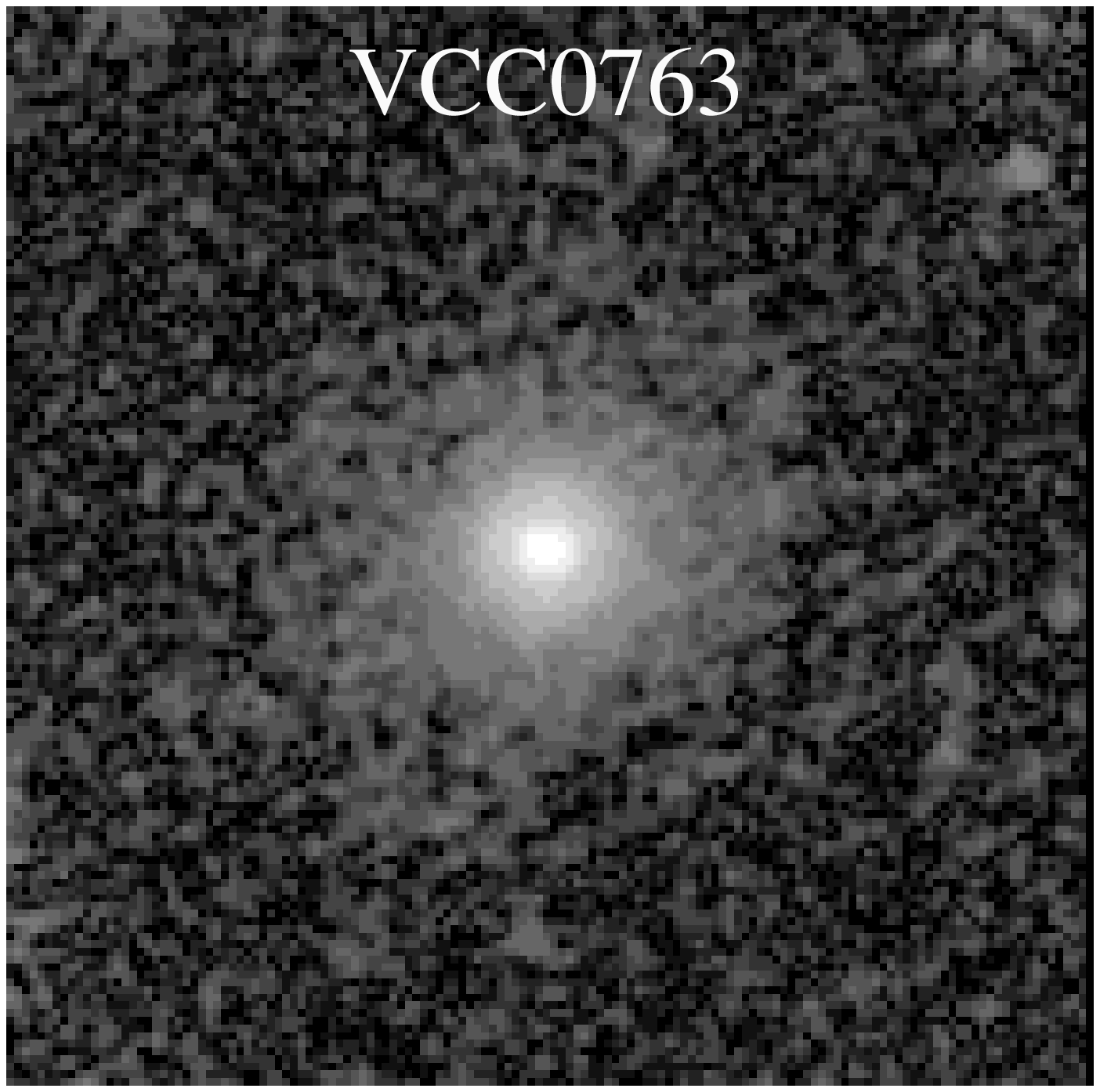}
\includegraphics[angle=0,scale=.28]{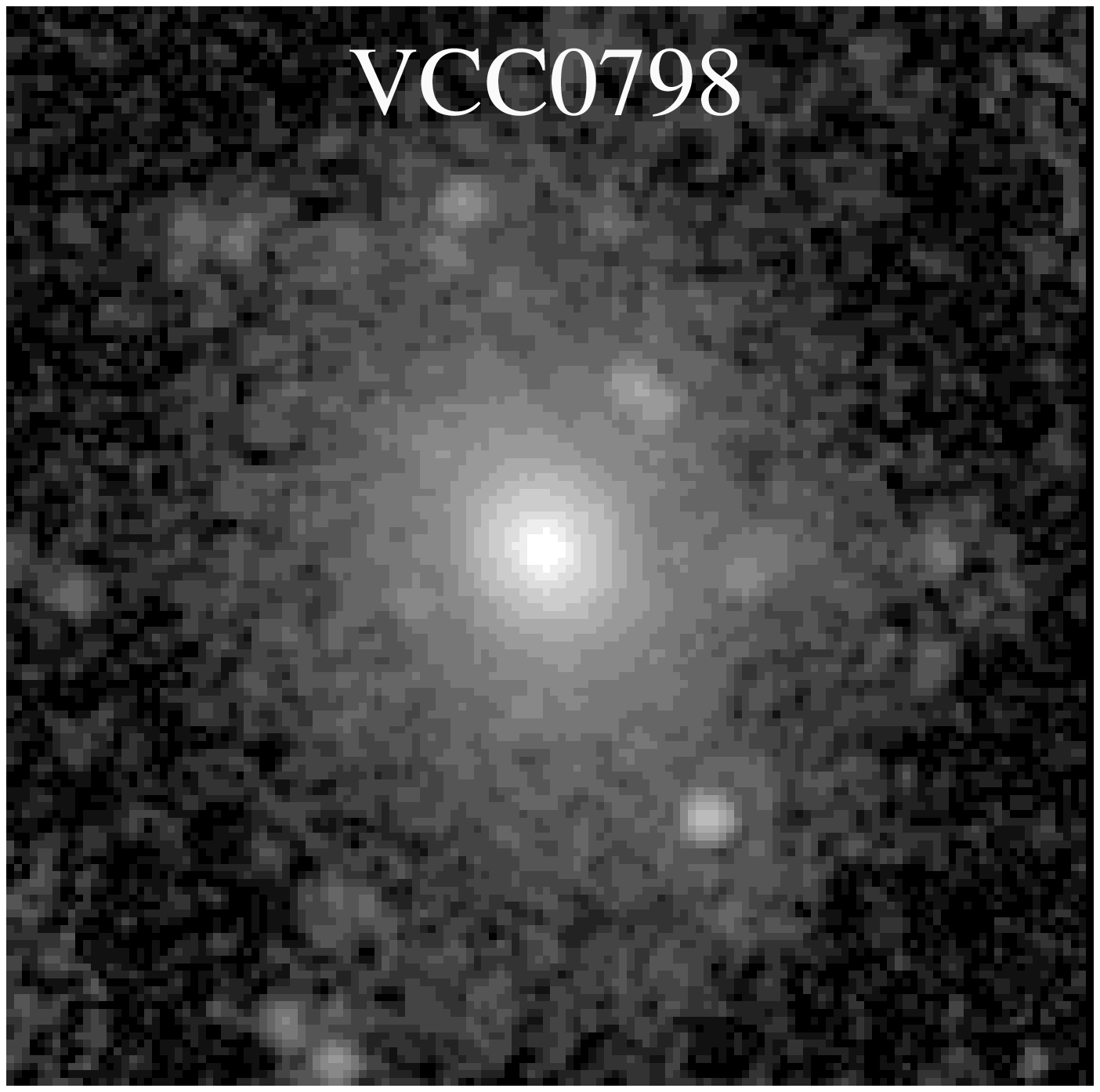}
\includegraphics[angle=0,scale=.28]{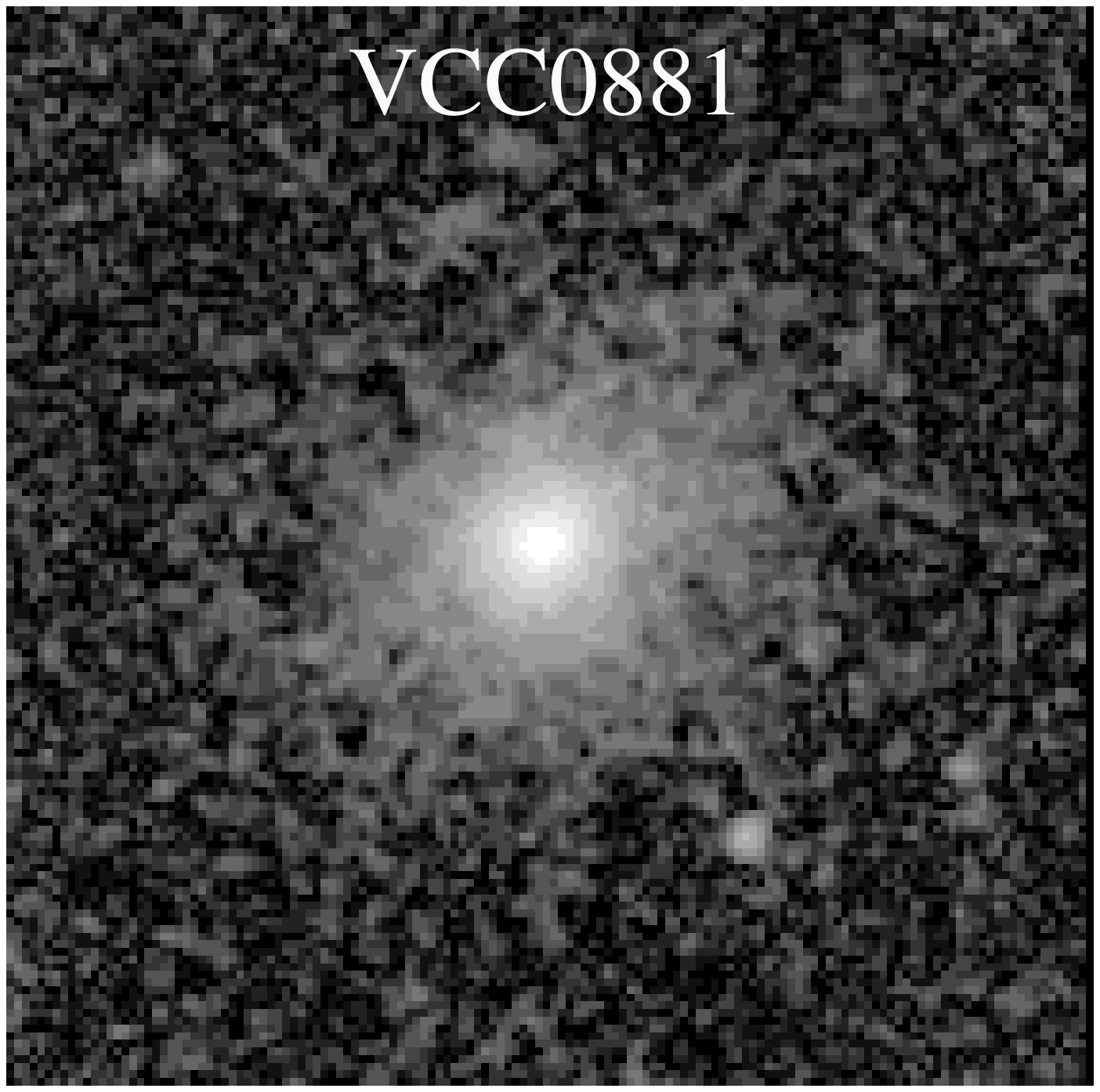}
\includegraphics[angle=0,scale=.28]{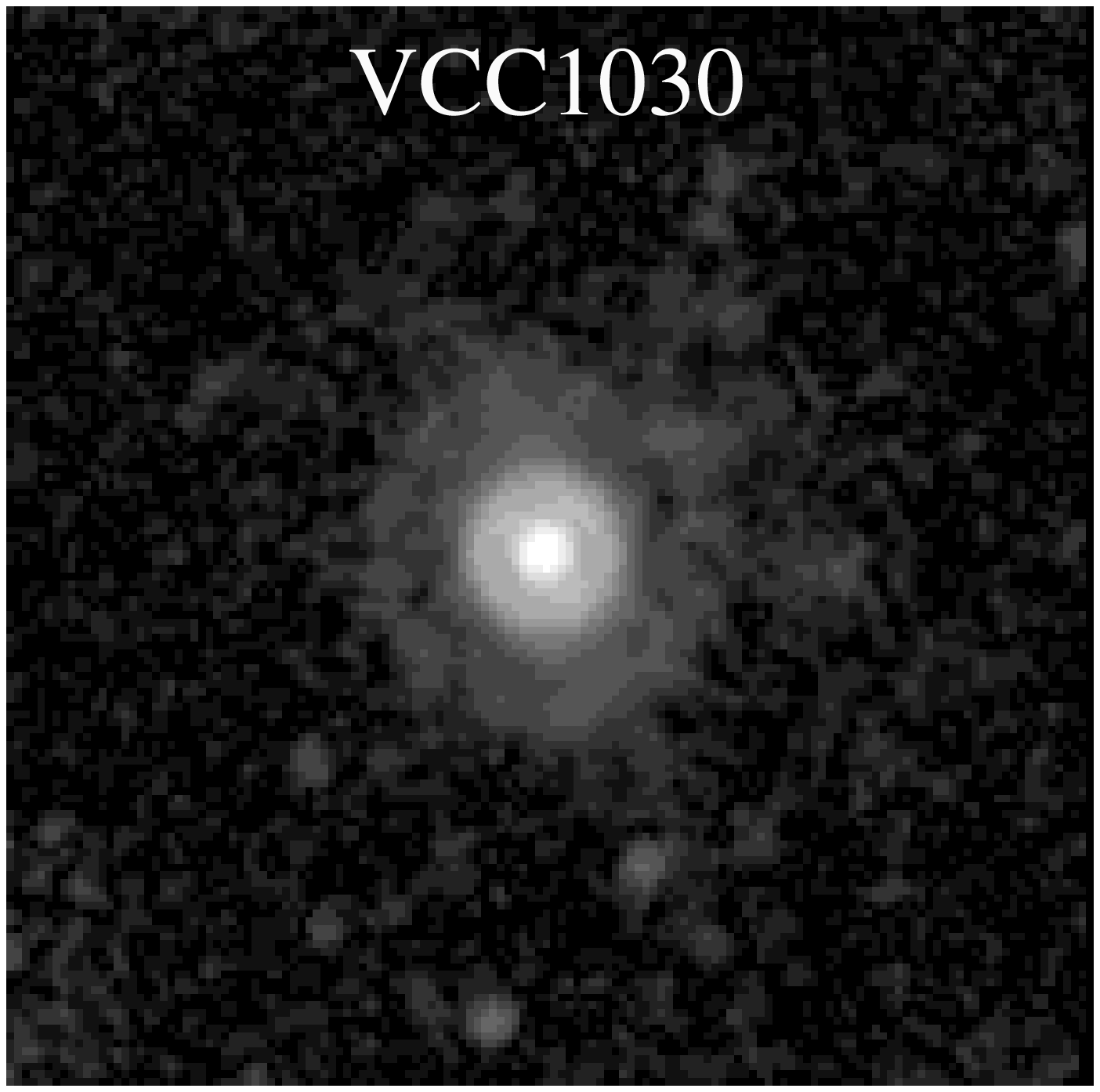}\vspace{0.025cm}
\includegraphics[angle=0,scale=.28]{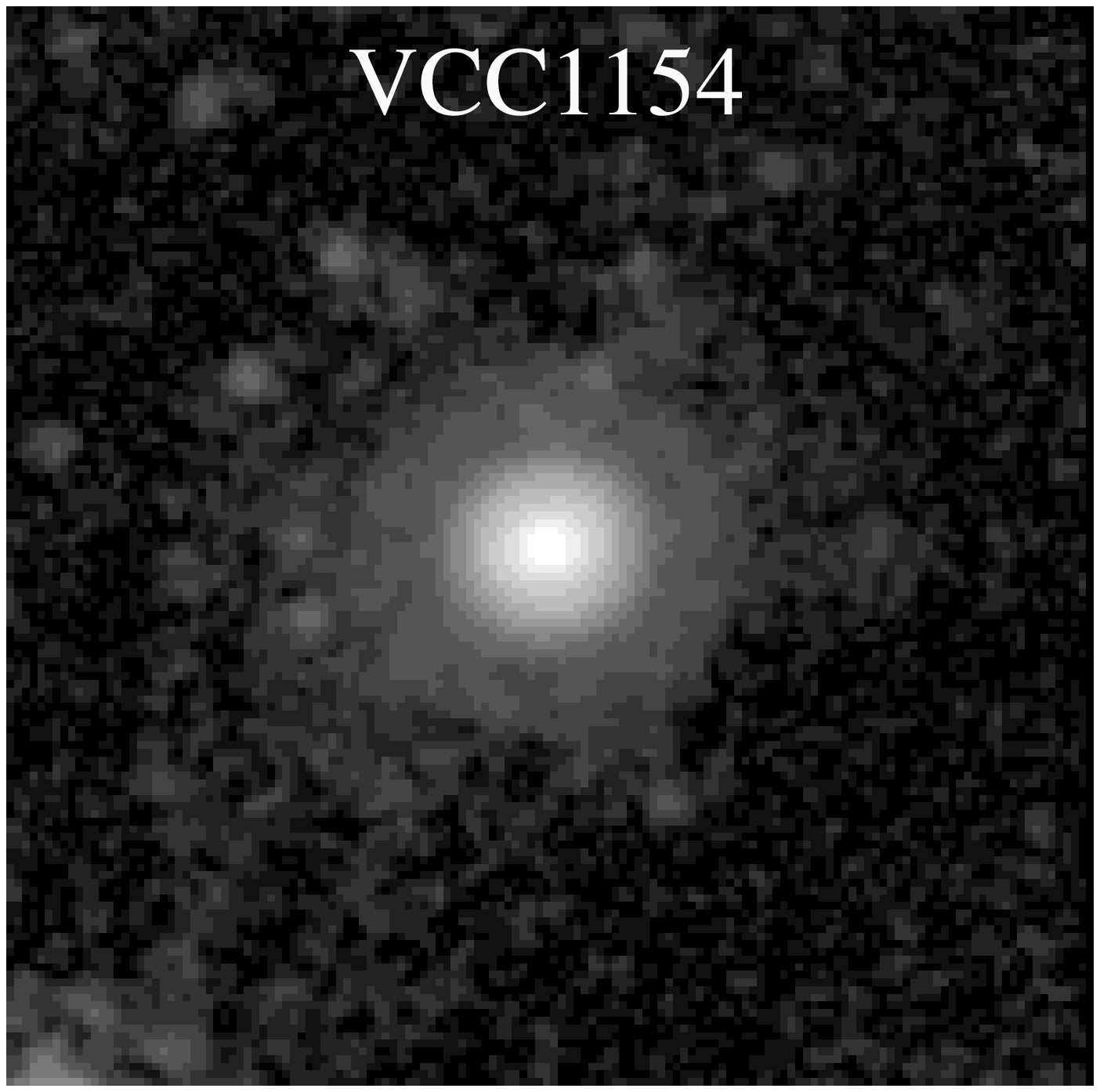}
\includegraphics[angle=0,scale=.28]{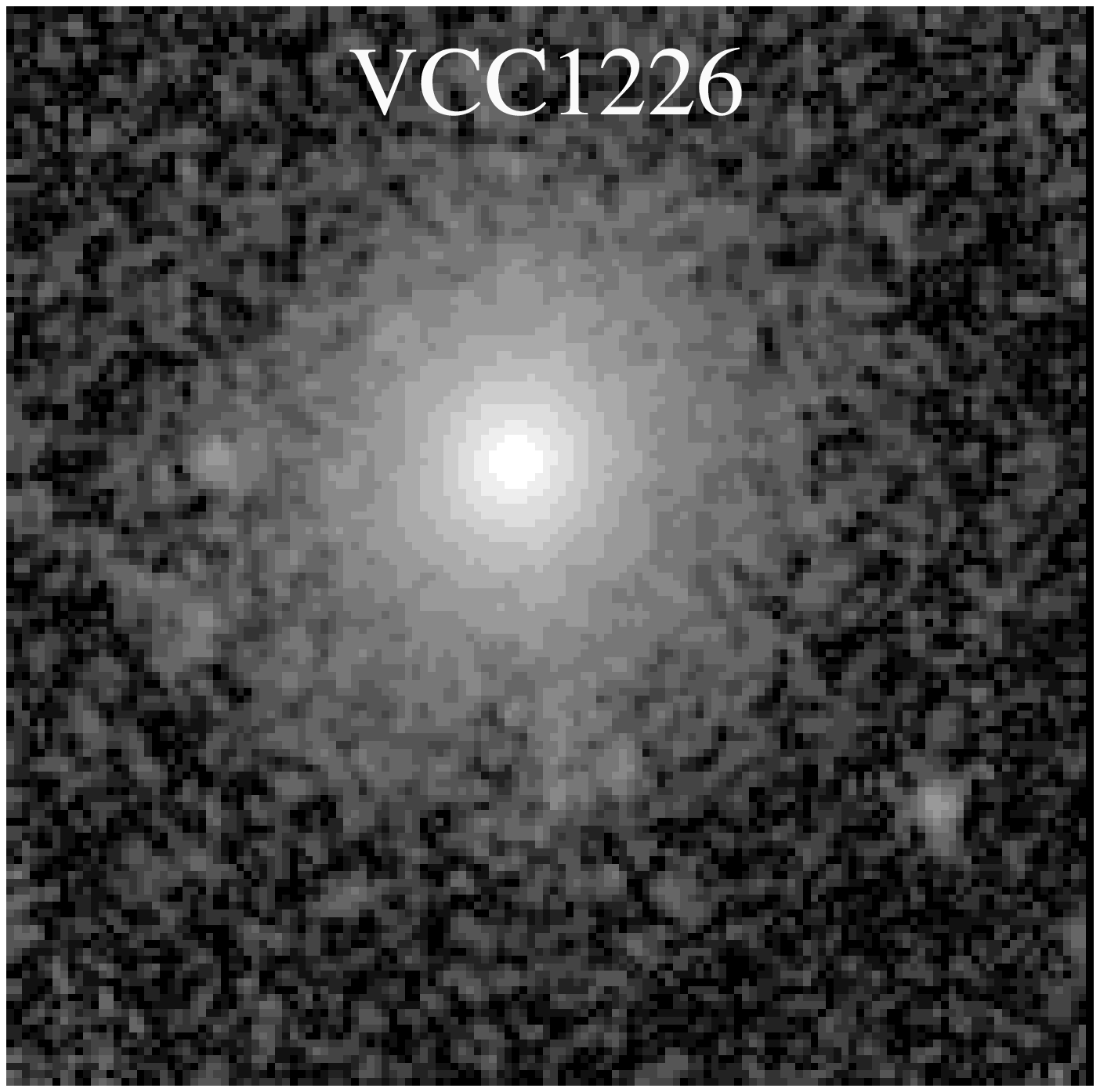}
\includegraphics[angle=0,scale=.28]{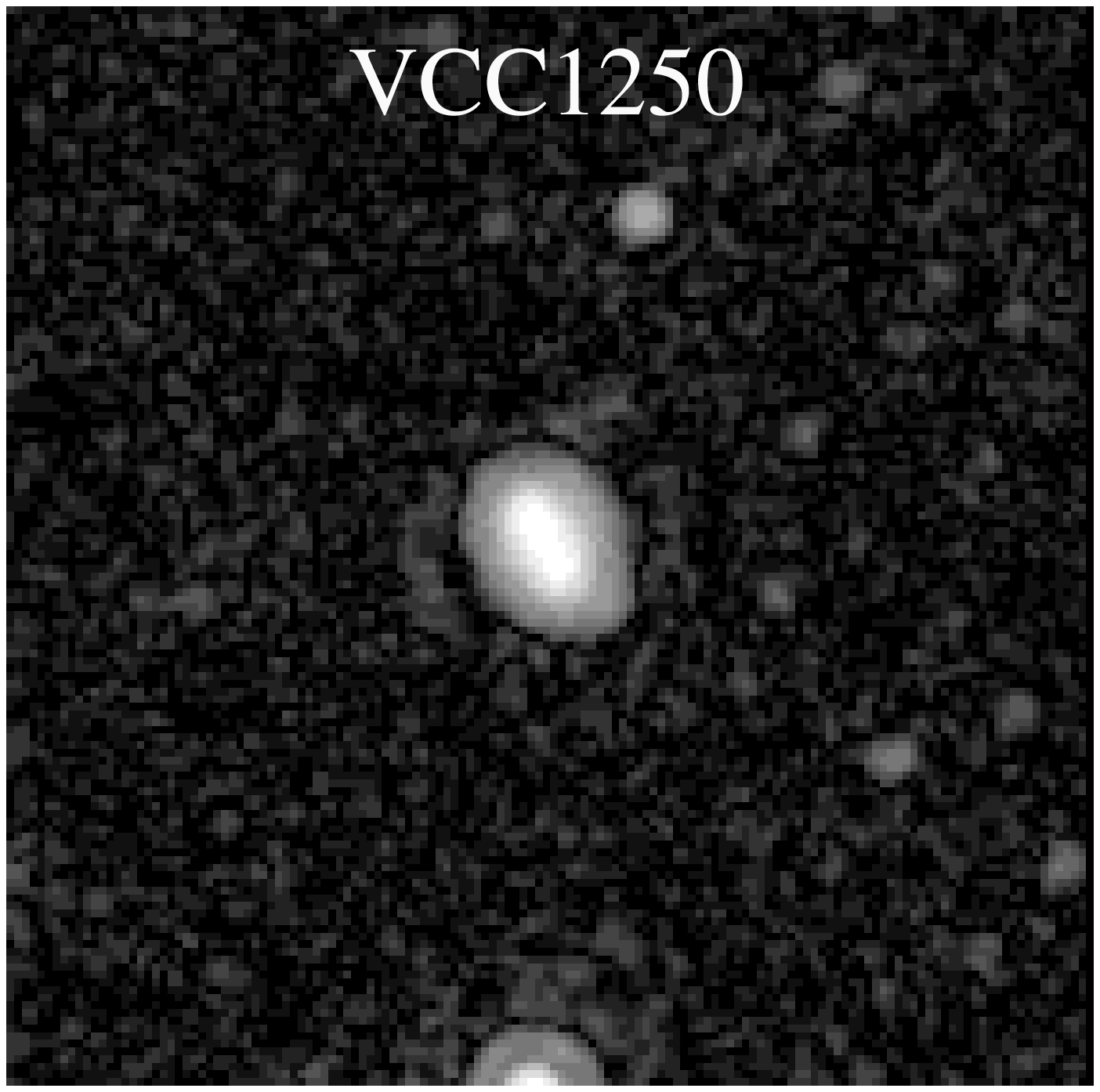}
\includegraphics[angle=0,scale=.28]{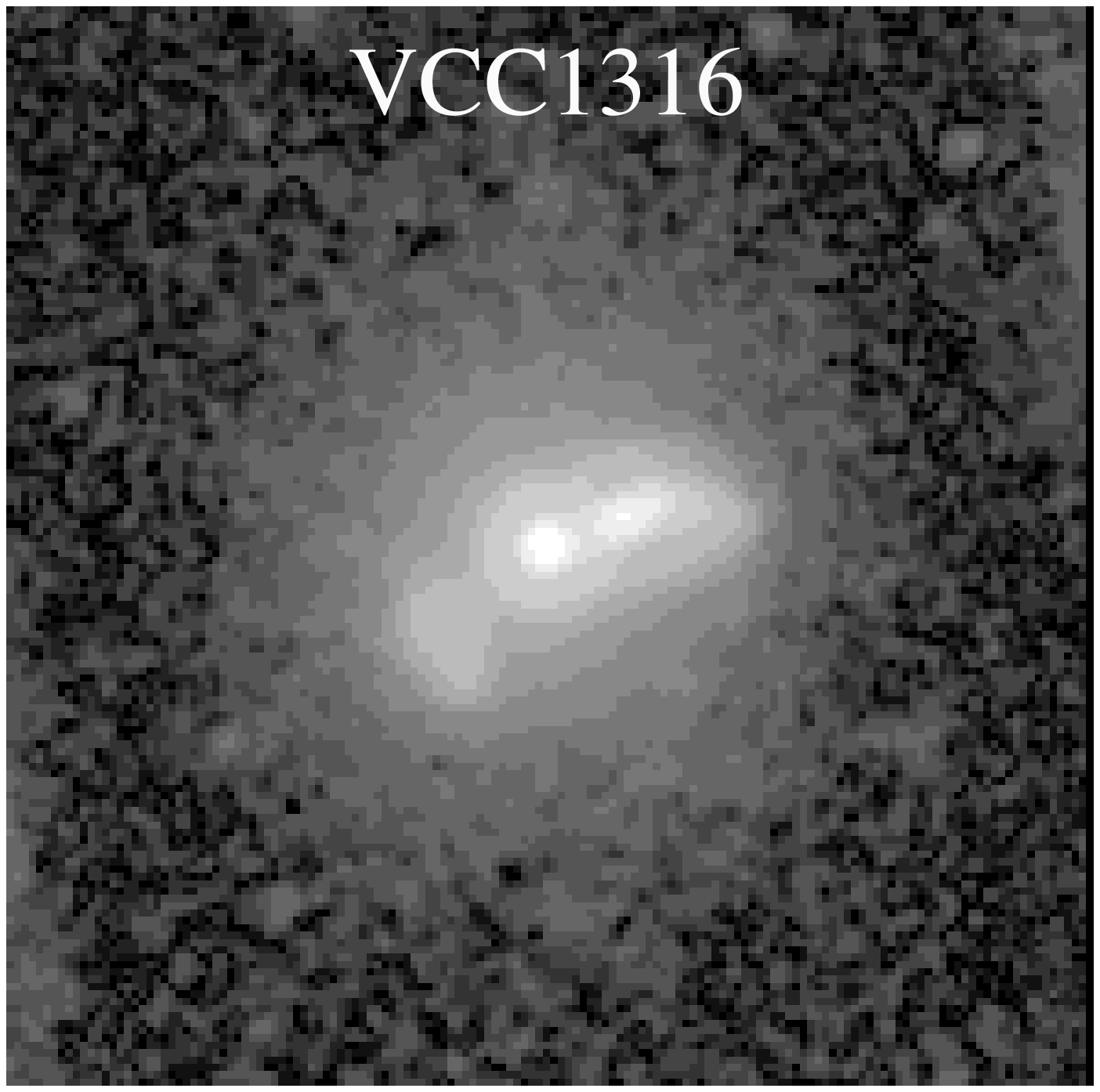}\vspace{0.025cm}
\includegraphics[angle=0,scale=.28]{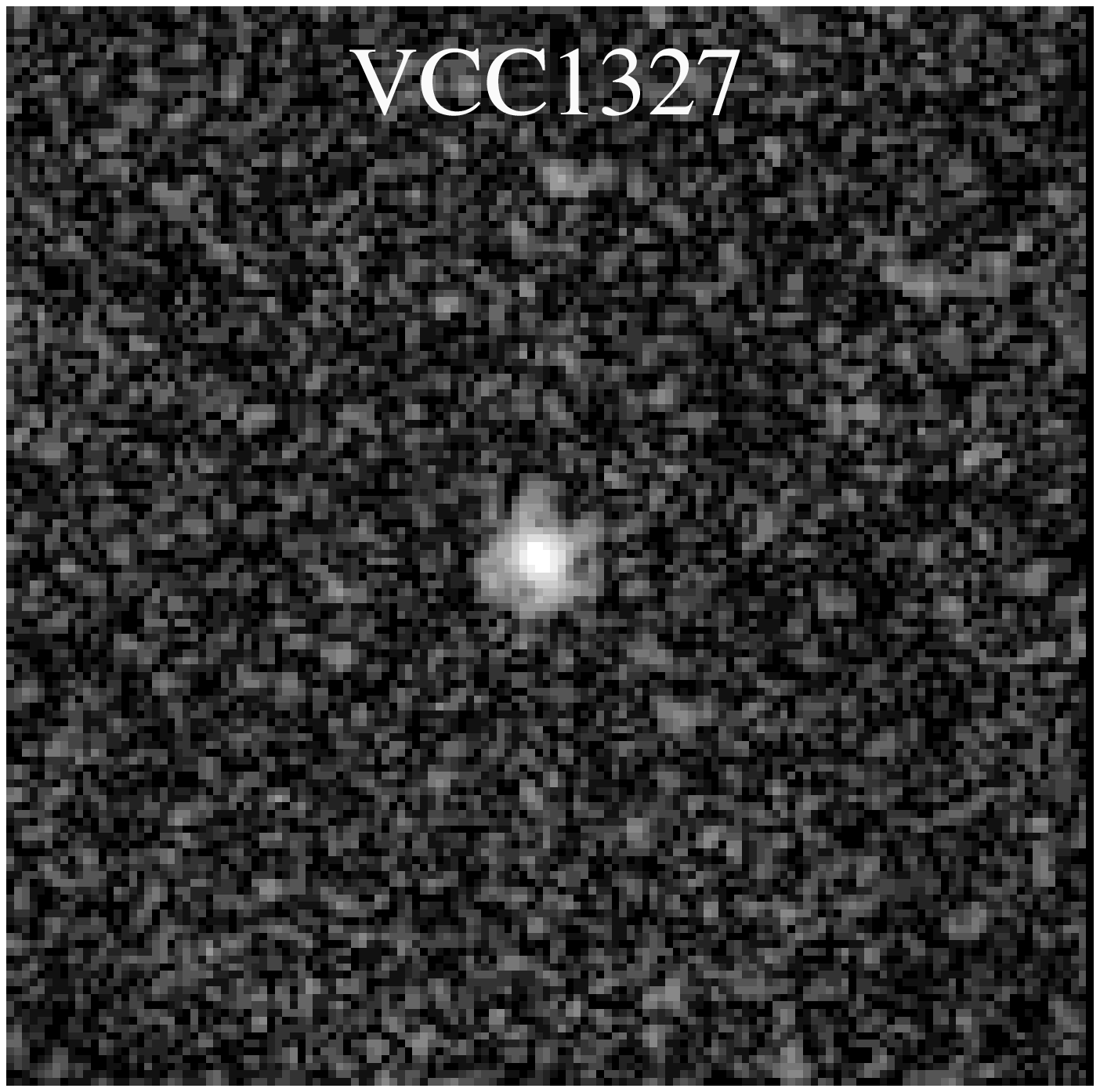}
\includegraphics[angle=0,scale=.28]{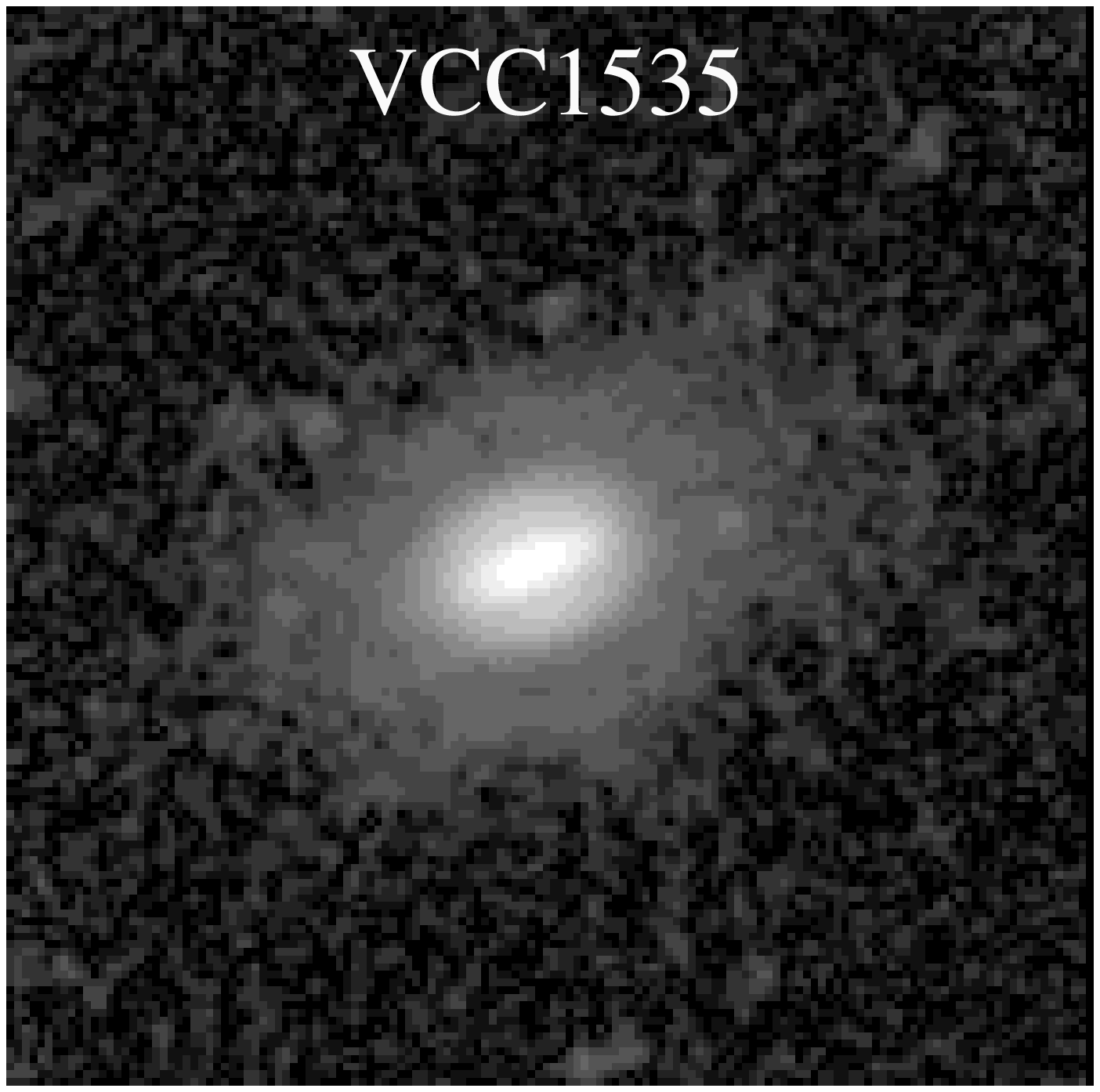}
\includegraphics[angle=0,scale=.28]{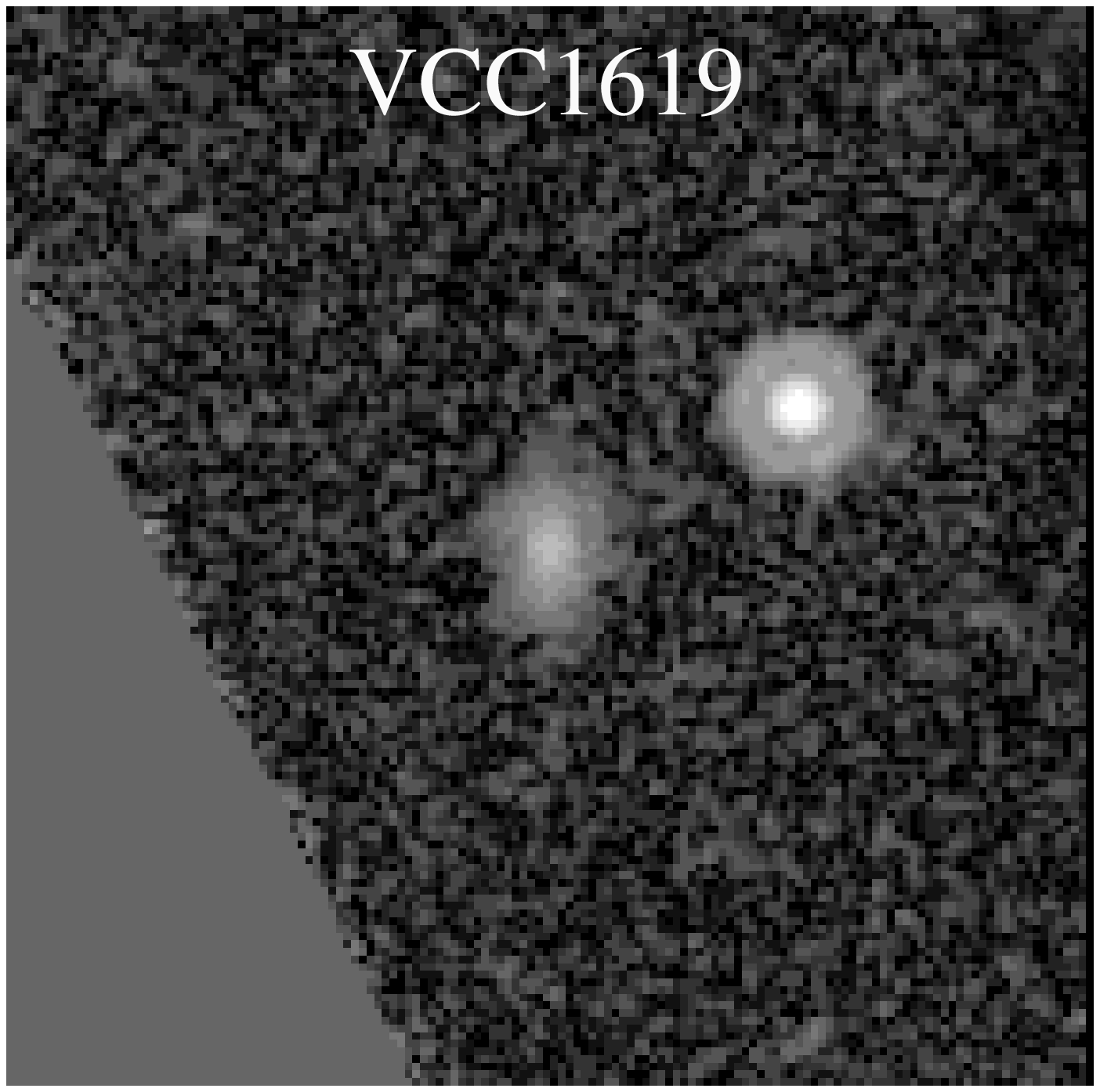}
\includegraphics[angle=0,scale=.28]{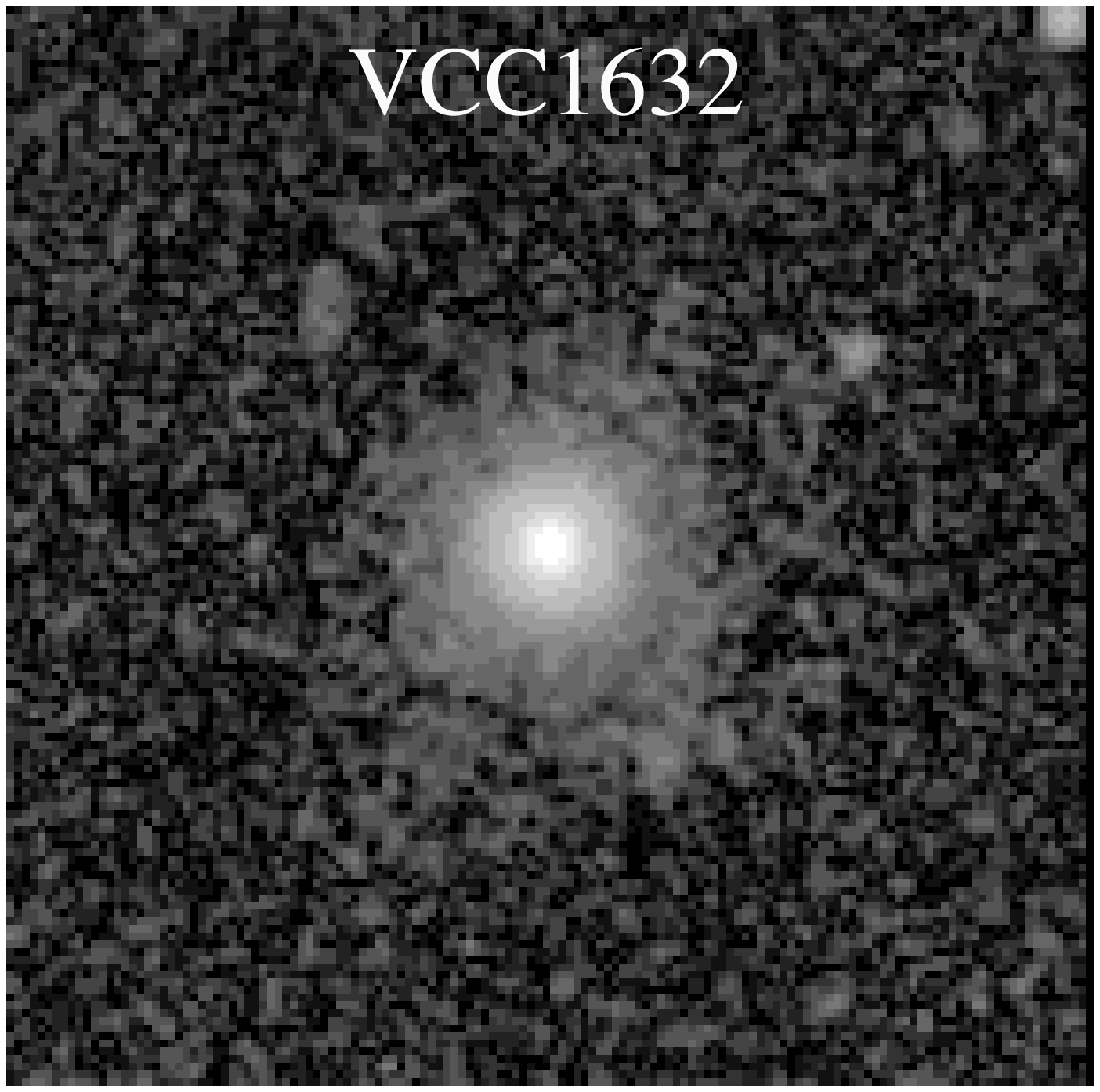}
\caption{MIPS 24\,$\mu$m images of sources with dust in their host galaxy 
as detected optically. The images are 3 arcminutes on a side, 
North is up, East is to the left.\label{images_hostdust} }
\end{figure*}

\section{Observations and data reduction}

Out of 100 ACS-VCS galaxies, 56 have been observed in photometry mode
as part of our survey.  Archival Spitzer images covering the galaxy nucleus 
were available for 39 out of the remaining 44 objects (the nuclear position 
fell just off the observed fields for VCC\,0140, VCC\,0654, VCC\,0751, 
VCC\,1192, and VCC\,1199 
-- none of which was detected in X-rays). This yields a total of 95 objects out of 100 composing the
original VCS sample for which MIPS 24\,$\mu$m data are available.

The new and archival \spi\ MIPS data have been reduced in the same
fashion: first, a second-order flat was created by median-combining
all frames (excluding the first frame). The individual files were then
corrected using this second-order flat. The flat-corrected frames were
then subject to an overlap correction (to adjust background levels)
and mosaicking, both performed using the {\sc Mopex} software
package. The final pixel size of the mosaics was chosen to be
1.275\arcsec/px.

Archival {\it HST}/ACS $z$-band images were re-analyzed using standard
data reduction procedures within {\sc Iraf}/STSDAS. The images were
drizzled onto a 0.049\arcsec/px grid: the plate scales for MIPS and
ACS were chosen to be integer multiples of each other, to allow for an
easy comparison of the galaxies' radial profiles while still
adequately representing the instrumental resolution.

\section{Analysis and Results}

\subsection{Aperture photometry}

First we performed aperture photometry to identify detections for
further study and determine upper flux limits for the
non-detections. We used an aperture with a diameter of 12\arcsec~(which
corresponds to about twice the size of the FWHM of MIPS
at 24\,$\mu$m) and a sky annulus between 20 and 32\arcsec. The noise
level was determined by the distribution of fluxes in 1000 apertures randomly placed 
in empty parts of the sky.

The flux distribution was then fitted by a Gaussian.  The resulting
standard deviation, appropriately corrected for PSF losses, was taken
as the $1\sigma$ photometric uncertainty for this mosaic. Sources
detected at less than 3$\sigma$ significance are considered
non-detections (with typical 3$\sigma$ upper limits of
$\sim$0.3\,mJy).  In total we detect 53/95 sources observed at
24\,$\mu$m. Upper limits for the non-detected sources are given in
Table\,\ref{tab_all}. We note
that the upper limits determined here are by design point-source upper
limits (i.e. for a nuclear MIR excess, see below) rather then upper
limits for the total galaxy emission.

Figure\,\ref{histogram} shows that all the objects undetected at
24$\mu$m have stellar masses $\log M_{\ast}/M_{\odot} < 10$. Stellar 
masses were taken from \citet[][see also Tab.\,\ref{tab_all}]{gal10}.

\subsection{Sources with extended dust}

One of the goals of this project is to detect possible nuclear MIR
emission which may point towards obscured low-level AGN
activity. However, in such a scenario, diffuse dust in the host galaxy
becomes a significant source of contamination. In order to isolate any
faint excess of nuclear dust emission, we must first account for
possible diffuse dust emission originating from the host galaxy
itself.

Some galaxies in our sample have extended (kpc-scale) dust features
clearly detectable from the optical images \citep[hereafter often
called ``optical dust'',][]{fer06}. The spatial resolution of {\it
Spitzer} is not sufficient to resolve MIR emission from this dust in
most of our objects.  In such cases we cannot disentangle how much of
the unresolved MIR flux comes from diffuse dusty structures or from a
putative nuclear source. Therefore we exclude any objects with
optically detected dust in the host galaxy from further analysis. For
reference, there are 18 objects with optically detected dust in our
sample and we detect most of them at 24\,$\mu$m, with the exception of
VCC\,1422 and VCC\,1779. Thus, for this sample sources with optical dust have a higher 
incidence of MIR detections than sources lacking optical dust features. 
The MIR images of the 16 detected sources are shown in Fig.\,\ref{images_hostdust}. 

For 11 of the 18 sources with optical dust, MIR spectra are available 
in the literature \citep{bre06,breg06,smi07,lei09} or in the {\it Spitzer} 
archive. Six of these show clear signs for recent (star formation) activity as inferred 
from the presence of PAH features (VCC\,0759, VCC\,0763, VCC\,1030, VCC\,1154, 
VCC\,1535, VCC\,1619). In addition, and conceivably as a result of the on-going star
formation, those sources are typically also detected at FIR
wavelengths \citep[e.g.][]{tem09}. One additional source is M\,87 (VCC\,1316) 
for which the MIR is dominated by non-thermal emission \citep[e.g.][]{per07,bus09}. 
The remaining four sources do not show any particular features hinting at activity in 
their MIR spectra. We note that the fraction 
of X-ray detections for MIR-detected objects is roughly comparable for sources 
with and without optical dust ($\sim$\,60\,\%, see Tab.\,\ref{tab1}).

\begin{table}[t!]
\begin{center}
\caption{Combinations of detections.\label{tab1}}
\begin{tabular}{ccc|c}
\tableline\tableline
optical  & MIR       & X-ray   & Number of  \\
dust     & detection & nucleus & sources    \\
\tableline
 +  &  +  &  +  &  10 \\
 +  &  +  & $-$ &   6 \\
 +  & $-$ & $-$ &   2 \\
 +  & $-$ &  +  &   0 \\
$-$ &  +  & $-$ &  17 \\
$-$ &  +  &  +  &  20 \\
$-$ & $-$ &  +  &   2 \\
$-$ & $-$ & $-$ &  38 \\
\tableline
$-$ & no MIR data & $-$ & 5 \\
\tableline
\end{tabular}
\end{center}
\end{table}

\begin{figure*}[t!]
\centering
\includegraphics[angle=0,scale=.38]{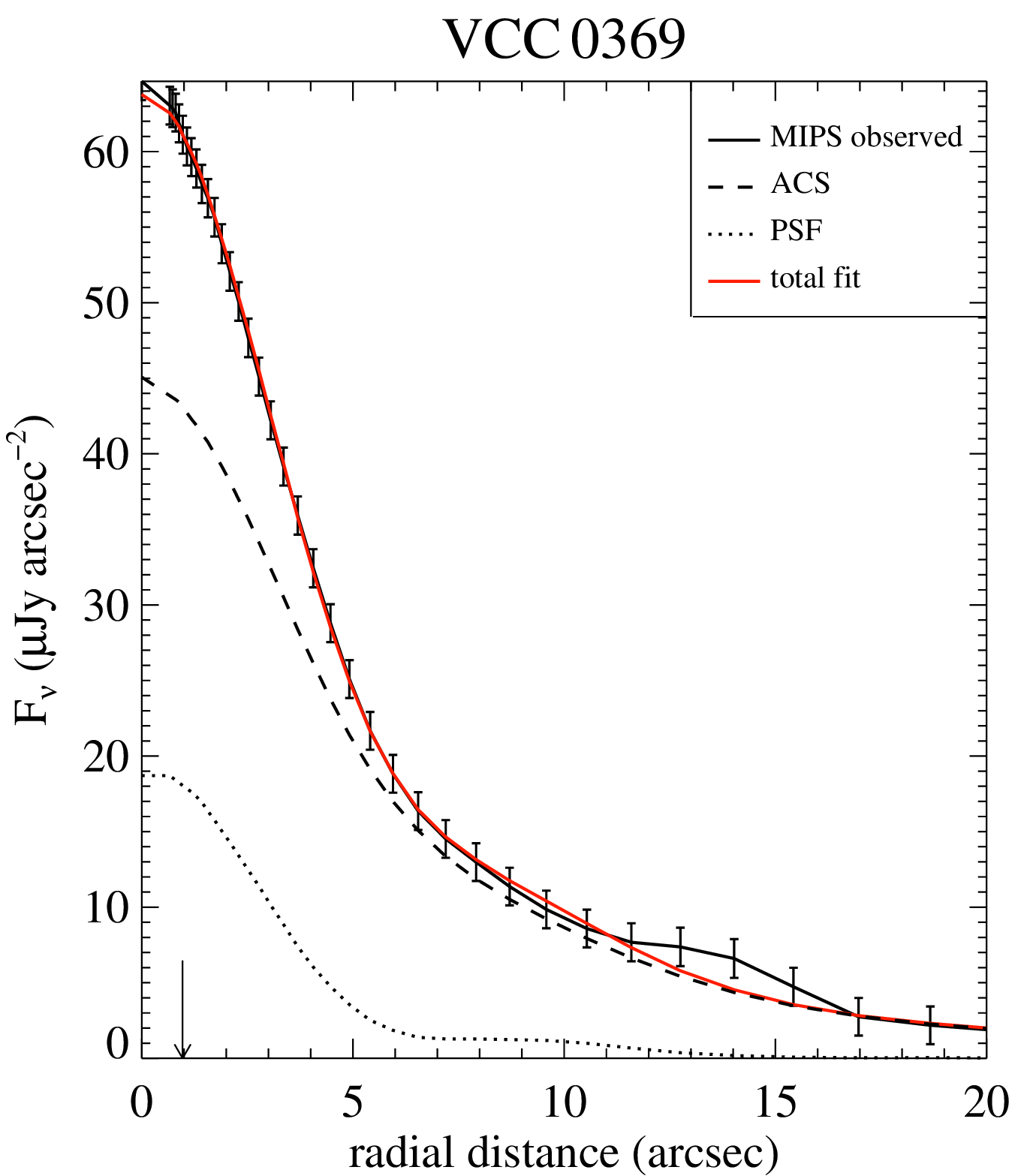}
\includegraphics[angle=0,scale=.38]{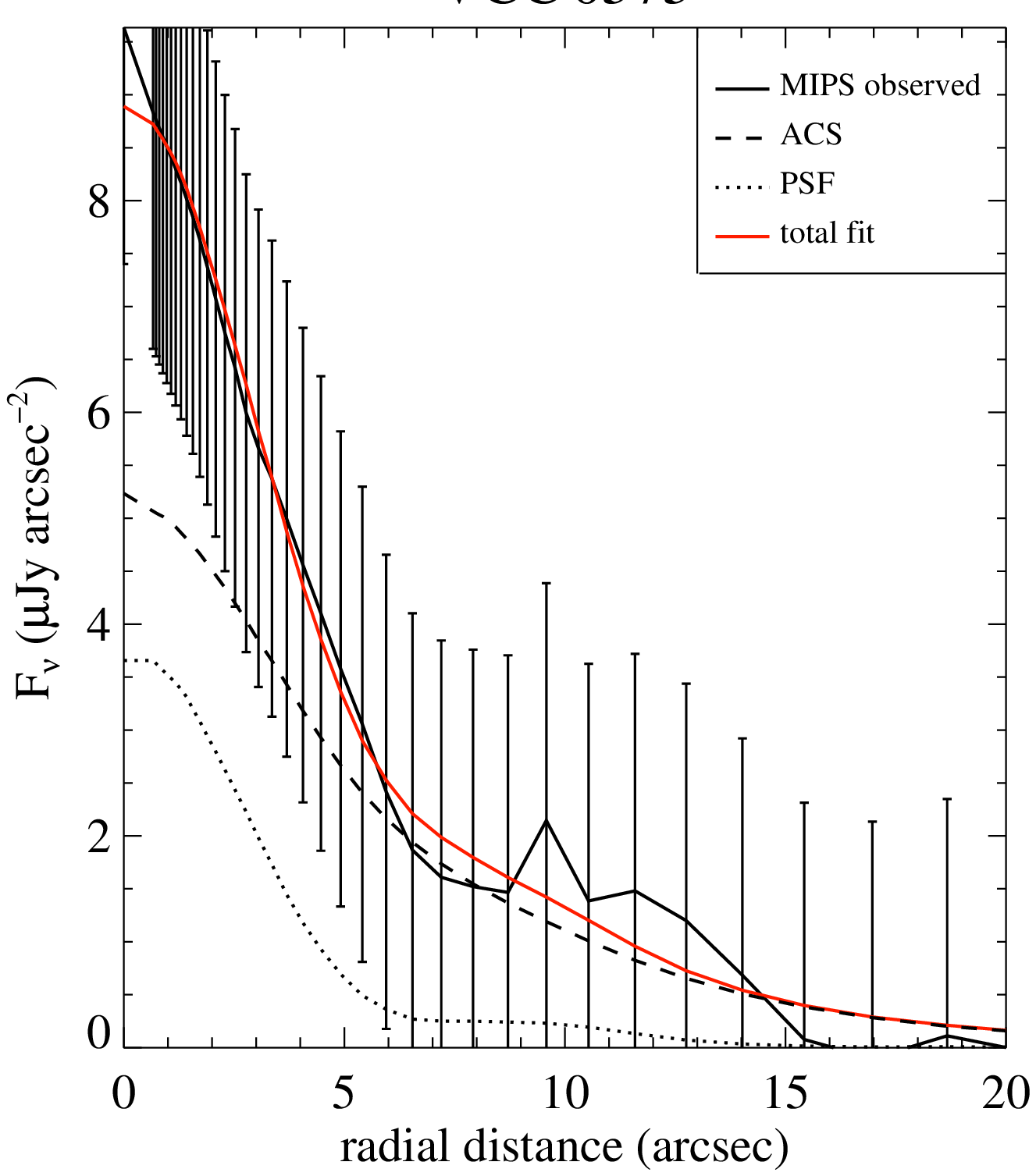}
\includegraphics[angle=0,scale=.38]{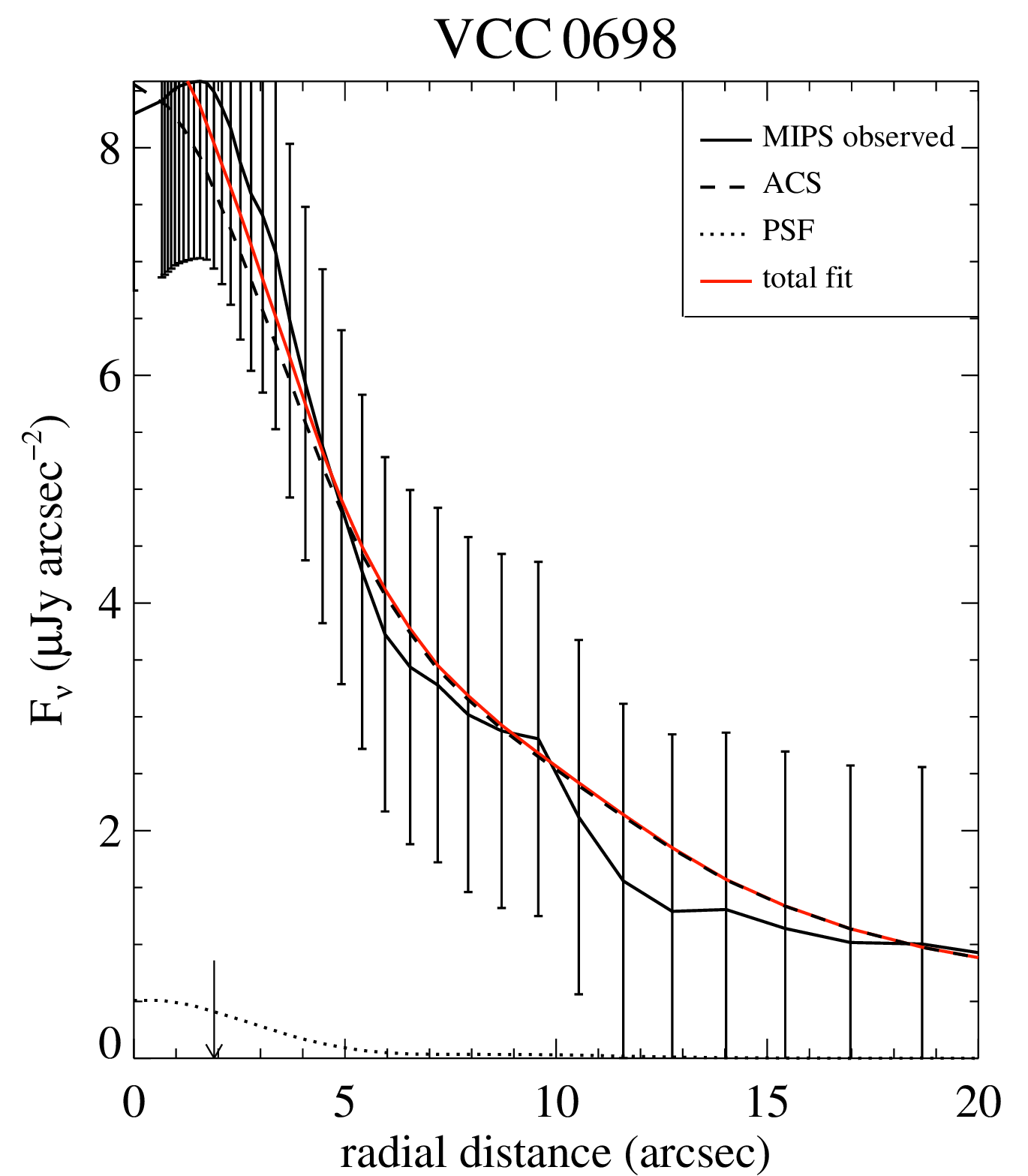}
\includegraphics[angle=0,scale=.38]{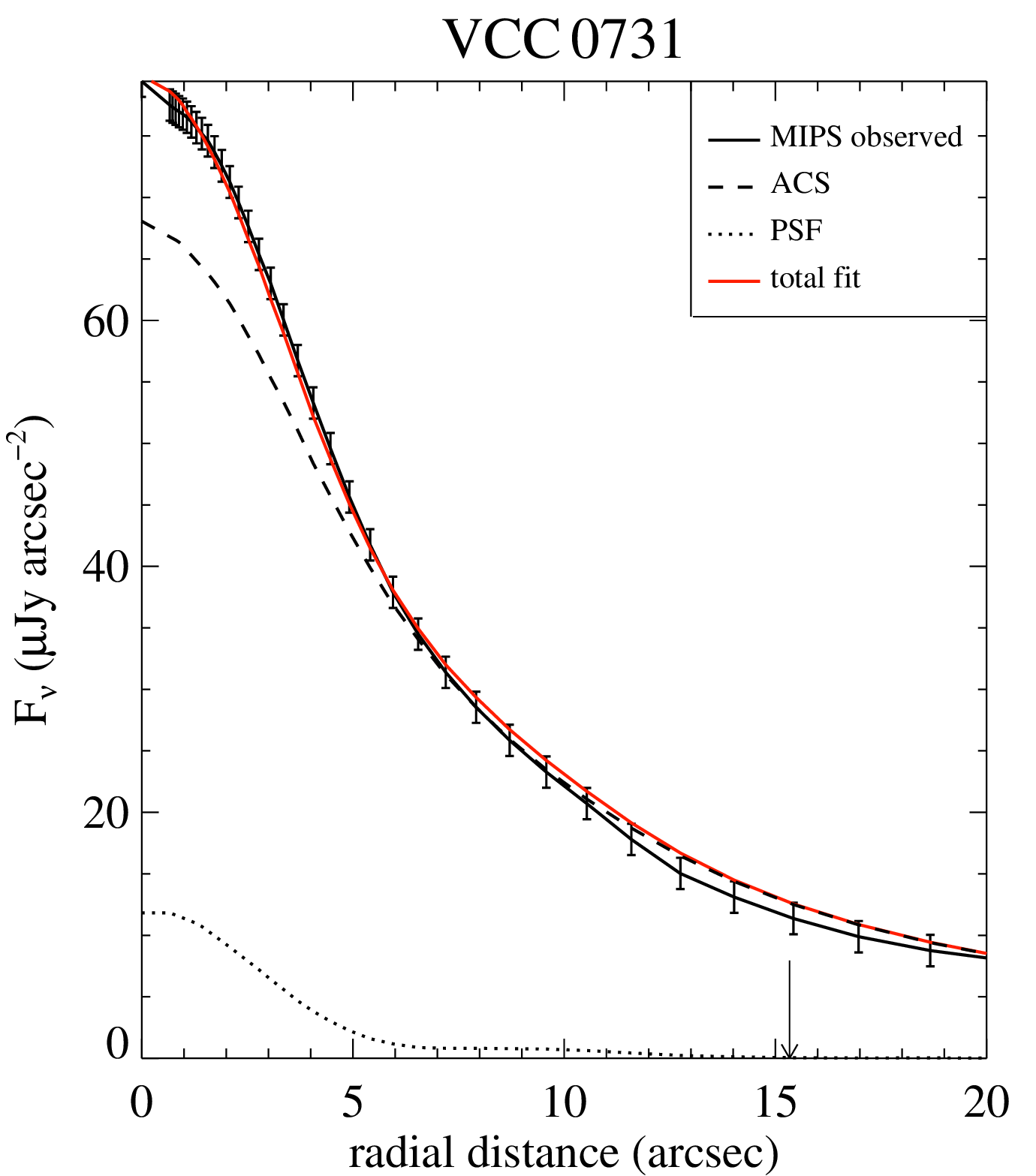}
\includegraphics[angle=0,scale=.38]{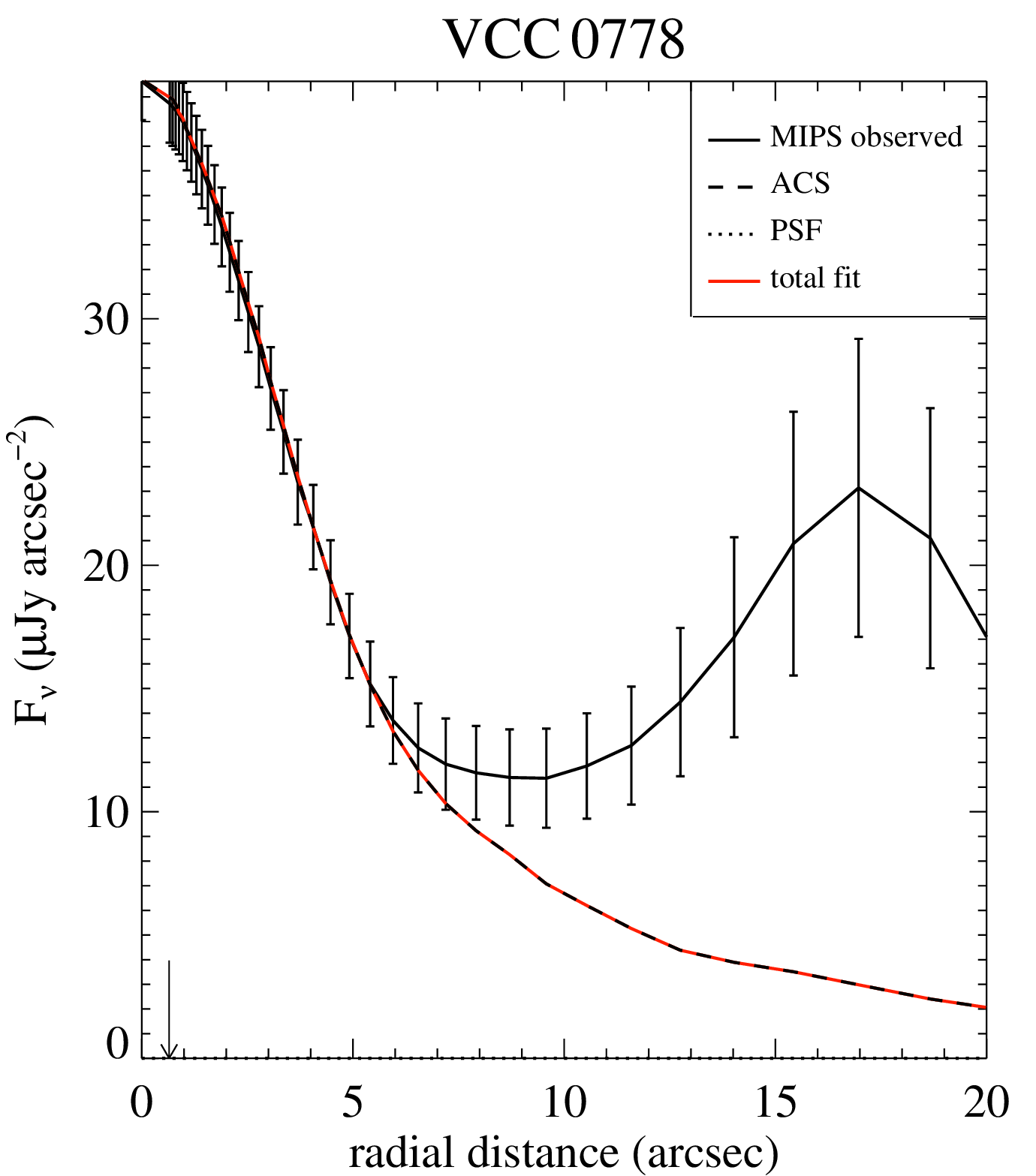}
\includegraphics[angle=0,scale=.38]{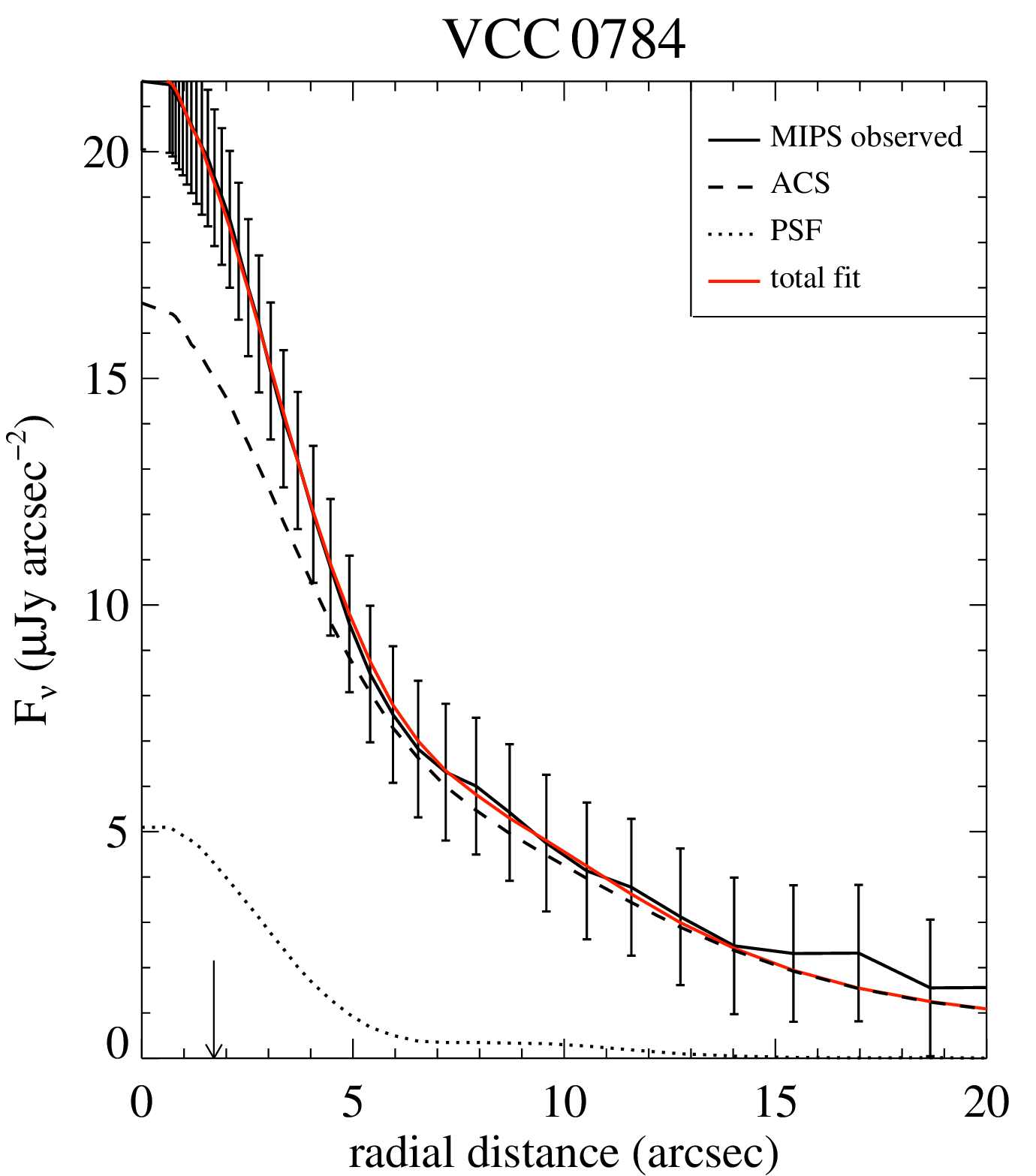}
\includegraphics[angle=0,scale=.38]{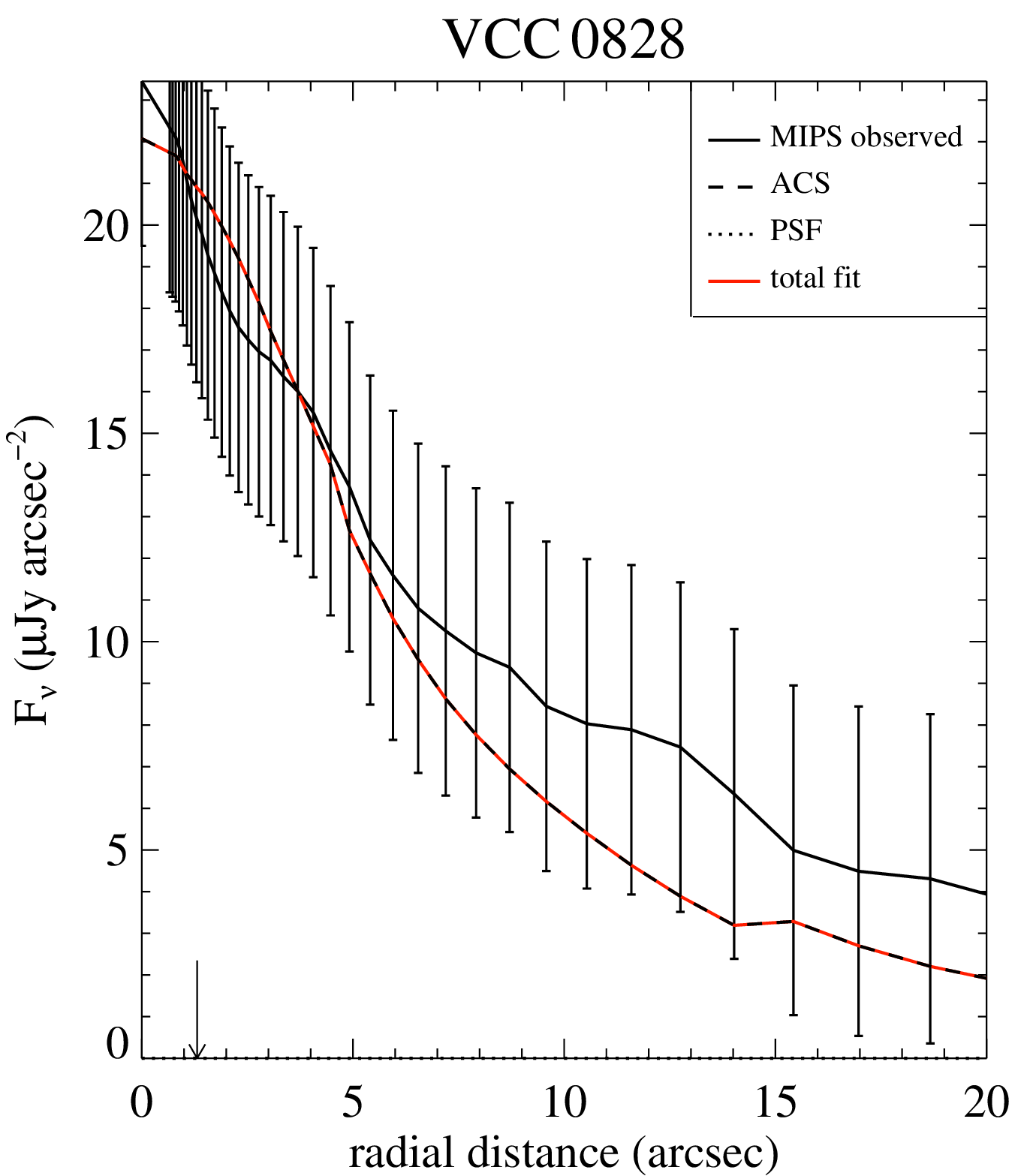}
\includegraphics[angle=0,scale=.38]{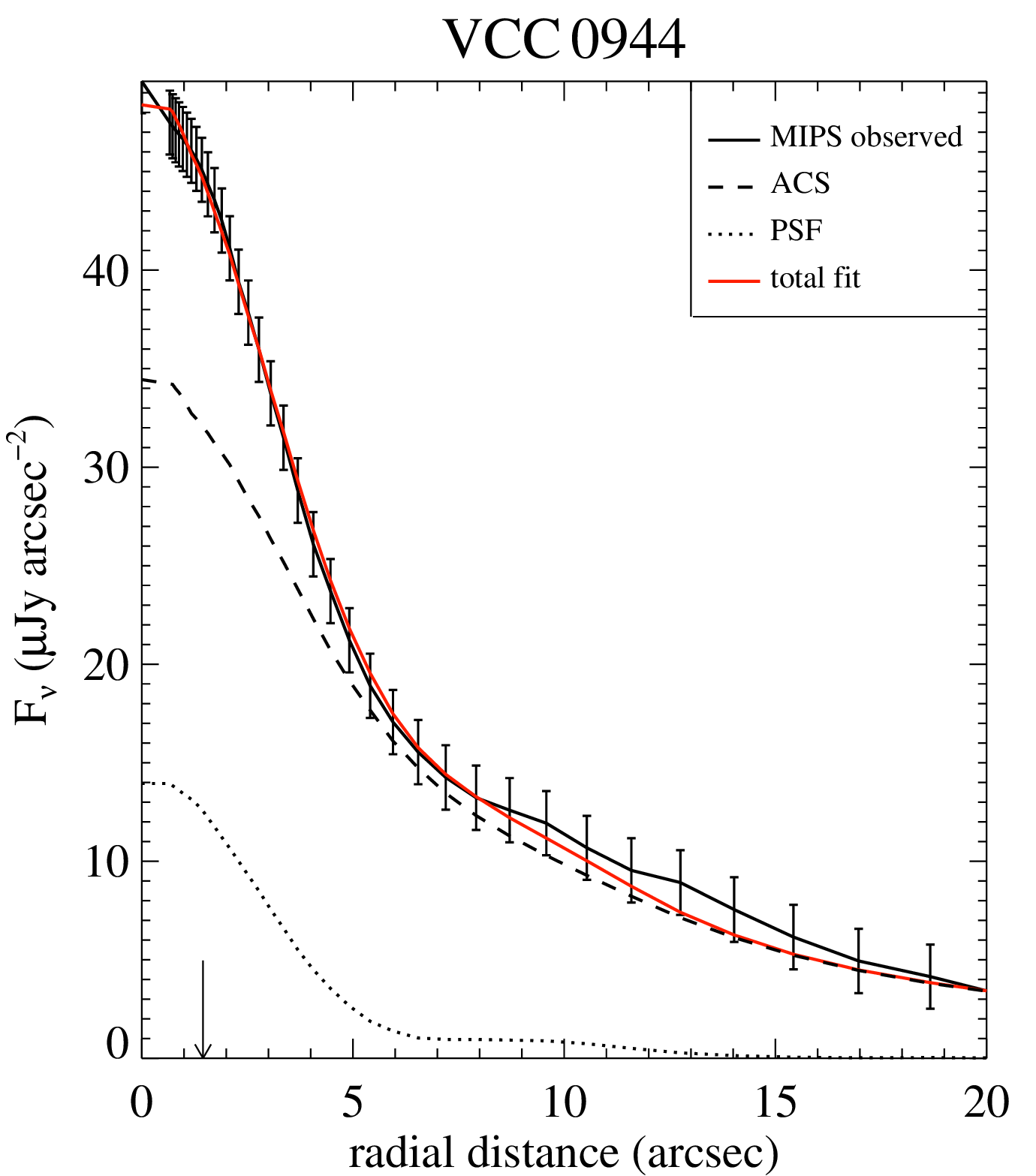}
\includegraphics[angle=0,scale=.38]{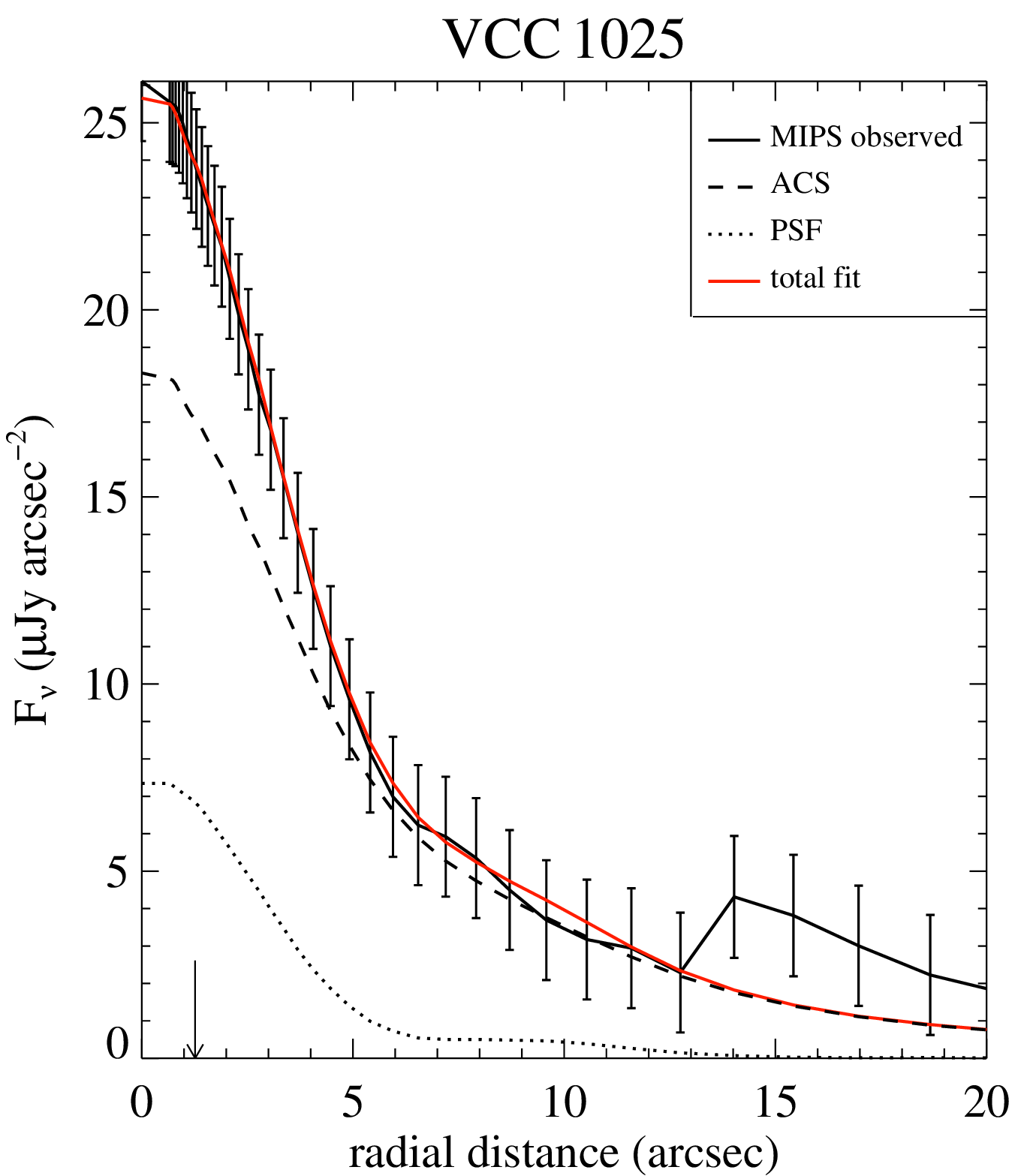}
\caption{Radial profiles of the 37 detected VCC early-type galaxies 
without optical signatures for host dust. The black solid line shows the 
observed MIR emission. The dashed and the dotted lines represent the 
scaled radial profiles of the MIR-PSF convolved $z$-band emission and of the 
additional point source contribution needed to fit the data, respectively. 
The red solid line gives the combination of these two components 
and corresponds to the total fit. The small arrow in each panel indicates 
R$_{\rm eff}$/8. Some radial profiles have larger errors due to the 
faintess of the sources. In the case of VCC\,0778 and VCC\,1743, the rise of the profile 
at radii $>$\,10\arcsec~is due to a bright nearby star (see Fig.\,\ref{images_galfit}).\label{images_radprof} }
\end{figure*}
\addtocounter{figure}{-1}
\begin{figure*}[t!]
\centering
\includegraphics[angle=0,scale=.38]{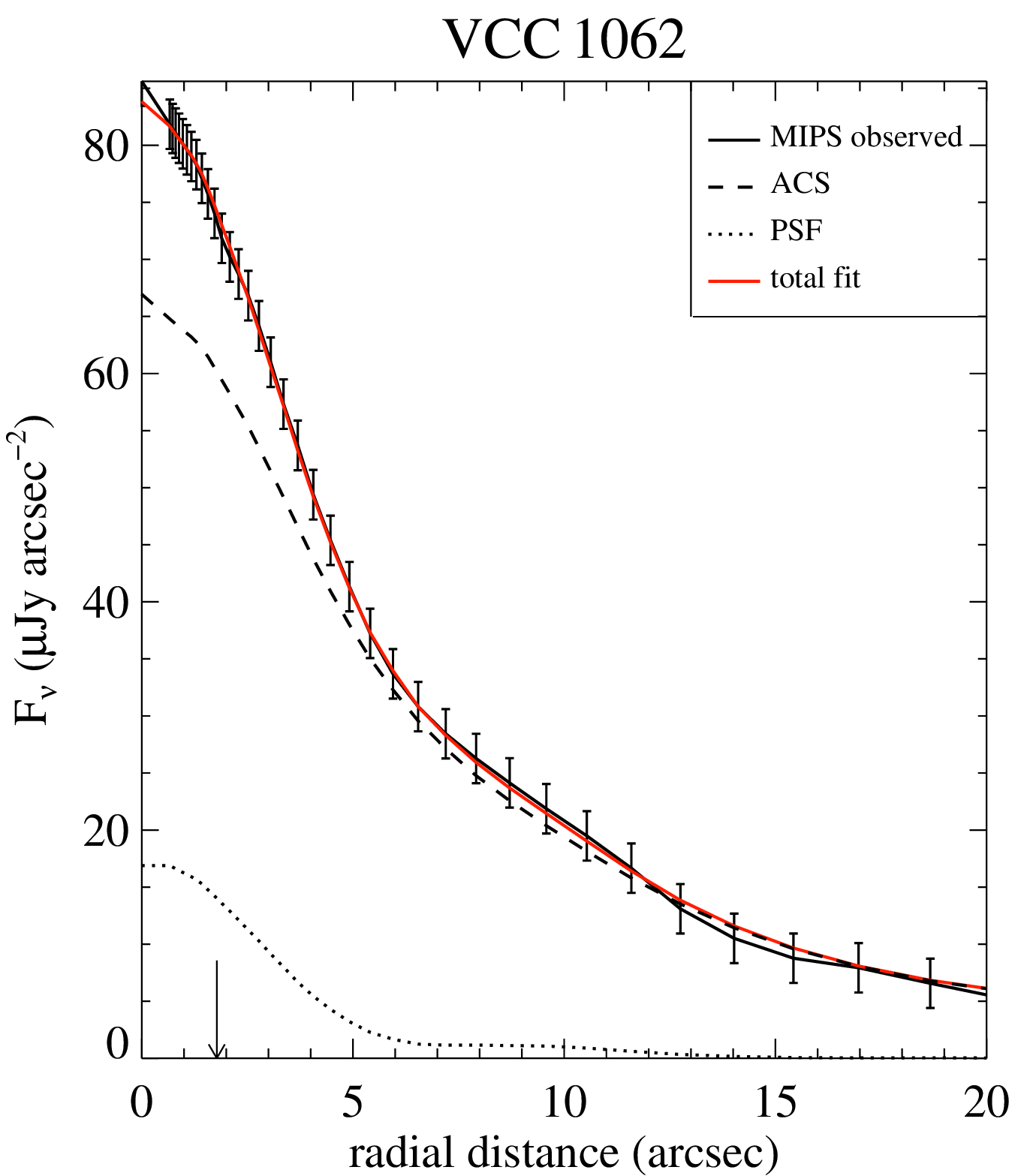}
\includegraphics[angle=0,scale=.38]{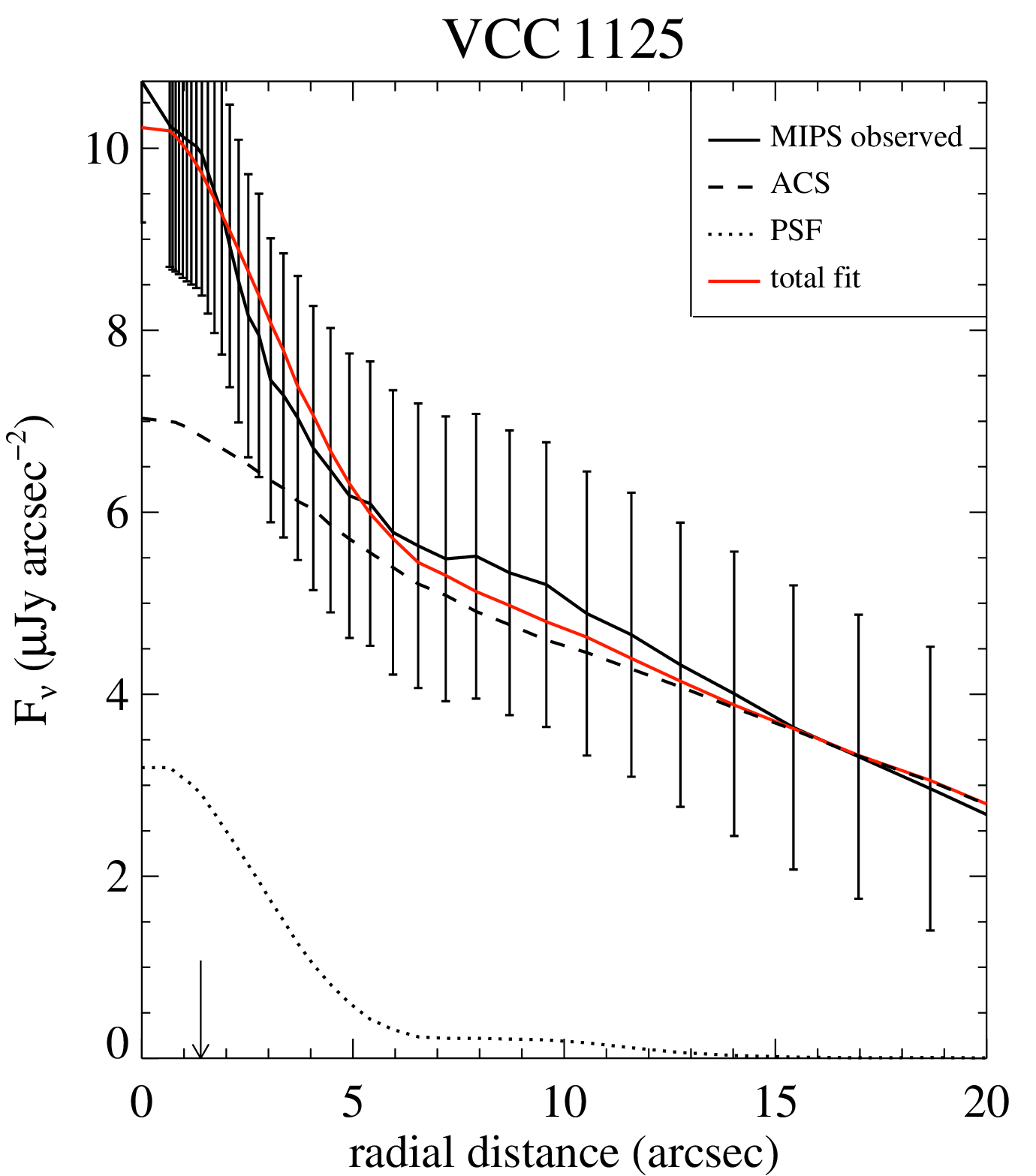}
\includegraphics[angle=0,scale=.38]{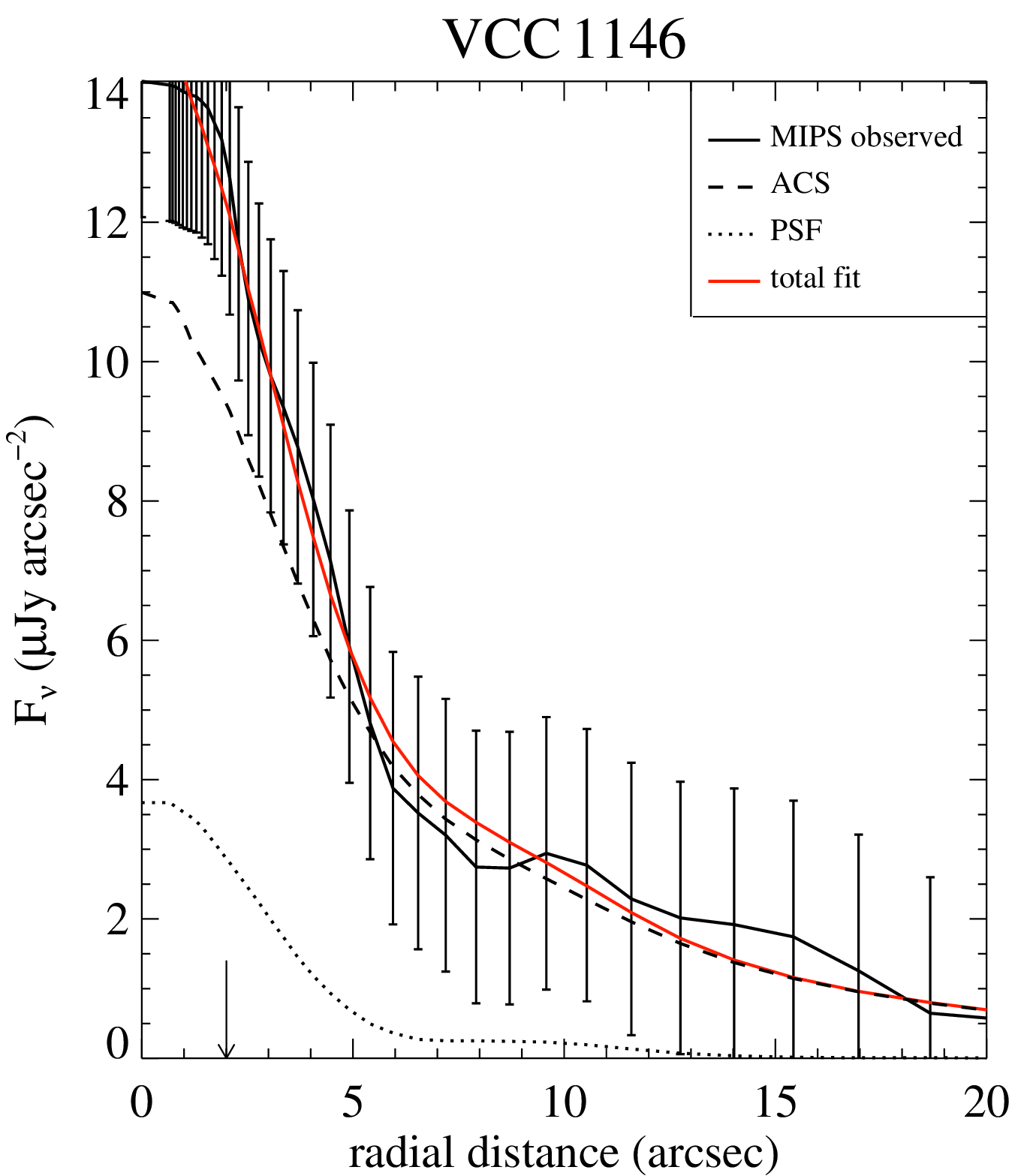}
\includegraphics[angle=0,scale=.38]{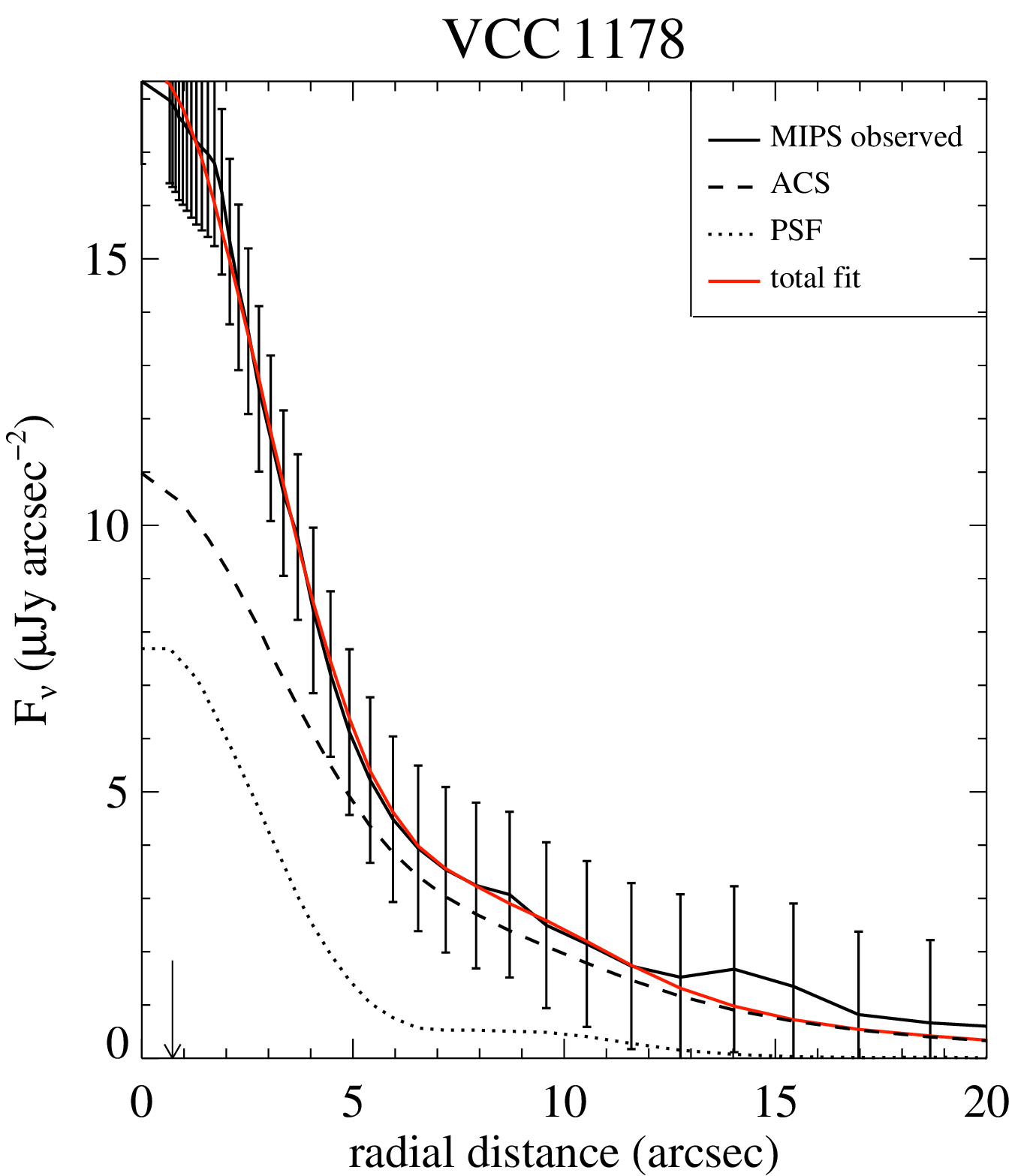}
\includegraphics[angle=0,scale=.38]{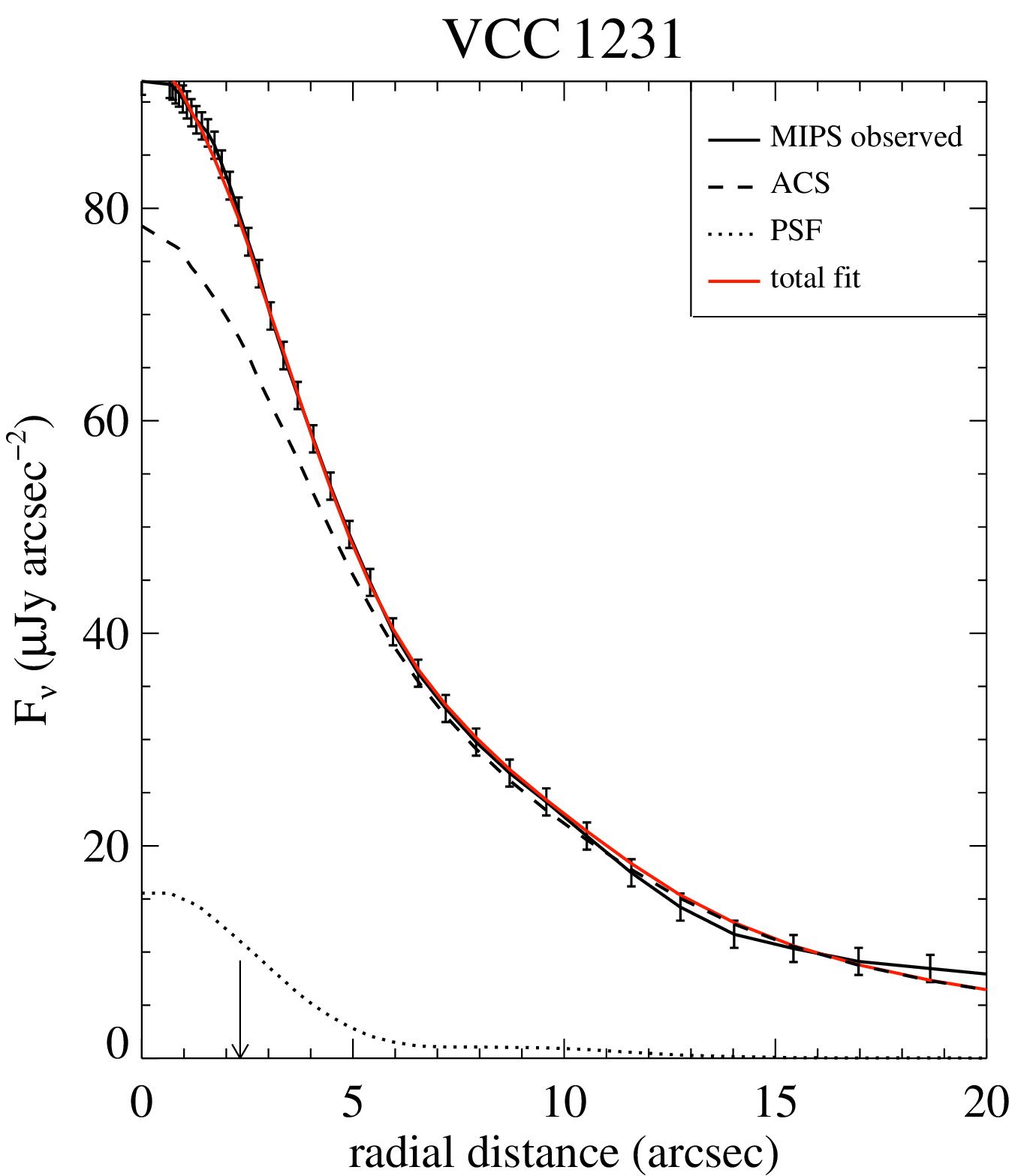}
\includegraphics[angle=0,scale=.38]{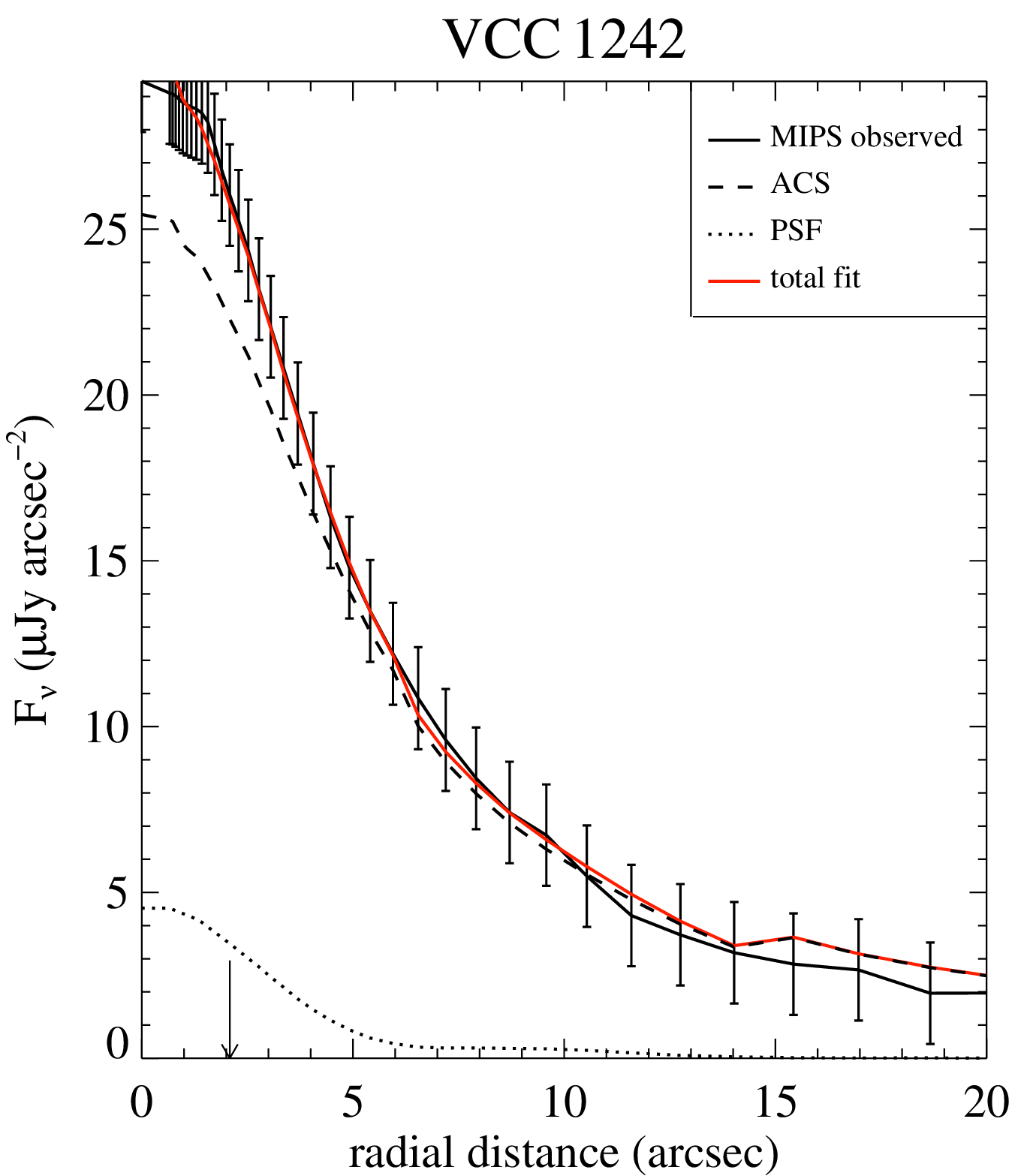}
\includegraphics[angle=0,scale=.38]{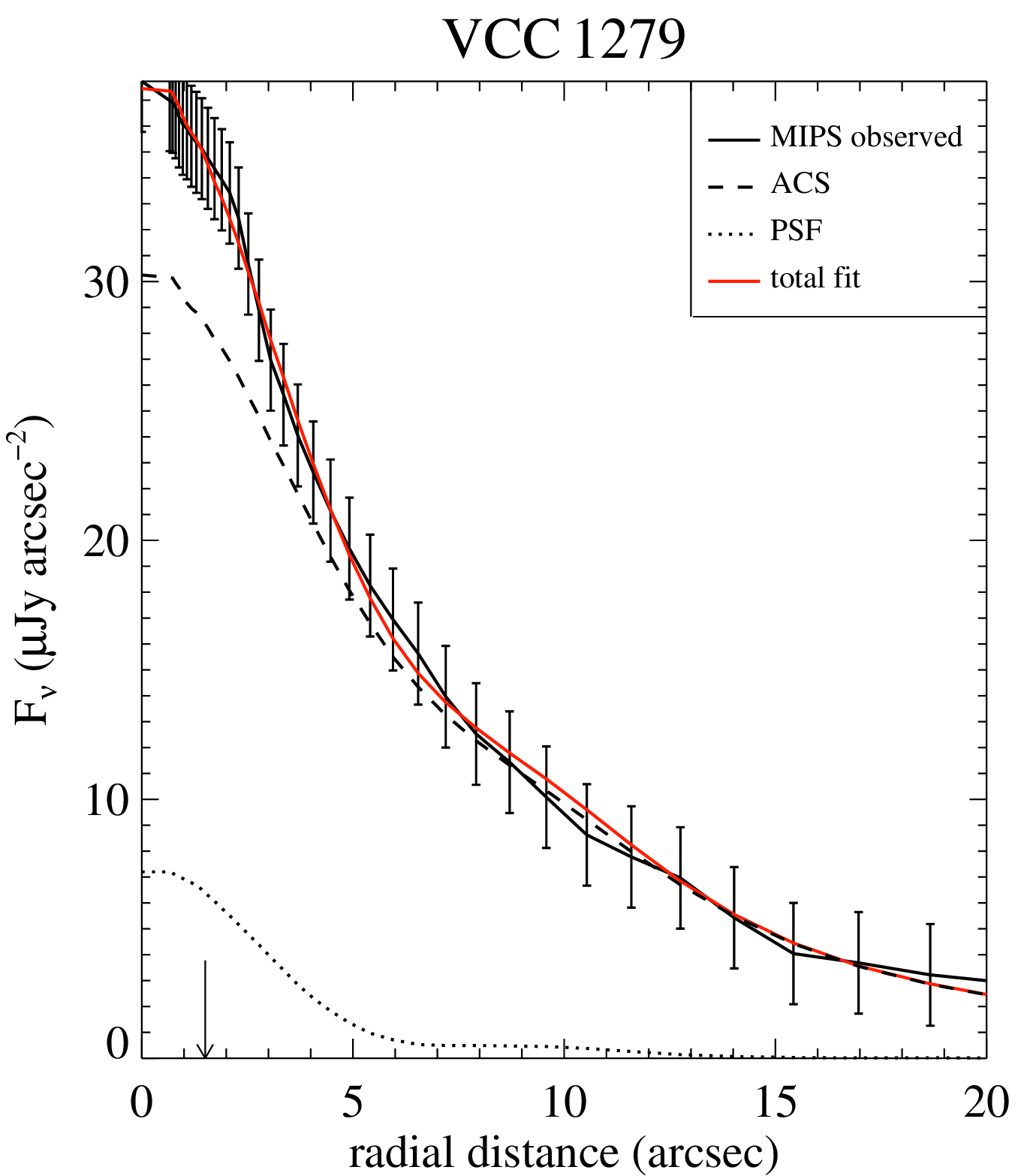}
\includegraphics[angle=0,scale=.38]{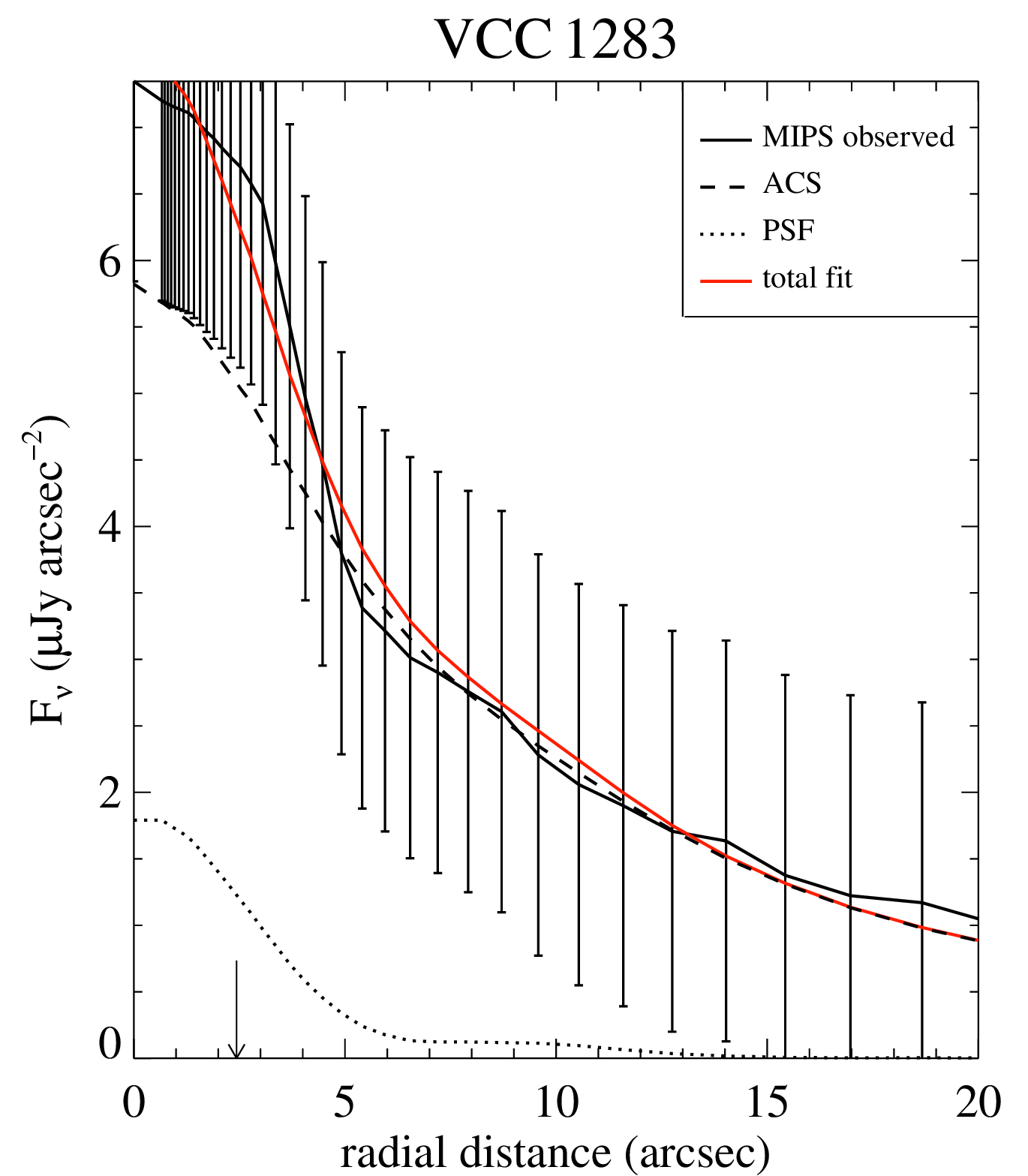}
\includegraphics[angle=0,scale=.38]{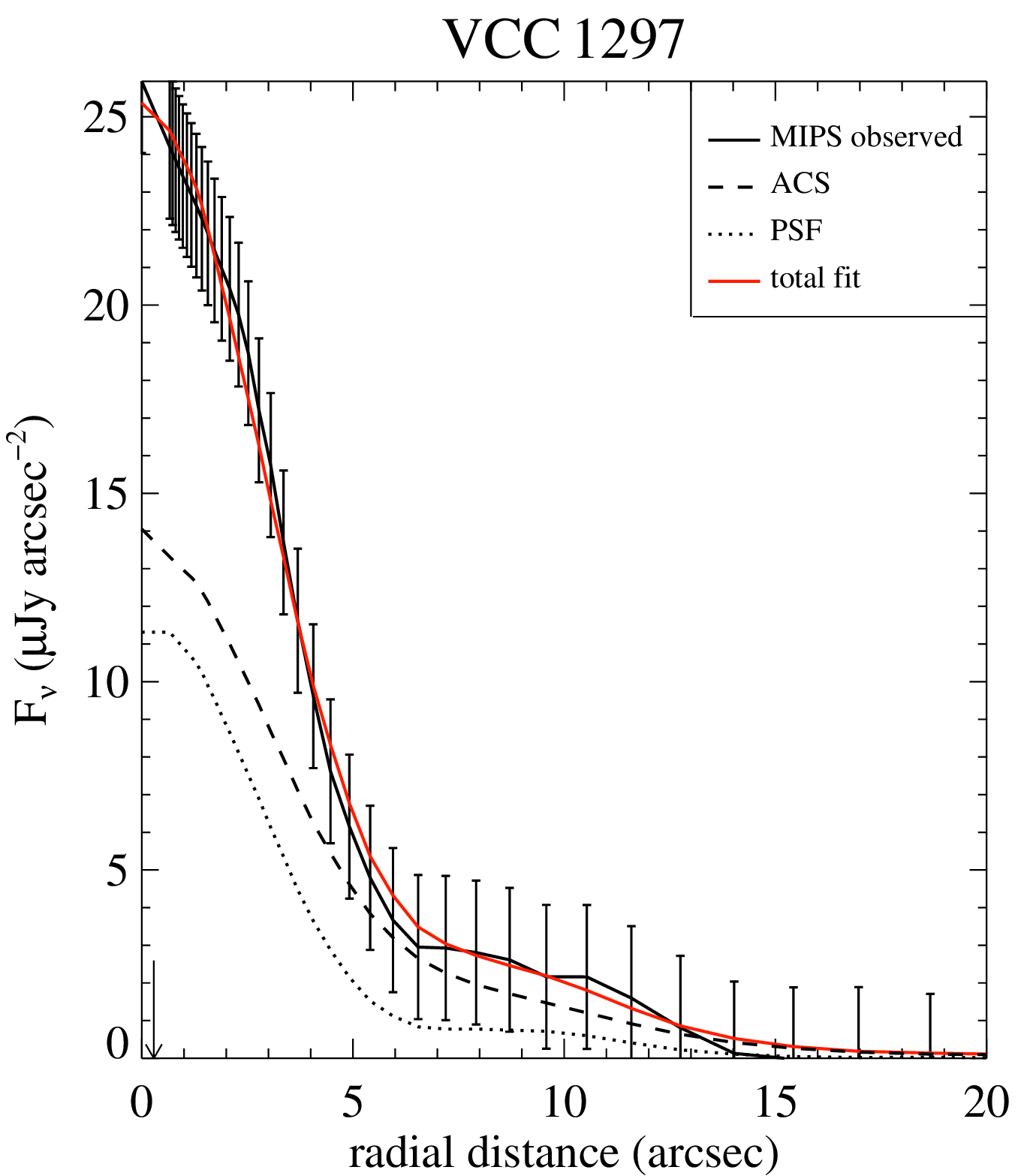}
\caption{{\it continued}}
\end{figure*}
\addtocounter{figure}{-1}
\begin{figure*}[t!]
\centering
\includegraphics[angle=0,scale=.38]{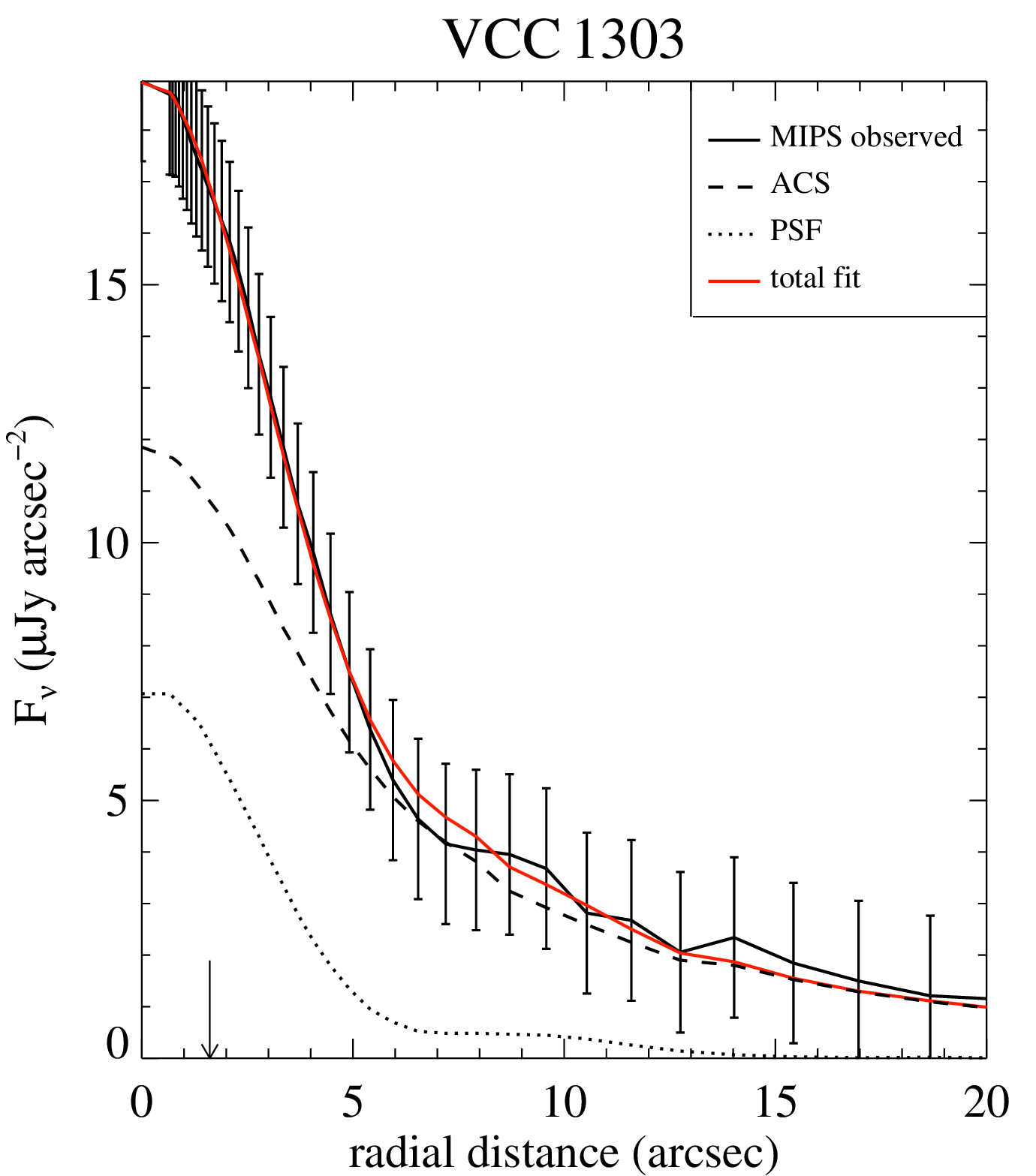}
\includegraphics[angle=0,scale=.38]{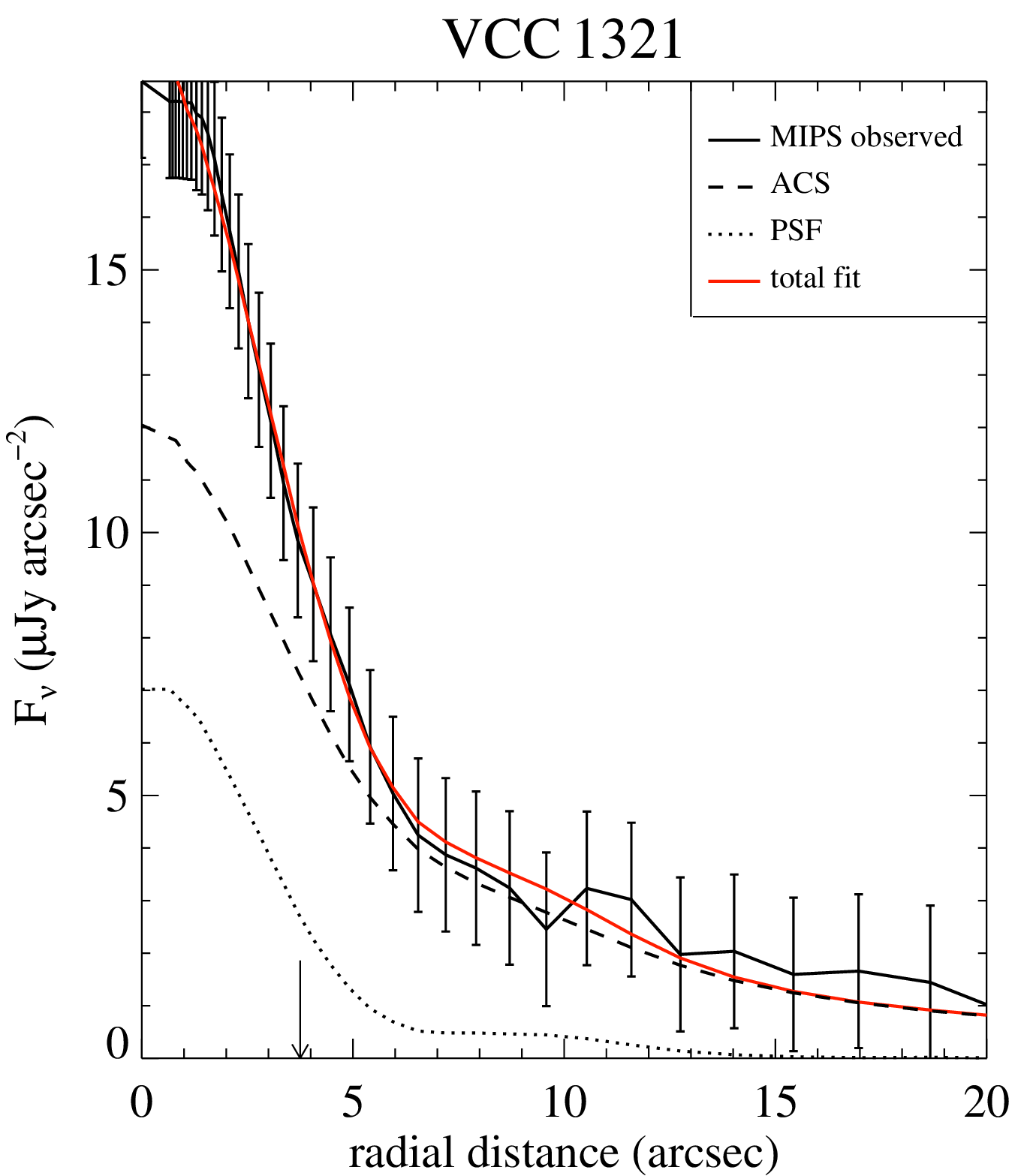}
\includegraphics[angle=0,scale=.38]{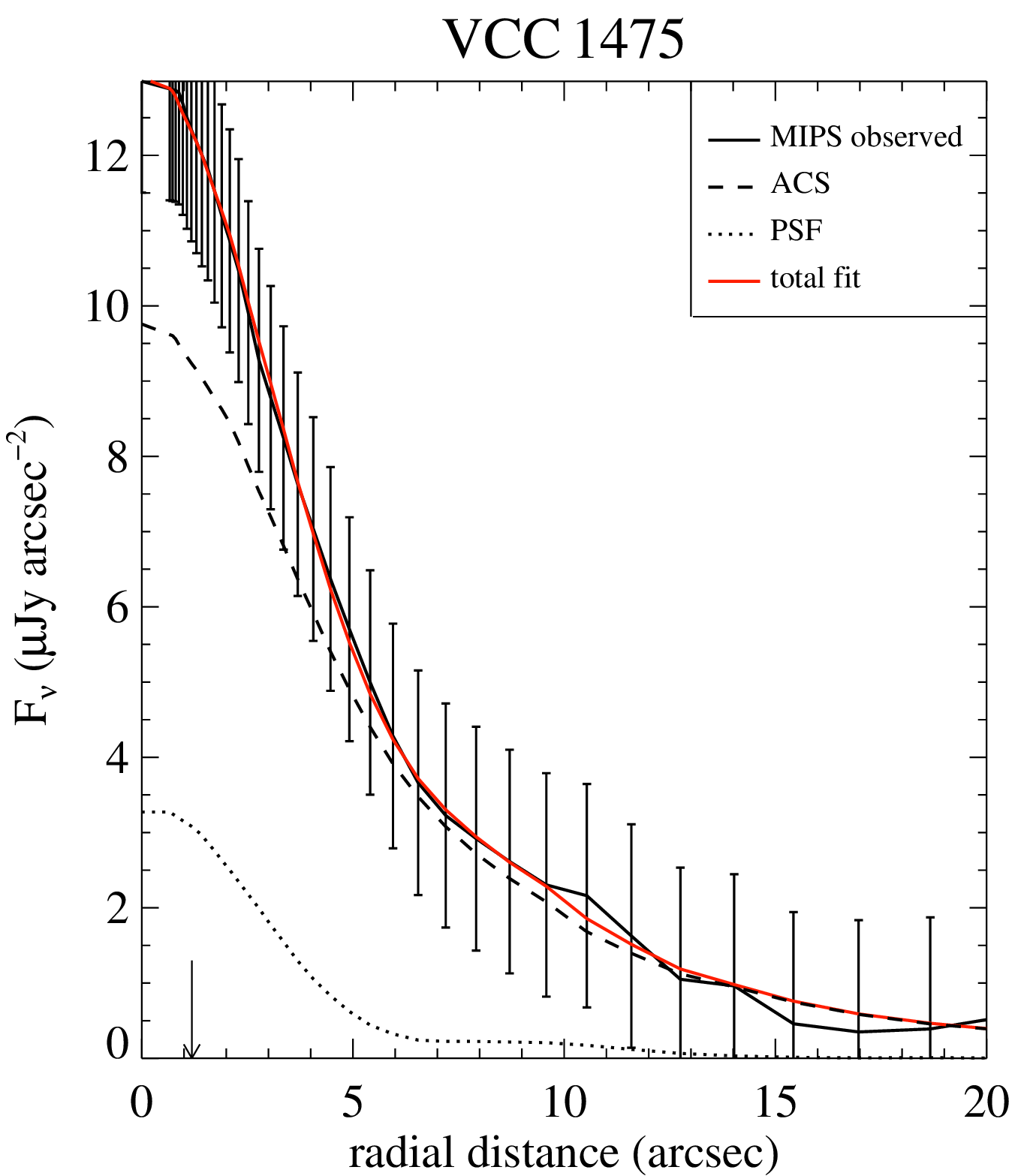}
\includegraphics[angle=0,scale=.38]{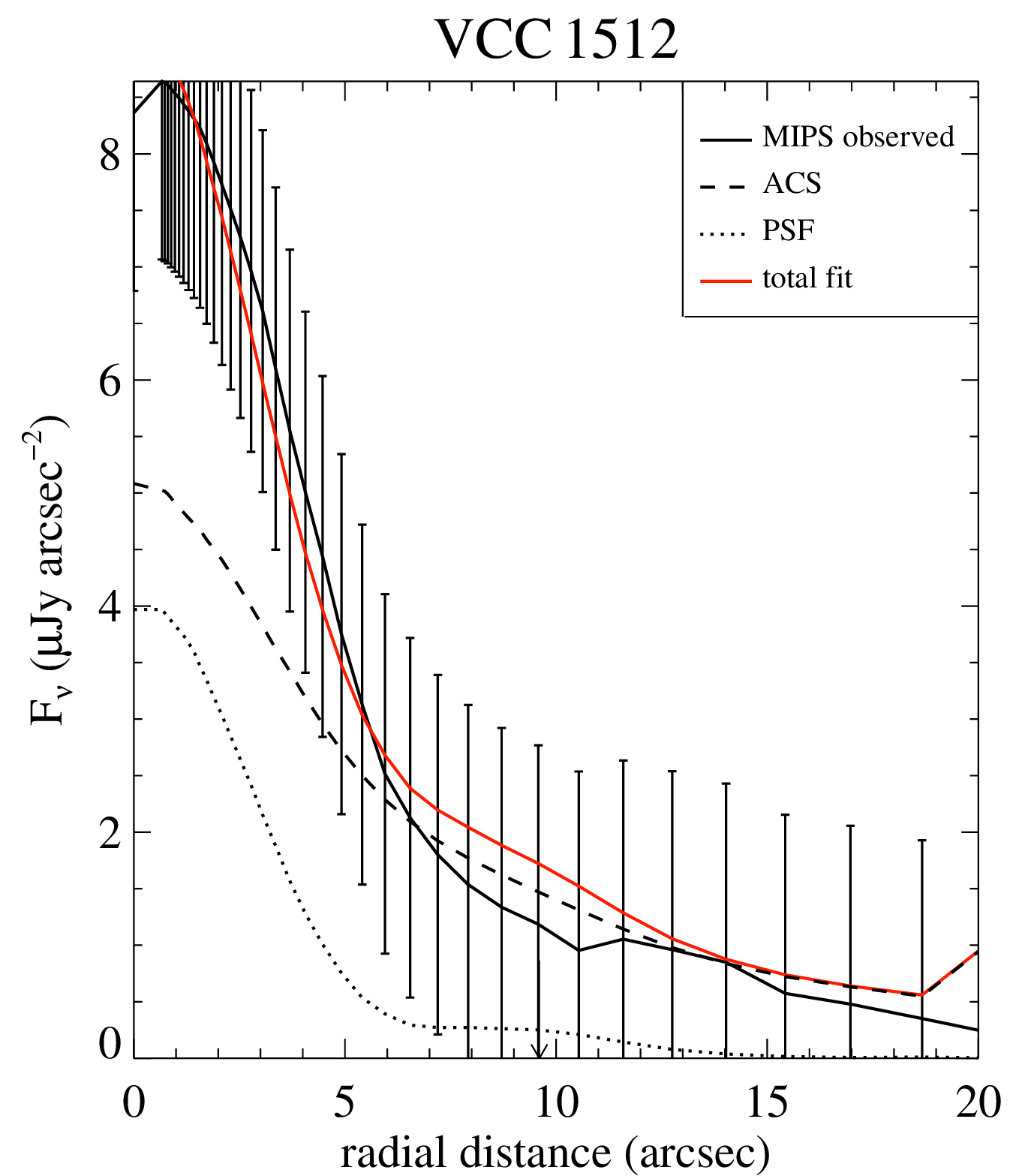}
\includegraphics[angle=0,scale=.38]{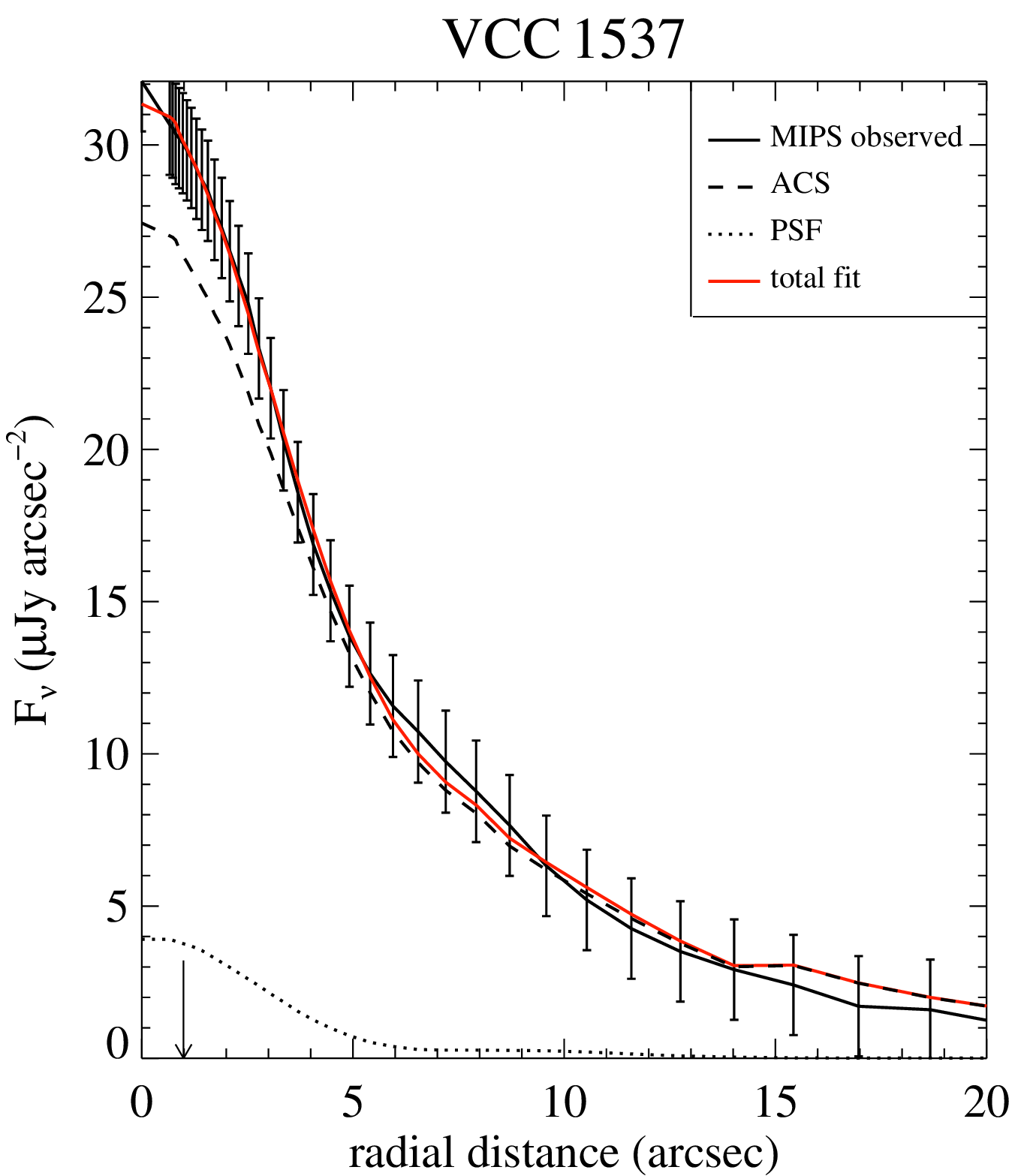}
\includegraphics[angle=0,scale=.38]{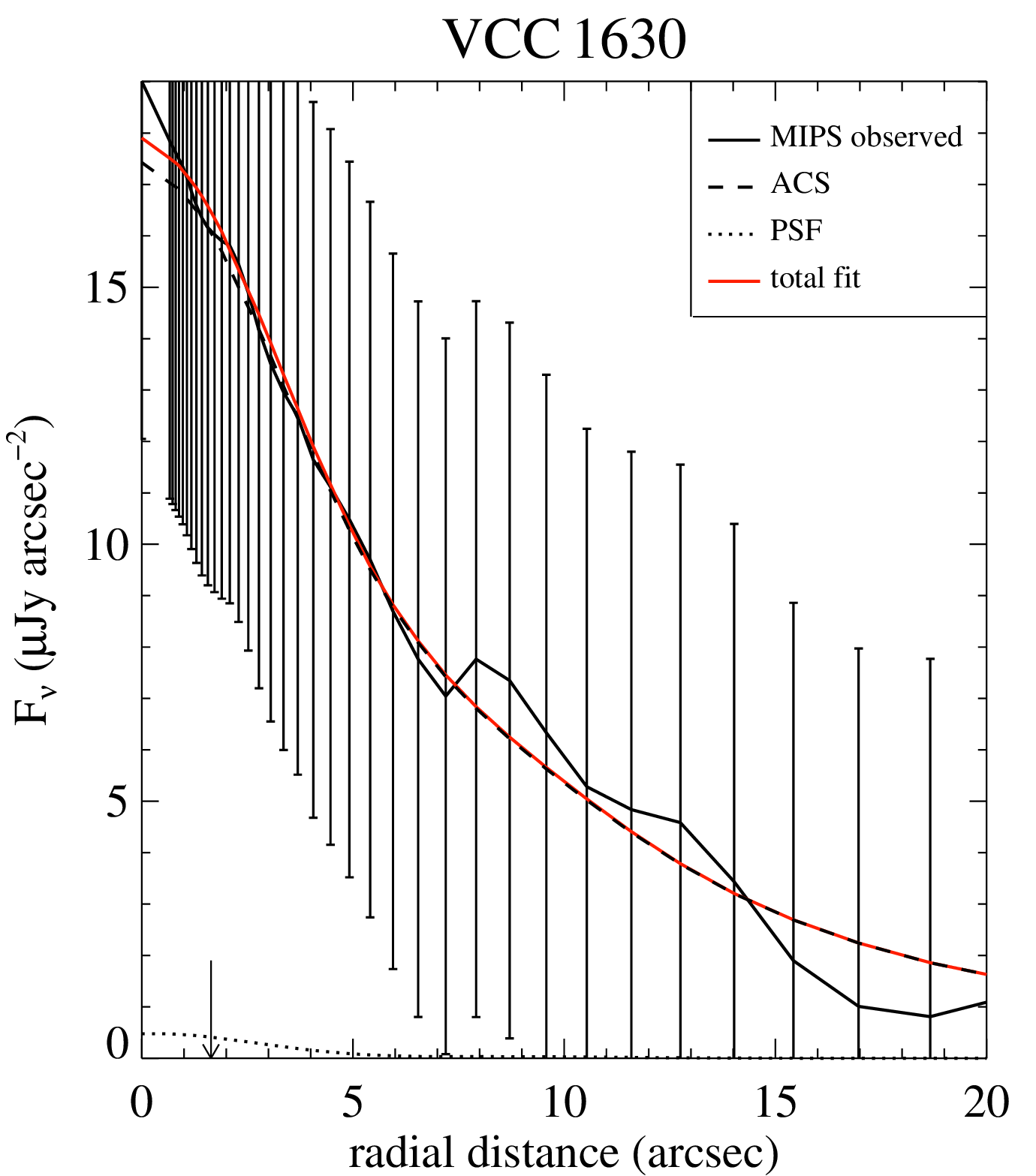}
\includegraphics[angle=0,scale=.38]{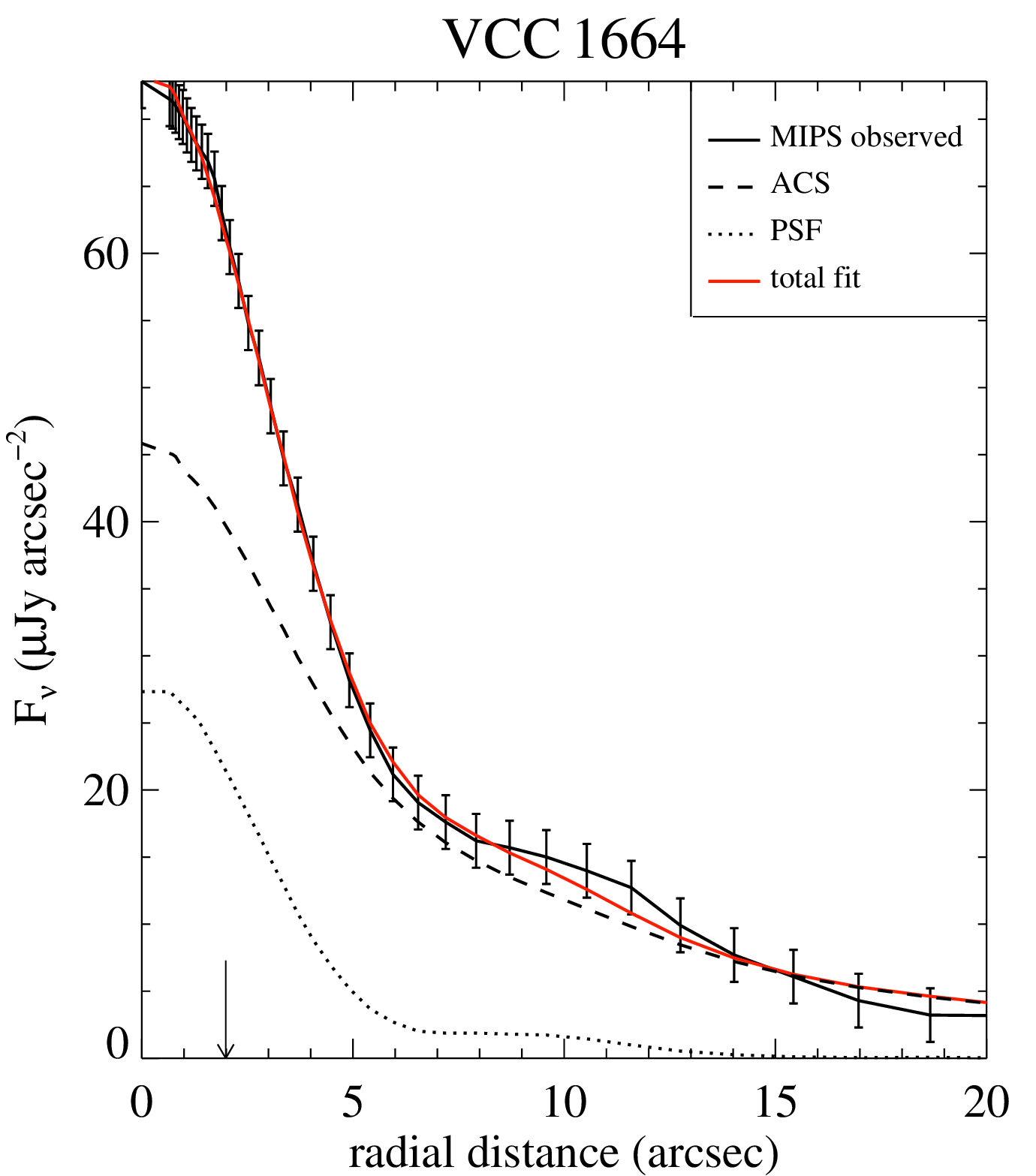}
\includegraphics[angle=0,scale=.38]{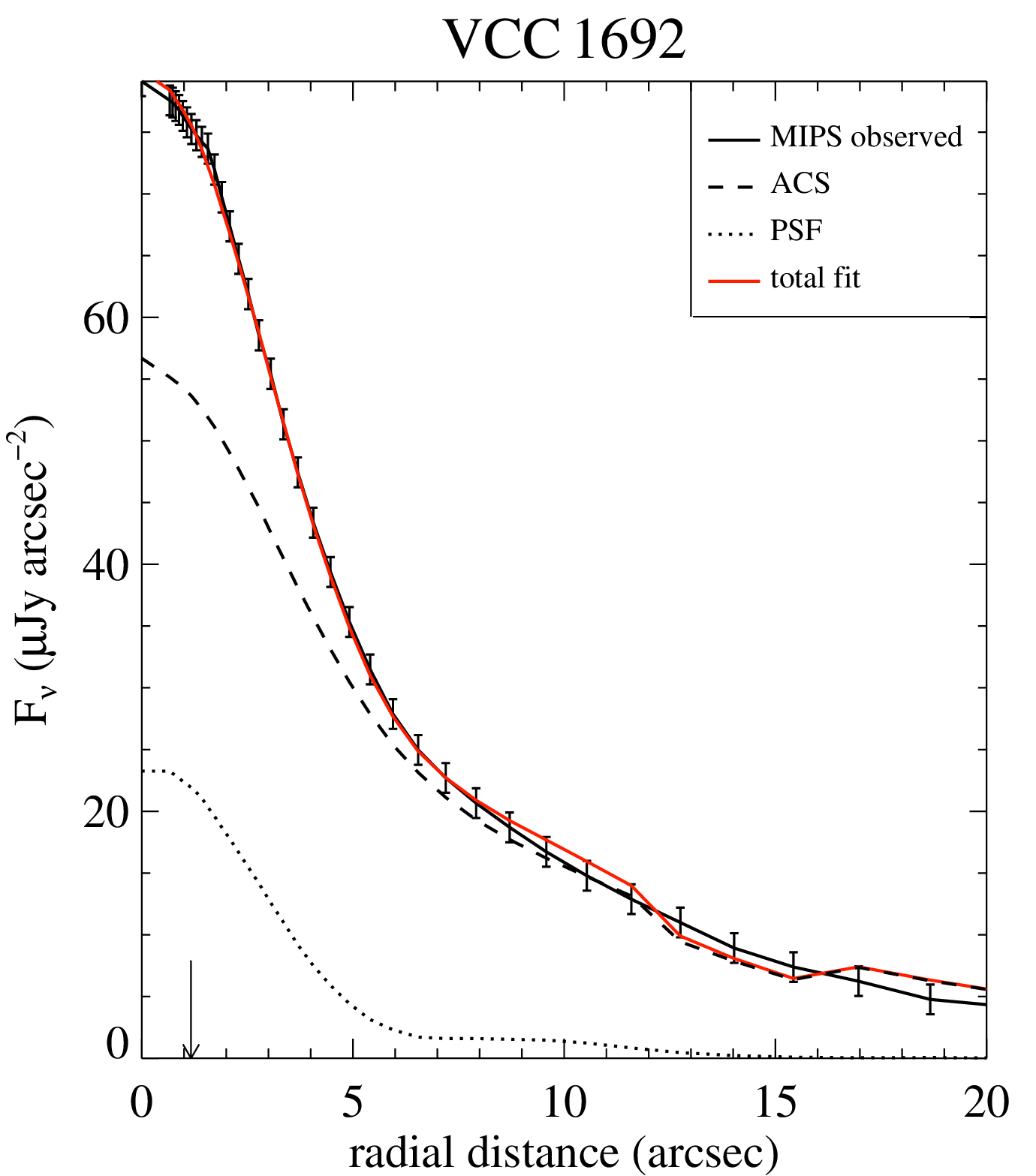}
\includegraphics[angle=0,scale=.38]{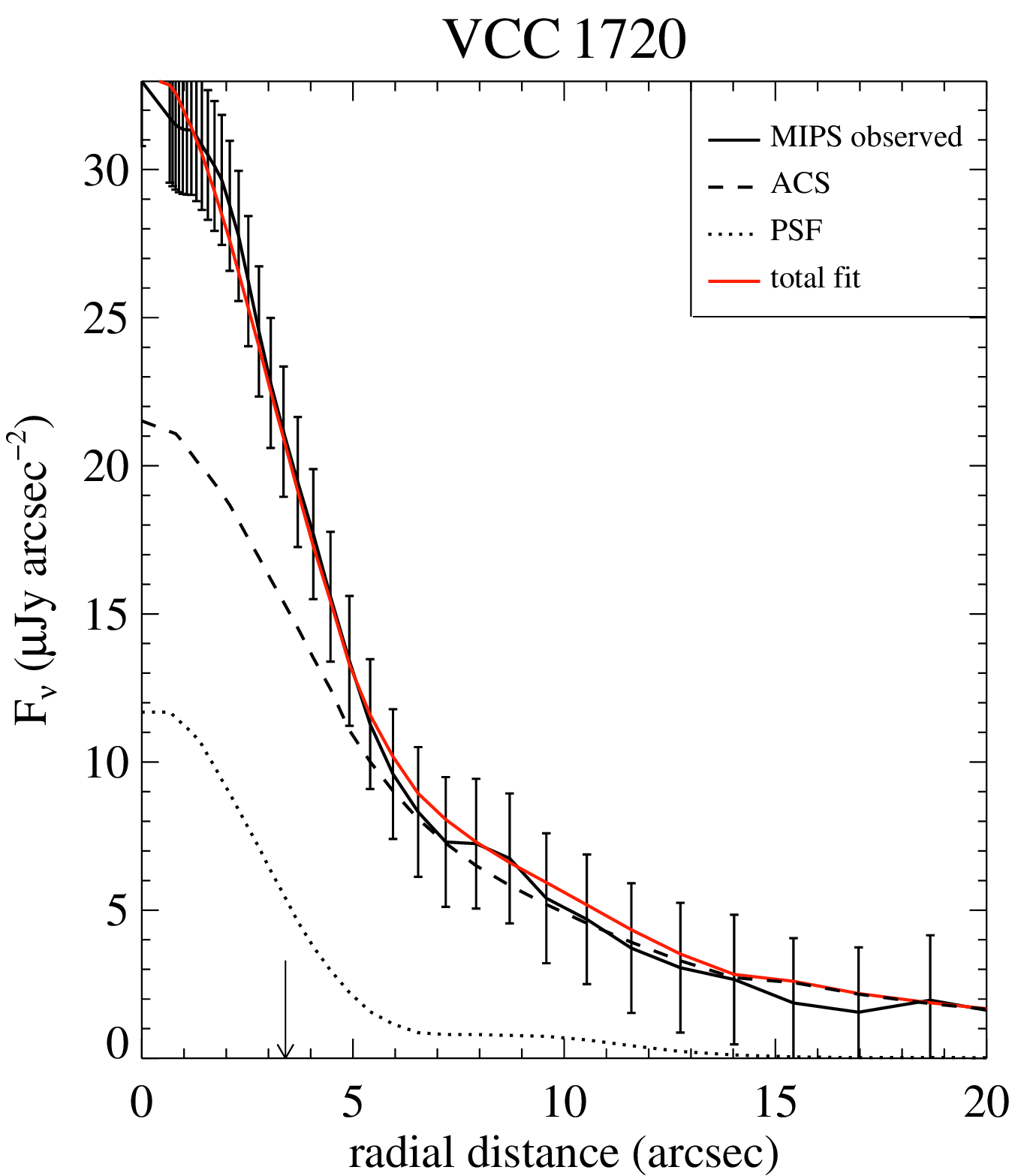}
\caption{{\it continued}}
\end{figure*}
\addtocounter{figure}{-1}
\begin{figure*}[t!]
\centering
\includegraphics[angle=0,scale=.38]{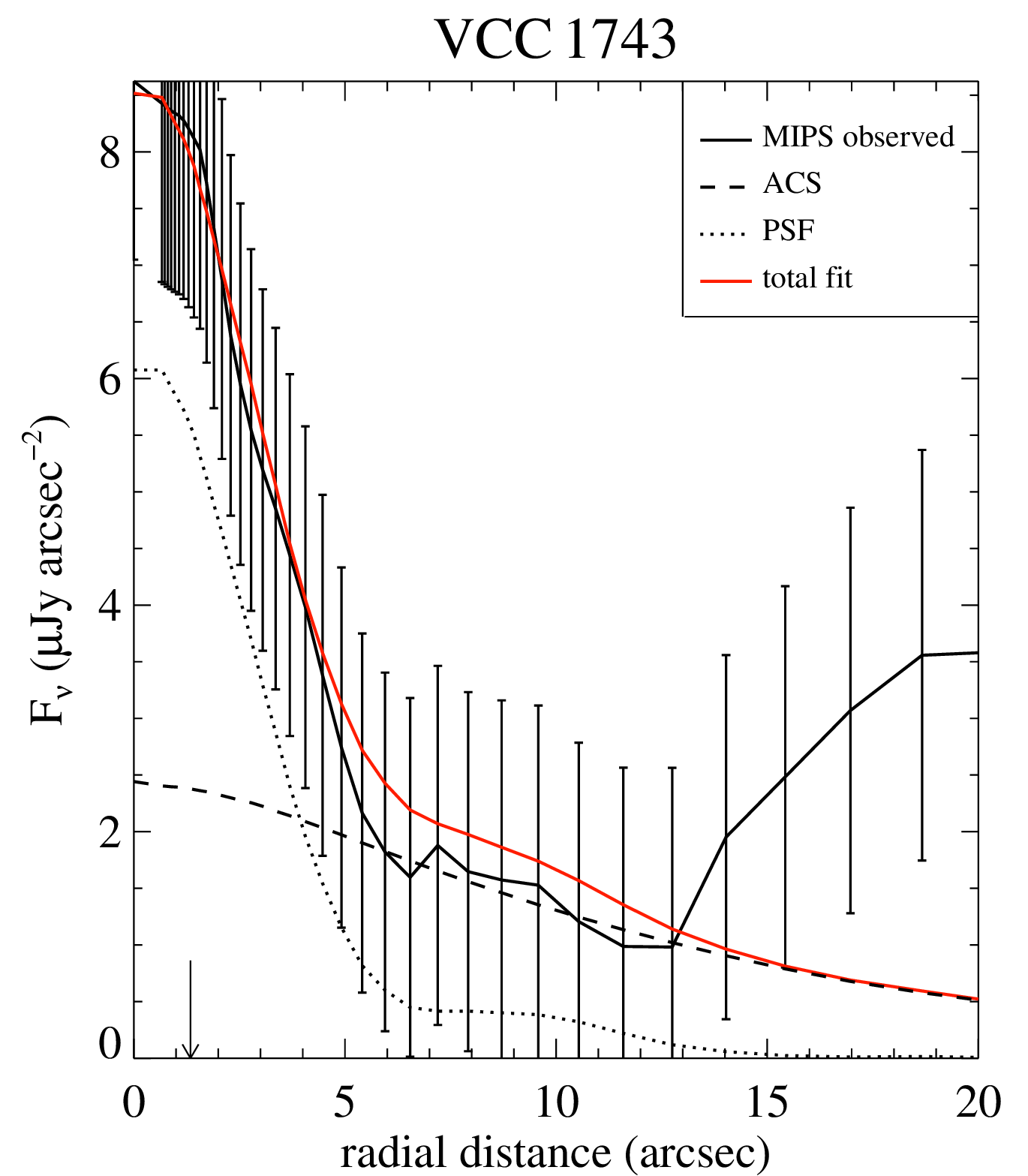}
\includegraphics[angle=0,scale=.38]{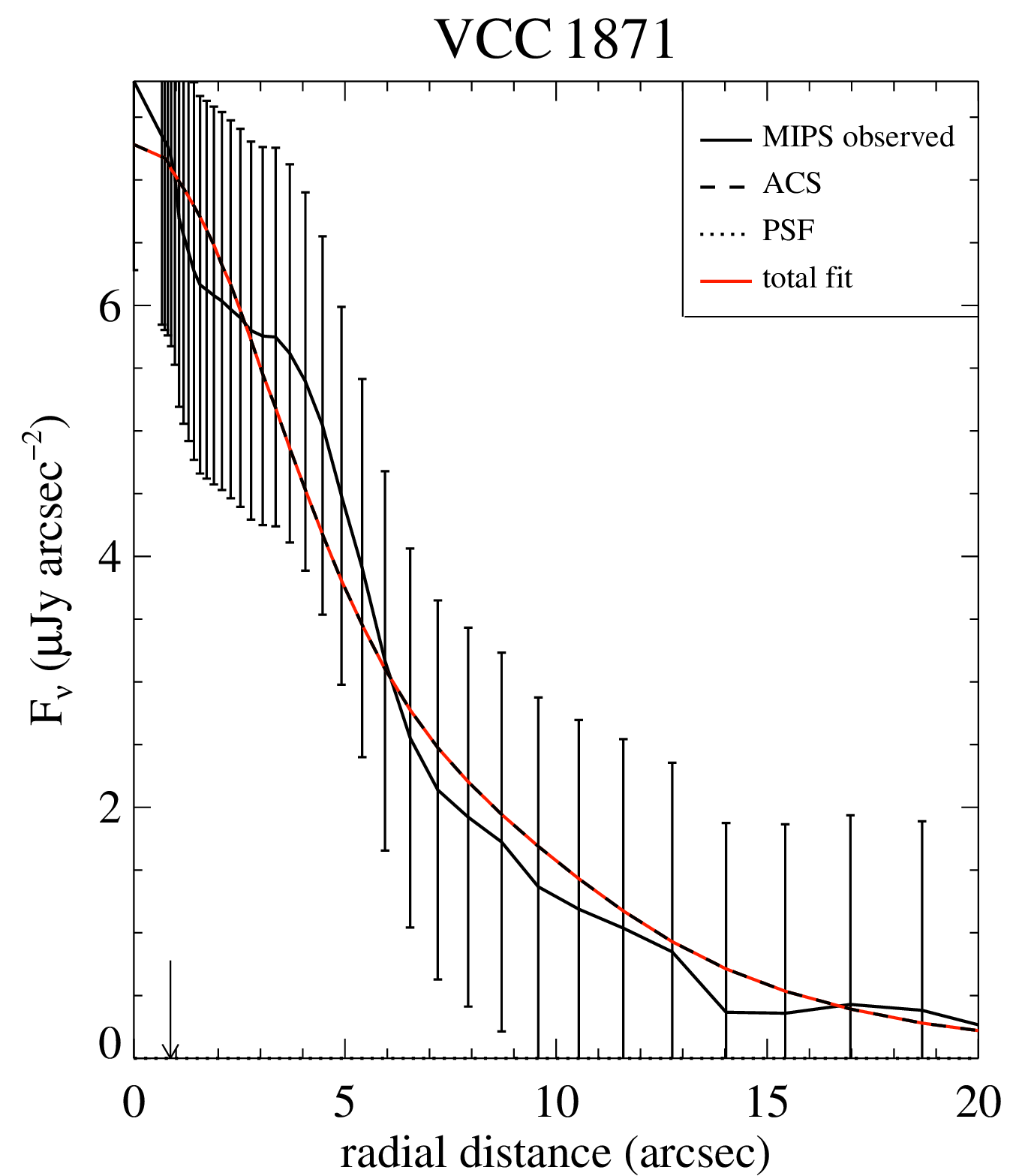}
\includegraphics[angle=0,scale=.38]{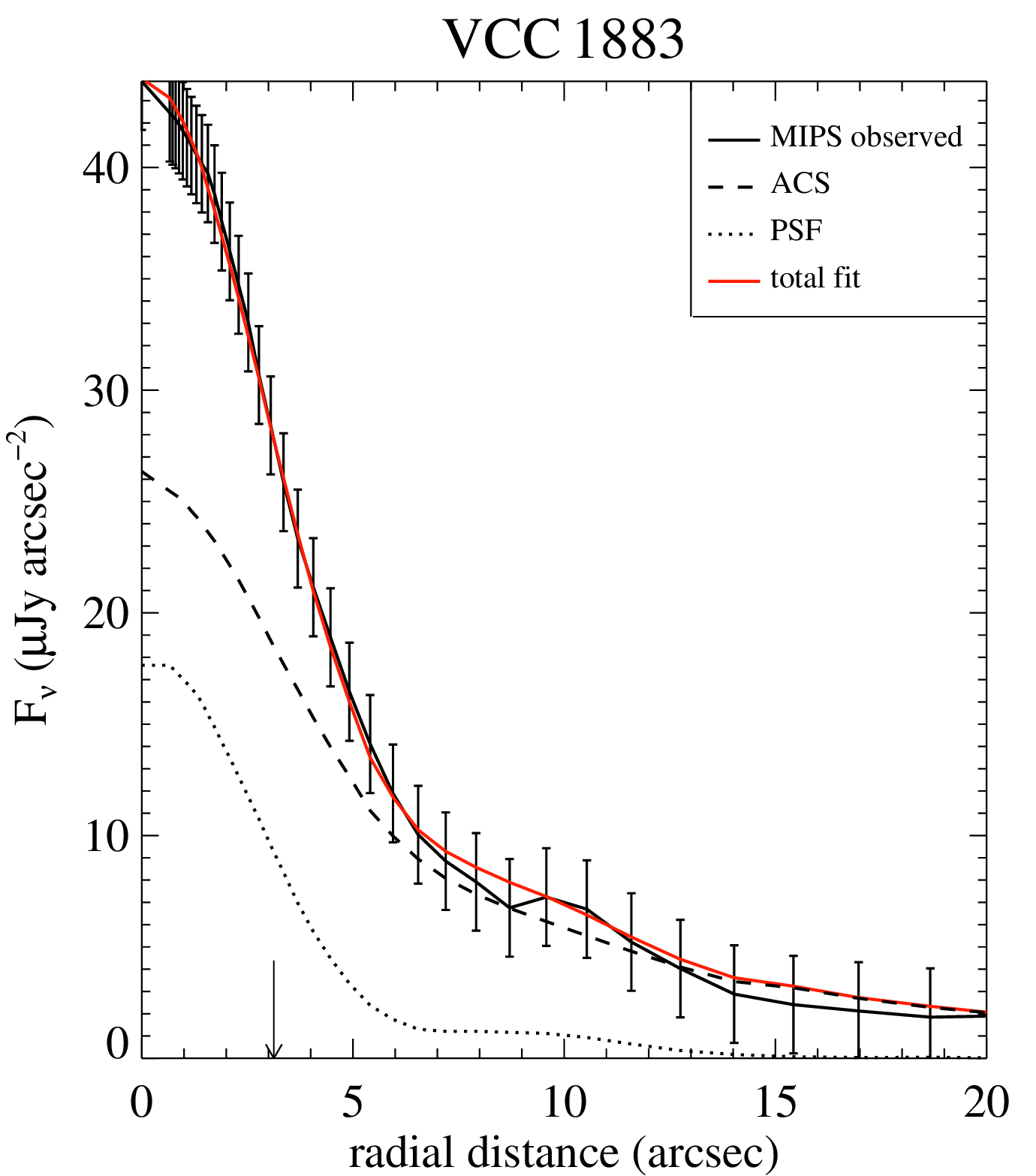}
\includegraphics[angle=0,scale=.38]{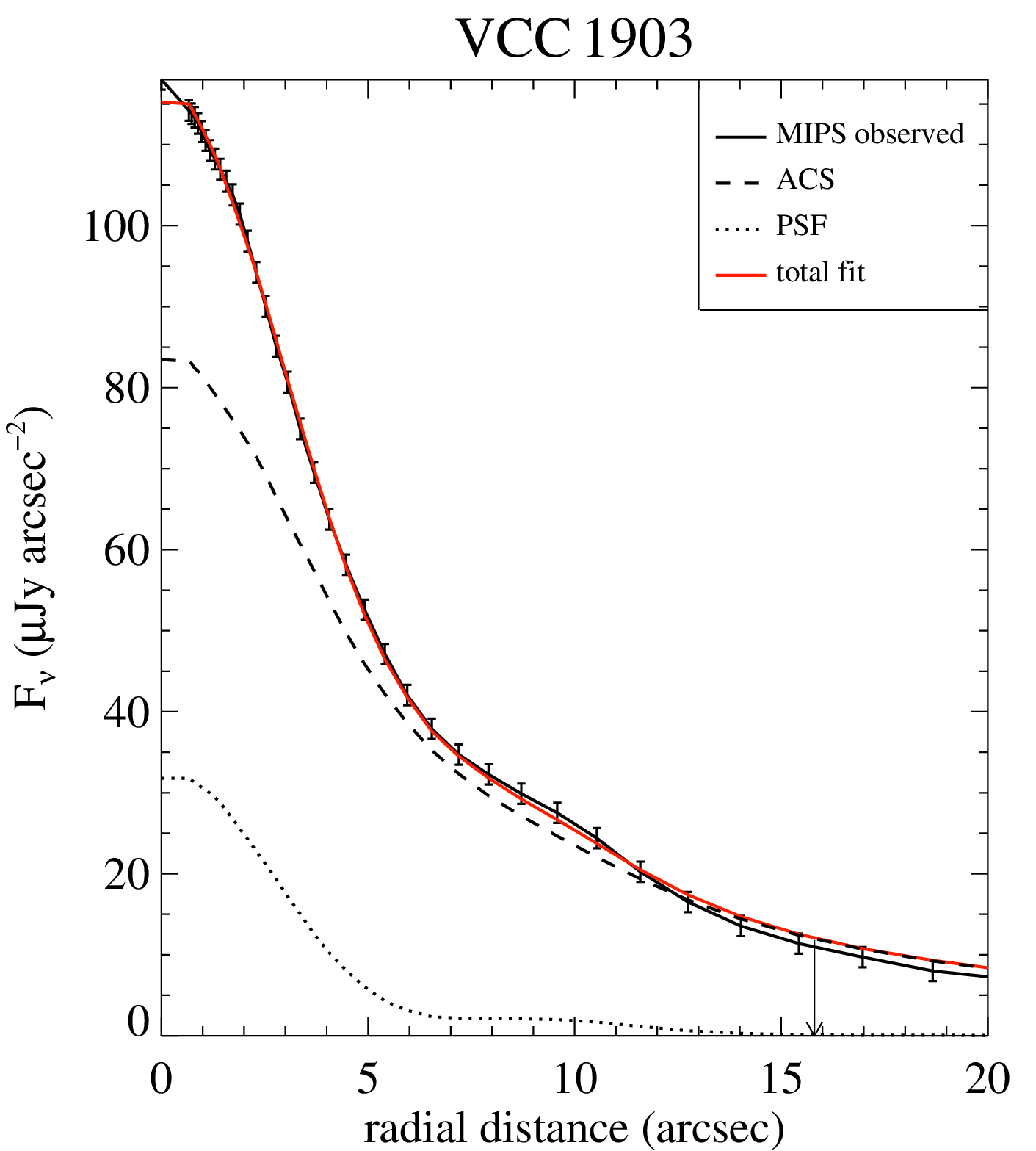}
\includegraphics[angle=0,scale=.38]{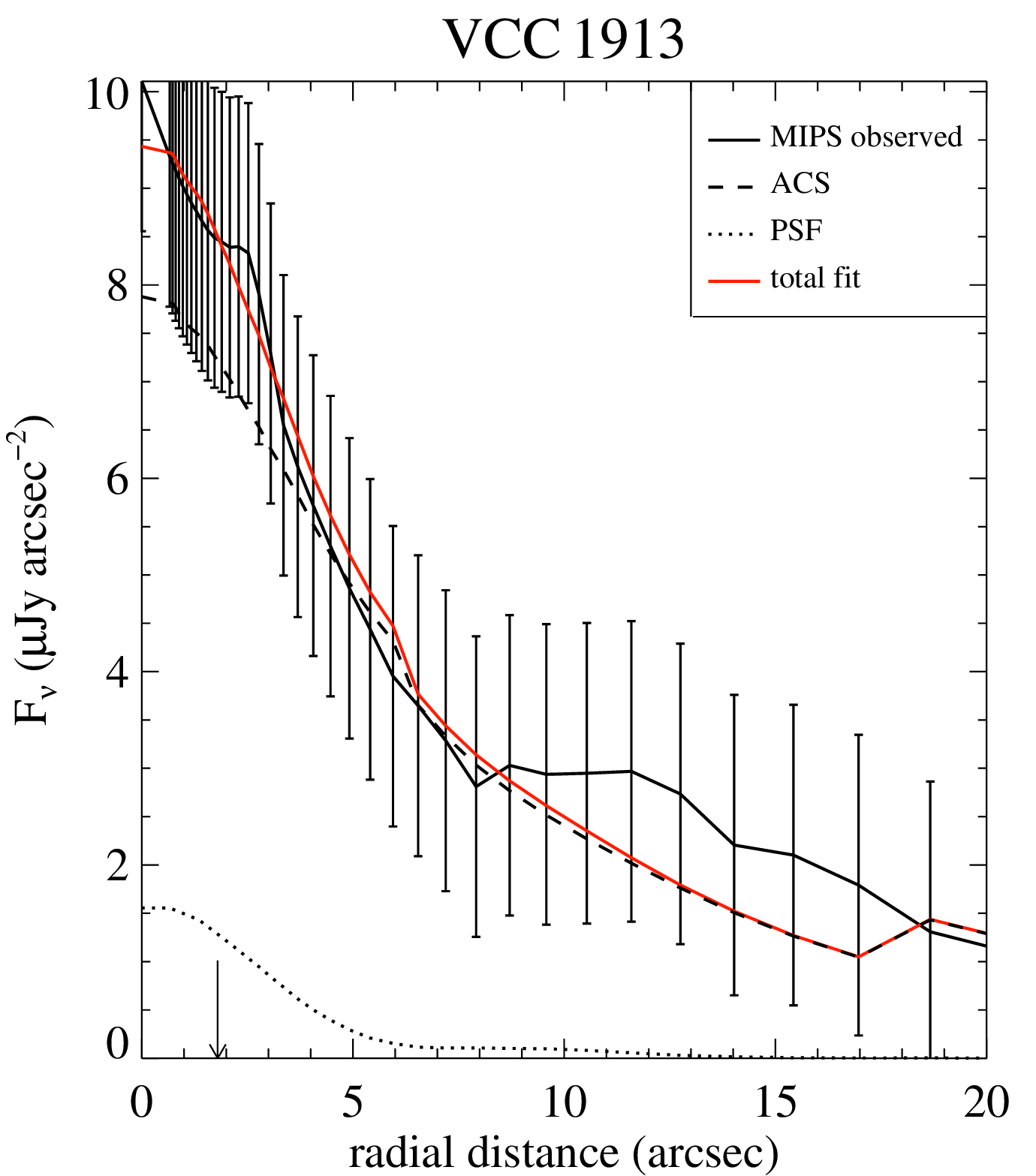}
\includegraphics[angle=0,scale=.38]{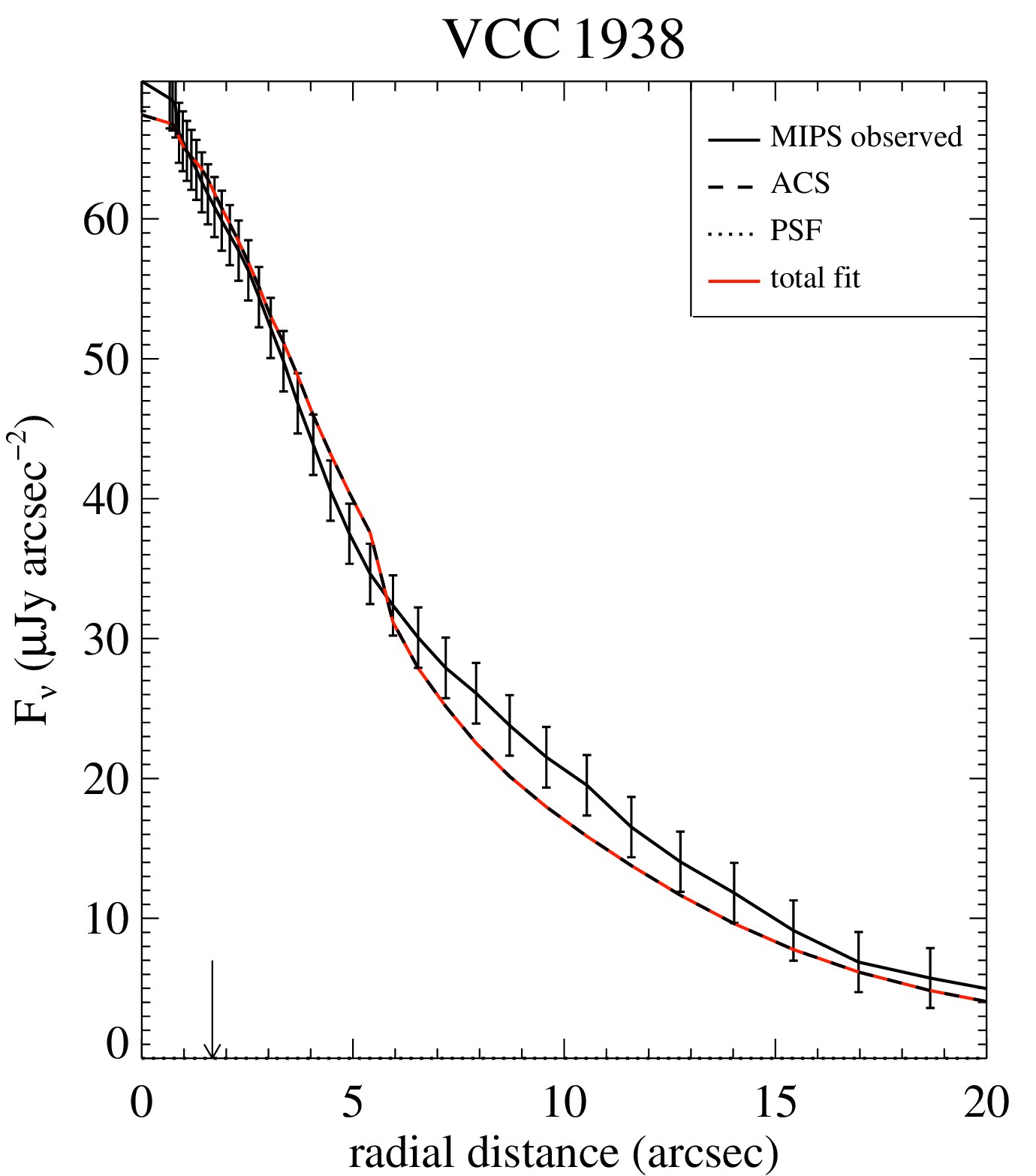}
\includegraphics[angle=0,scale=.38]{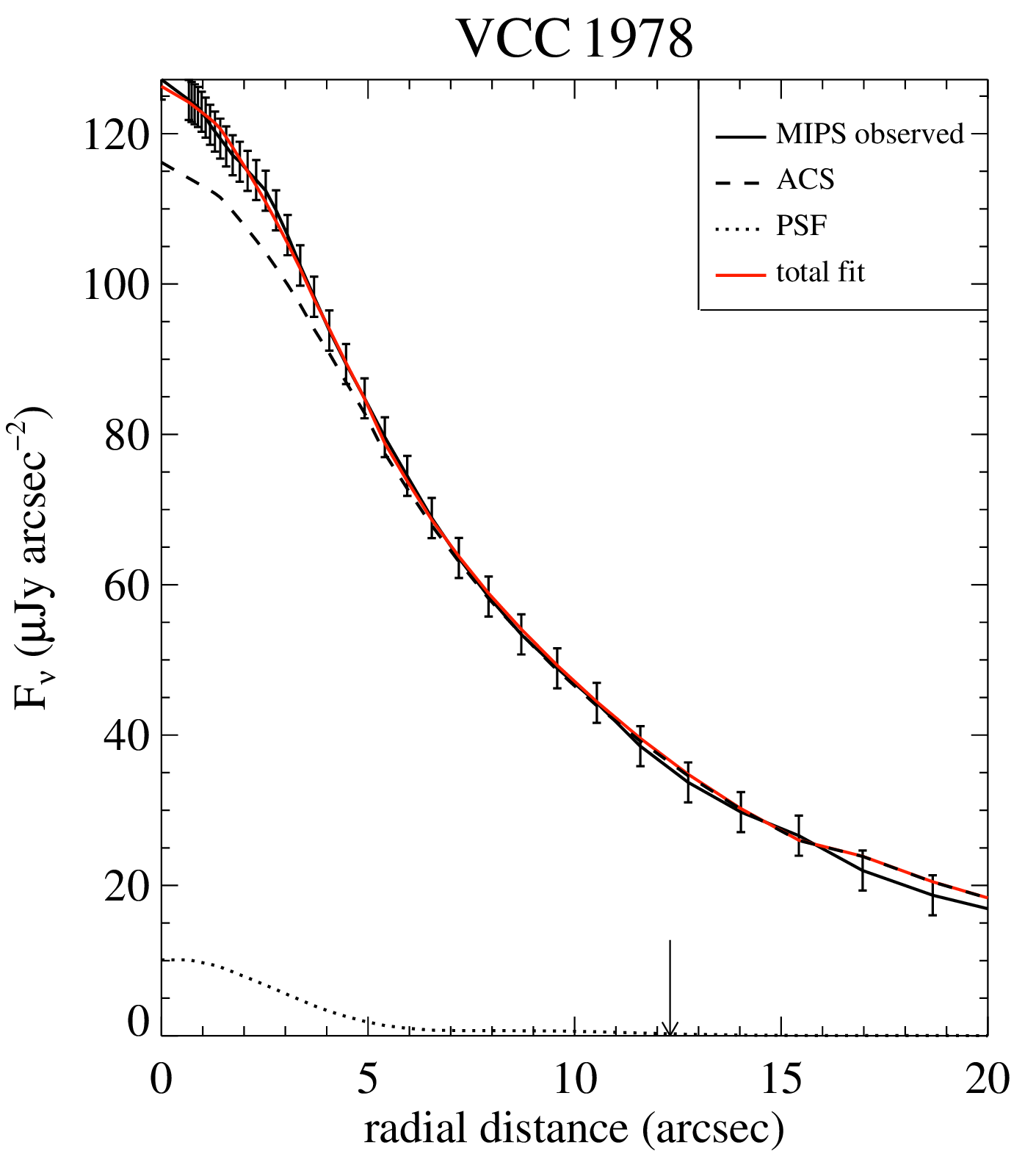}
\includegraphics[angle=0,scale=.38]{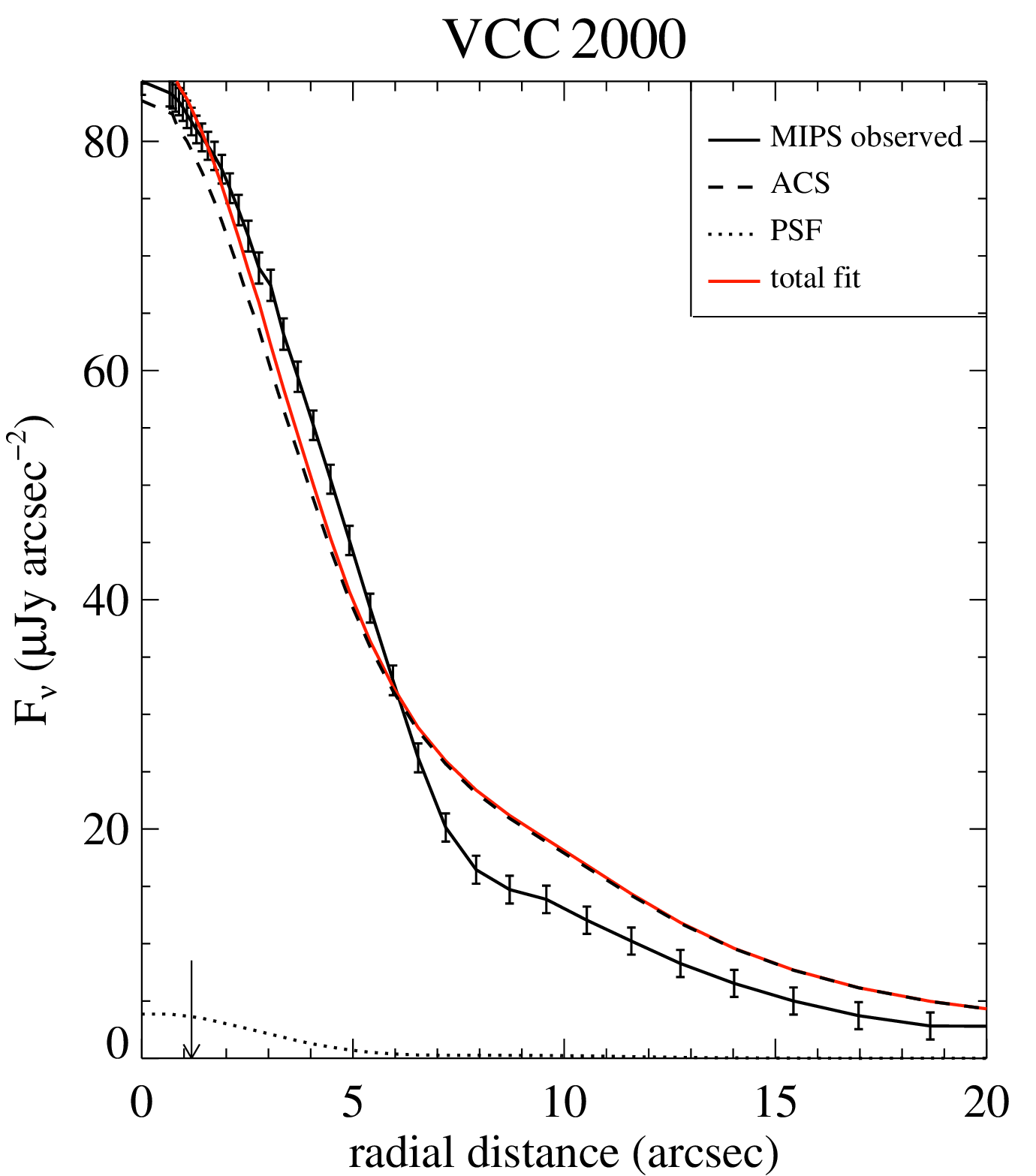}
\includegraphics[angle=0,scale=.38]{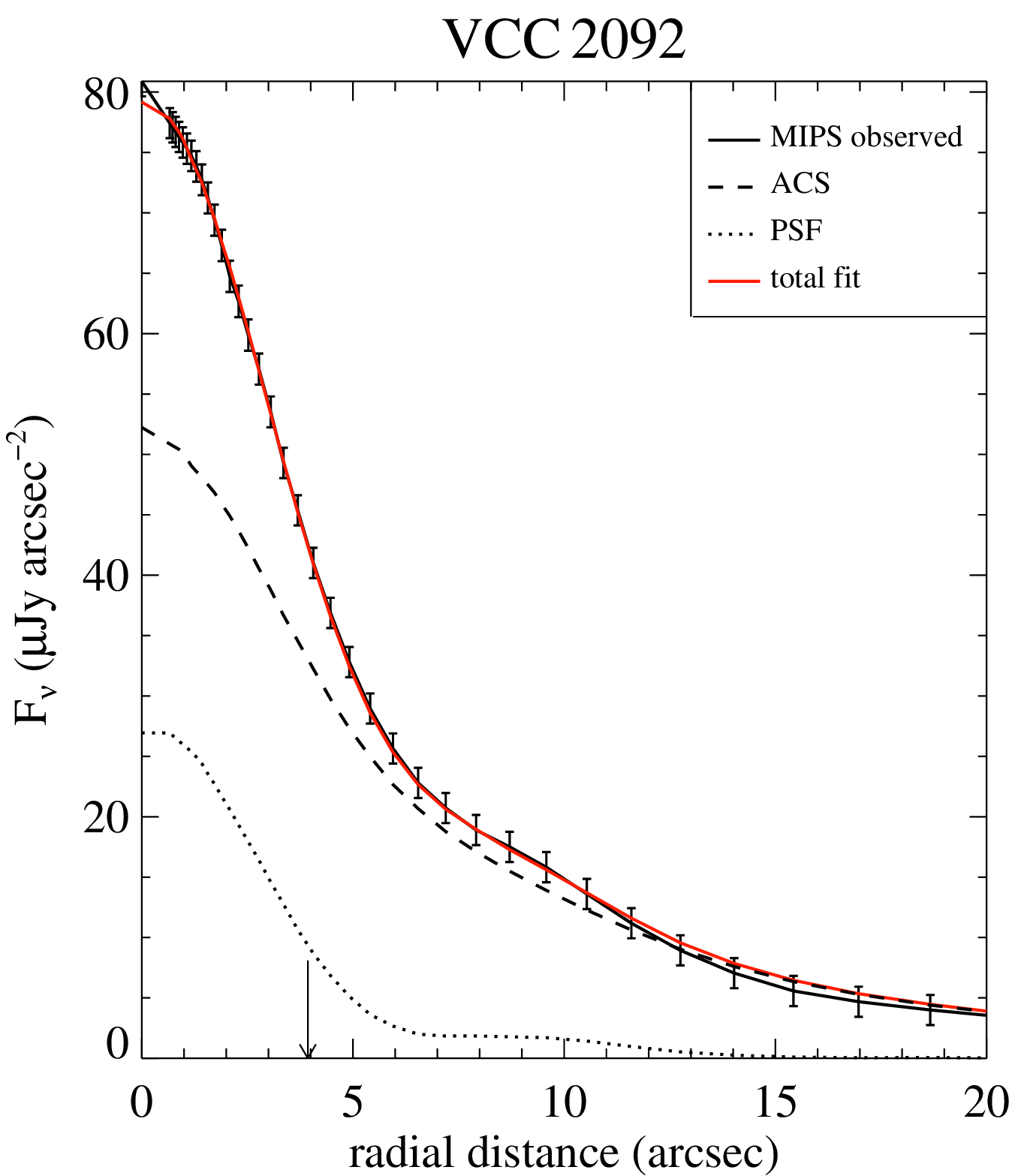}
\includegraphics[angle=0,scale=.38]{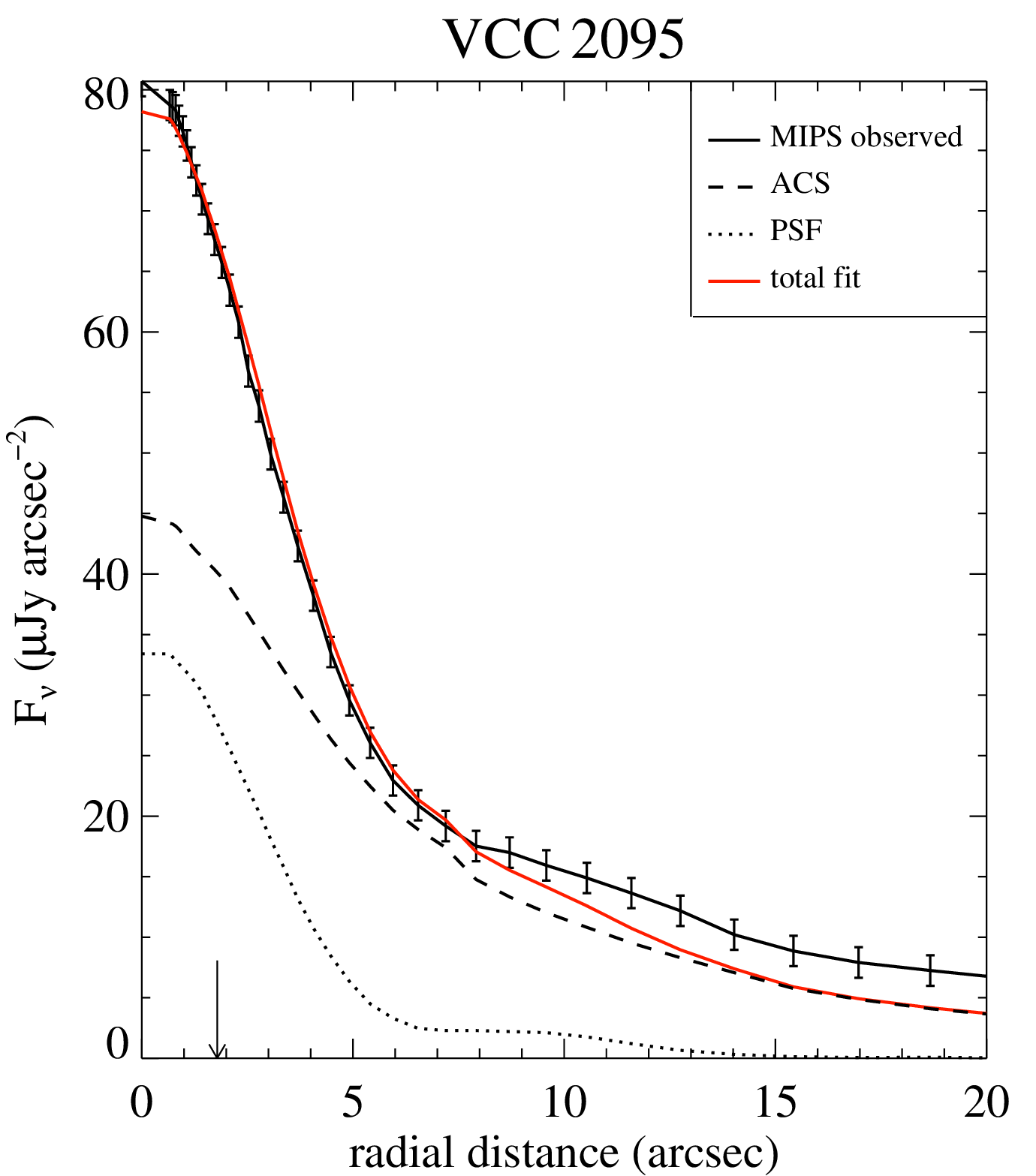}
\caption{{\it continued}}
\end{figure*}

\subsection{Radial profiles}
\label{sec:radprof}

After excluding the 16 sources mentioned above, the remaining 37
sources with clear MIR detections show no indications of host galaxy
dust in the optical from their {\it HST} images. Twenty of these also 
show a nuclear X-ray source as detected by \cxo\ \citep[][see also Table\,\ref{tab1}]{gal10}.

It has been shown that for early-type galaxies the total luminosity at
24\,$\mu$m correlates well with the optical luminosity
\citep[e.g.][]{tem07}. Furthermore, the mid-infrared radial light
profiles follow the optical light profiles very closely
\citep[e.g.][]{kna92,tem07,tem08}.  In addition, the MIR spectra of
quiescent early-type galaxies reveal emission in excess of
photospheric Rayleigh-Jeans predictions \citep[e.g.][]{bre06}. This
is generally interpreted as emission from dust produced in the
photospheres and circum-stellar regions of evolved mass-losing stars,
mainly on the asymptotic giant branch (AGB). These findings argue that
most of the MIR emission in early-type galaxies is directly associated
with the stellar population. This provides a valuable tool in this
study, because the optical light profiles (scaled appropriately) can
be used as a proxy for the bulk of MIR emission which is associated
with the stellar population. 

For the 37 MIR-detected objects without optical dust detection we
produced and analyzed azimuthally averaged radial surface brightness
profiles at 24\,$\mu$m.  Radial light profiles were created using the
task {\tt ellipse} within {\sc Iraf}.  The central question is whether
these galaxies require an additional nuclear component to explain
their MIR light profiles or if the observed MIR emission is in
accordance with emission from the host galaxy (stellar light and dust
emission associated directly with the stellar population).

In order to use the optical light profiles as a proxy for the MIR
emission expected from the stellar populations, we also created radial
profiles from the re-reduced ACS $z$-band images, which we convolved
with an observed MIPS PSF, after aligning them to the MIPS images to
ACS-pixel accuracy.  Because only few MIR maps included a sufficiently
bright star to serve as a MIPS PSF for our analysis, we chose to
create an average PSF from twelve individual stars on our maps which
were reasonably bright, not close to the edges of the maps, and also
fairly isolated. The radial surface brightness profile for the MIR PSF
was derived in the same manner as for the ACS and MIPS images.

The observed MIR surface brightness profiles were then fitted as the 
linear combination of the radial profile of the MIR-PSF convolved 
optical images (i.e. stellar host galaxy contributions) and the radial 
profile of the average MIR PSF (i.e. a possible nuclear excess of unresolved
dust). We show the radial profile fits in
Fig.\,\ref{images_radprof}. In most objects an additional PSF component is
required to match the optical light profile with the MIR light
profile. This nuclear MIR excess can be small (e.g. VCC\,1537), of
medium strength (e.g. VCC\,0731), or quite substantial
(e.g. VCC\,2092). We will discuss the likely origin of this nuclear
excess emission in Section\,\ref{sec:results}.

\subsection{Image fitting}
\label{sec:imfit}

As a second step in our analysis we modeled the observed MIR images
using {\sc Galfit} \citep[][]{pen02}. As we have seen in the previous
section, a two component model is well suited to describe the MIR
radial profiles of the early-type galaxies in our sample. Therefore we
chose to fit the two-dimensional images also with two components: one
component representing the host contributions, plus a PSF component
representing a possible nucleus. As tracers for the host galaxy light
distribution we again used the {\it HST} $z$-band images, which have
been modeled previously by \citet[][]{fer06}. We fixed the shape
parameters (e.g. effective radius, Sersic index, axis ratio,
orientation) describing the host component to those determined by
\citet[][]{fer06}, leaving the absolute scaling and centroid as the only free
parameters. For the PSF component we provided the same average PSF
used in section \ref{sec:radprof}. For the vast majority of objects
this two-component model provided a very good description of the MIR
images. In a few individual cases we added an additional PSF to
account for nearby stars. In the cases of VCC\,1125 and VCC\,2095 an
additional disk component was needed to achive a satisfactory model of
the galaxy. The model for VCC\,1938 was more complex and required a
second sersic profile and an additional (off nuclear) PSF. The MIPS
24\,$\mu$m images and the residuals after subtracting the best fit
model for all 37 sources for which the fitting was performed are
presented in Fig.\,\ref{images_galfit}.

\begin{figure*}
\centering
\includegraphics[angle=0,scale=.28]{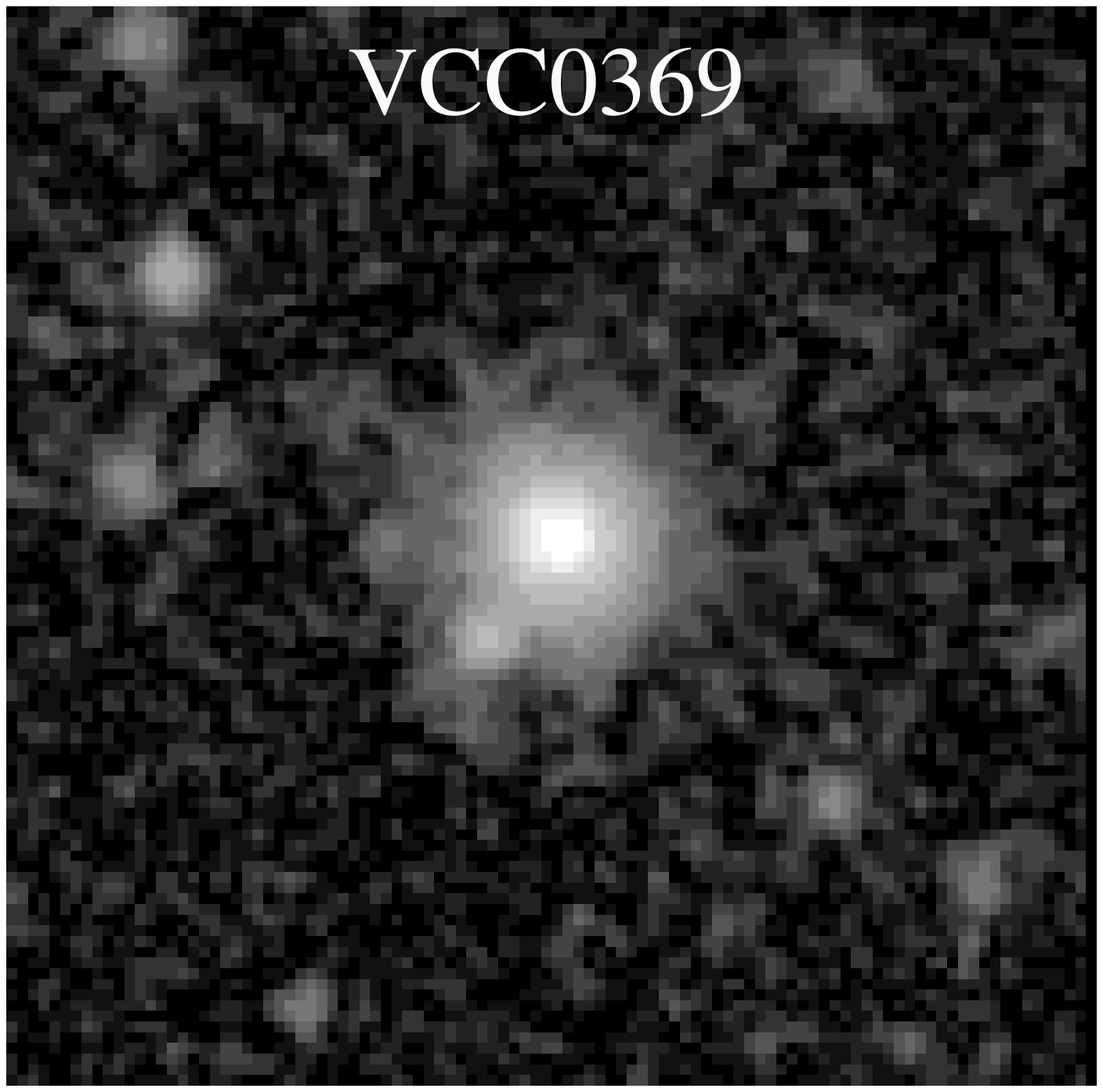}
\includegraphics[angle=0,scale=.28]{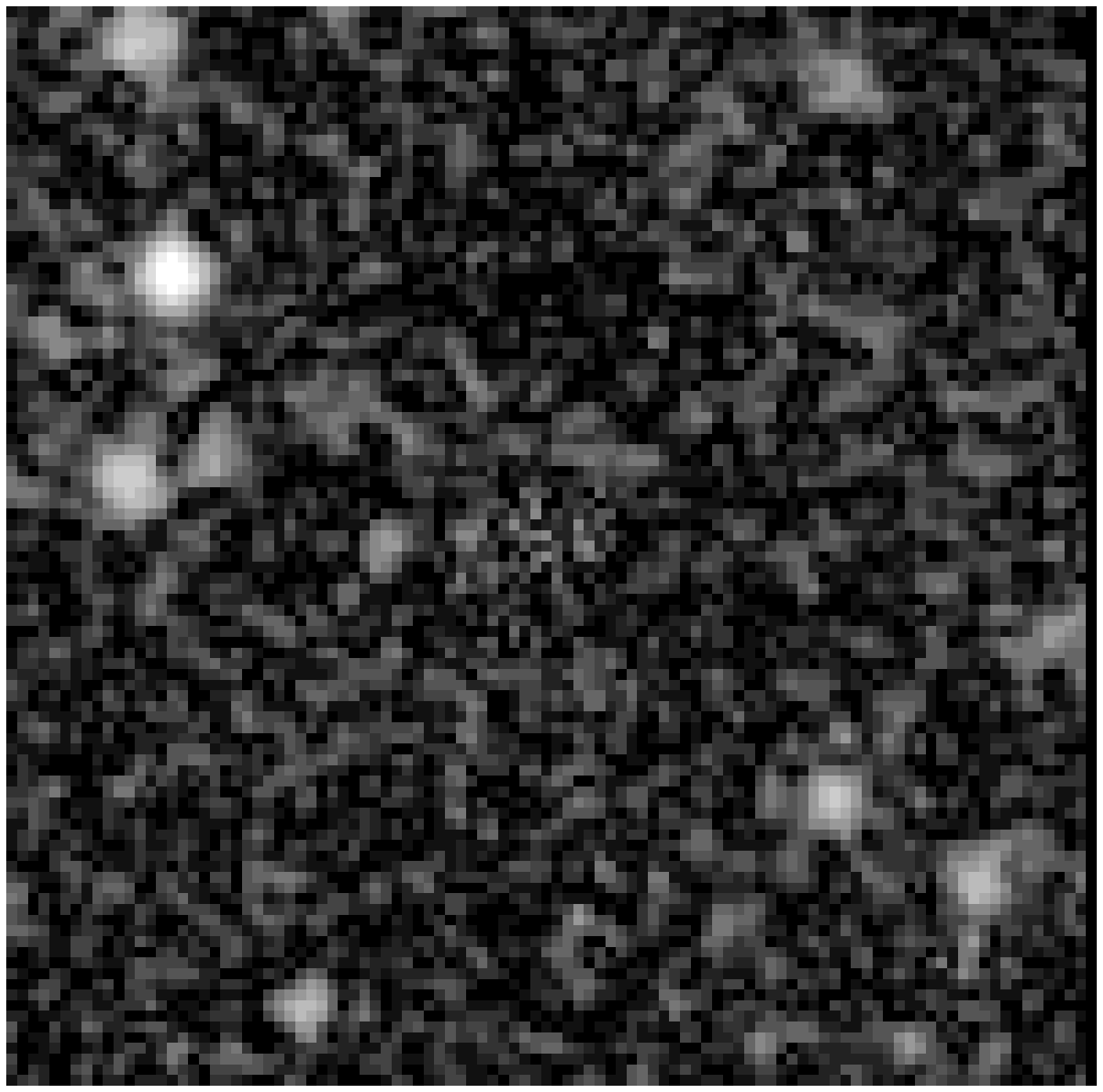}\hspace{0.06cm}
\includegraphics[angle=0,scale=.28]{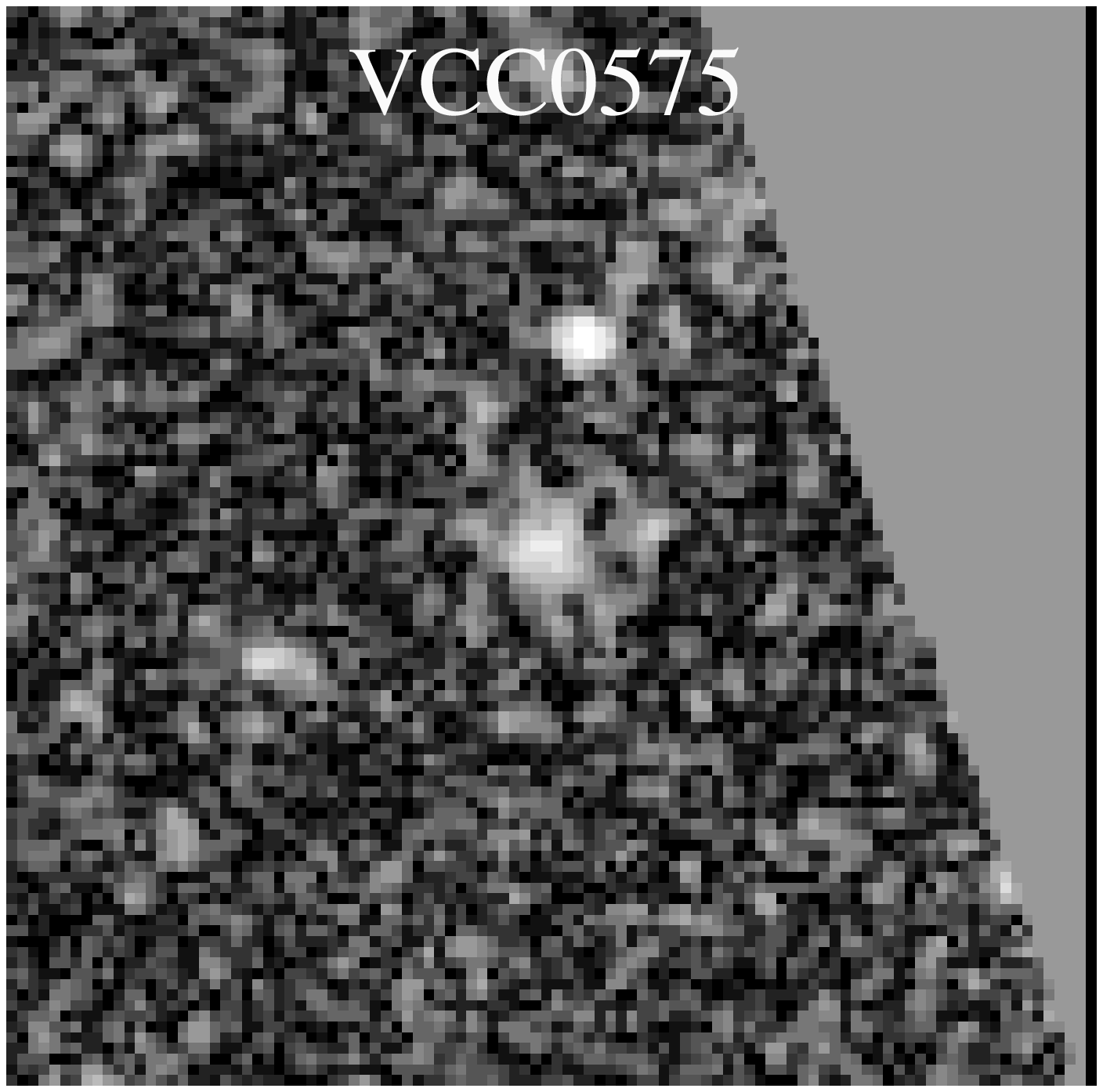}
\includegraphics[angle=0,scale=.28]{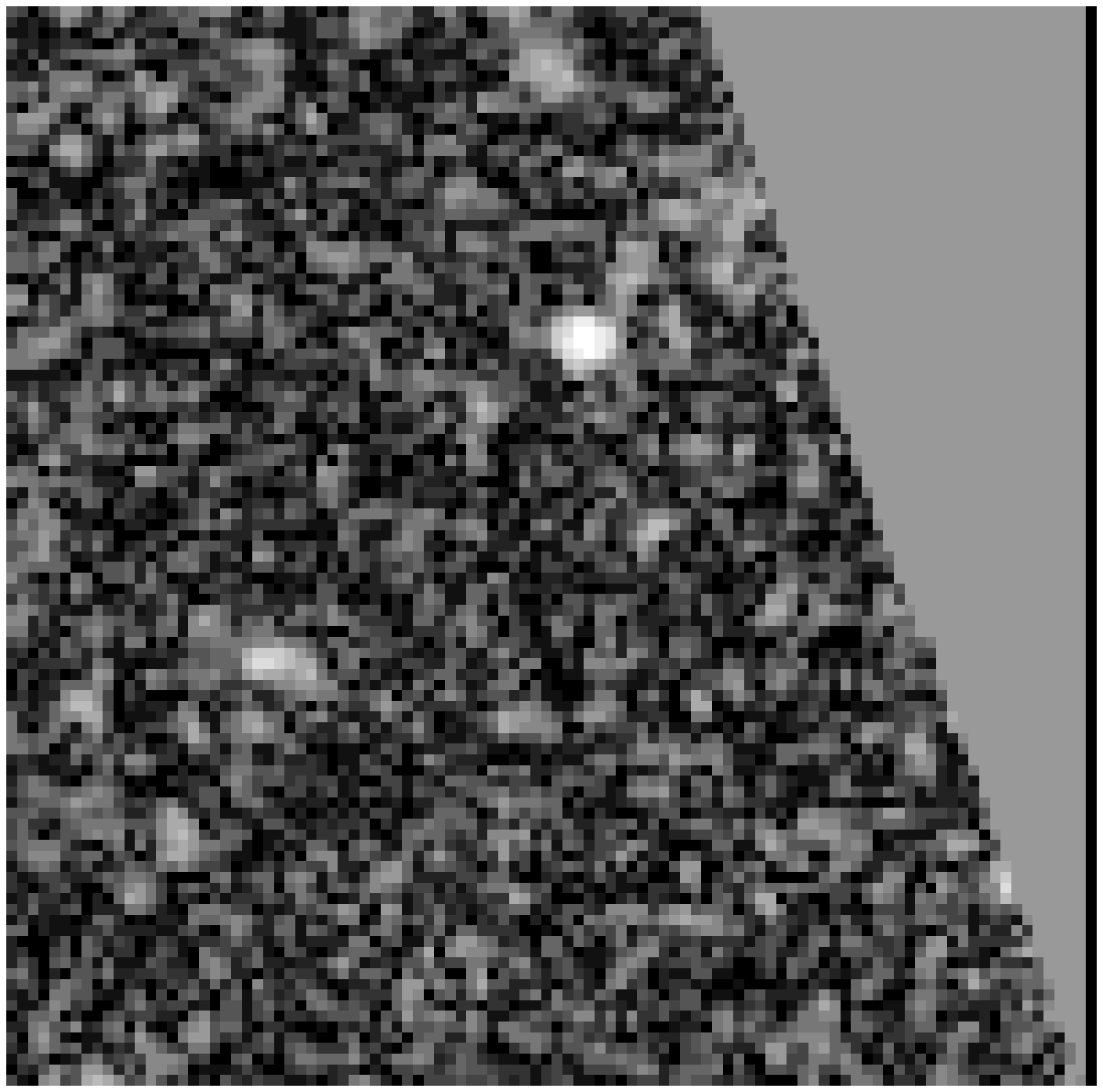}\vspace{0.025cm}
\includegraphics[angle=0,scale=.28]{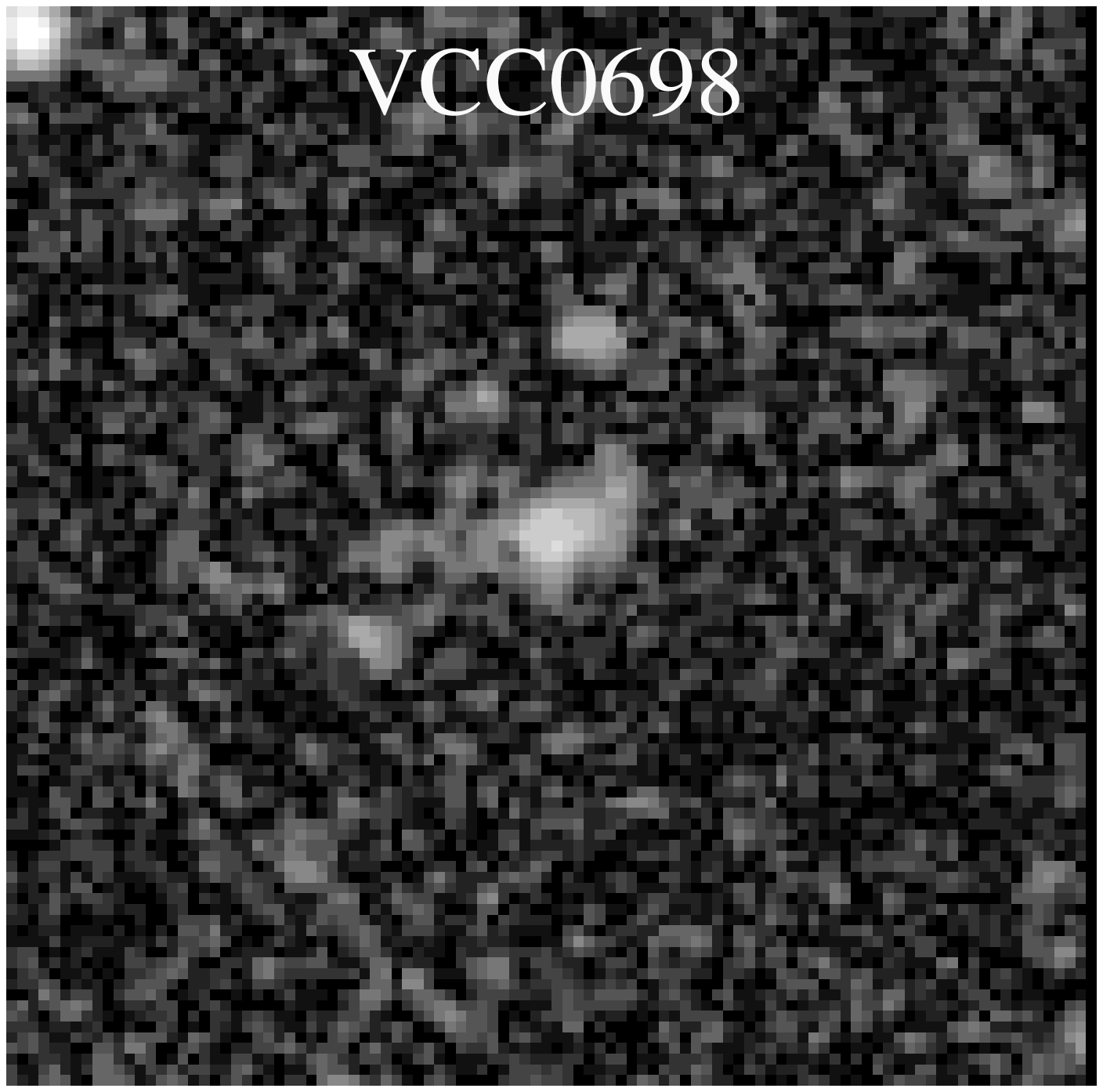}
\includegraphics[angle=0,scale=.28]{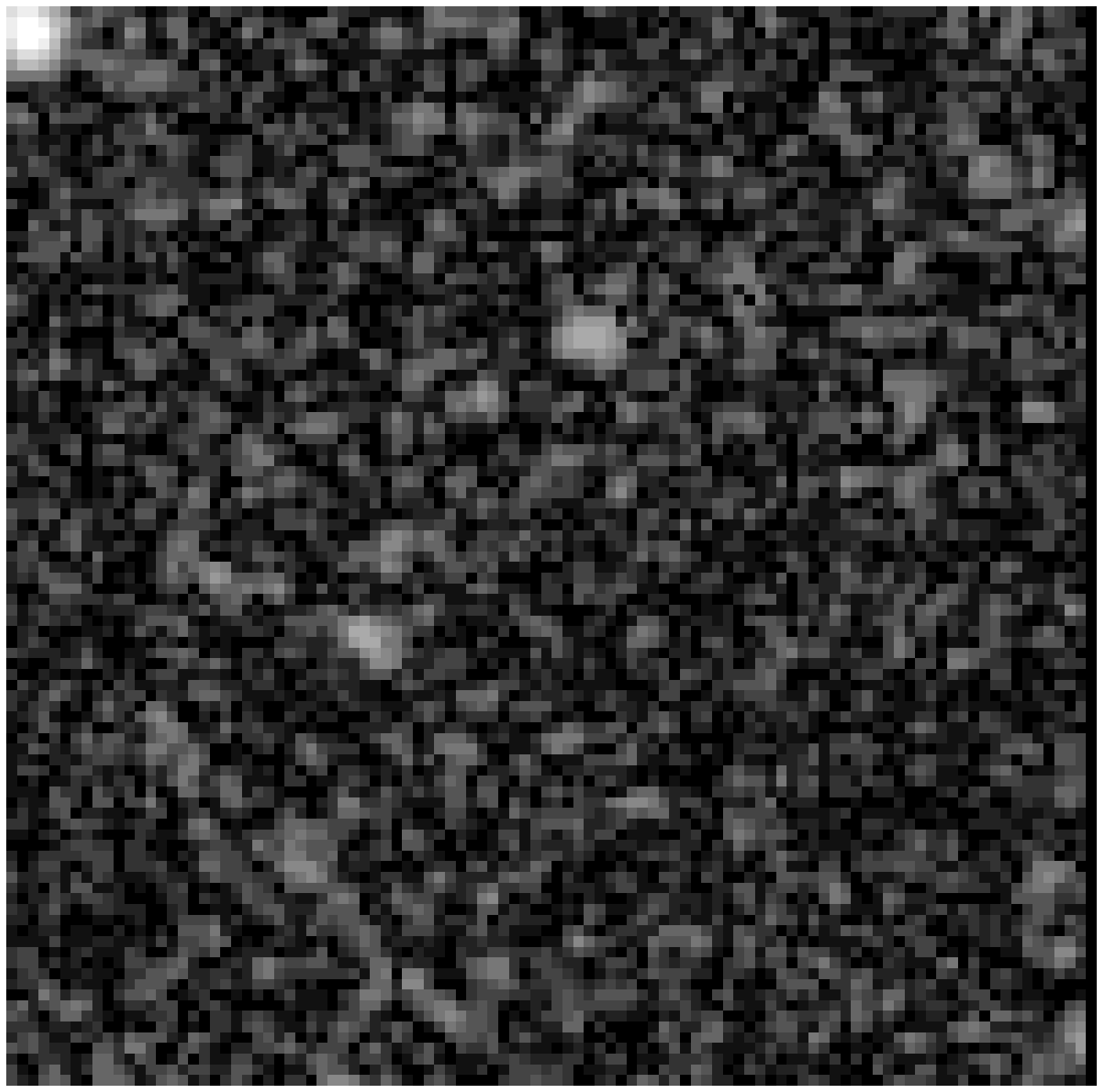}\hspace{0.06cm}
\includegraphics[angle=0,scale=.28]{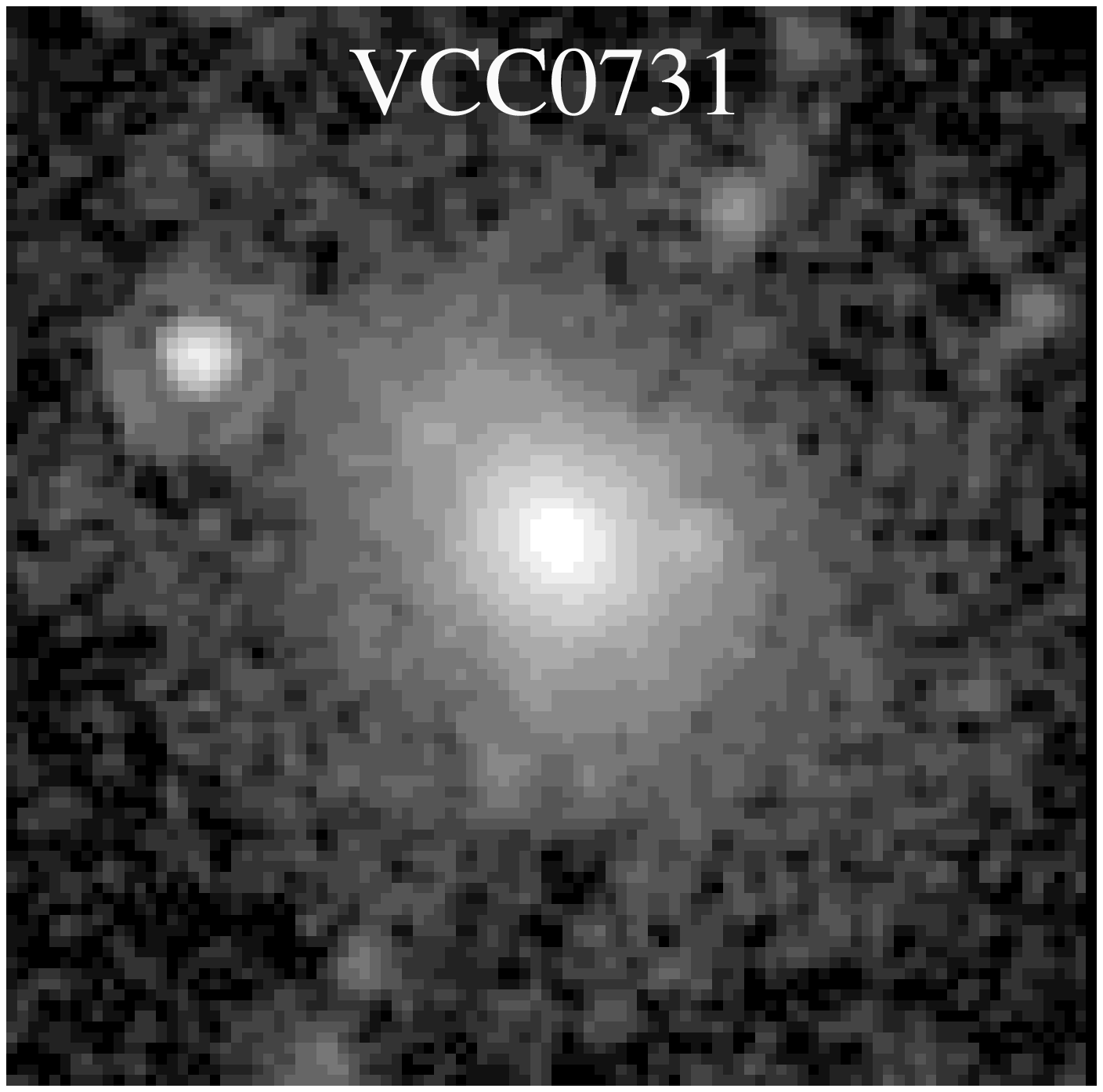}
\includegraphics[angle=0,scale=.28]{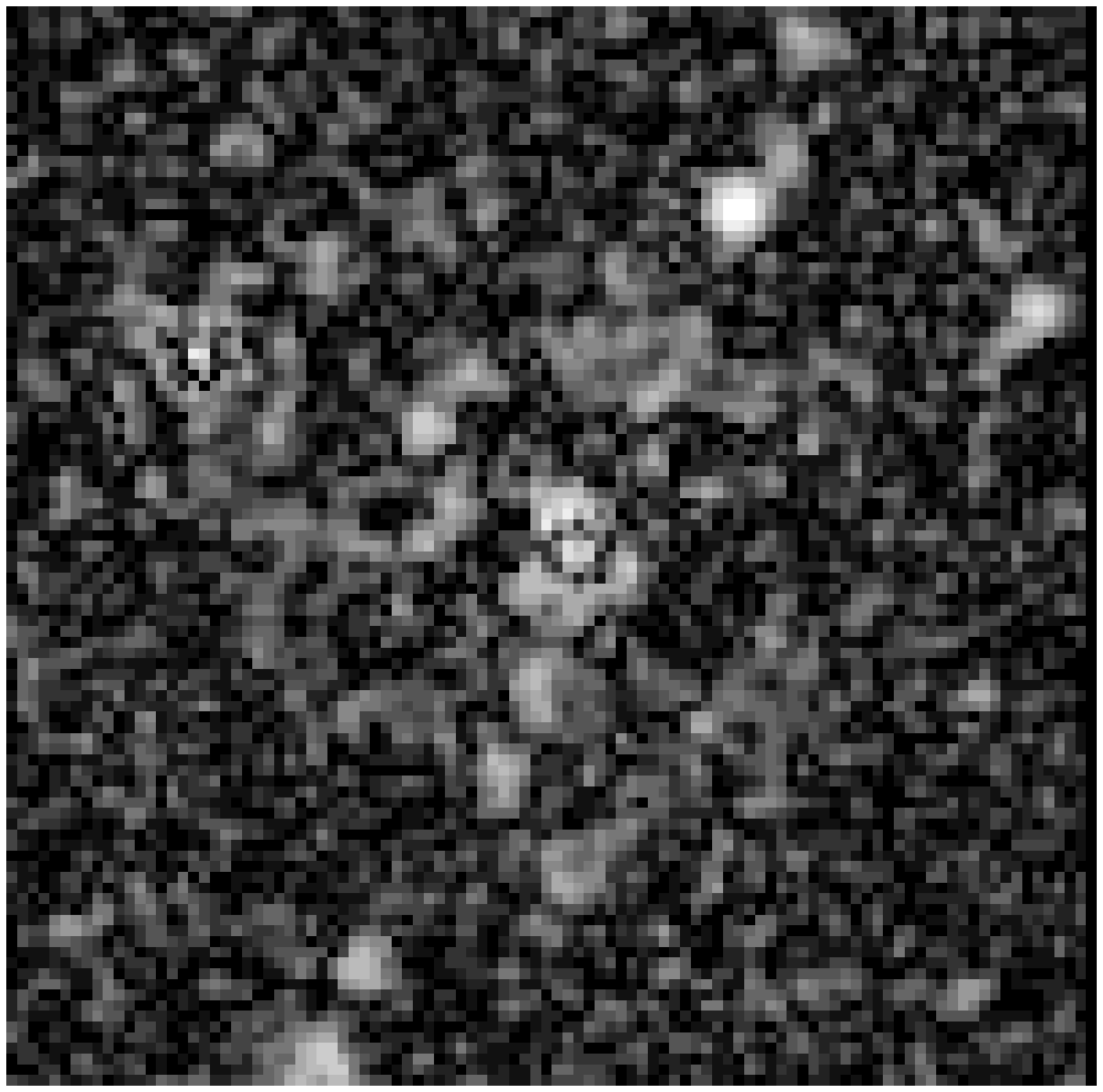}\vspace{0.025cm}
\includegraphics[angle=0,scale=.28]{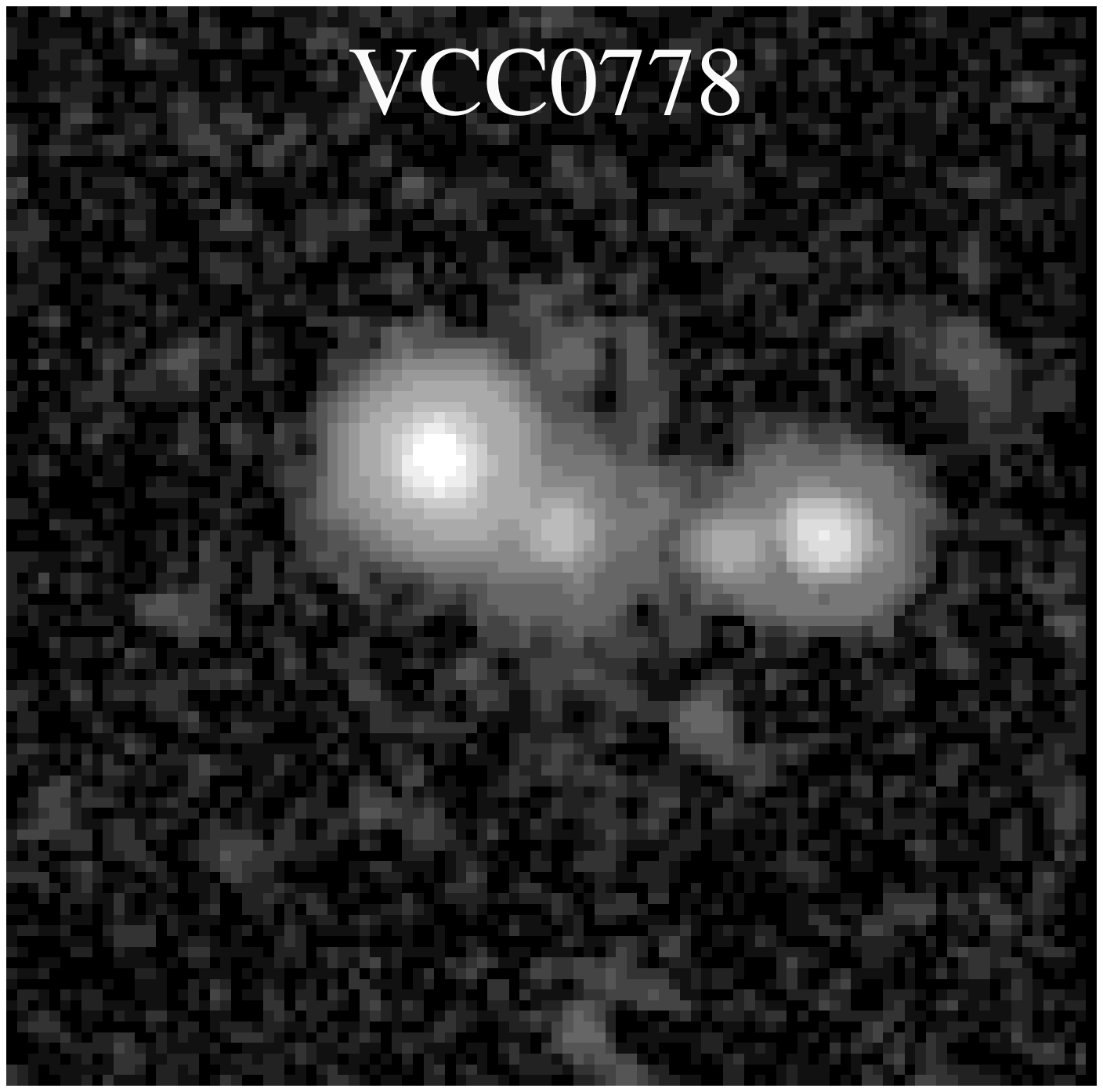}
\includegraphics[angle=0,scale=.28]{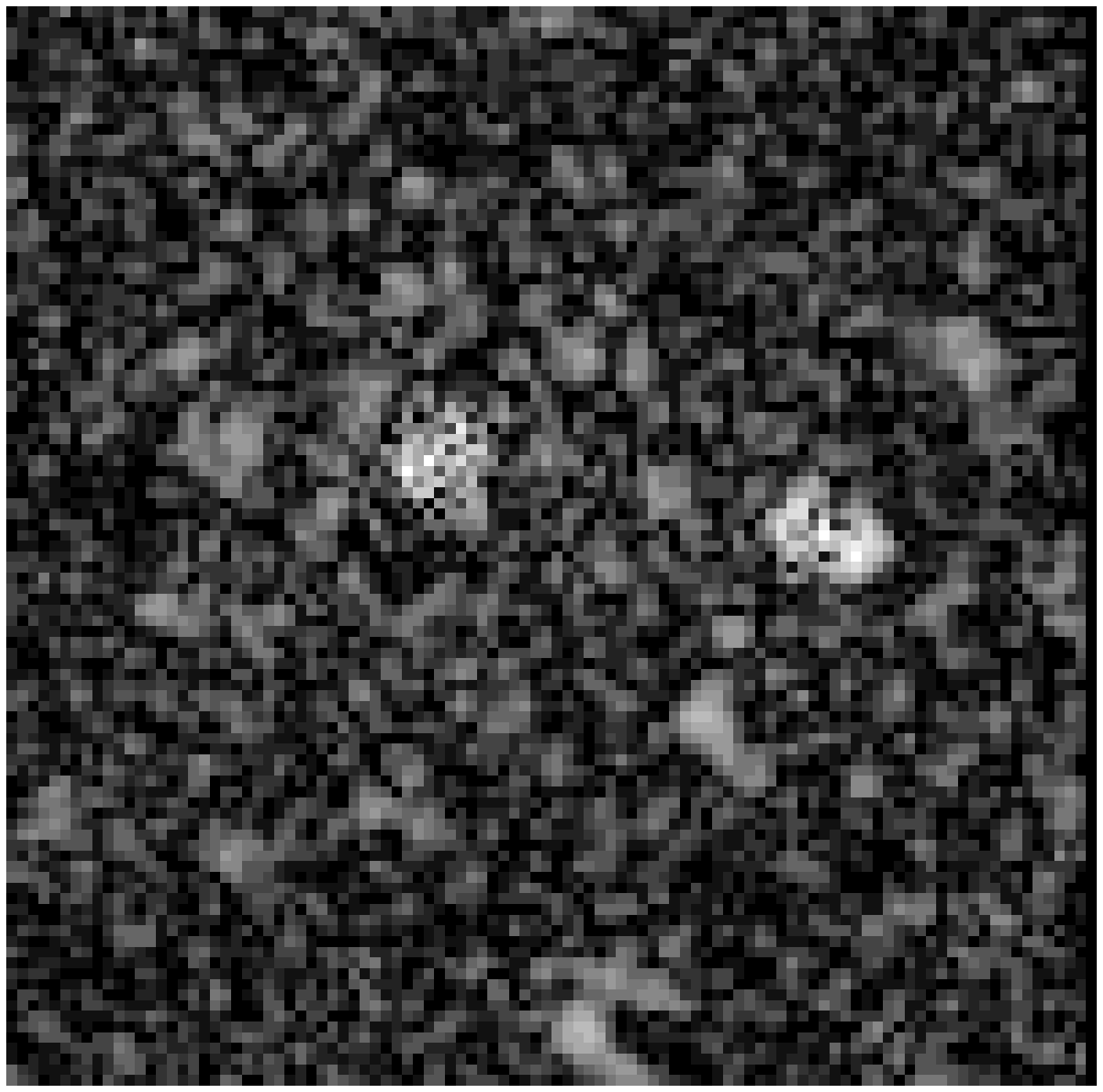}\hspace{0.06cm}
\includegraphics[angle=0,scale=.28]{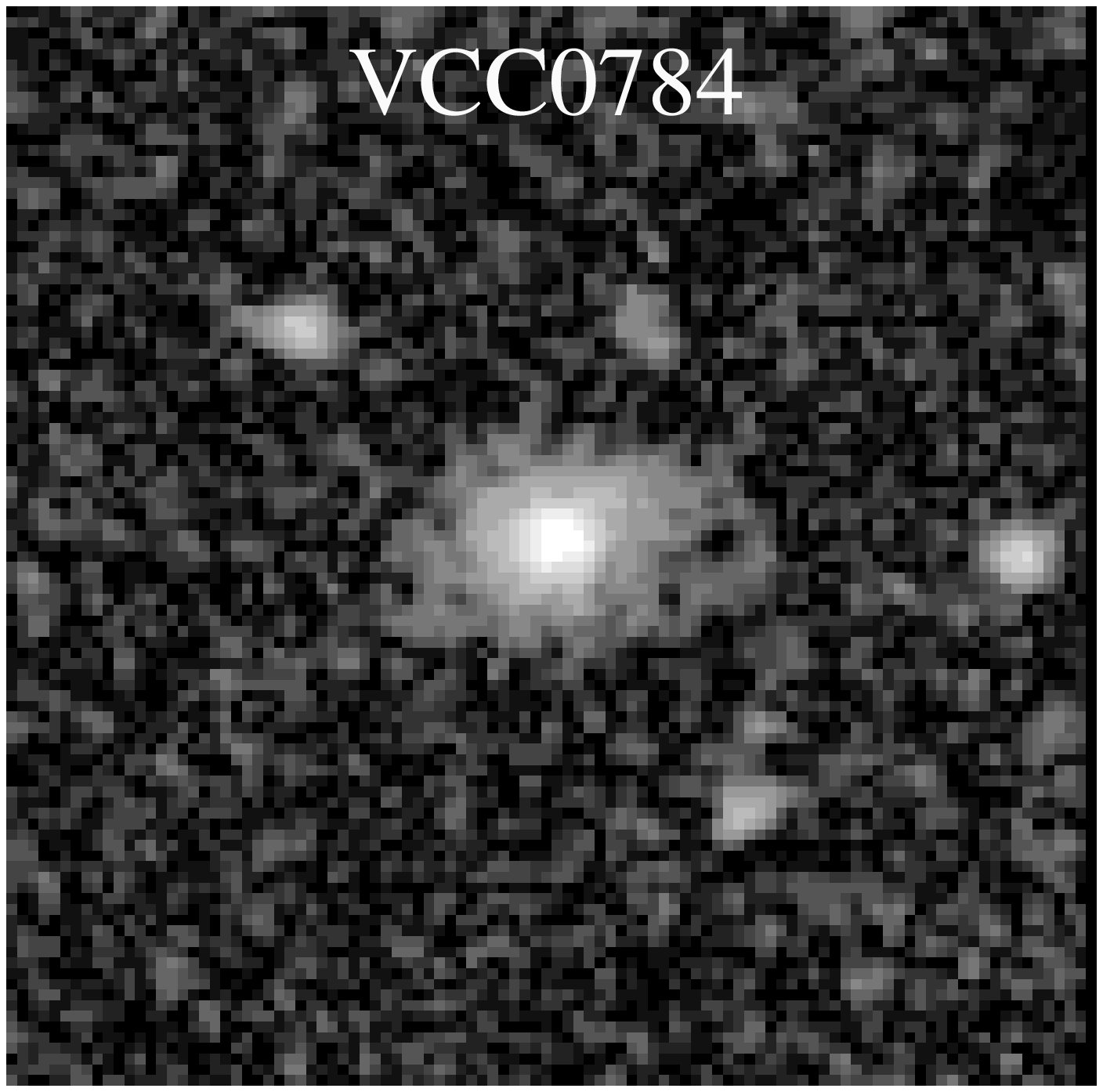}
\includegraphics[angle=0,scale=.28]{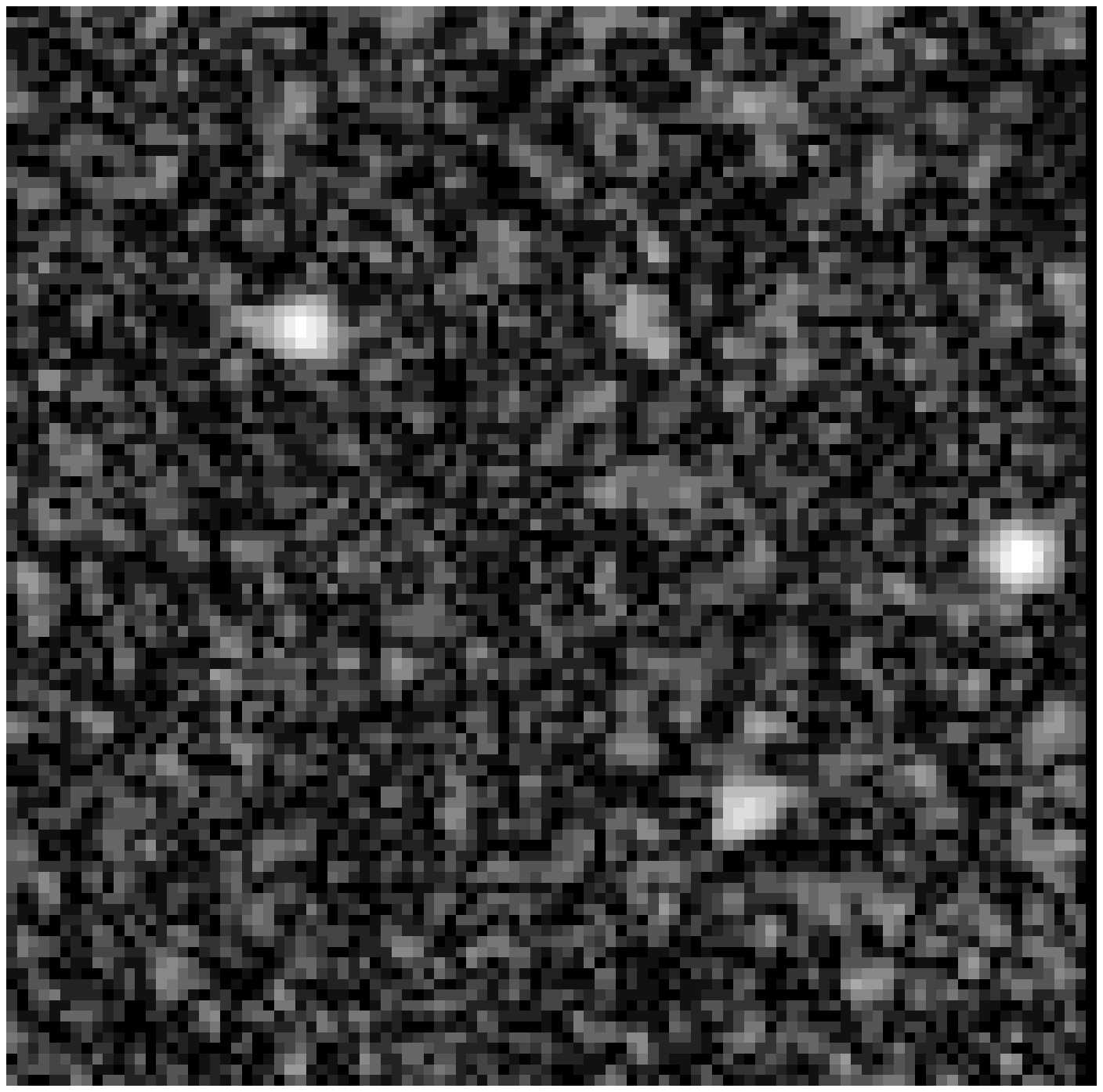}\vspace{0.025cm}
\includegraphics[angle=0,scale=.28]{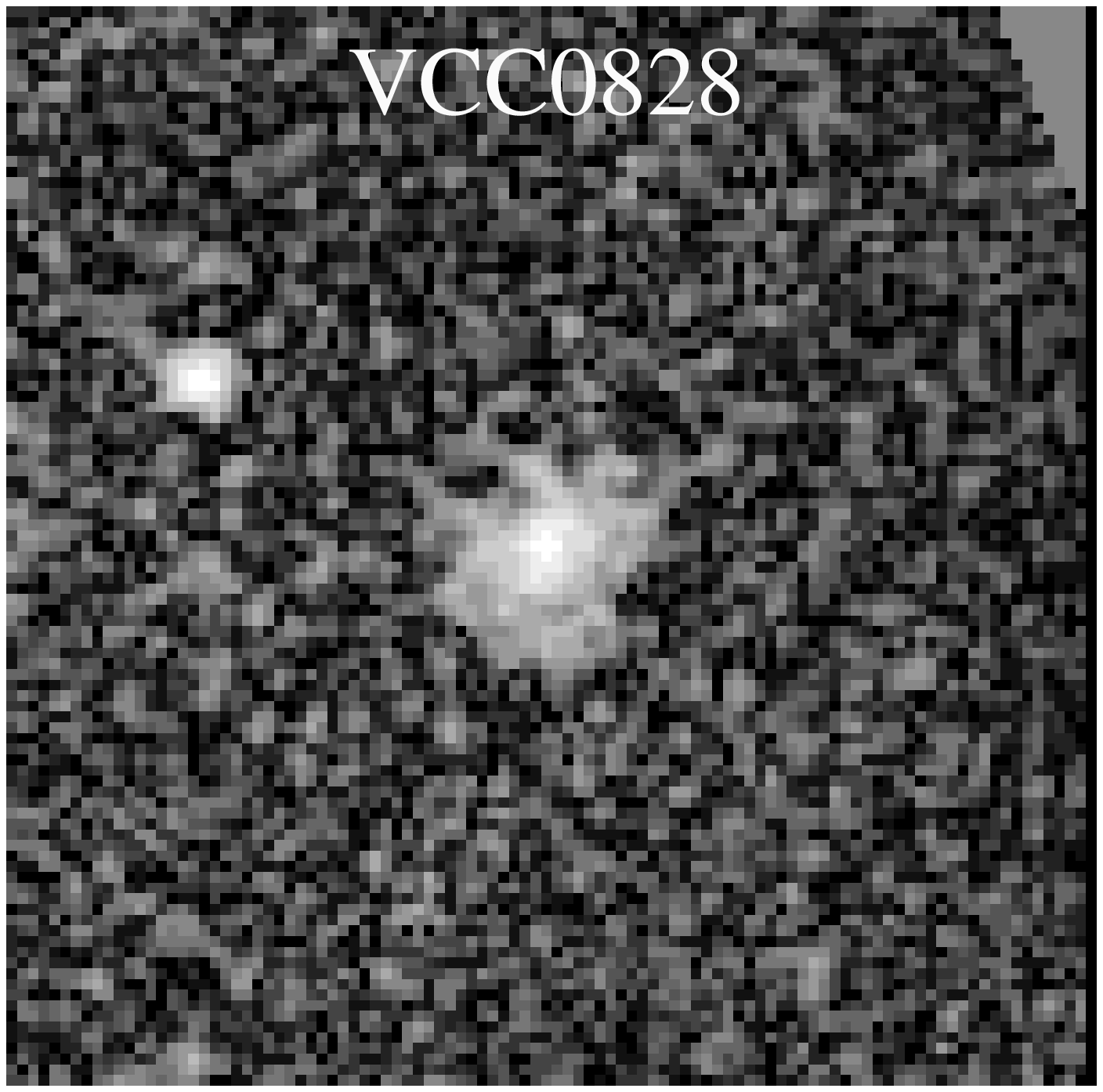}
\includegraphics[angle=0,scale=.28]{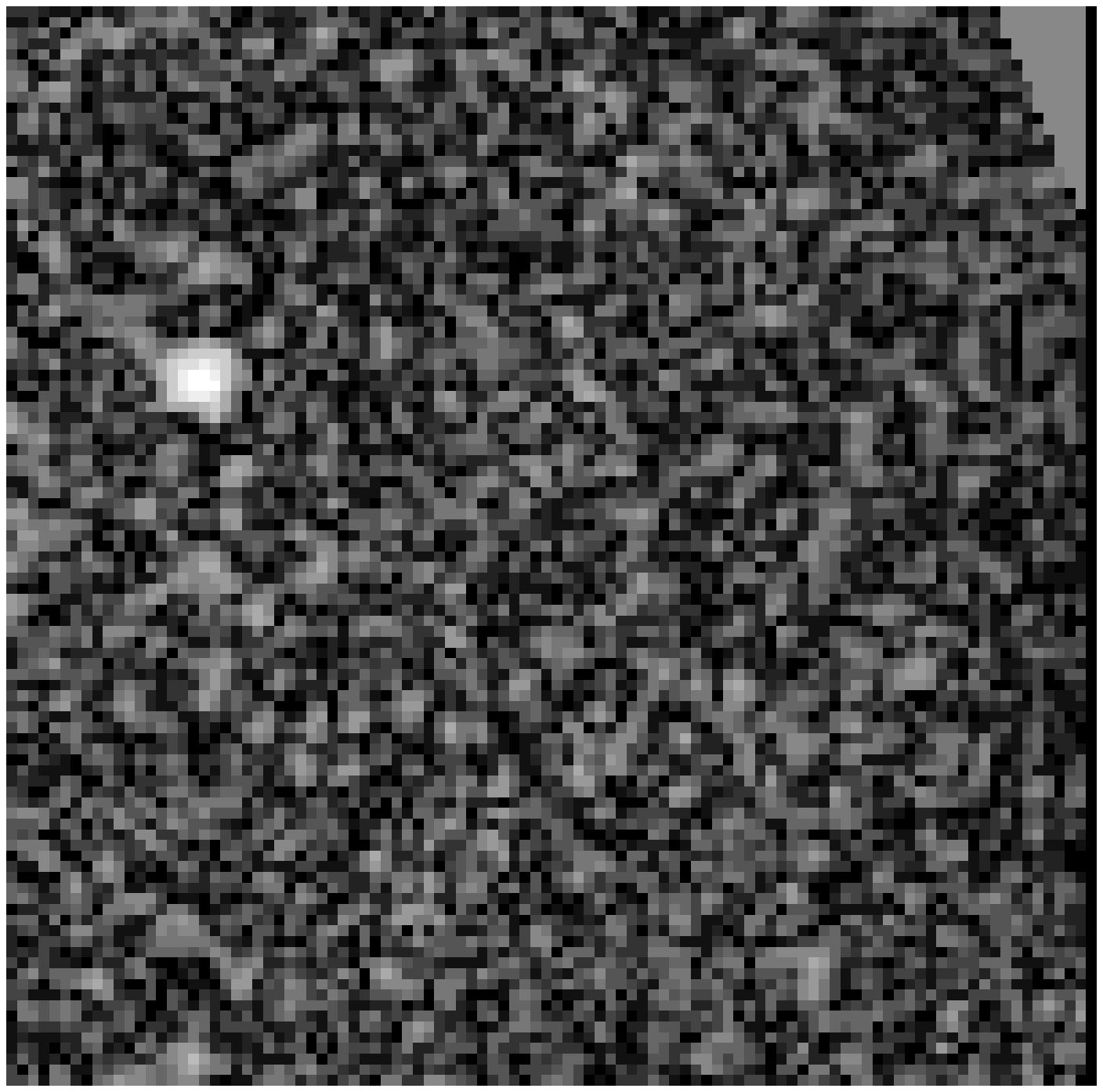}\hspace{0.06cm}
\includegraphics[angle=0,scale=.28]{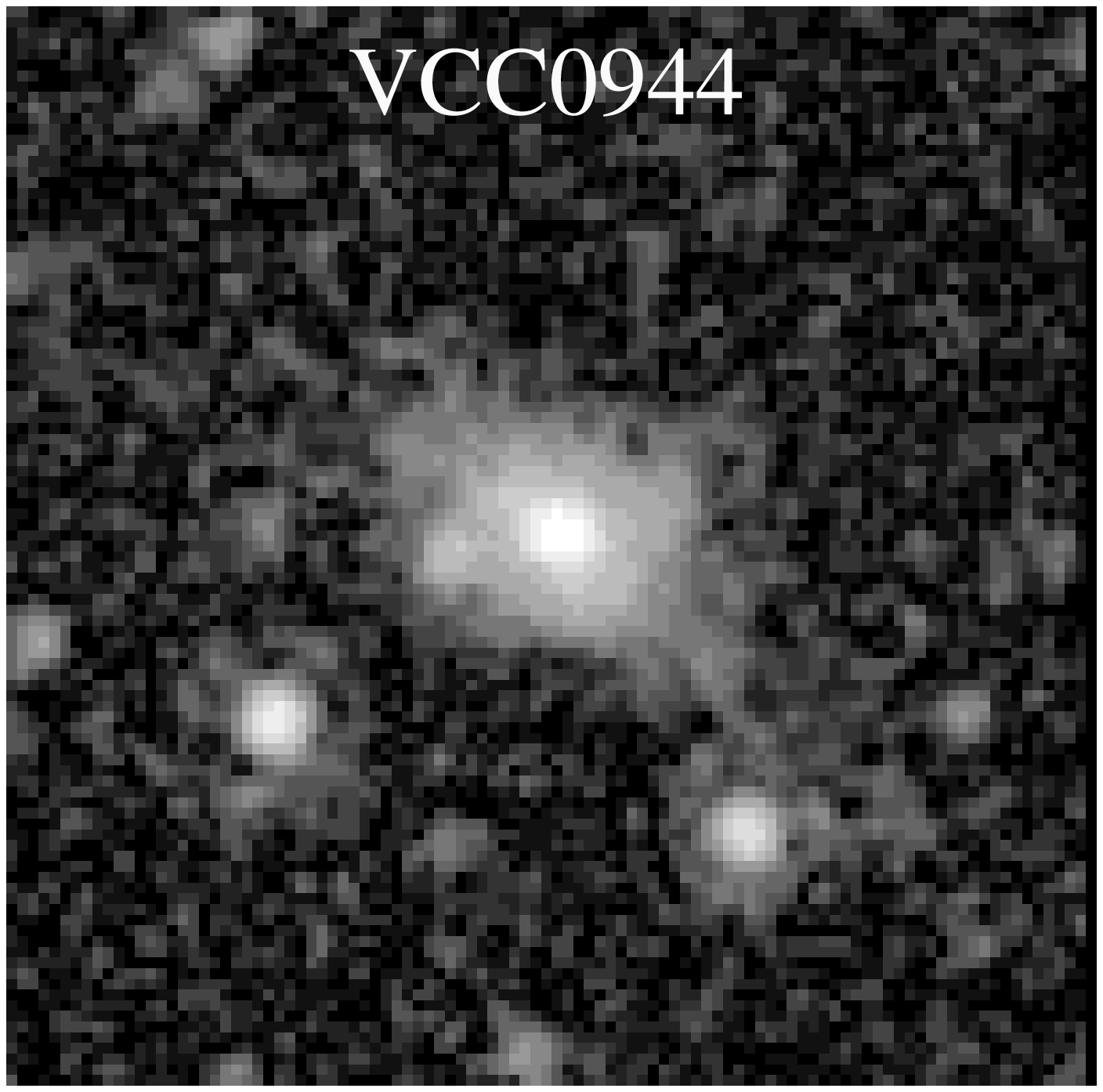}
\includegraphics[angle=0,scale=.28]{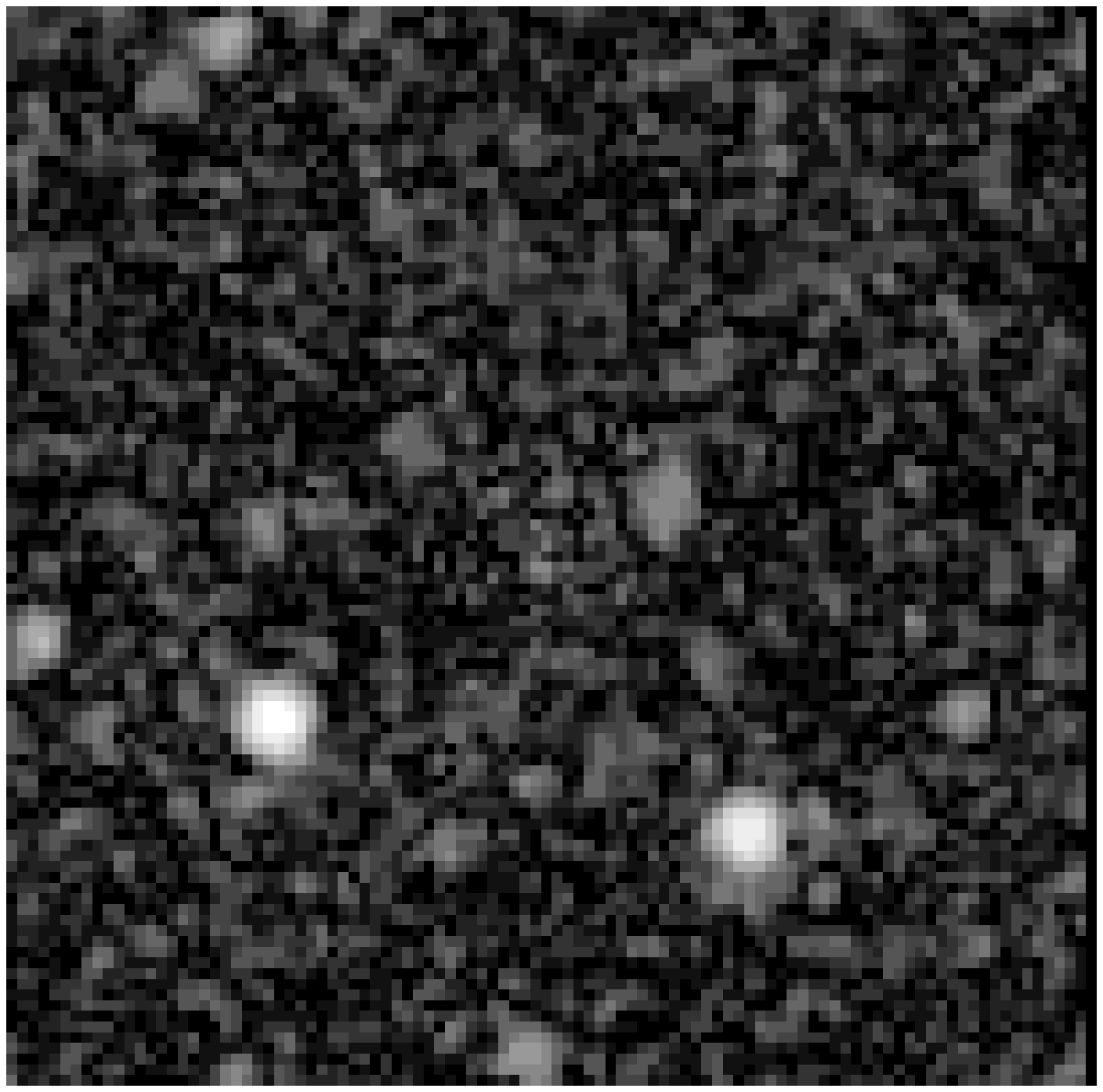}\vspace{0.025cm}
\includegraphics[angle=0,scale=.28]{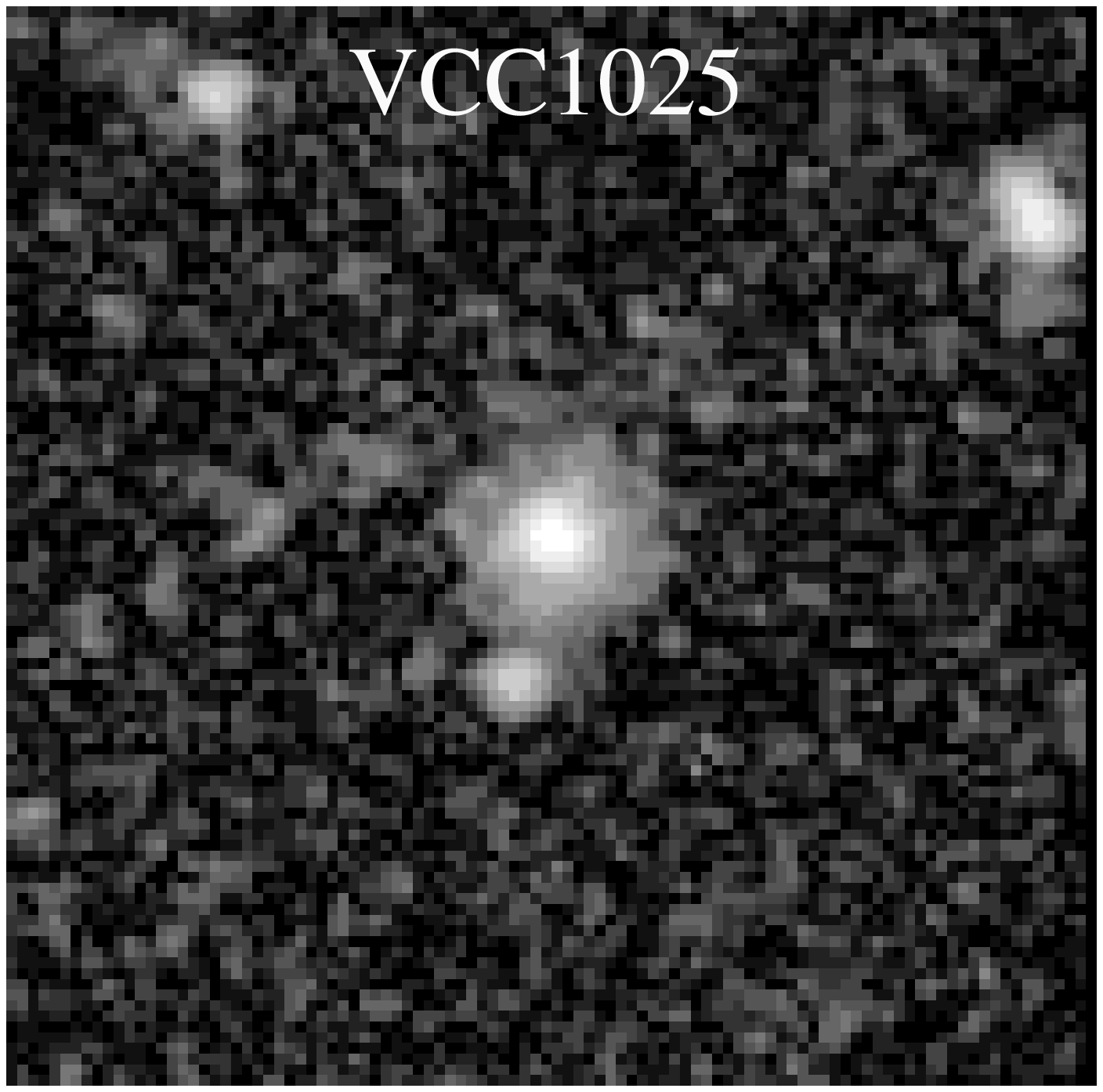}
\includegraphics[angle=0,scale=.28]{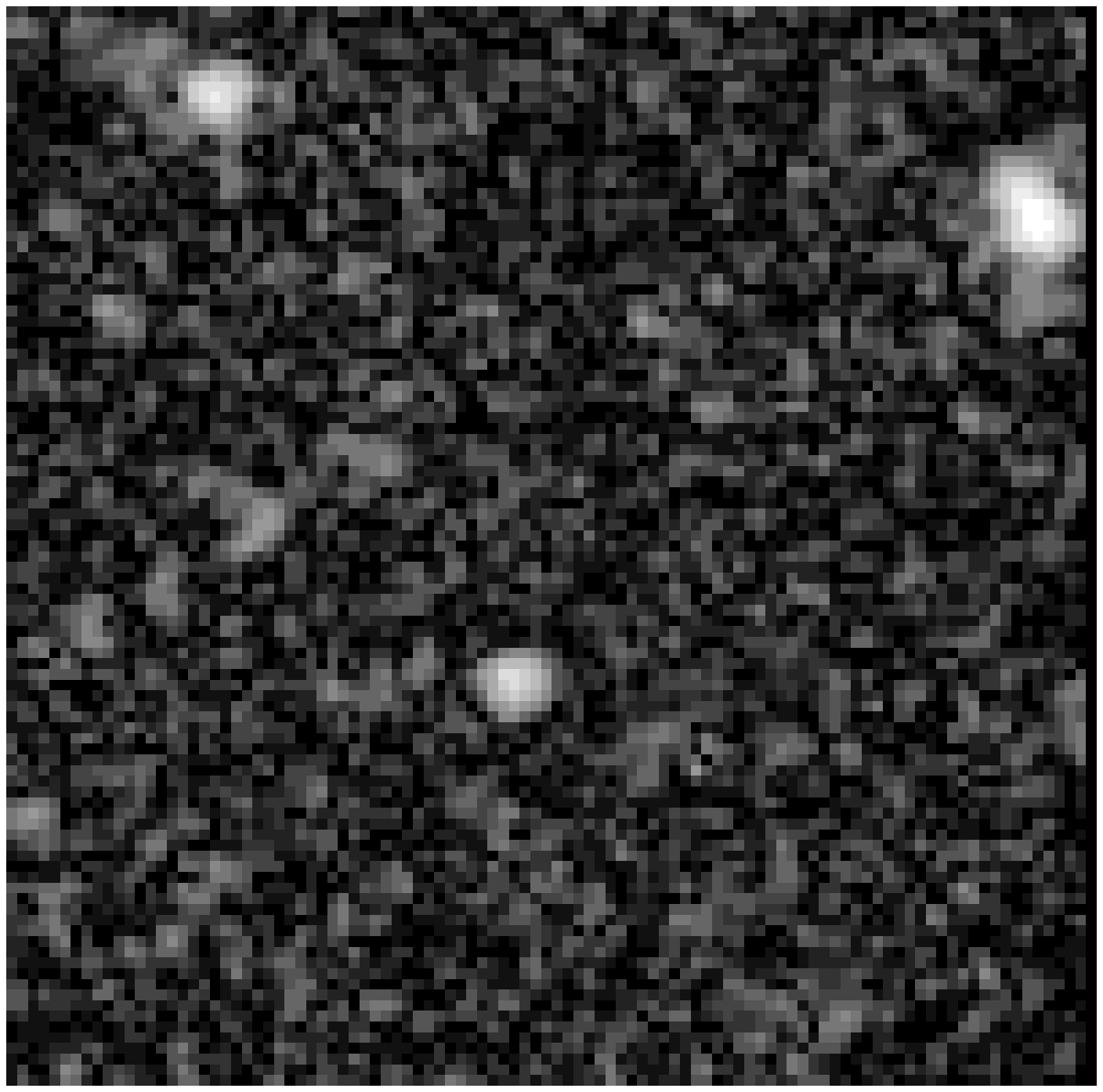}\hspace{0.06cm}
\includegraphics[angle=0,scale=.28]{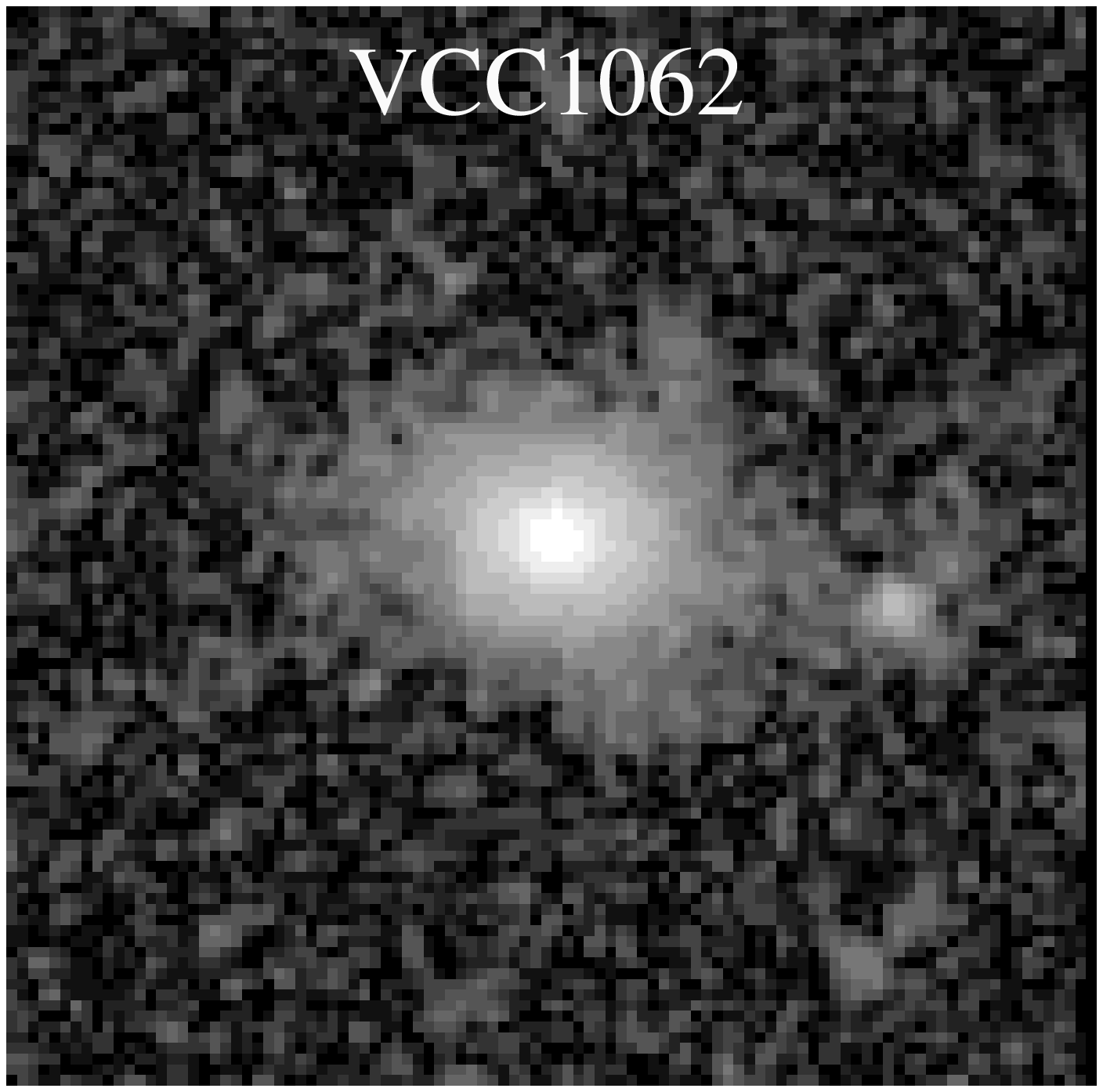}
\includegraphics[angle=0,scale=.28]{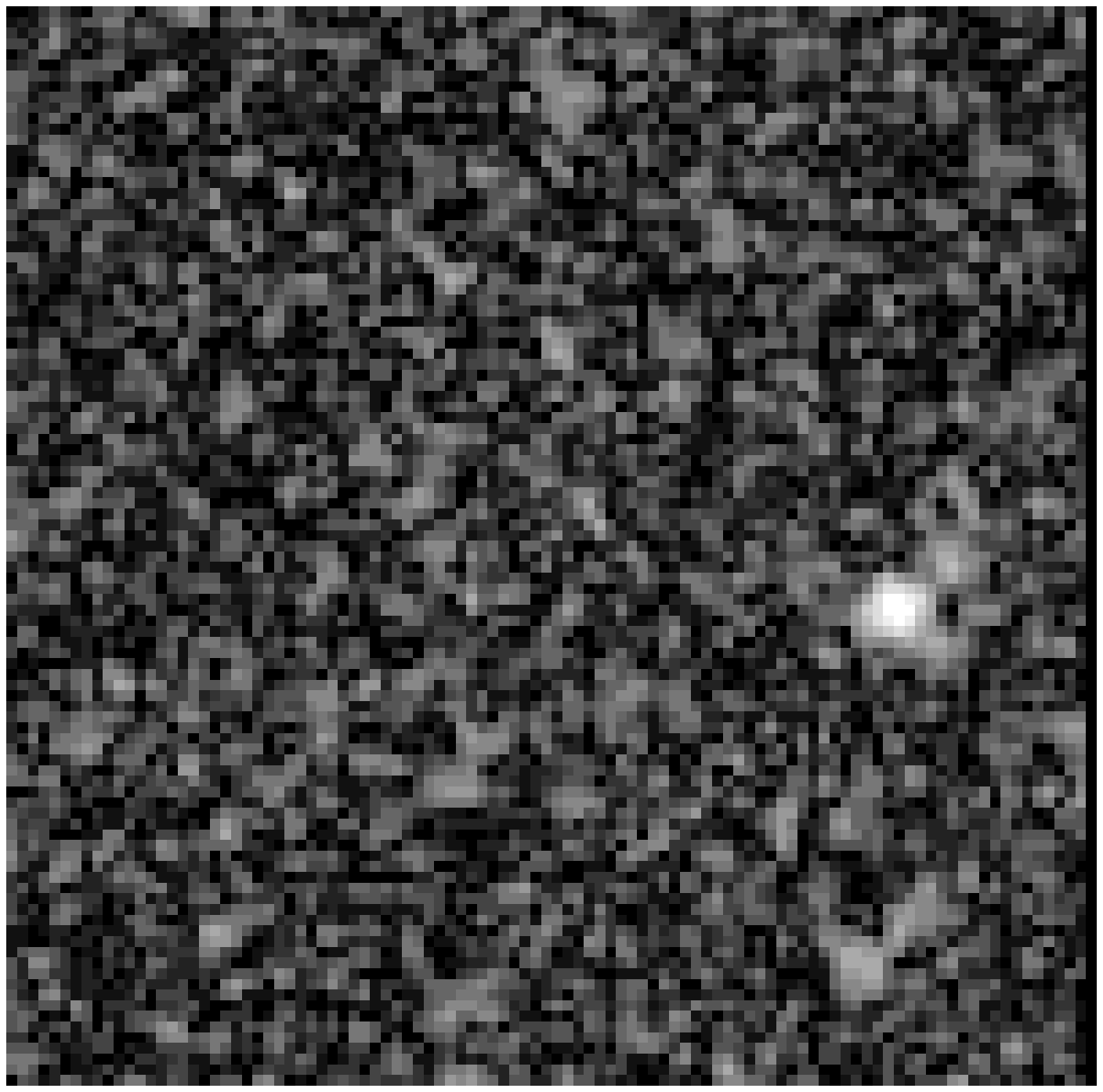}
\vspace{1cm}
\caption{Observed 24\,$\mu$m images of the 37 detected VCC early-type galaxies 
without optical signatures for host dust. To the right of the image we show 
the residual after subtracting the galaxy model (see section \ref{sec:imfit}). 
The images are 2 arcminutes on the side and North is up with East to the 
left.\label{images_galfit} }
\end{figure*}
\addtocounter{figure}{-1}
\begin{figure*}
\centering
\includegraphics[angle=0,scale=.28]{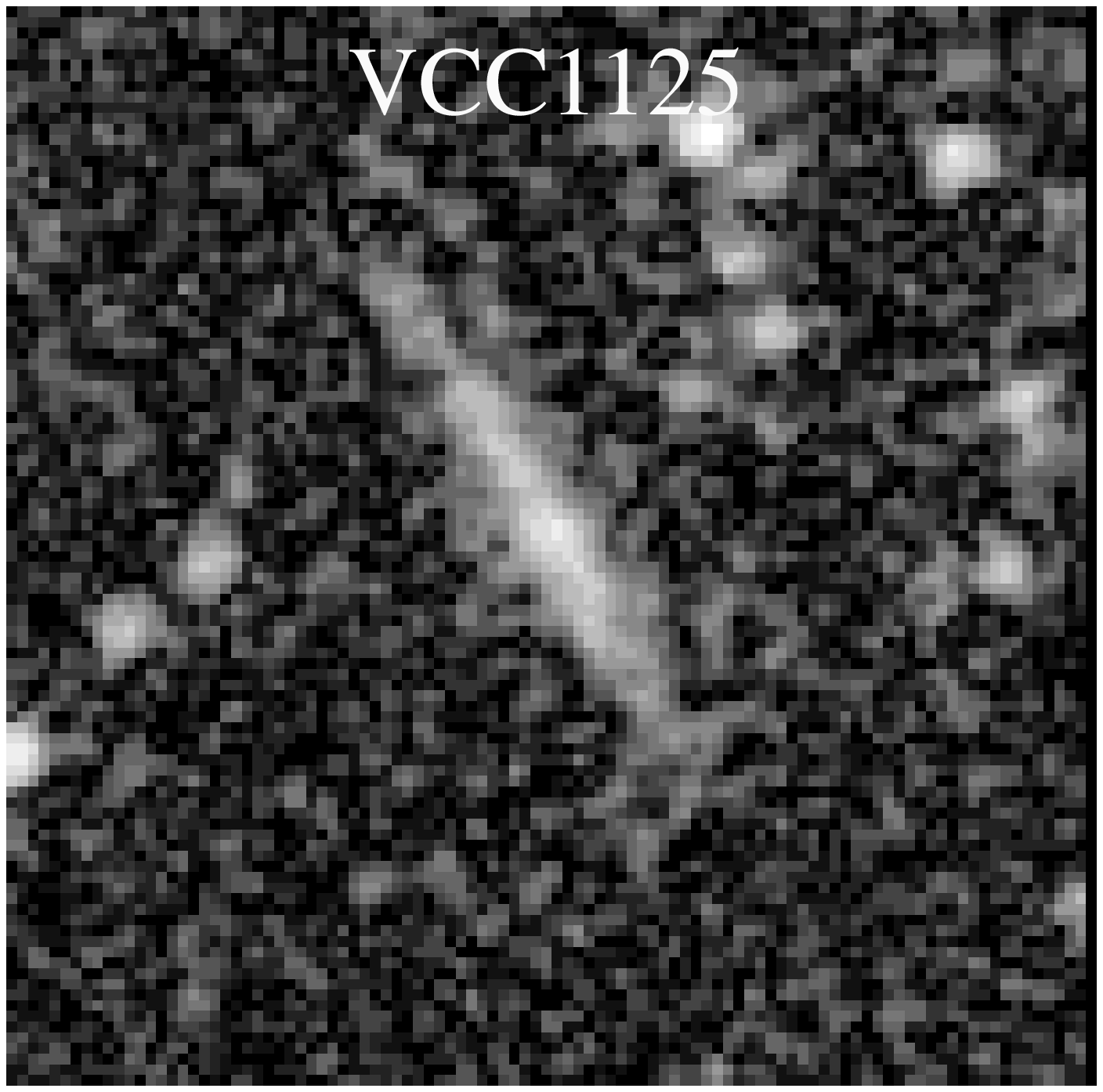}
\includegraphics[angle=0,scale=.28]{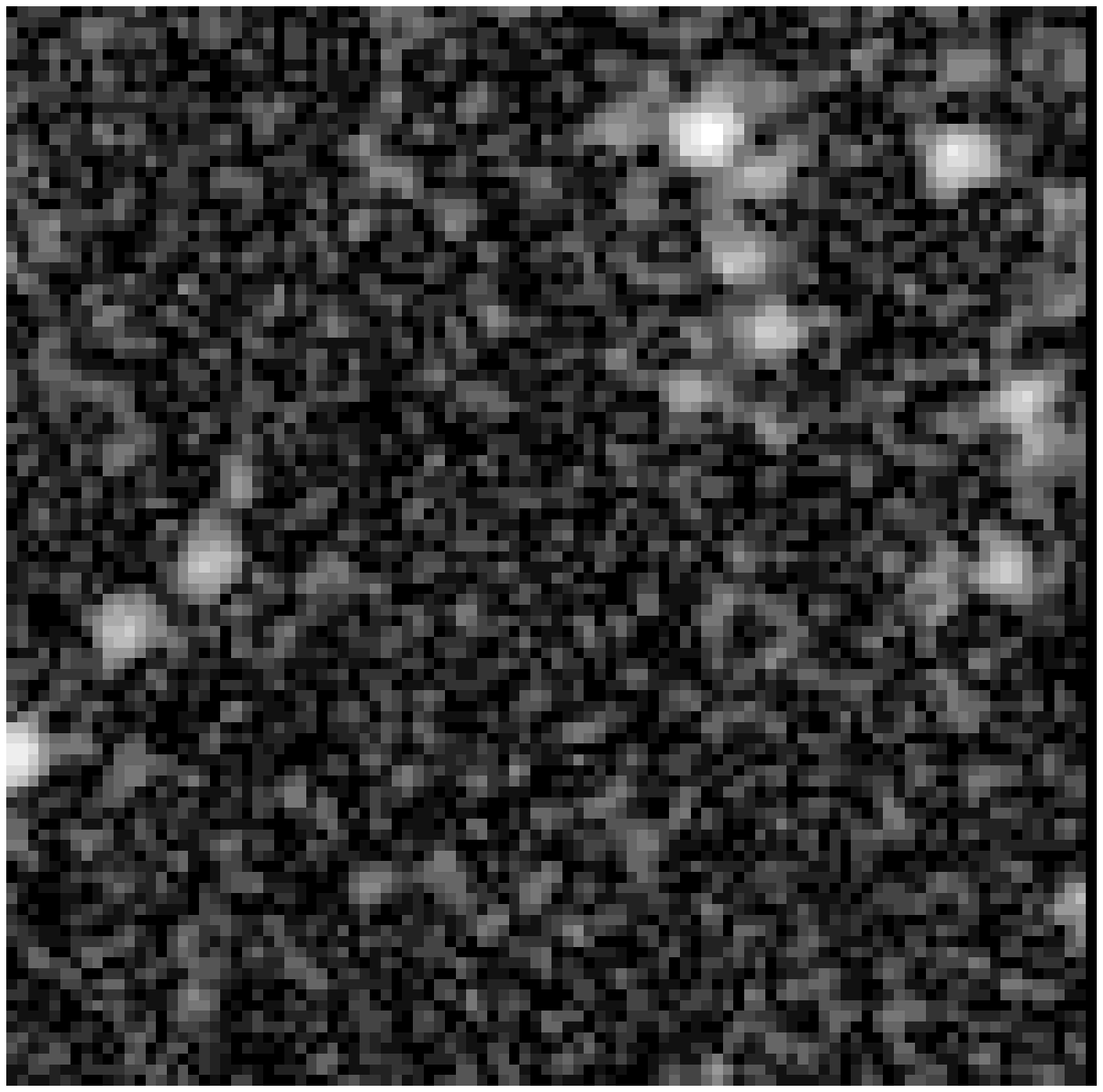}\hspace{0.06cm}
\includegraphics[angle=0,scale=.28]{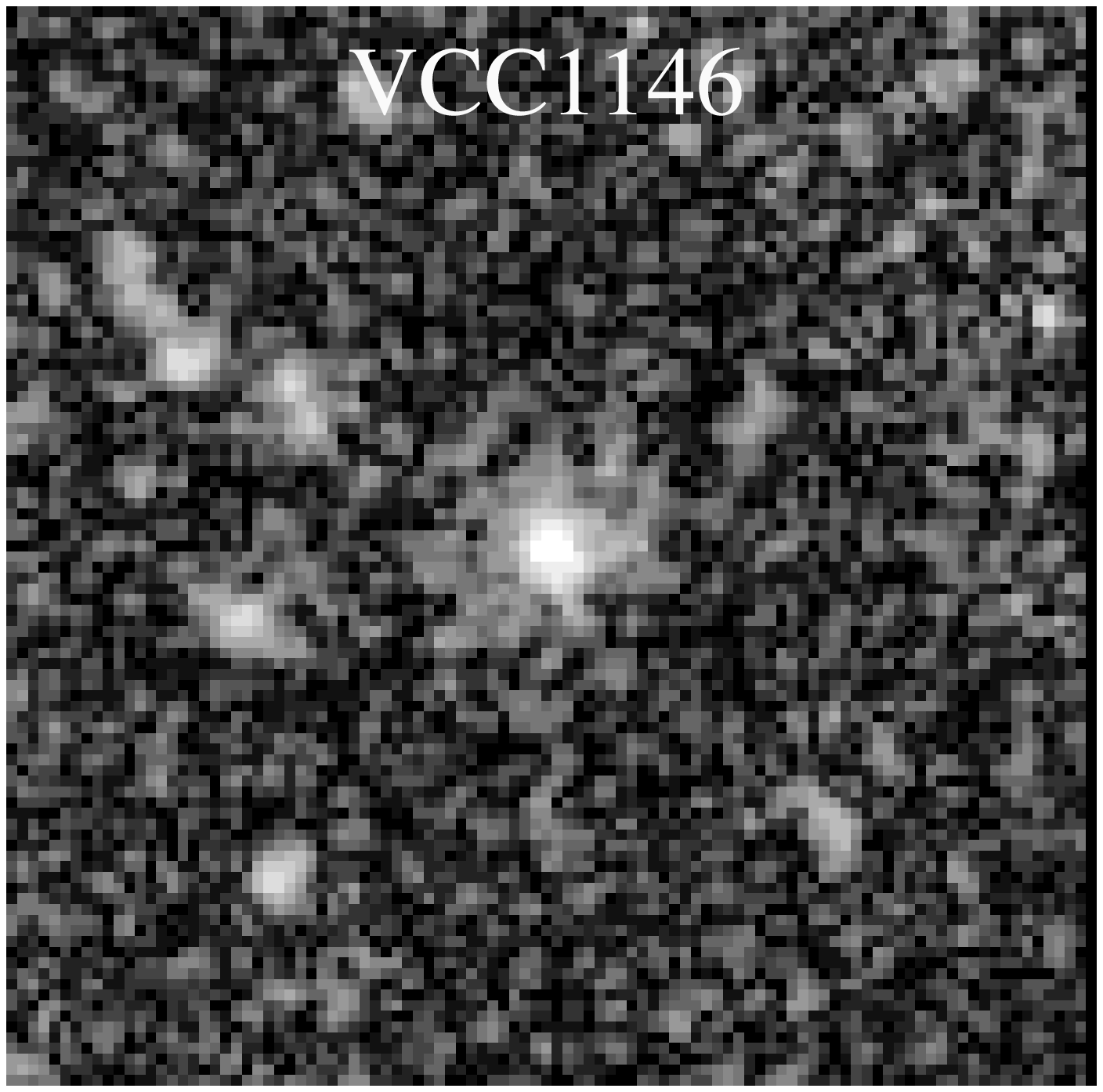}
\includegraphics[angle=0,scale=.28]{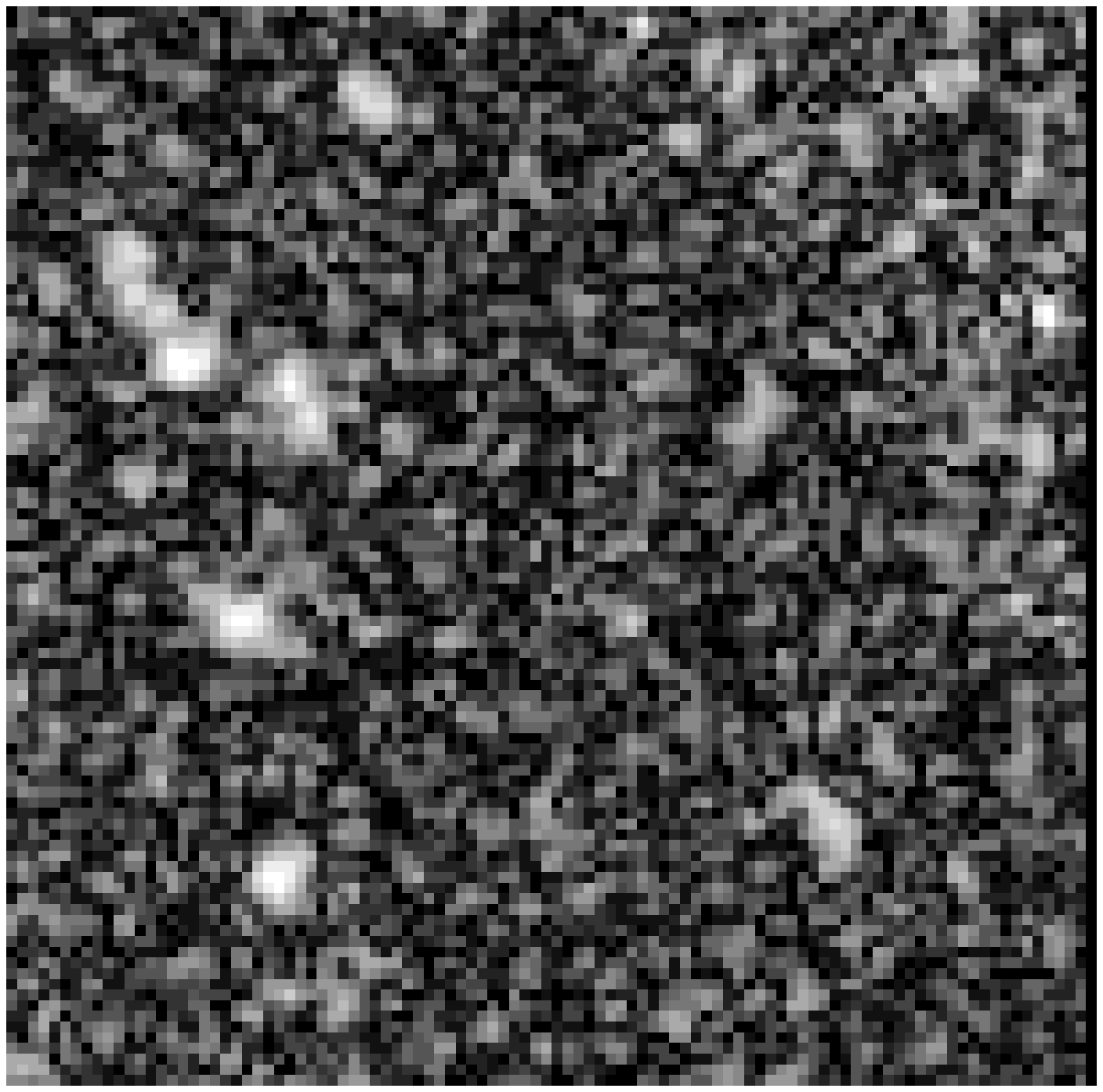}\vspace{0.025cm}
\includegraphics[angle=0,scale=.28]{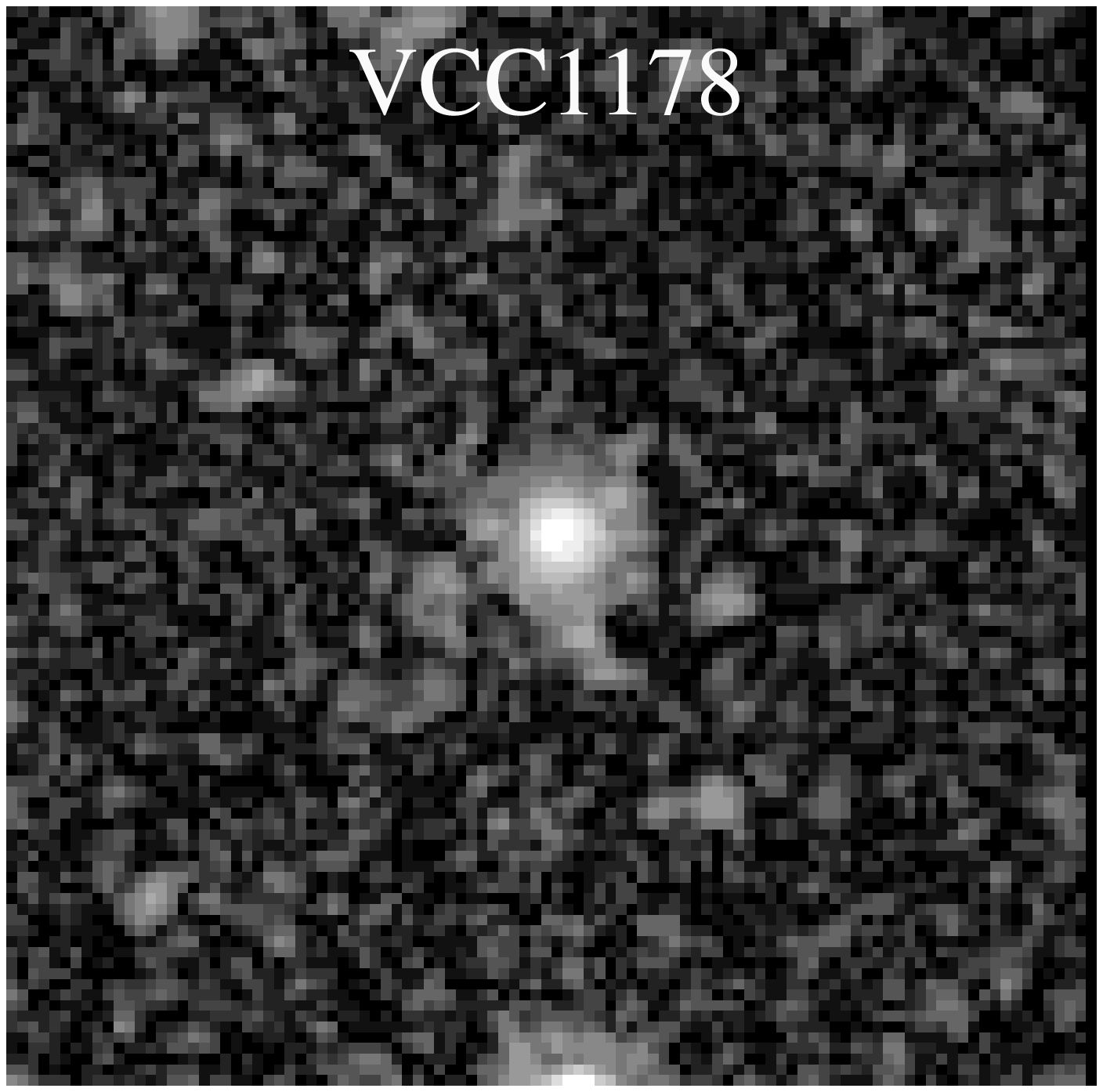}
\includegraphics[angle=0,scale=.28]{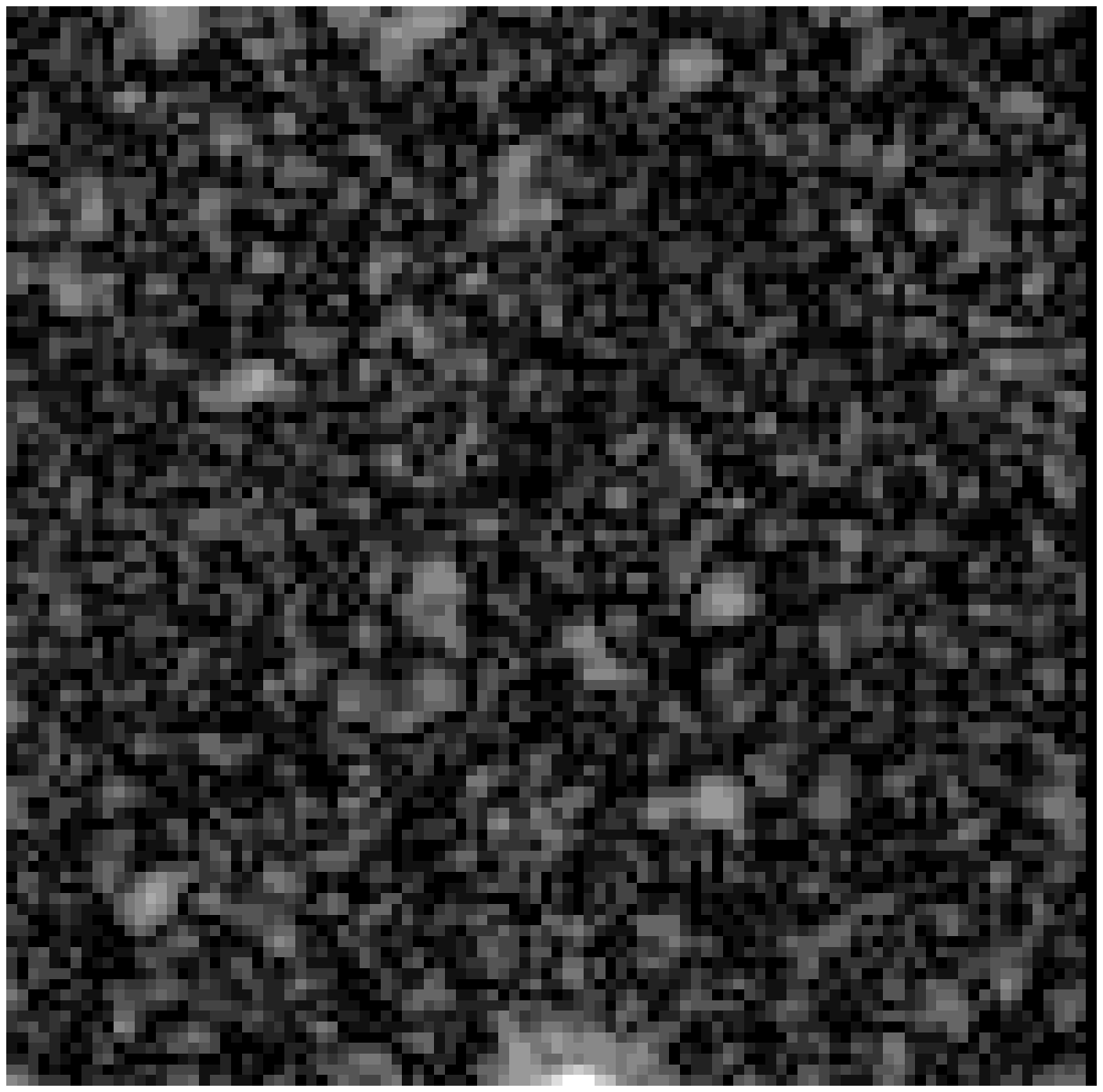}\hspace{0.06cm}
\includegraphics[angle=0,scale=.28]{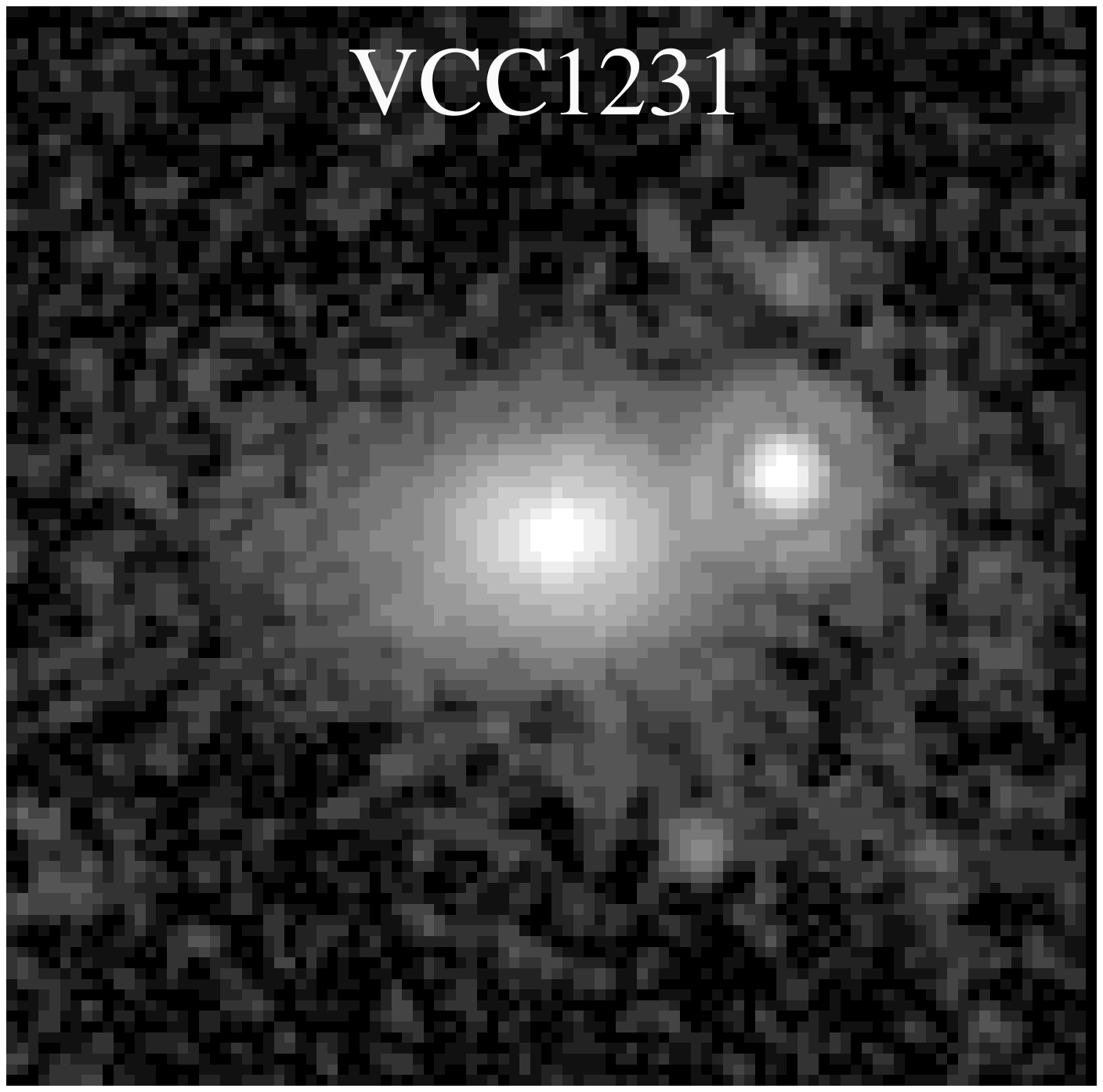}
\includegraphics[angle=0,scale=.28]{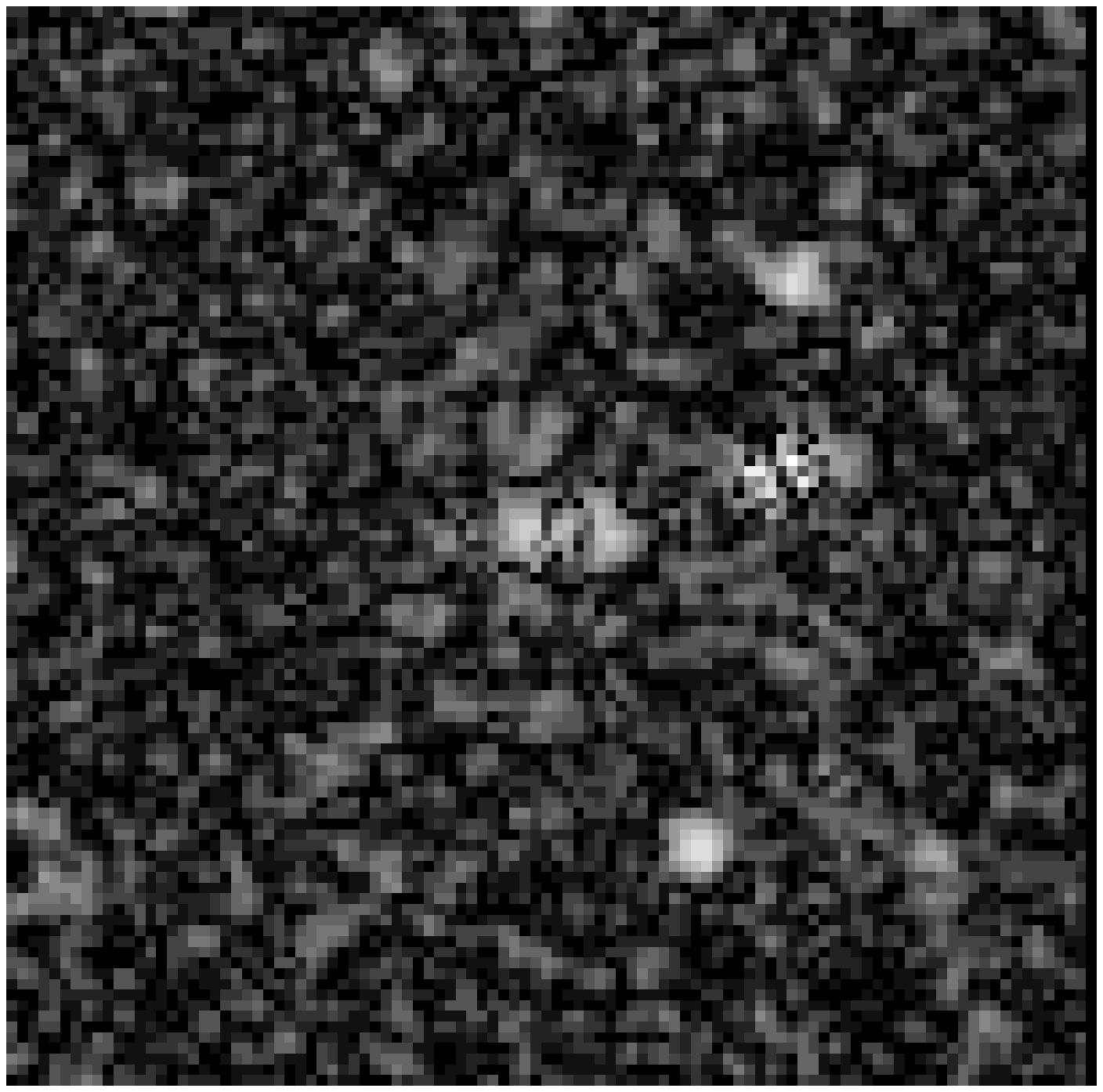}\vspace{0.025cm}
\includegraphics[angle=0,scale=.28]{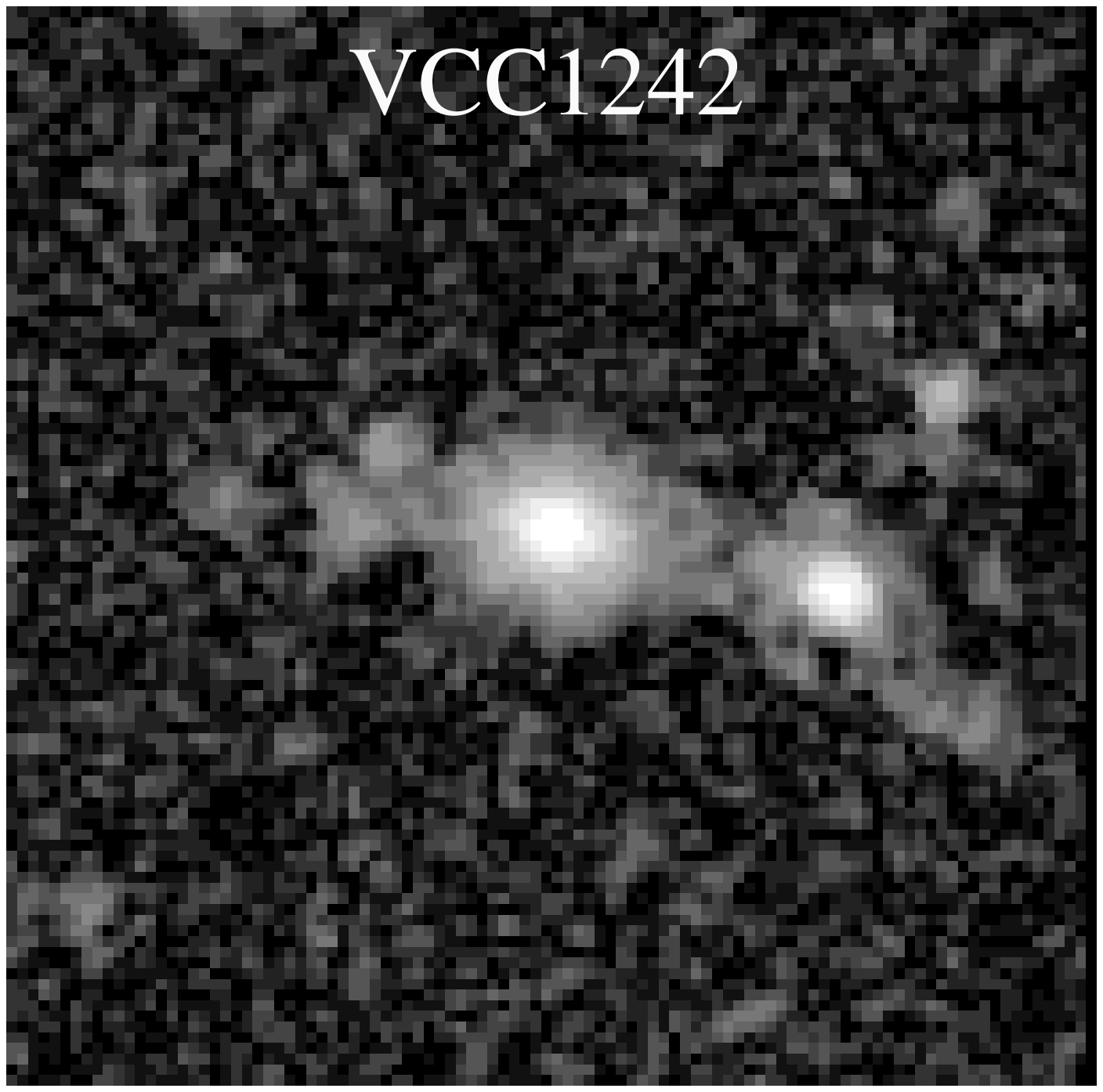}
\includegraphics[angle=0,scale=.28]{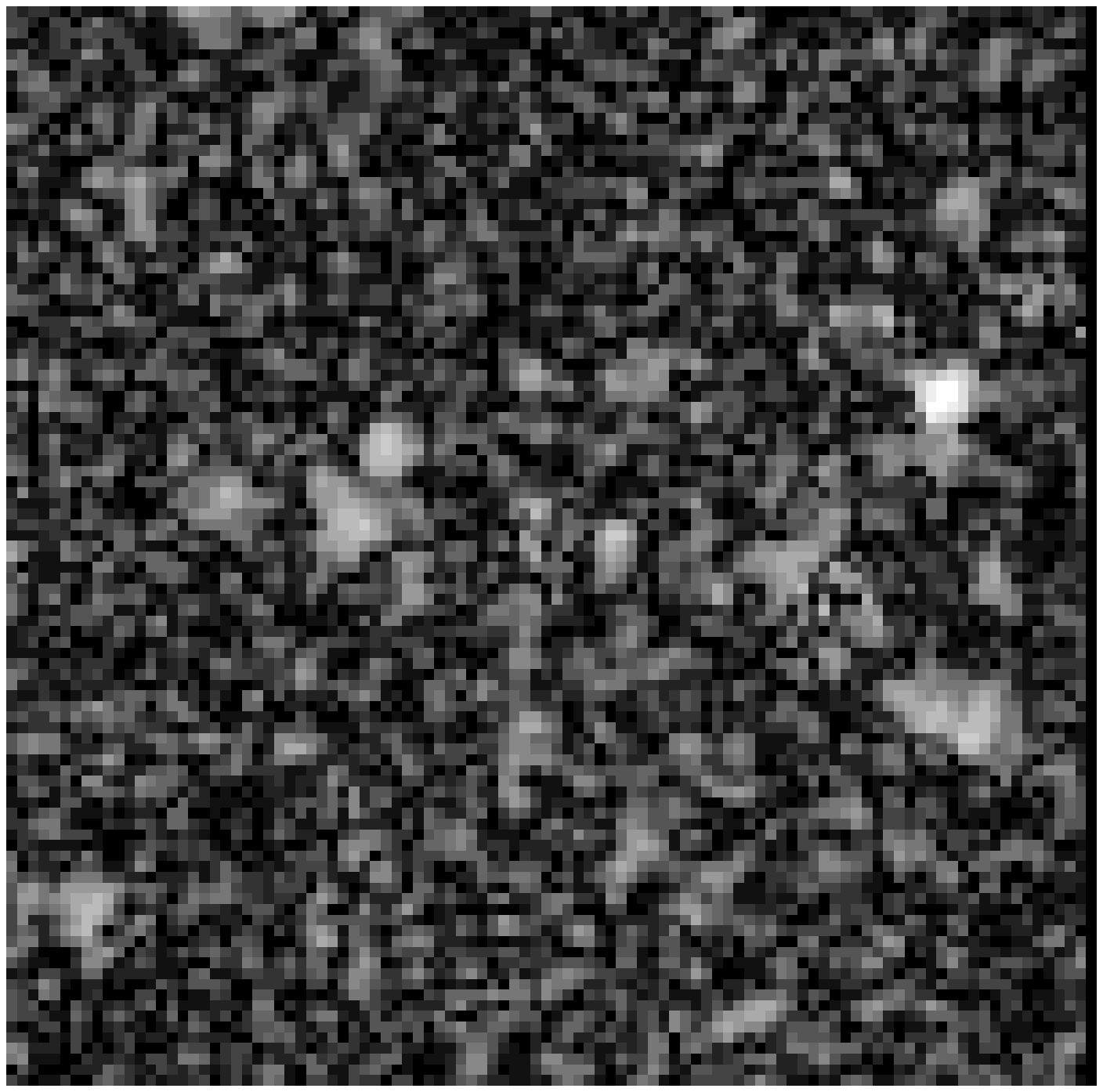}\hspace{0.06cm}
\includegraphics[angle=0,scale=.28]{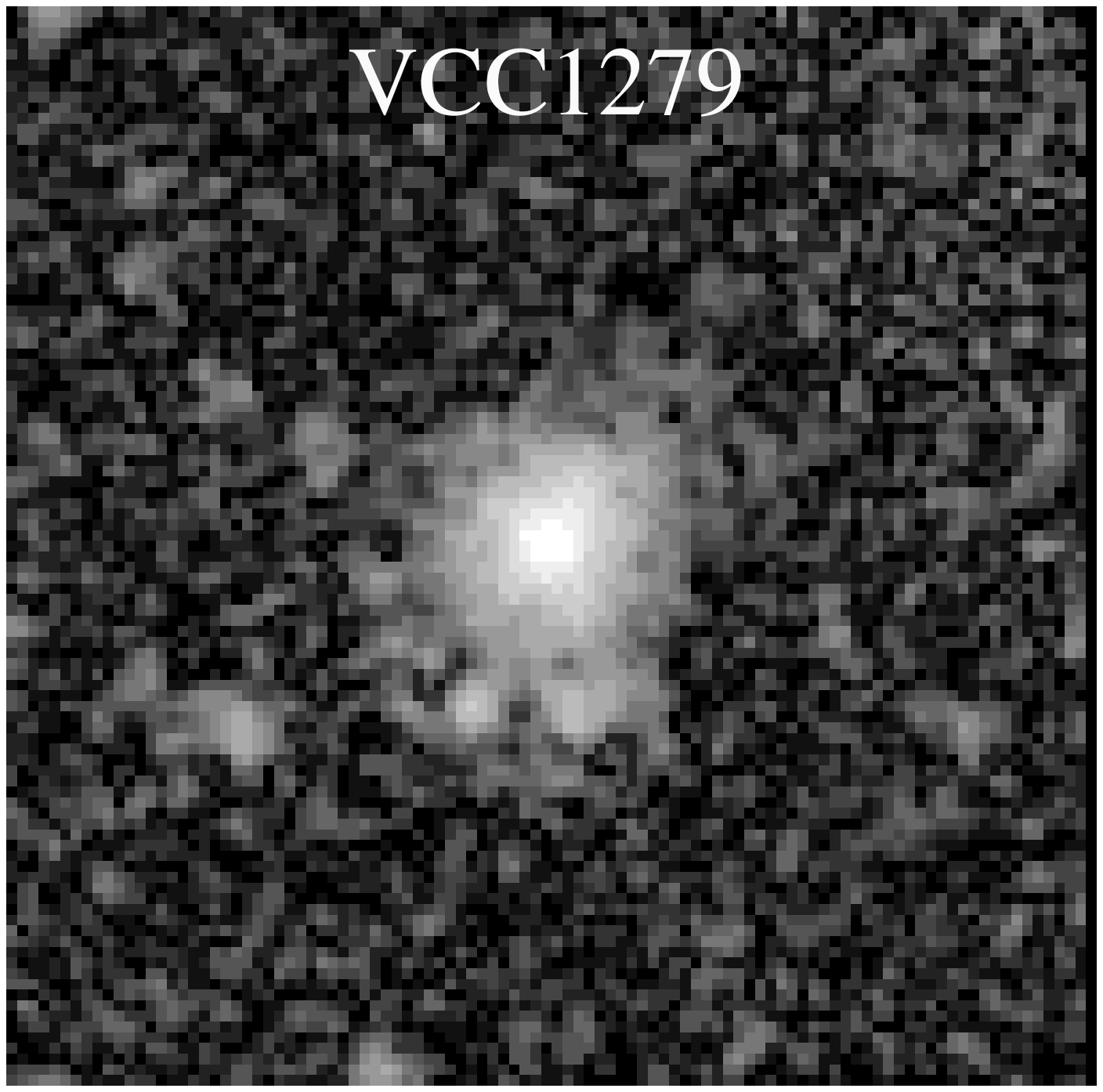}
\includegraphics[angle=0,scale=.28]{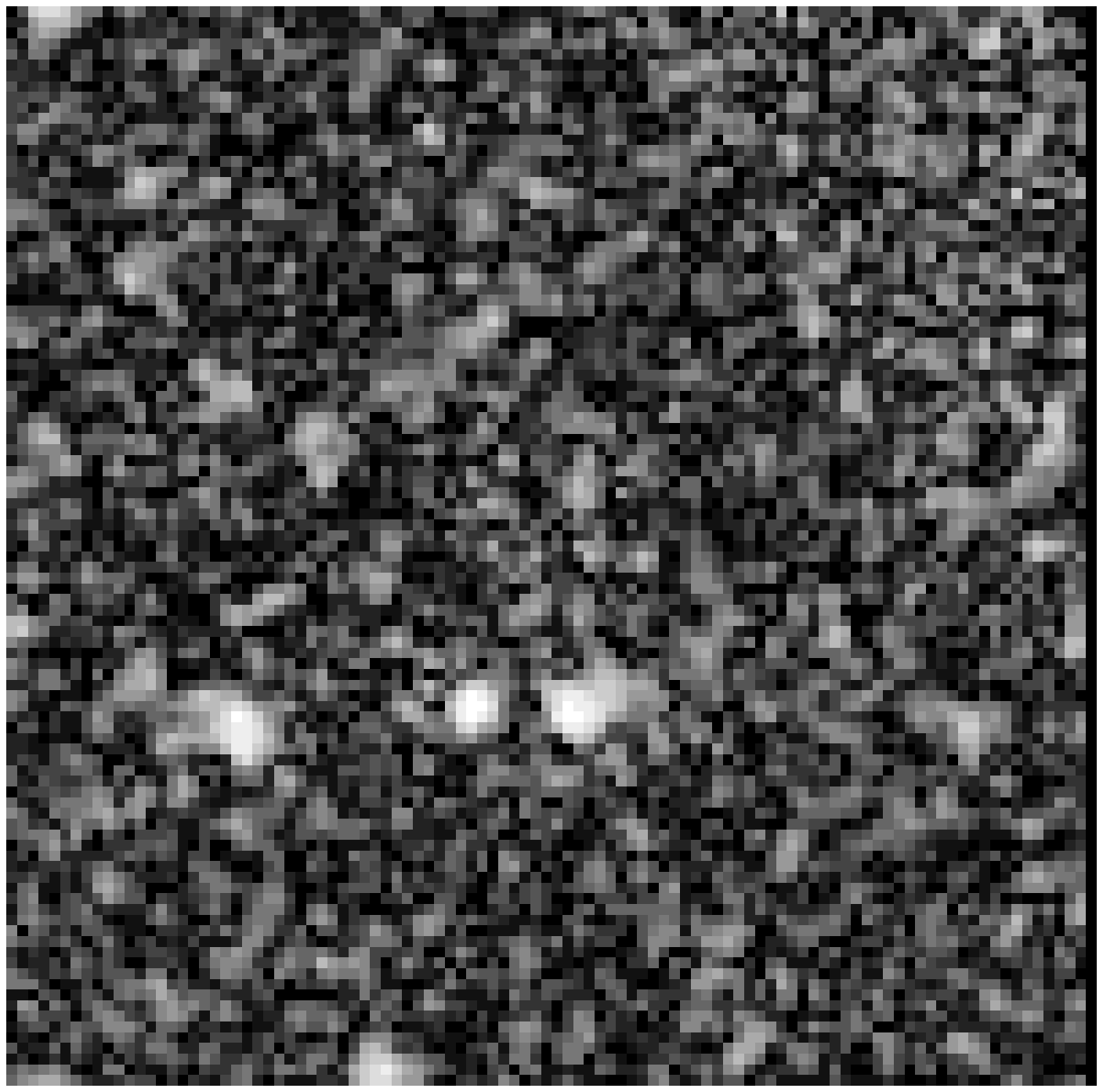}\vspace{0.025cm}
\includegraphics[angle=0,scale=.28]{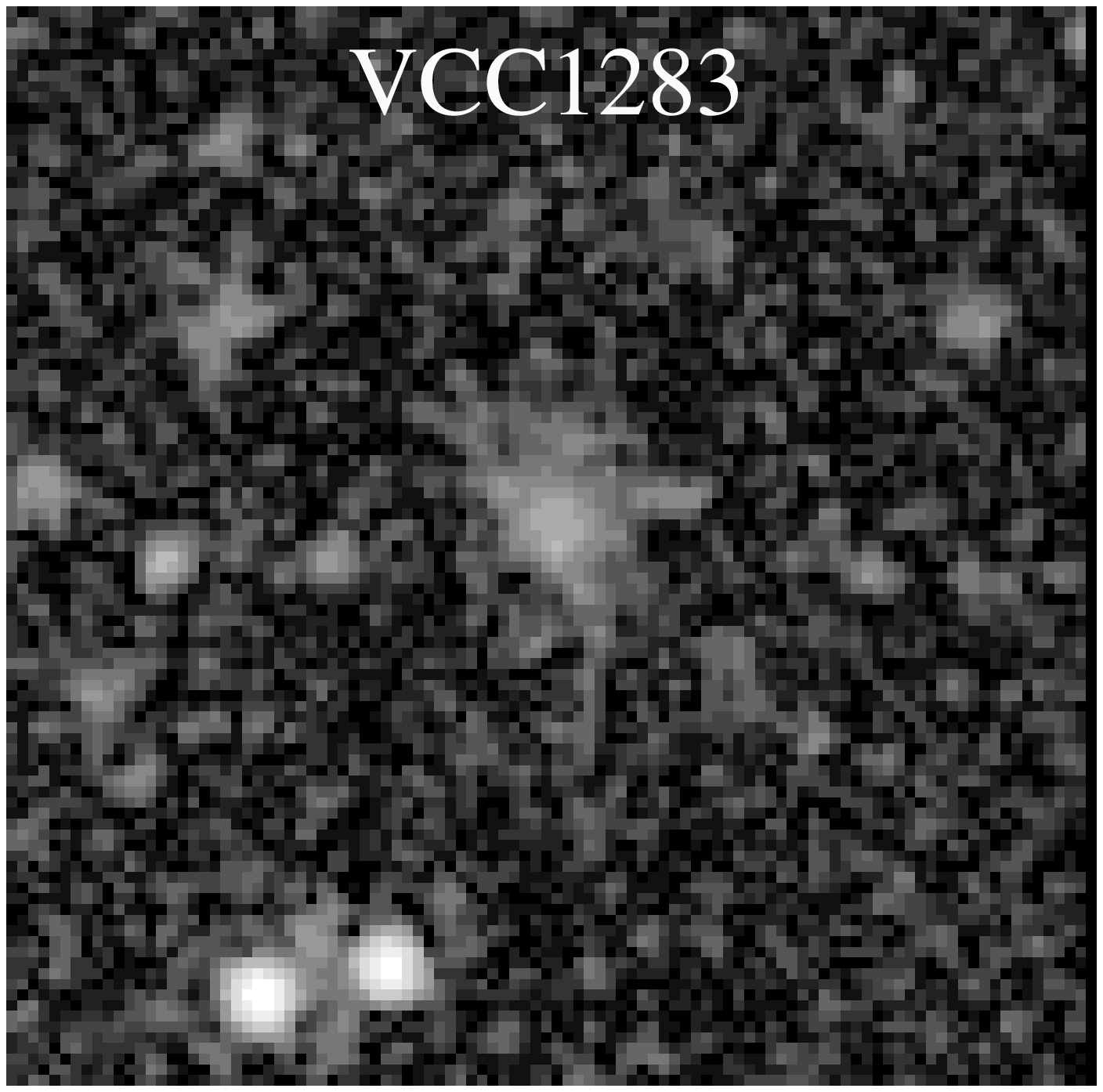}
\includegraphics[angle=0,scale=.28]{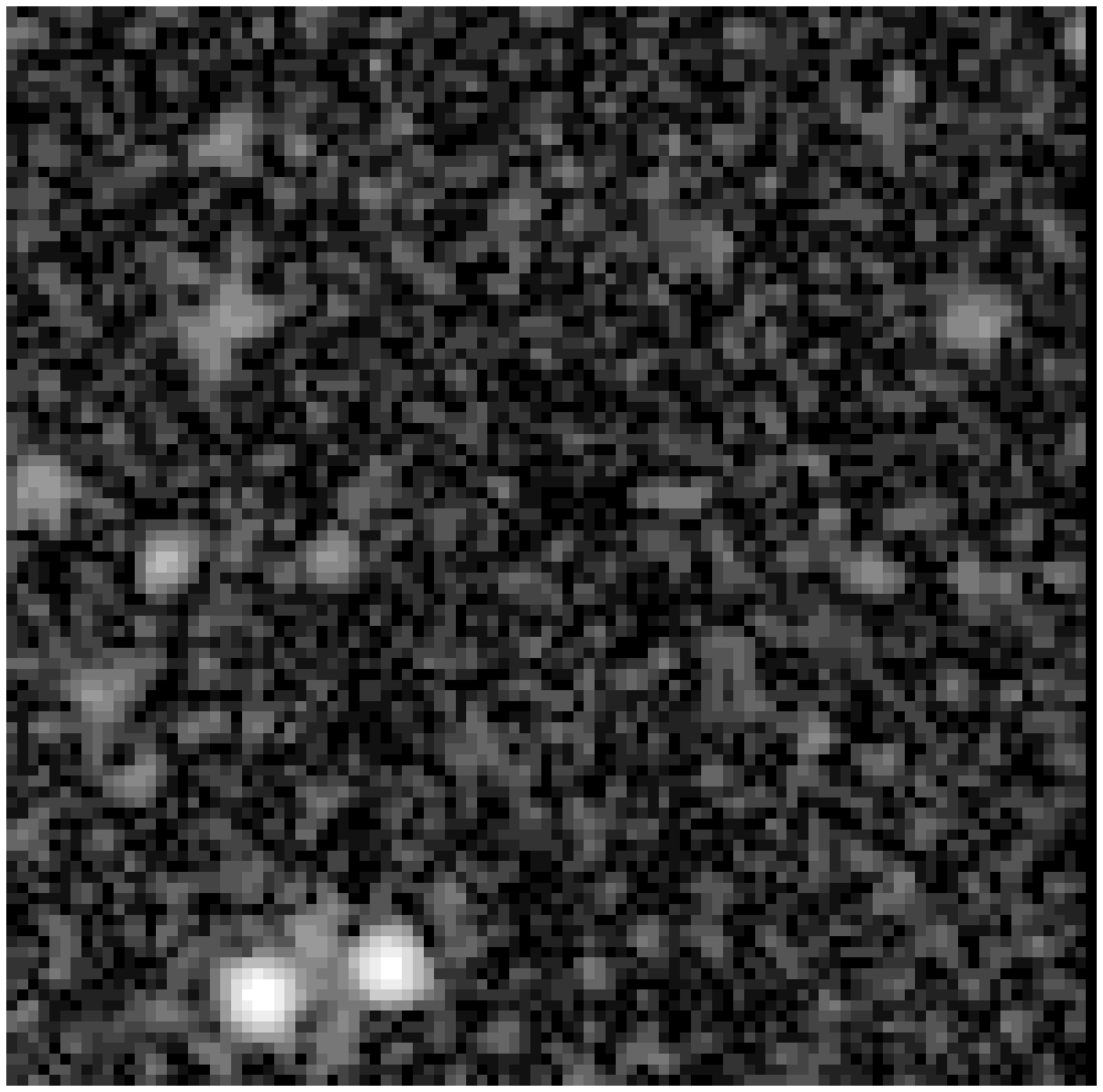}\hspace{0.06cm}
\includegraphics[angle=0,scale=.28]{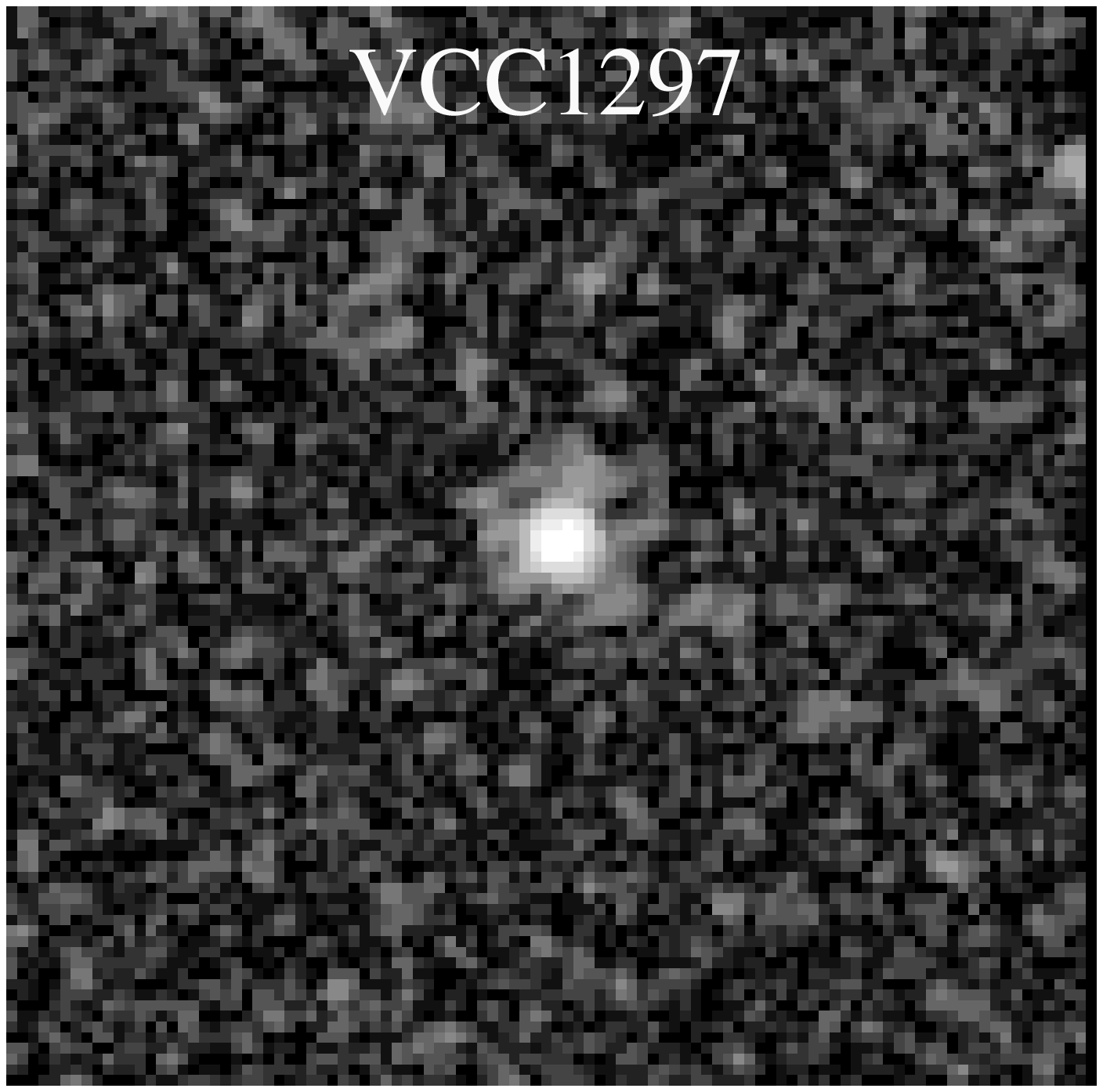}
\includegraphics[angle=0,scale=.28]{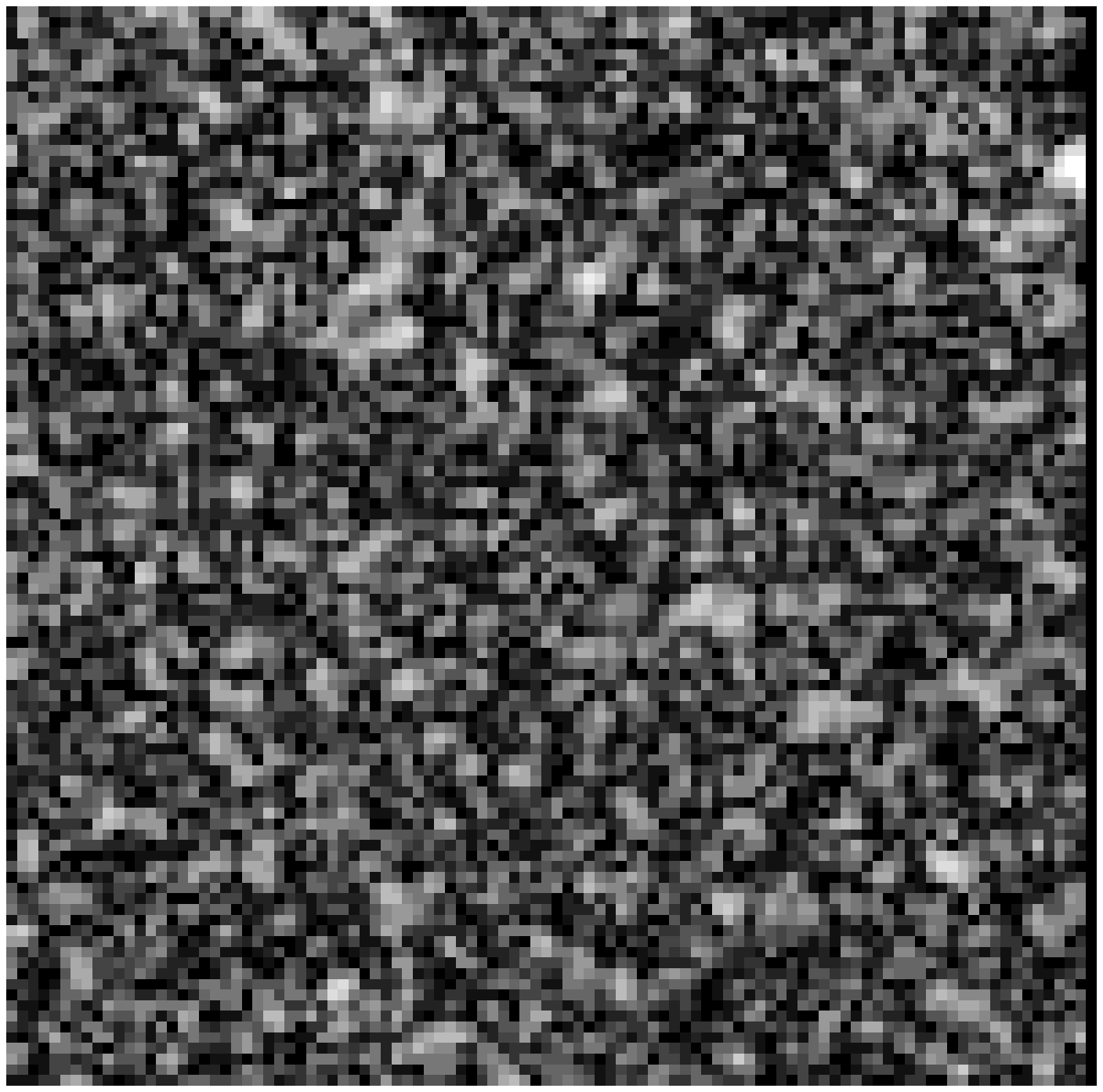}\vspace{0.025cm}
\includegraphics[angle=0,scale=.28]{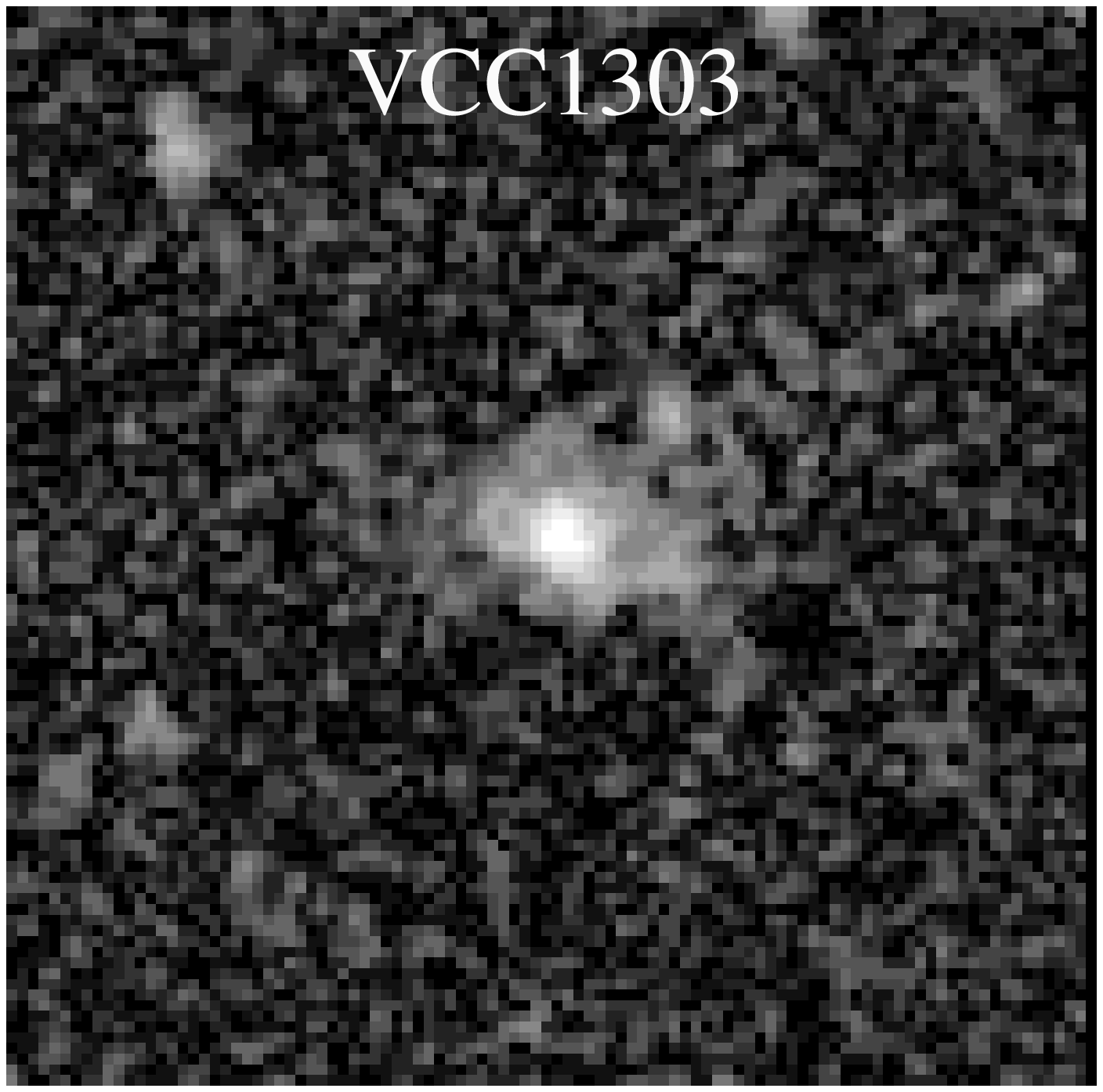}
\includegraphics[angle=0,scale=.28]{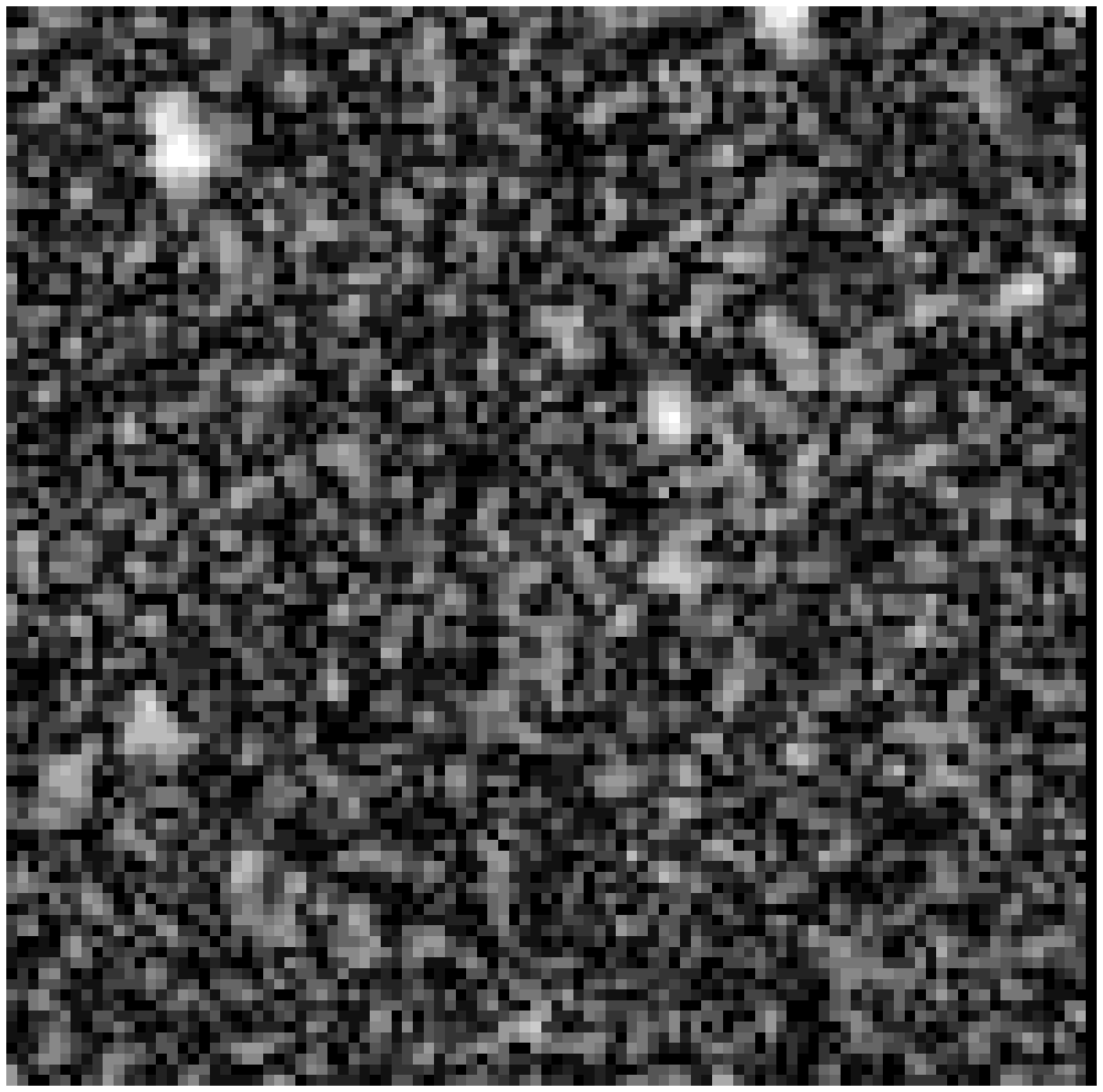}\hspace{0.06cm}
\includegraphics[angle=0,scale=.28]{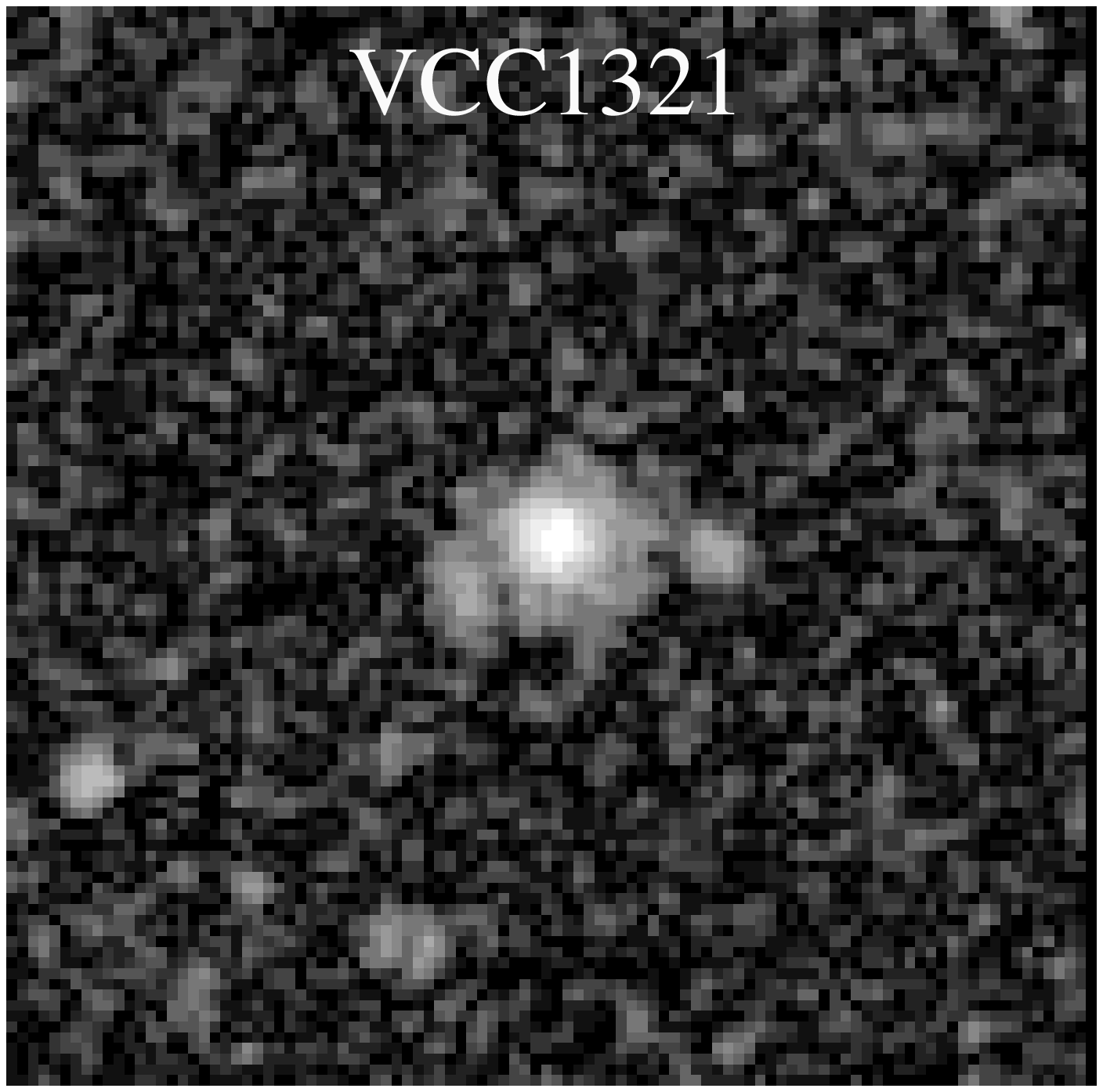}
\includegraphics[angle=0,scale=.28]{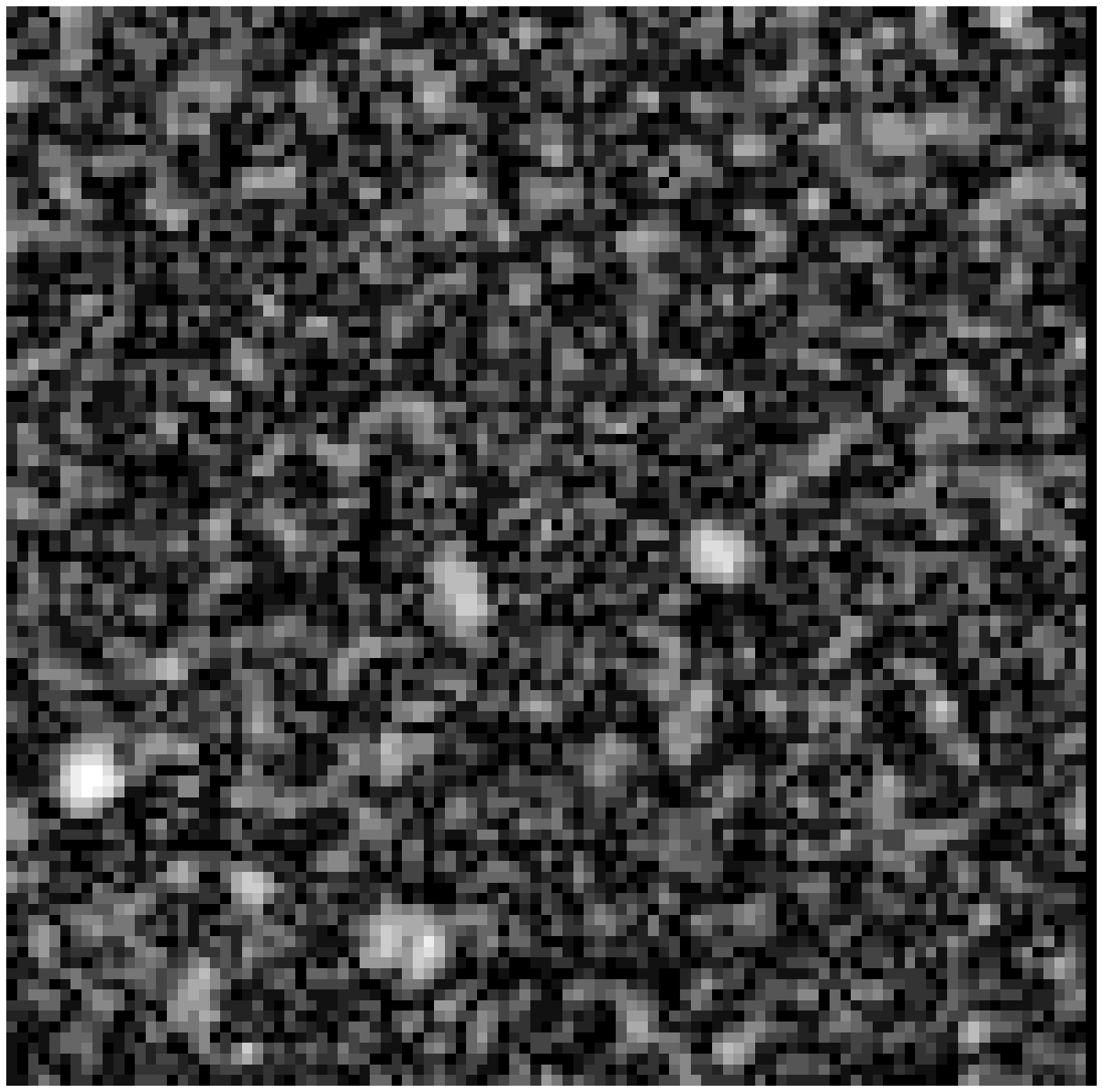}
\caption{{\it continued}}
\end{figure*}
\addtocounter{figure}{-1}

\begin{figure*}
\centering
\includegraphics[angle=0,scale=.28]{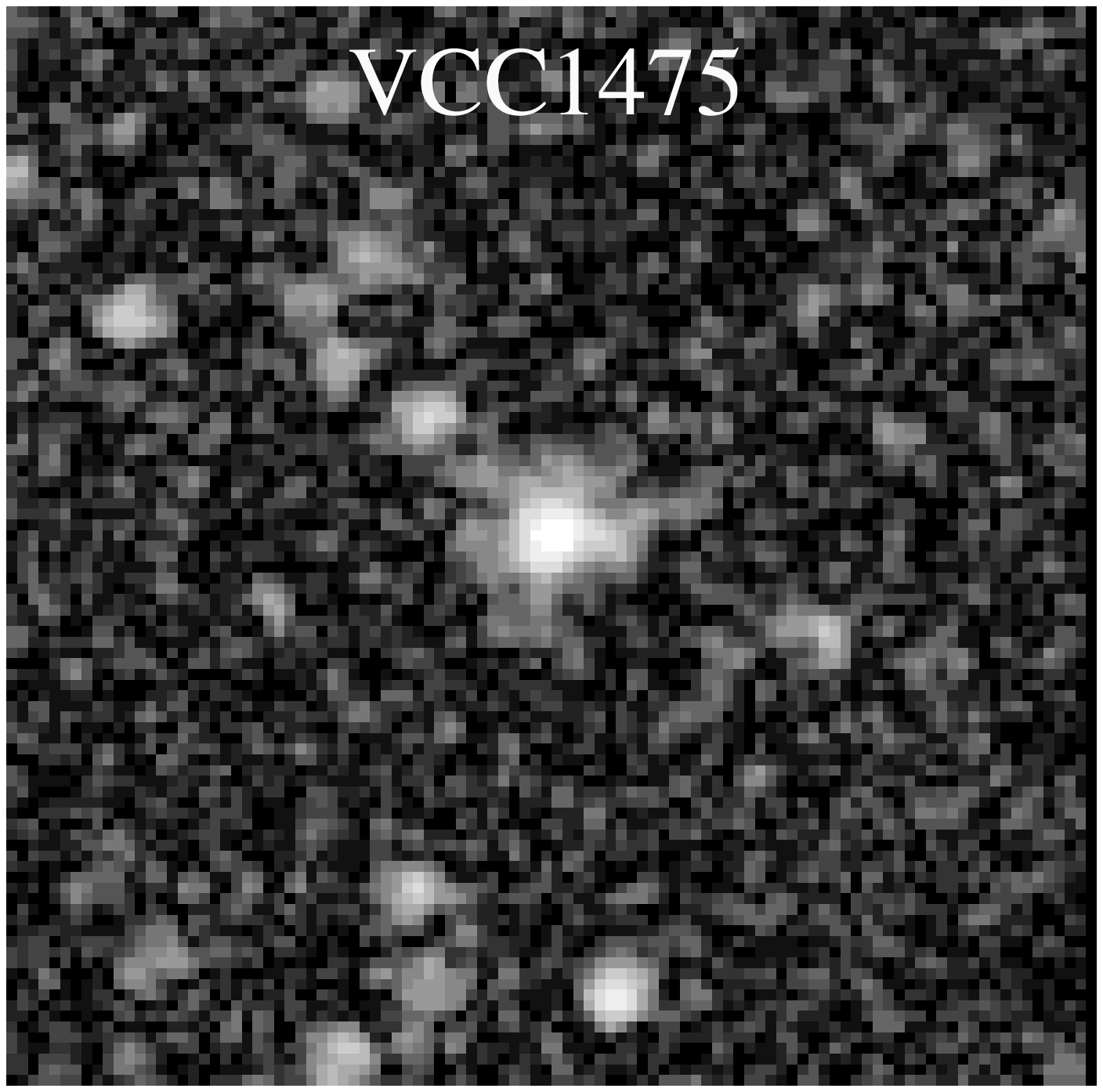}
\includegraphics[angle=0,scale=.28]{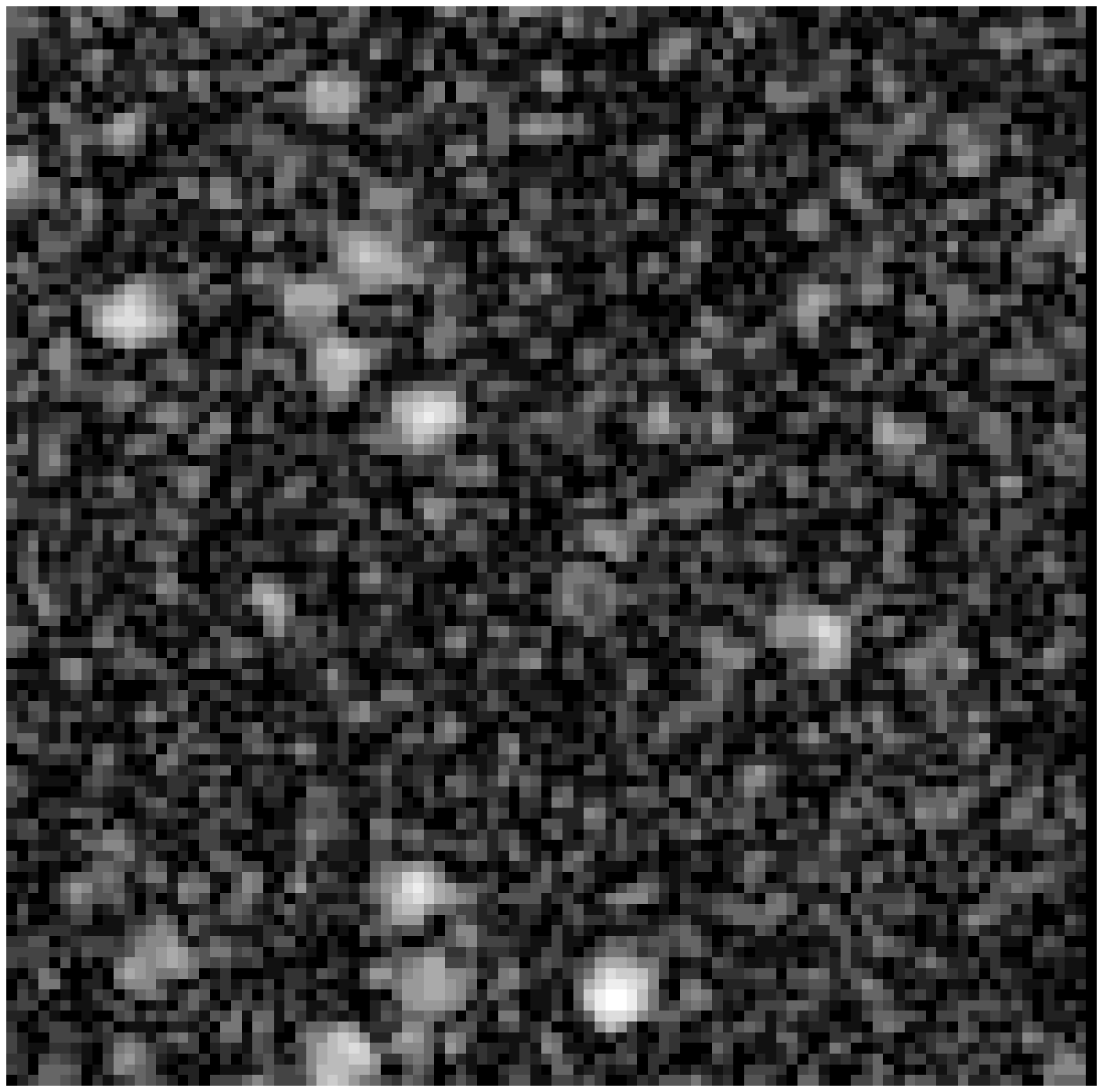}\hspace{0.06cm}
\includegraphics[angle=0,scale=.28]{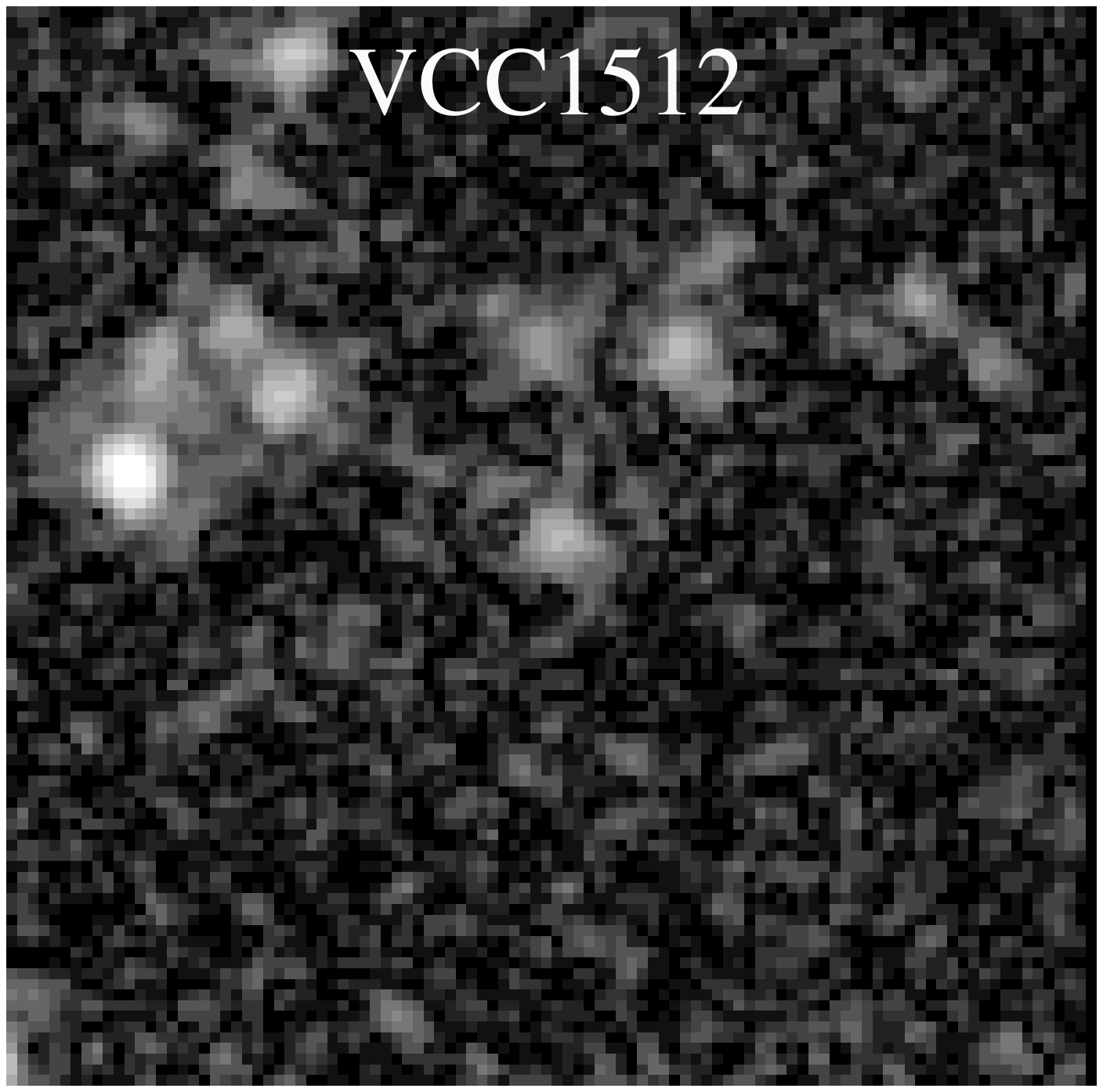}
\includegraphics[angle=0,scale=.28]{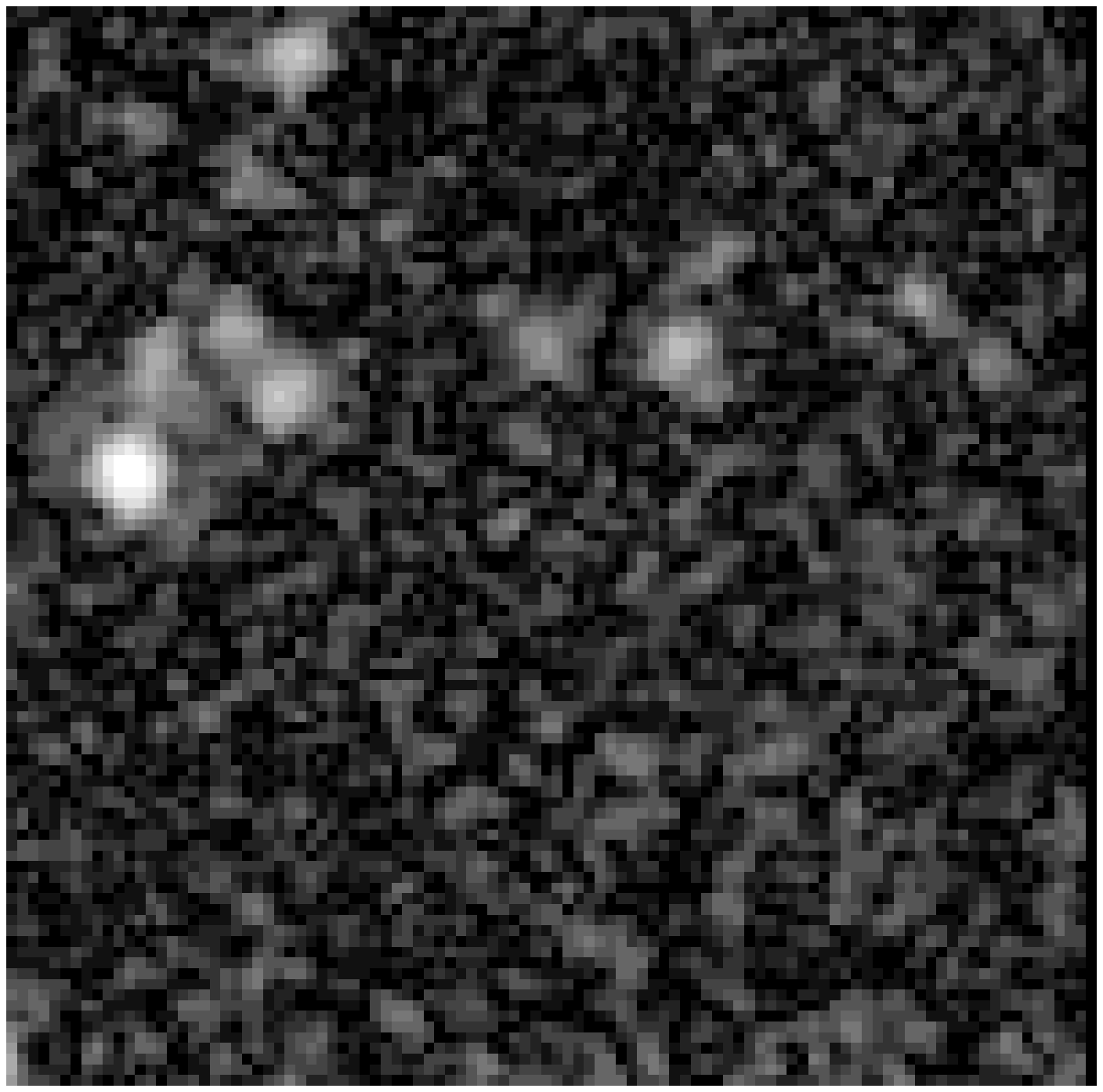}\vspace{0.025cm}
\includegraphics[angle=0,scale=.28]{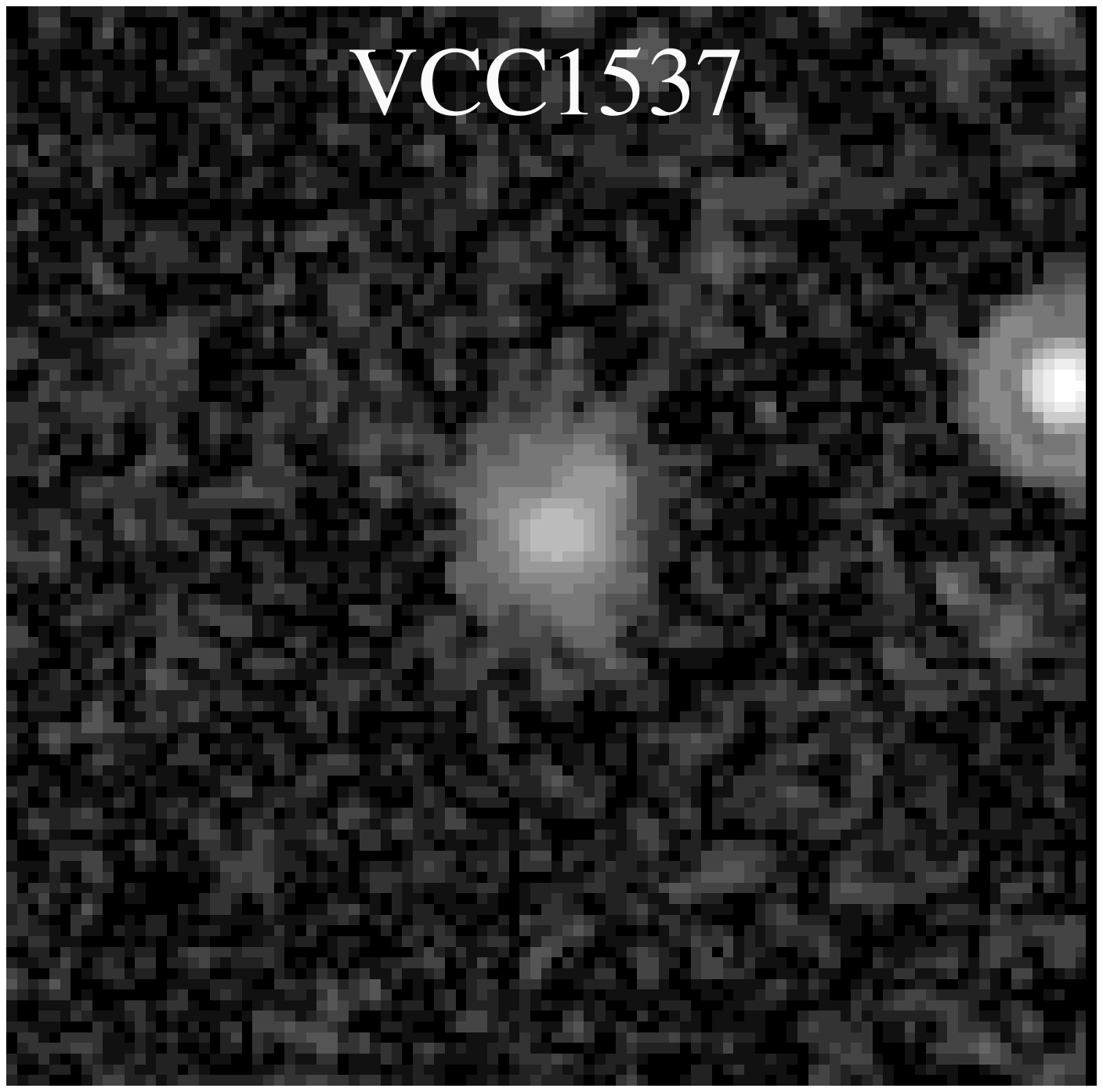}
\includegraphics[angle=0,scale=.28]{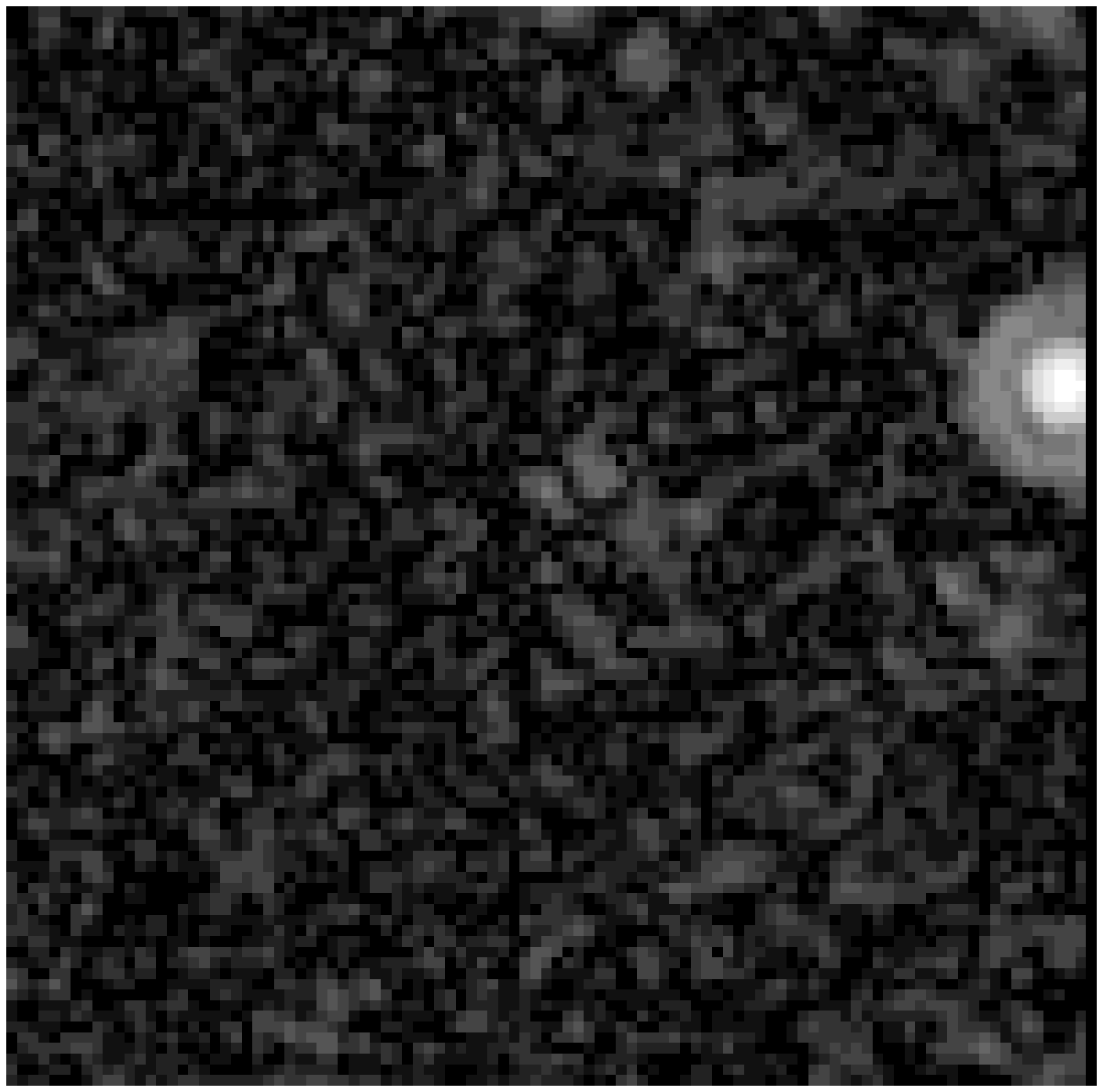}\hspace{0.06cm}
\includegraphics[angle=0,scale=.28]{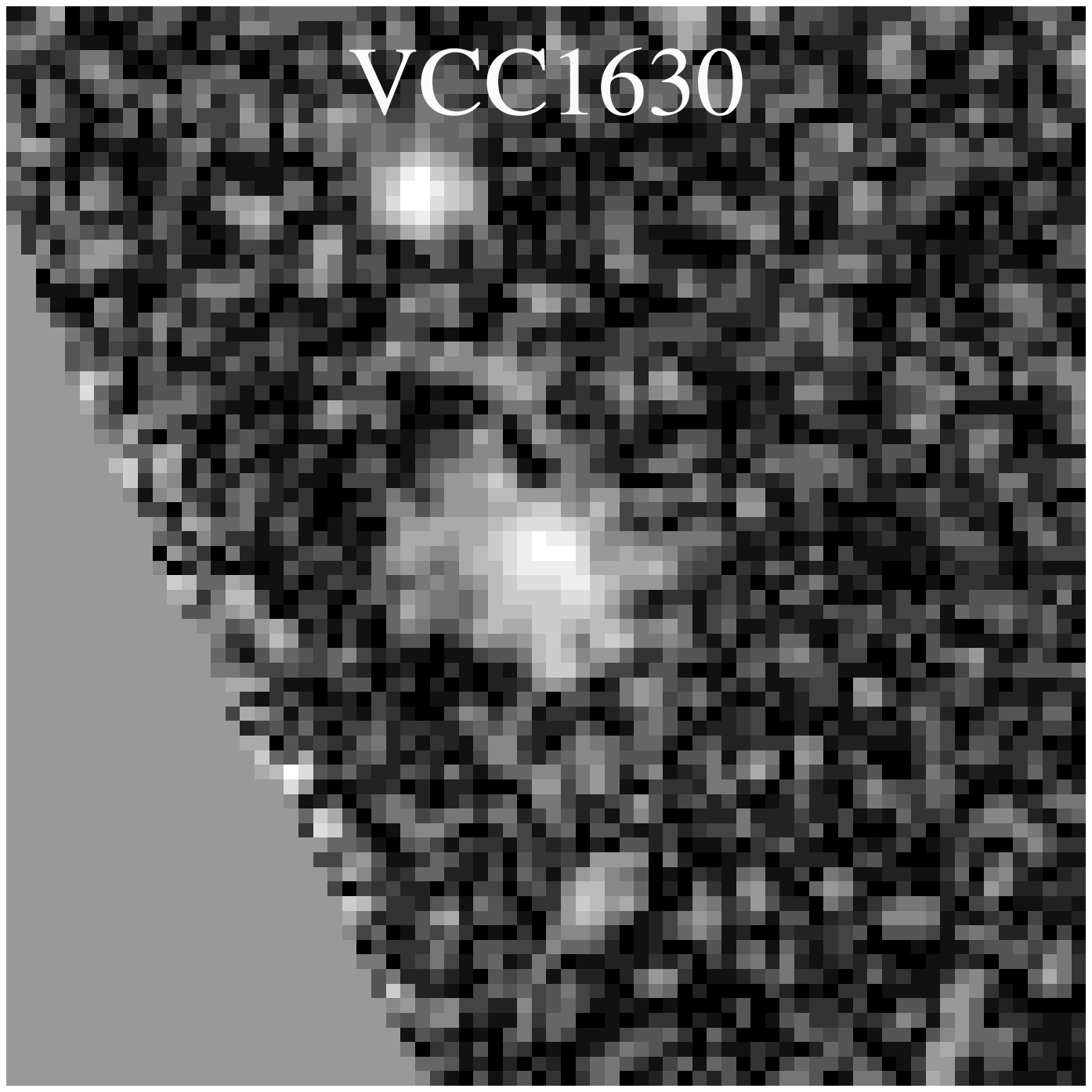}
\includegraphics[angle=0,scale=.28]{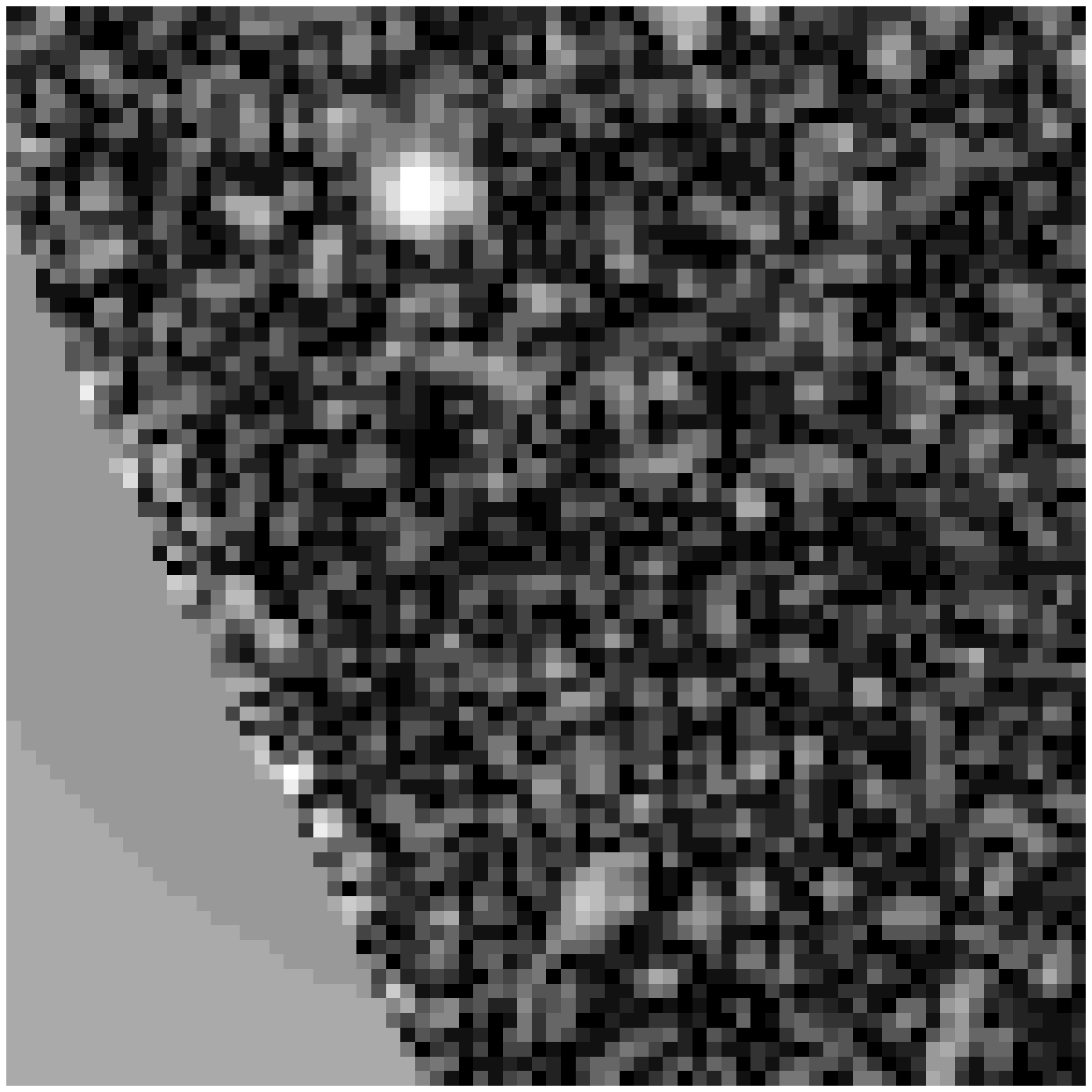}\vspace{0.025cm}
\includegraphics[angle=0,scale=.28]{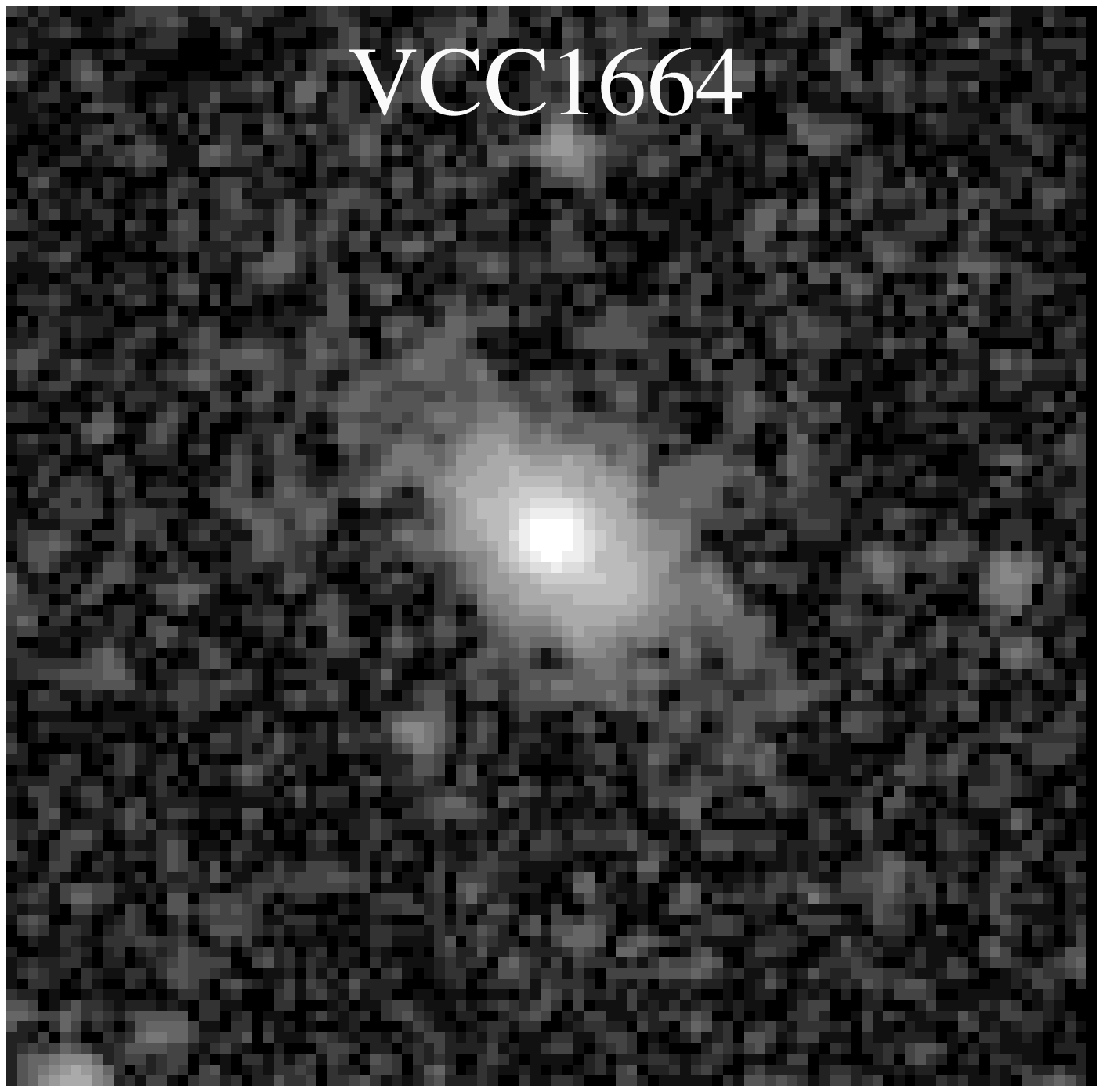}
\includegraphics[angle=0,scale=.28]{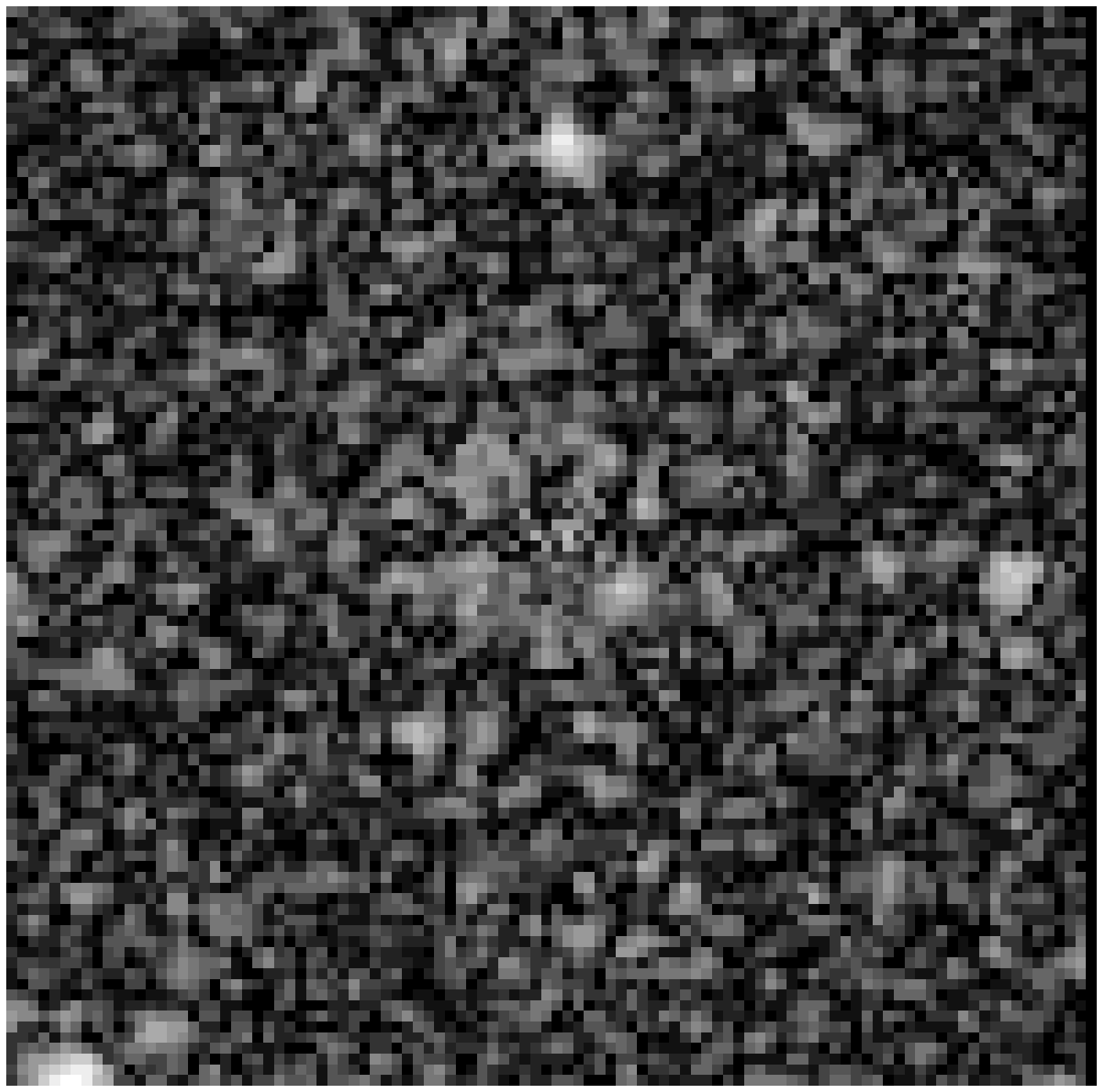}\hspace{0.06cm}
\includegraphics[angle=0,scale=.28]{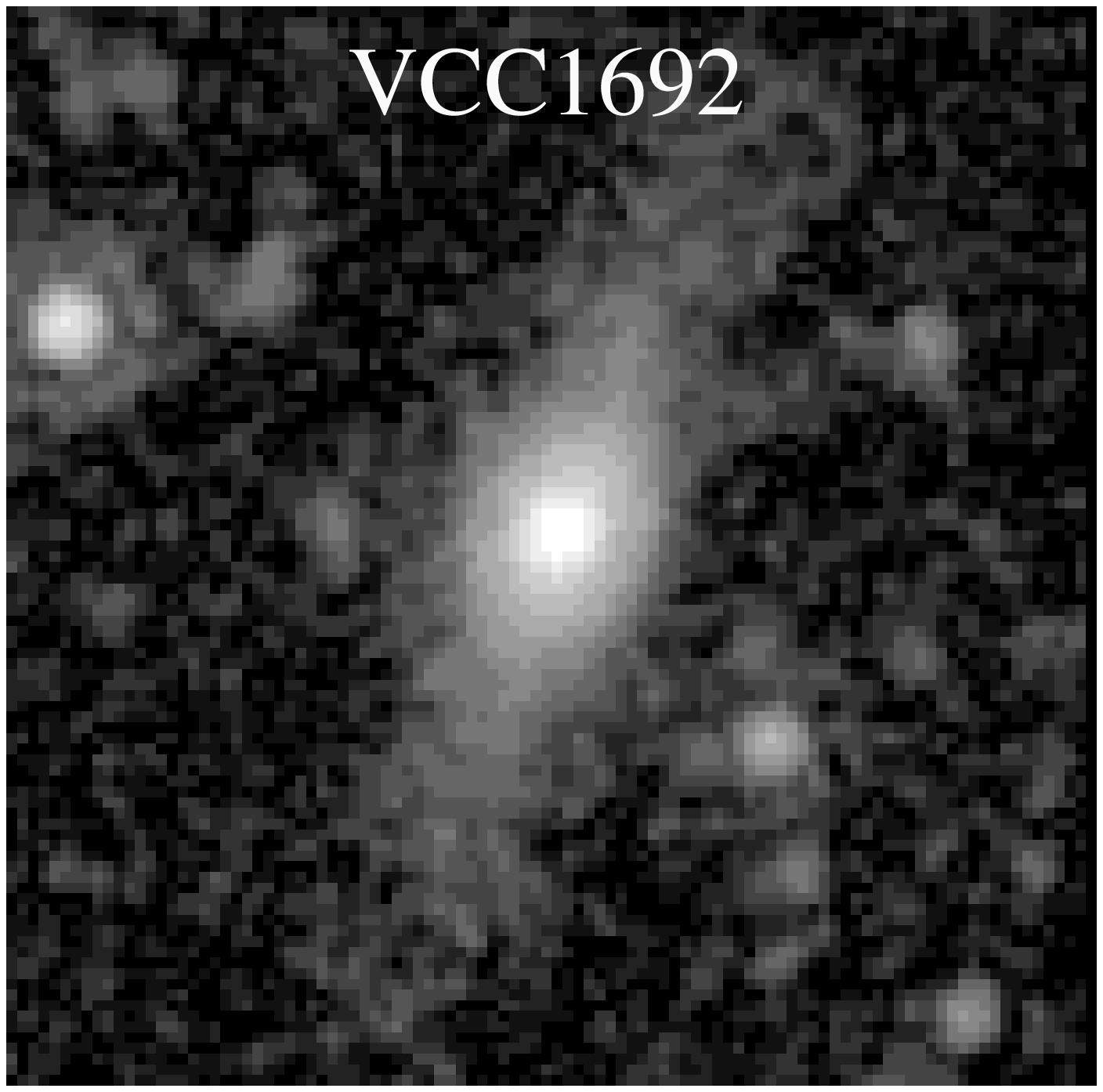}
\includegraphics[angle=0,scale=.28]{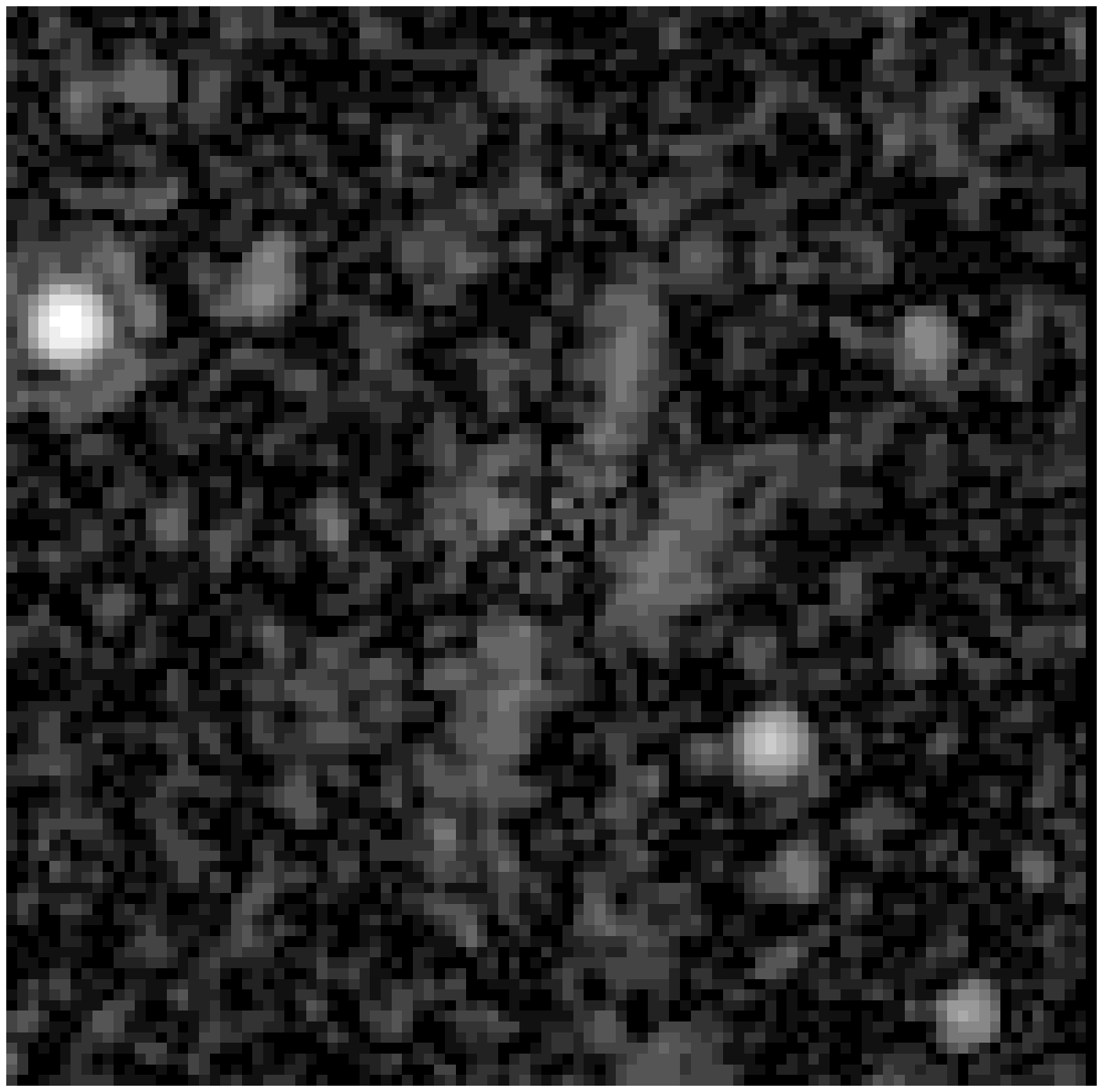}\vspace{0.025cm}
\includegraphics[angle=0,scale=.28]{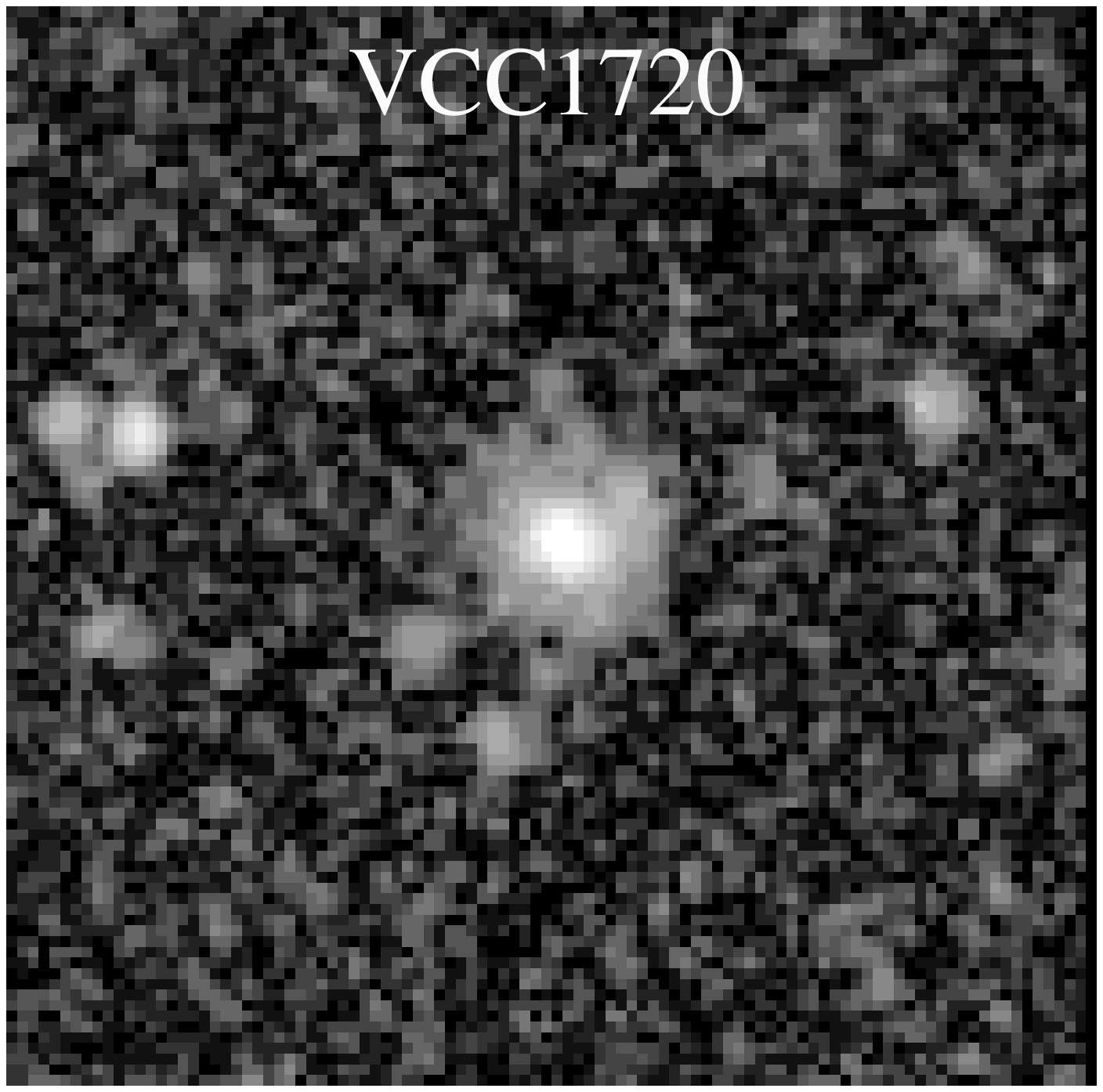}
\includegraphics[angle=0,scale=.28]{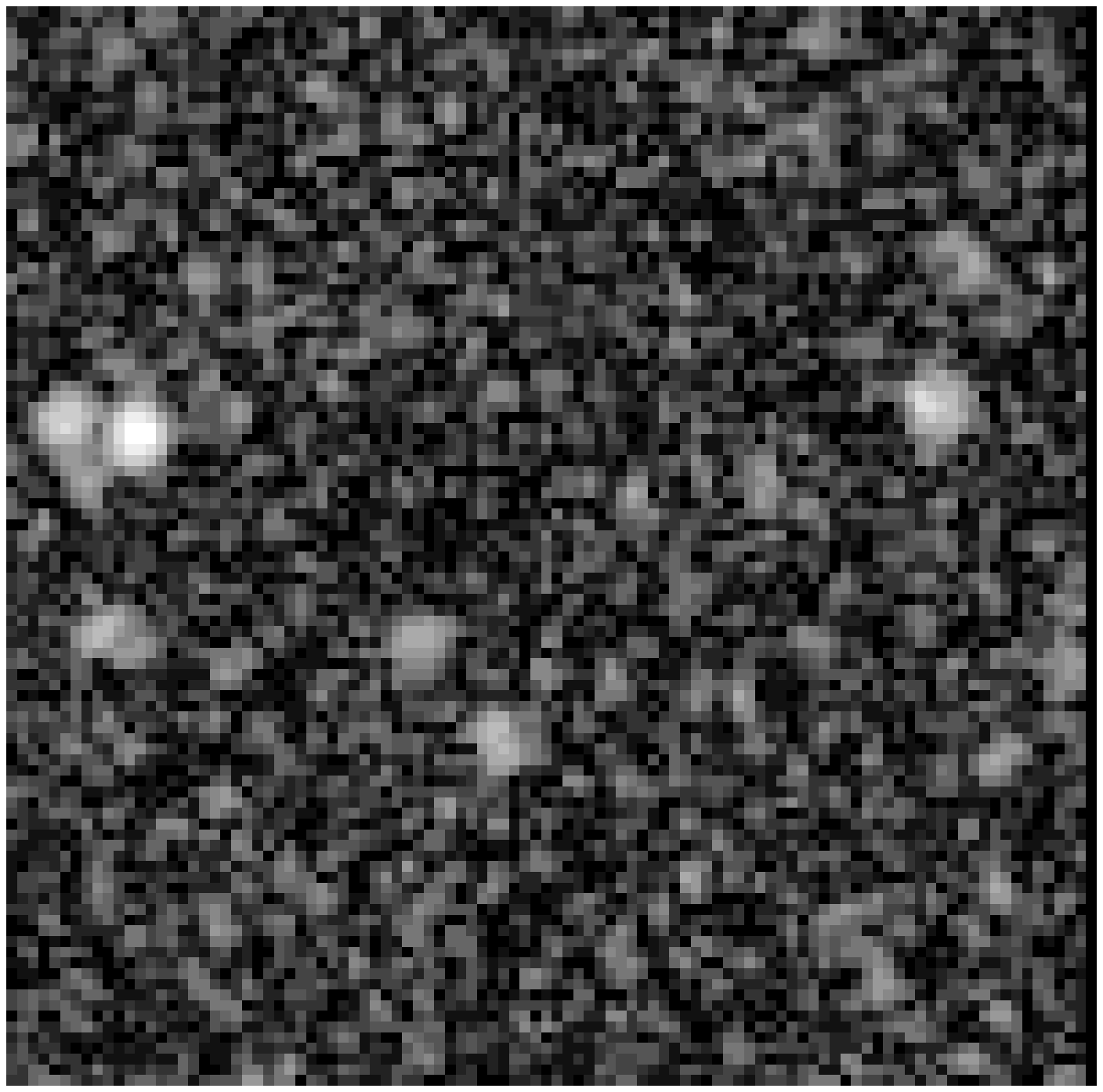}\hspace{0.06cm}
\includegraphics[angle=0,scale=.28]{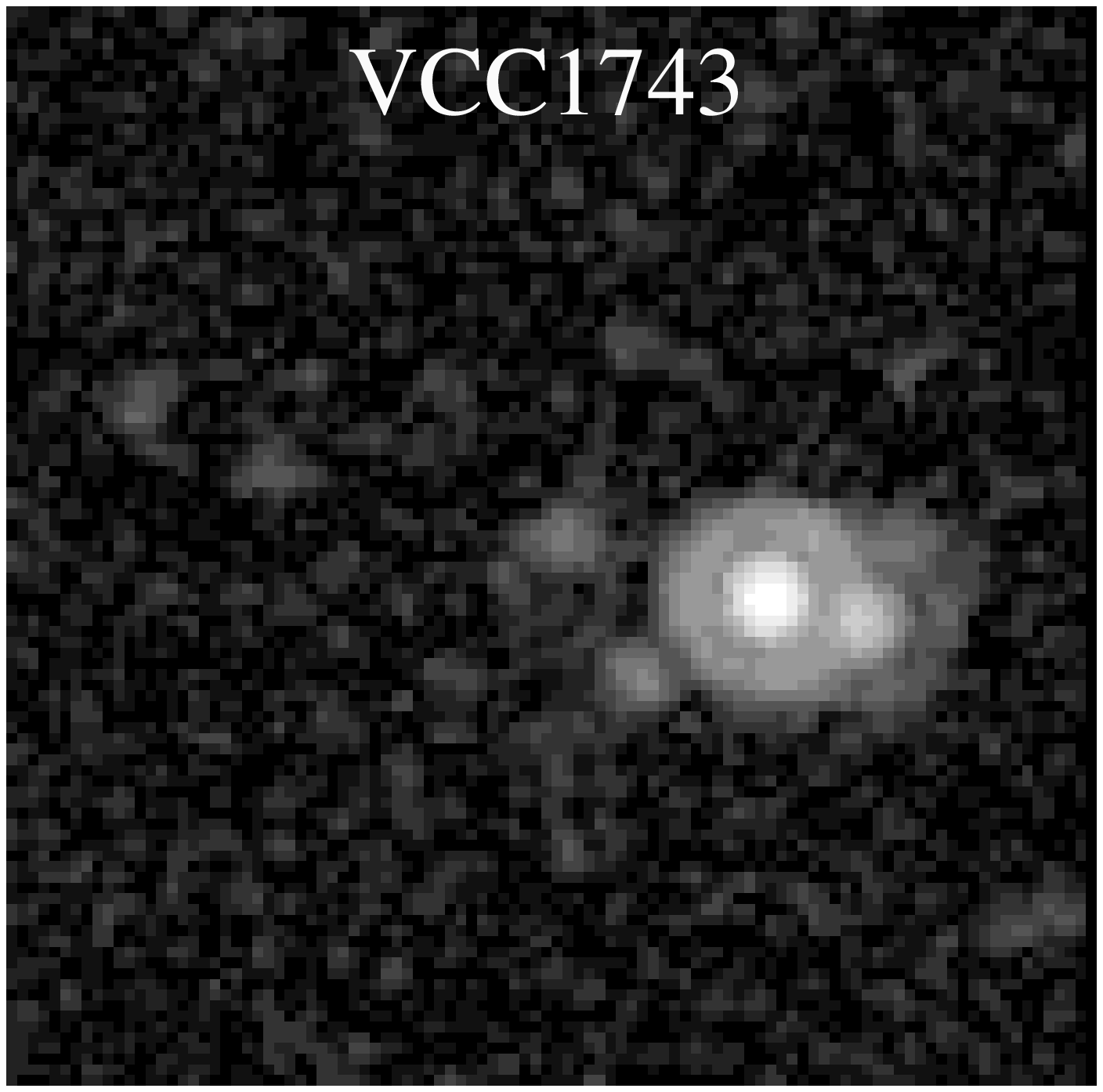}
\includegraphics[angle=0,scale=.28]{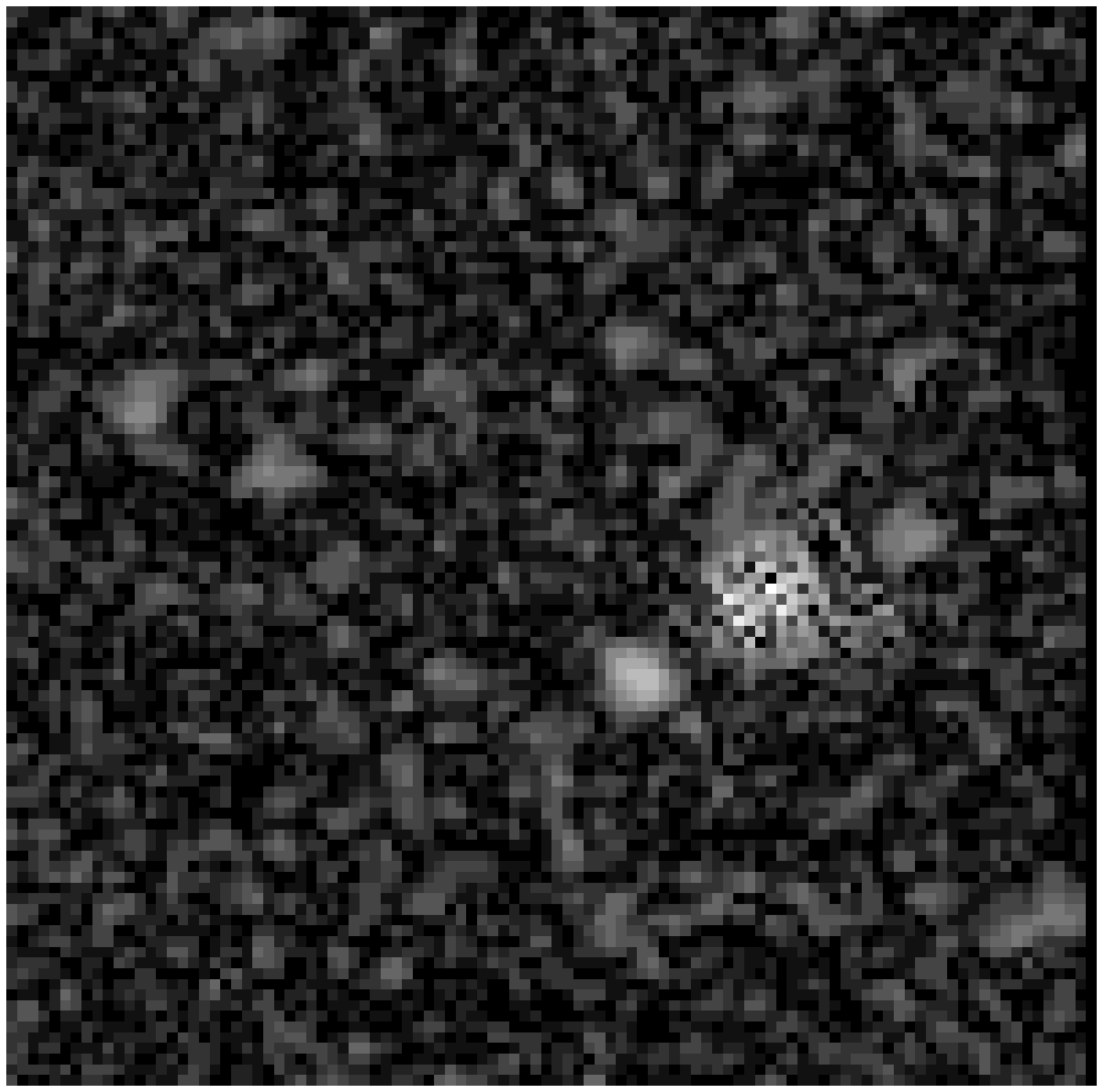}\vspace{0.025cm}
\includegraphics[angle=0,scale=.28]{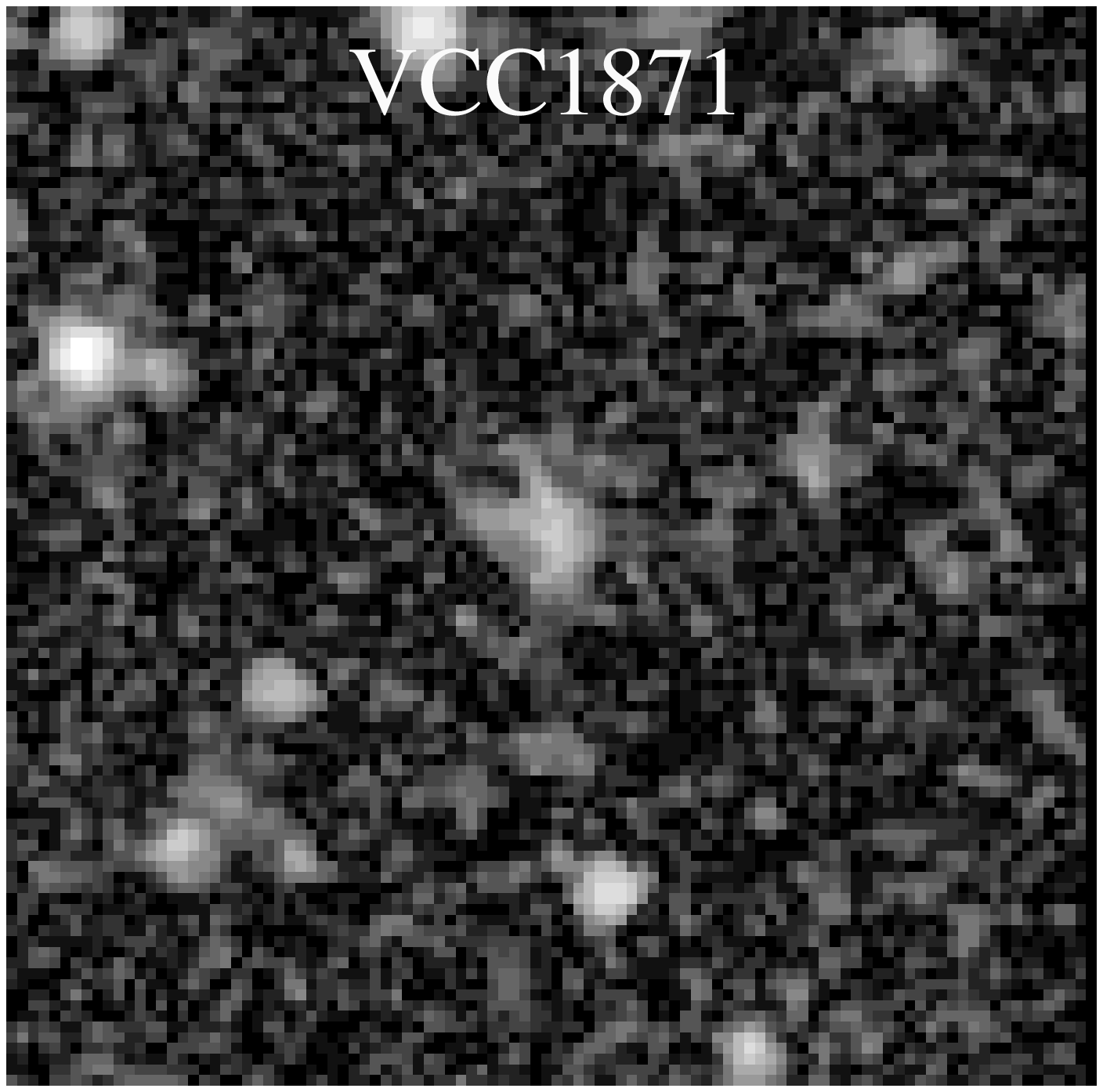}
\includegraphics[angle=0,scale=.28]{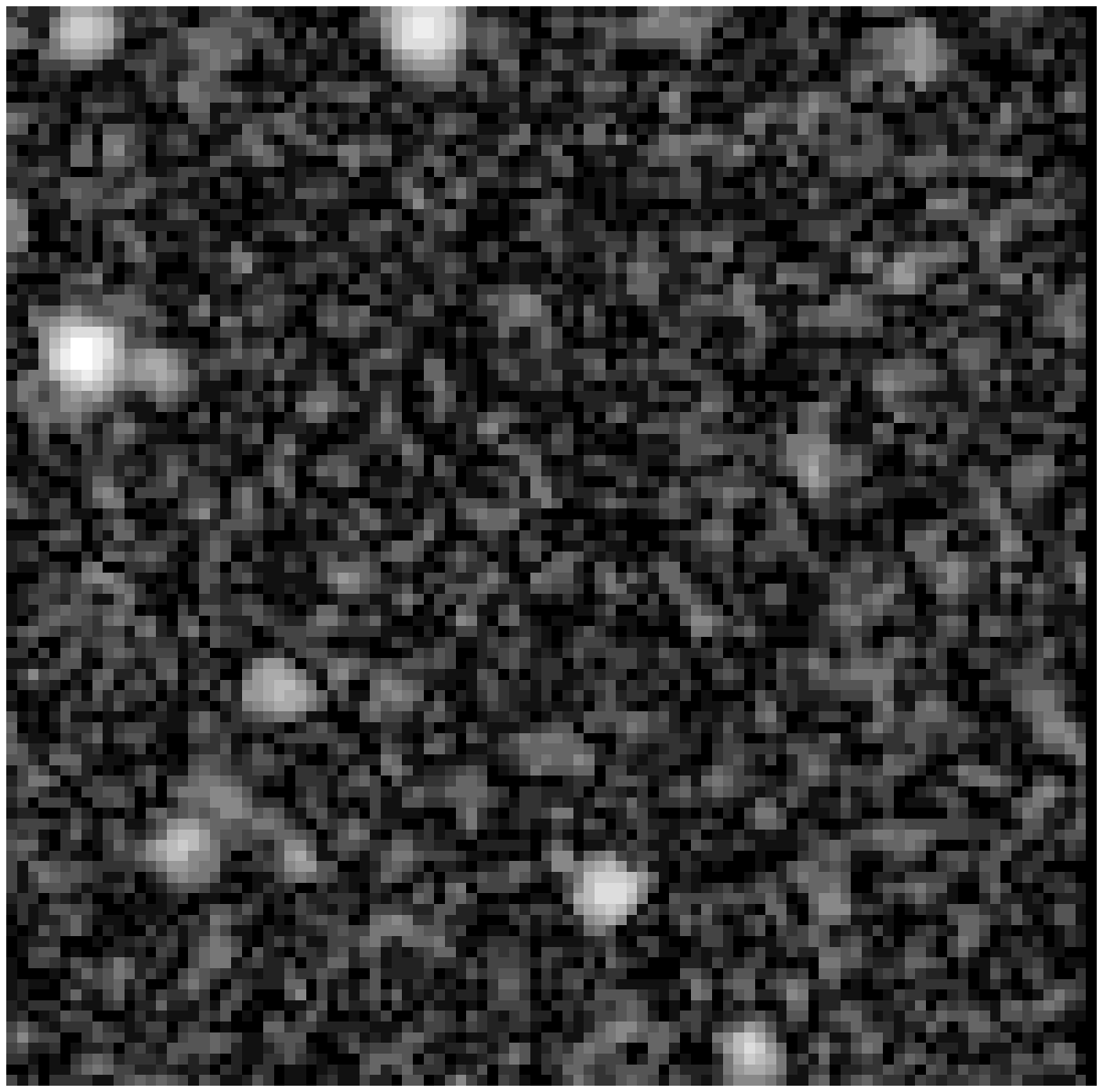}\hspace{0.06cm}
\includegraphics[angle=0,scale=.28]{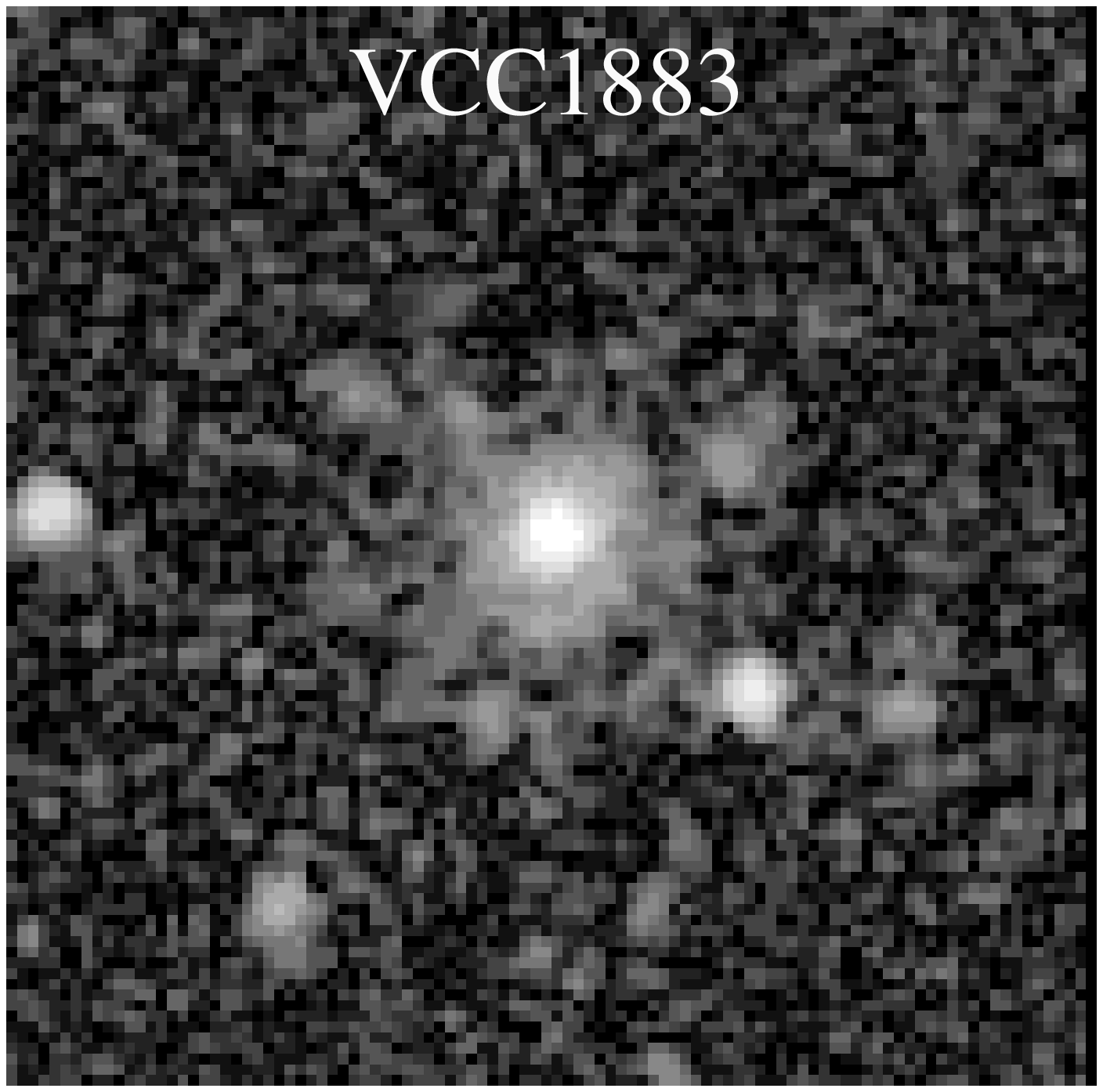}
\includegraphics[angle=0,scale=.28]{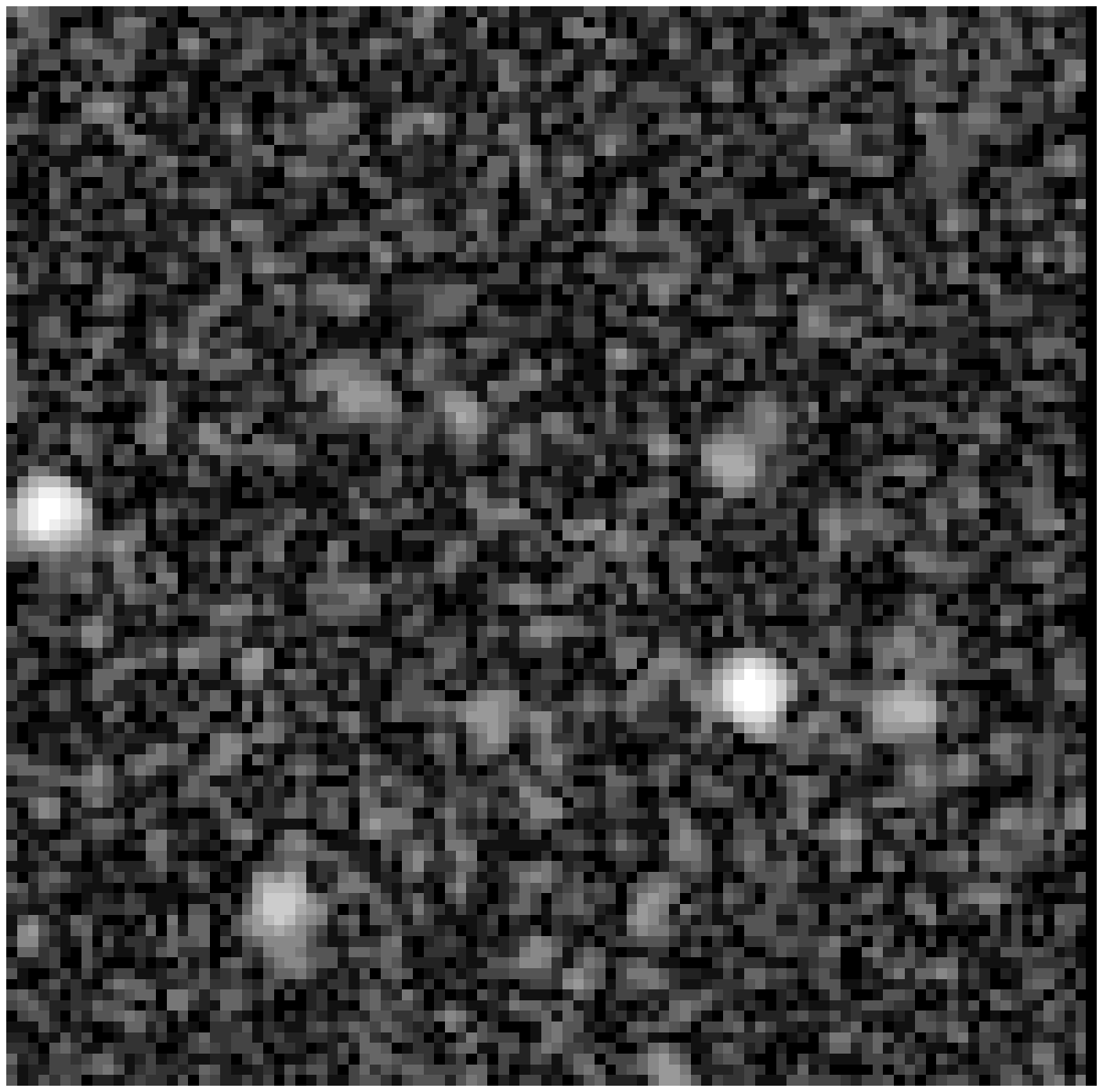}
\caption{{\it continued}}
\end{figure*}
\addtocounter{figure}{-1}

\begin{figure*}
\centering
\includegraphics[angle=0,scale=.28]{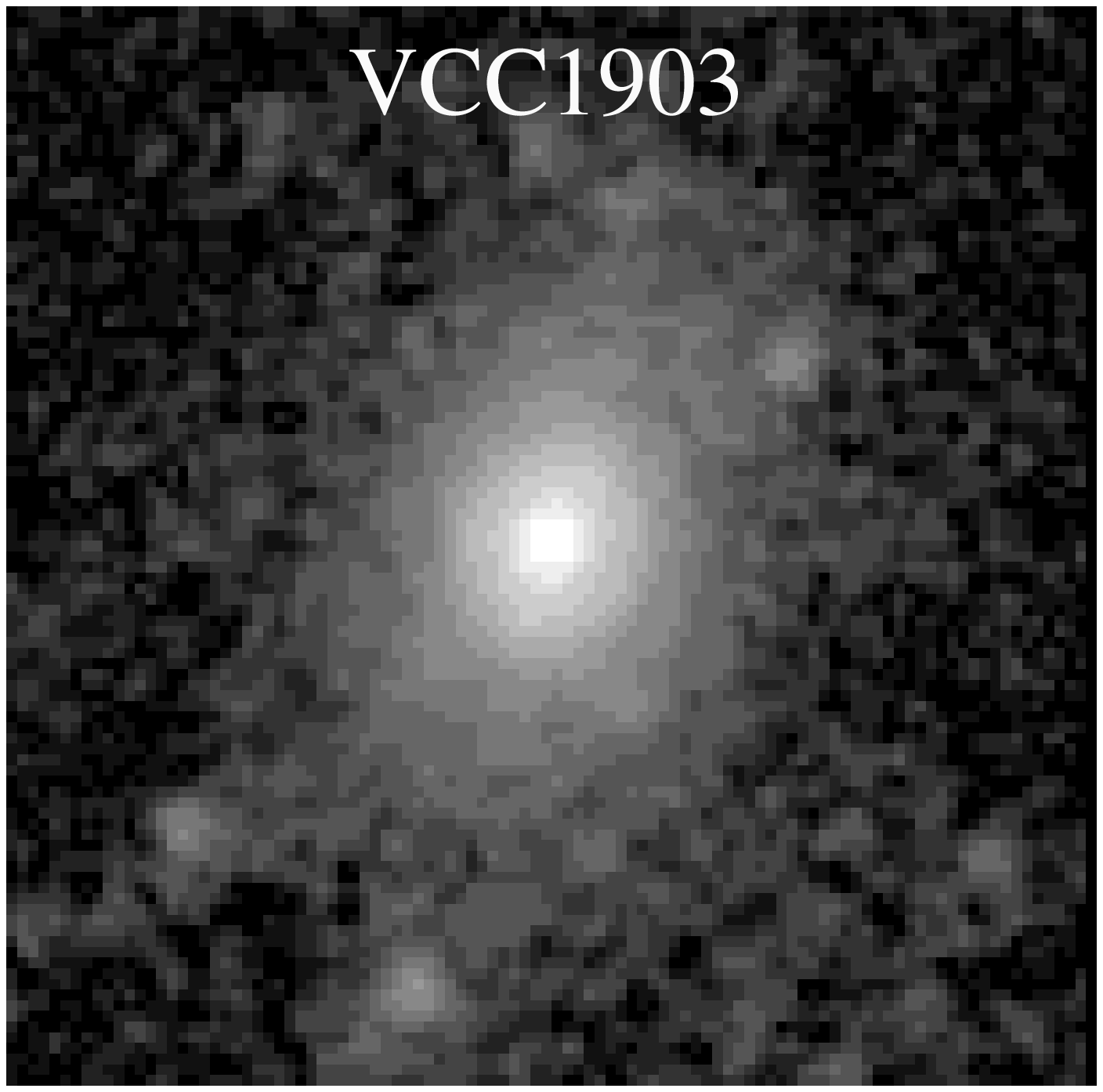}
\includegraphics[angle=0,scale=.28]{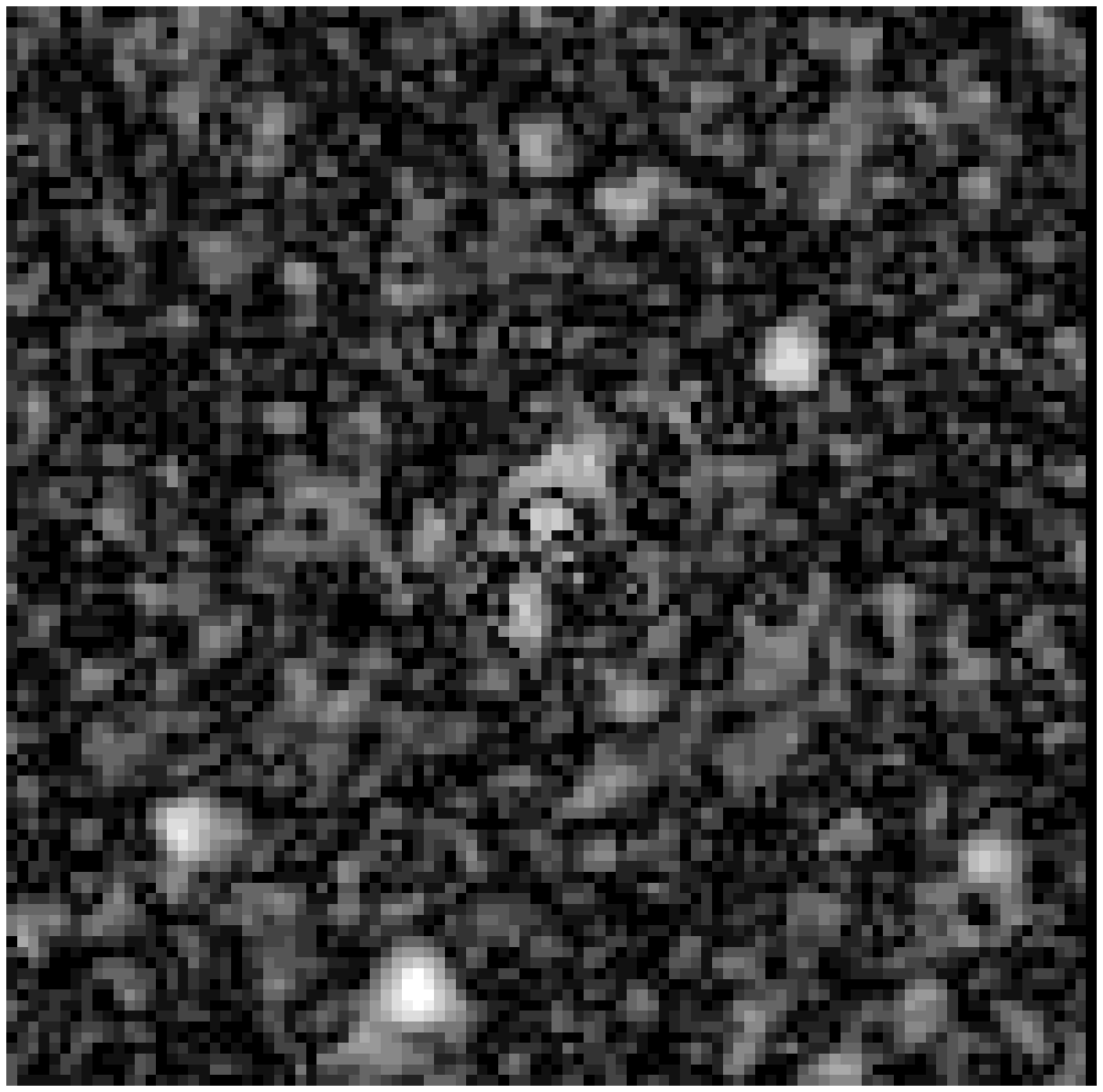}\hspace{0.06cm}
\includegraphics[angle=0,scale=.28]{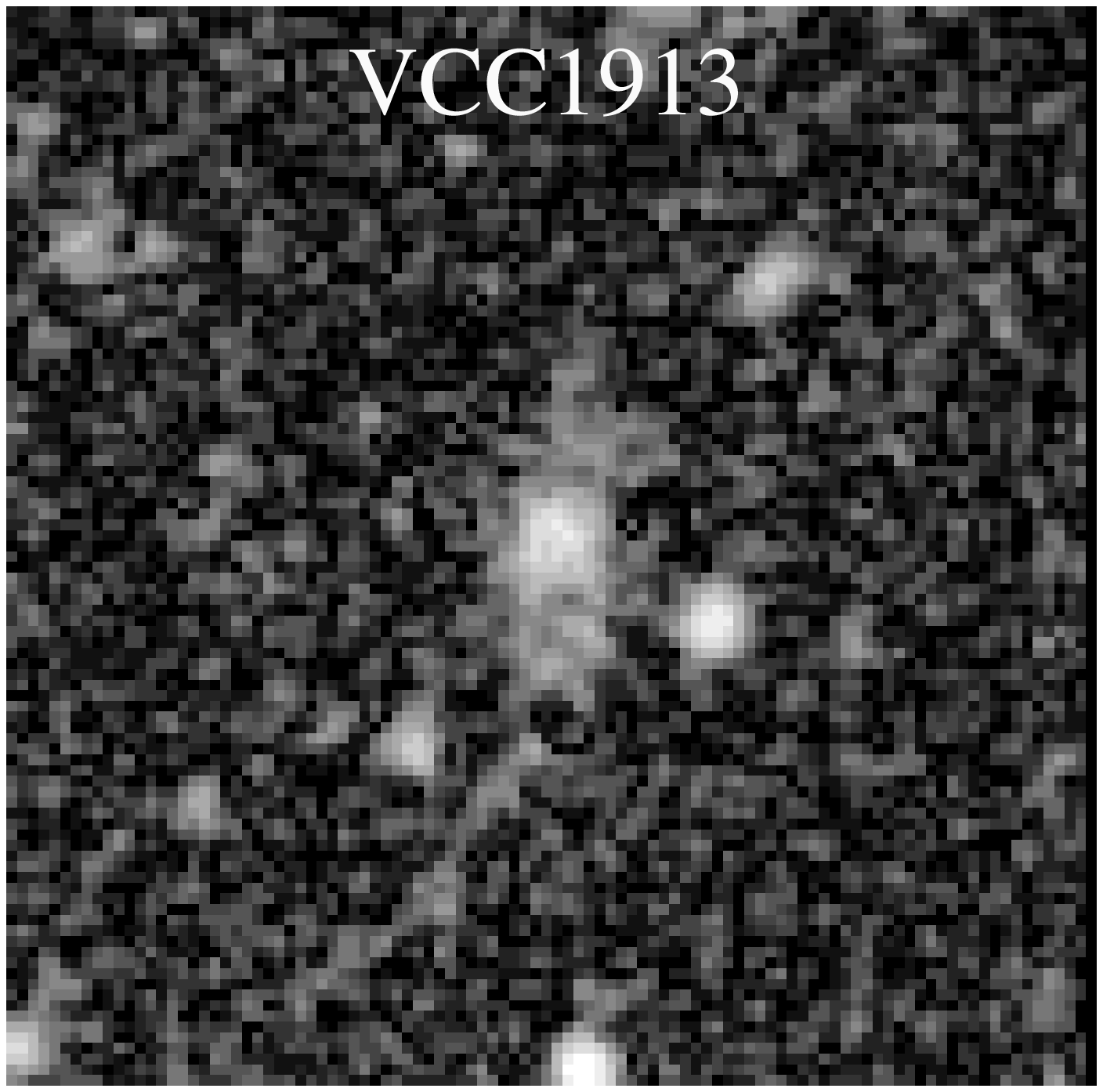}
\includegraphics[angle=0,scale=.28]{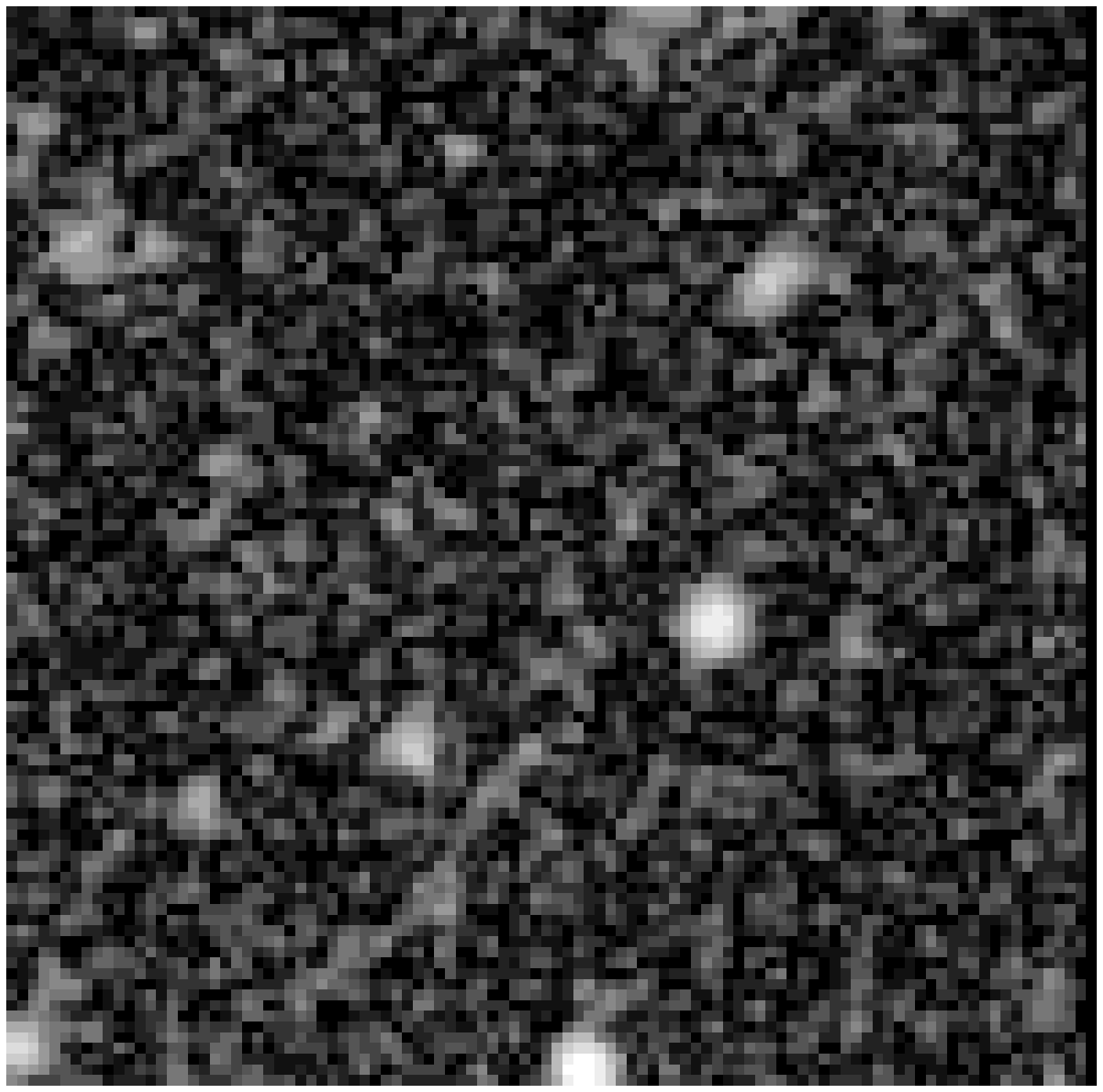}\vspace{0.025cm}
\includegraphics[angle=0,scale=.28]{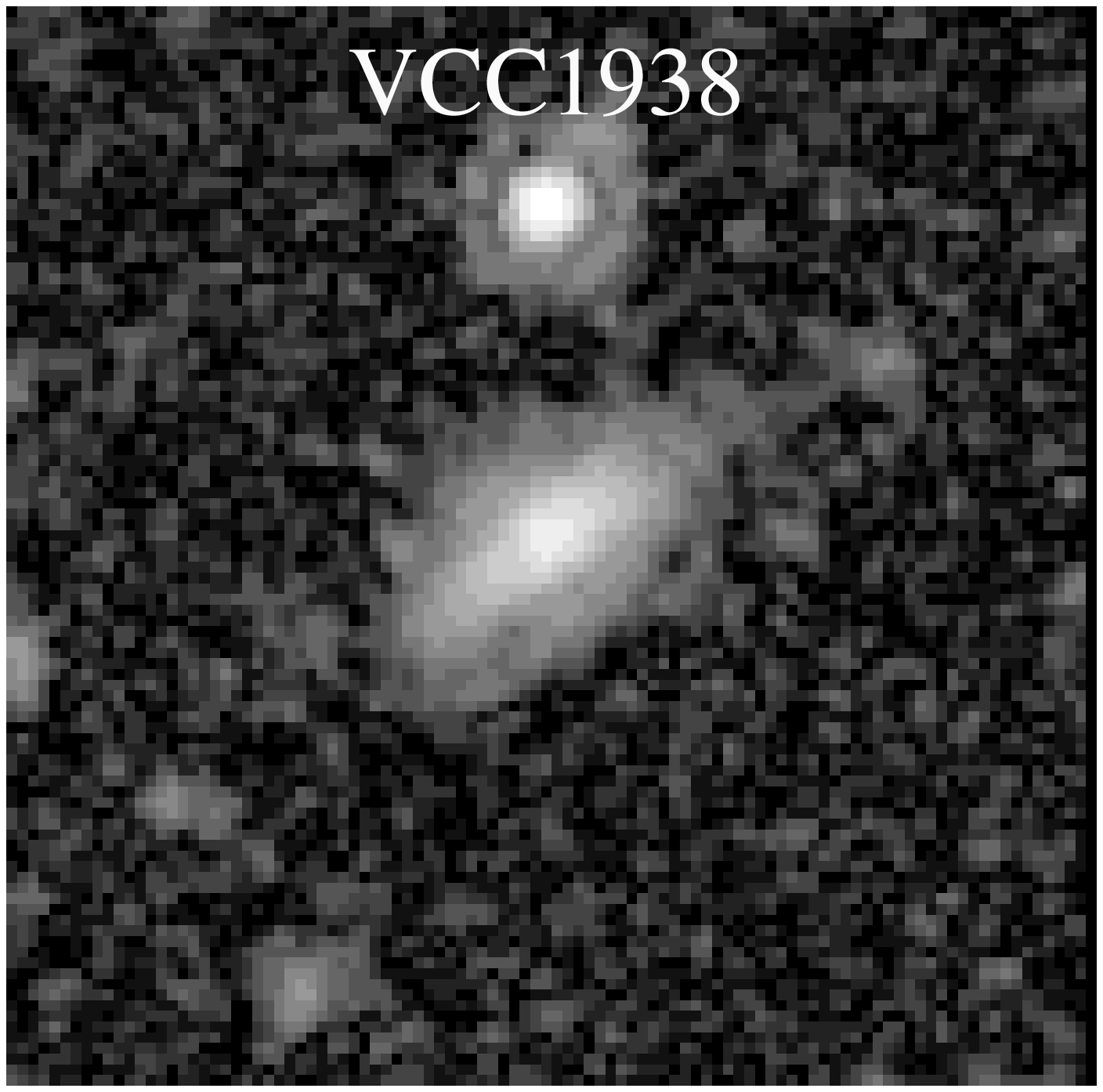}
\includegraphics[angle=0,scale=.28]{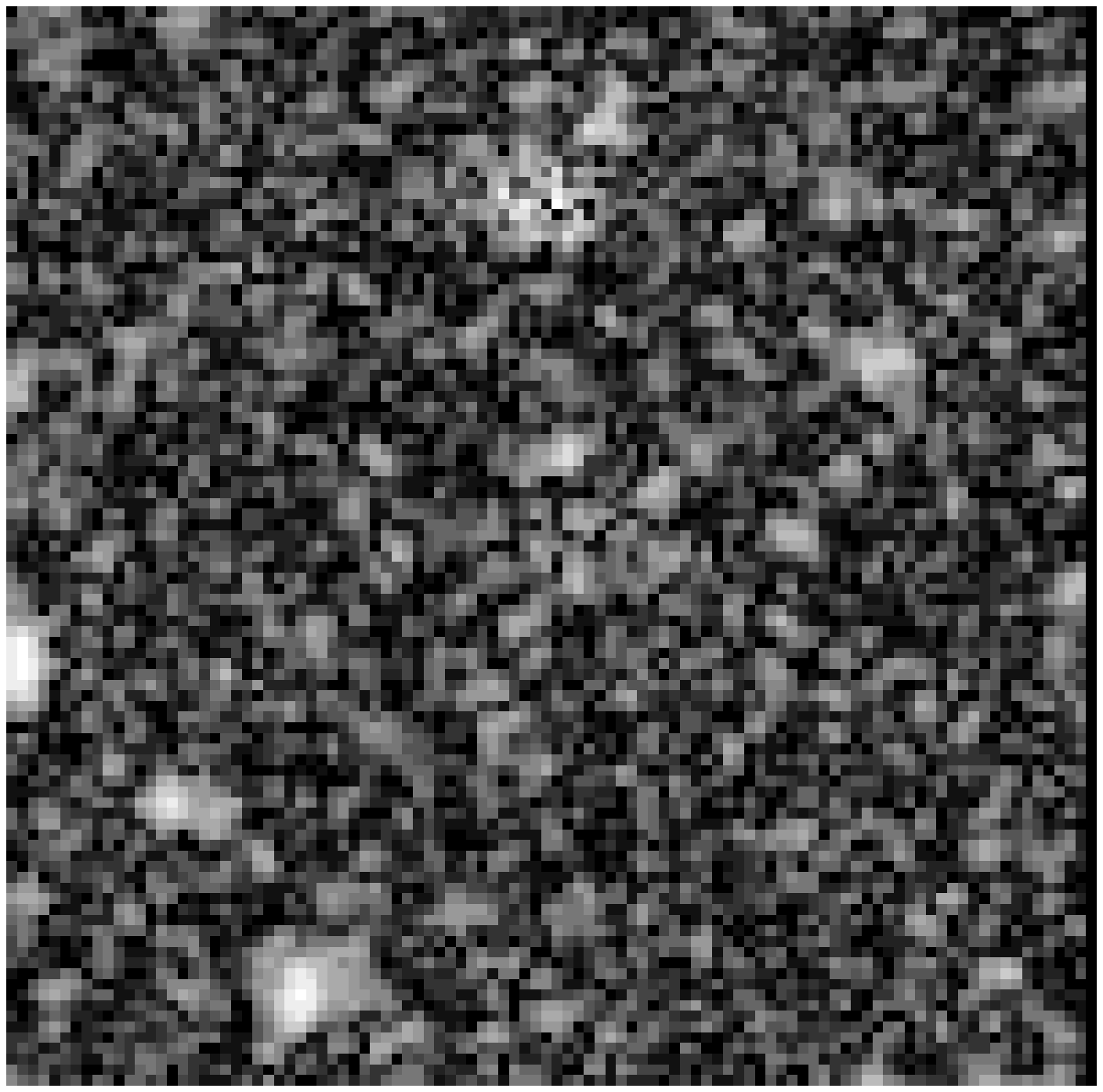}\hspace{0.06cm}
\includegraphics[angle=0,scale=.28]{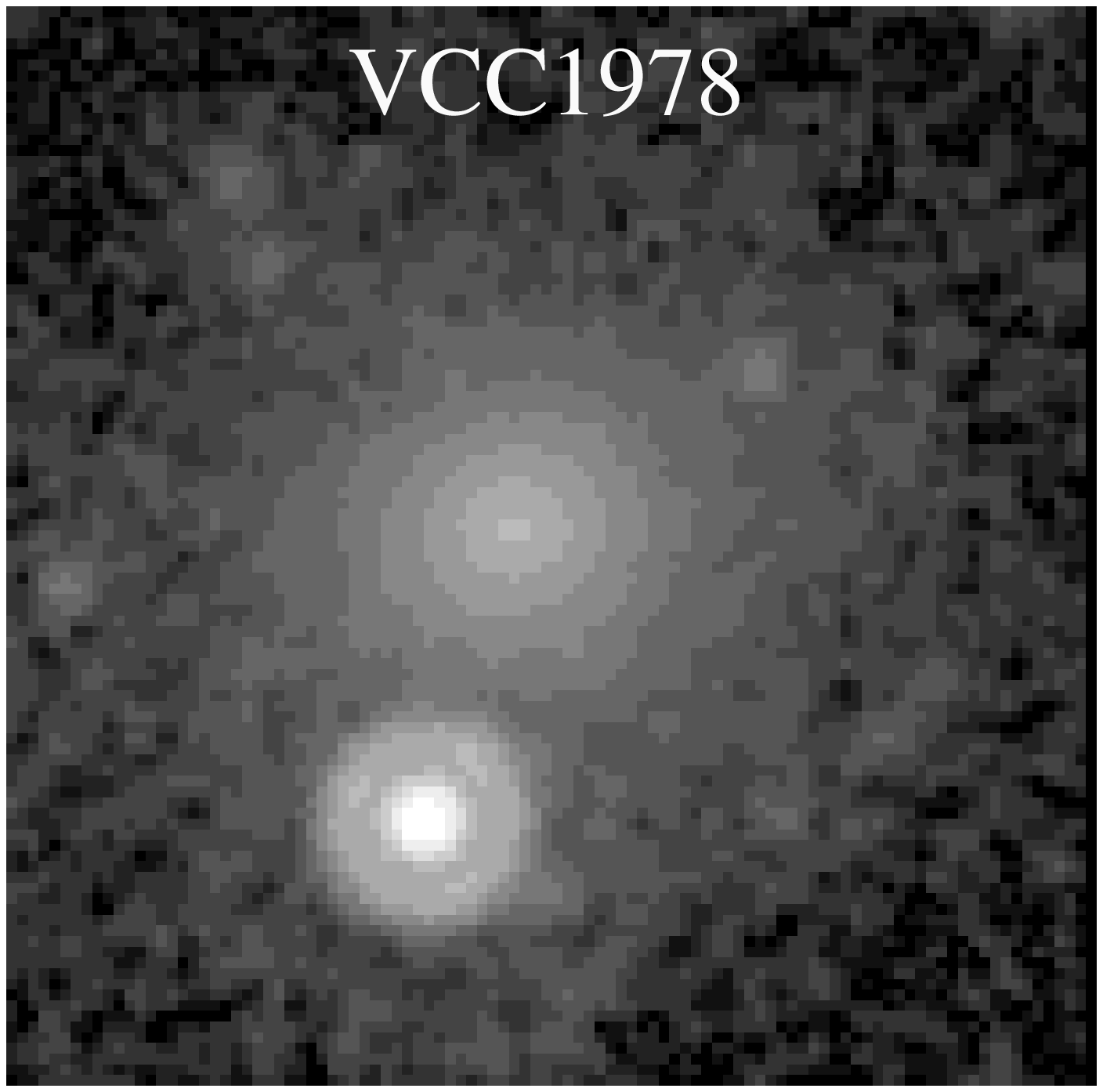}
\includegraphics[angle=0,scale=.28]{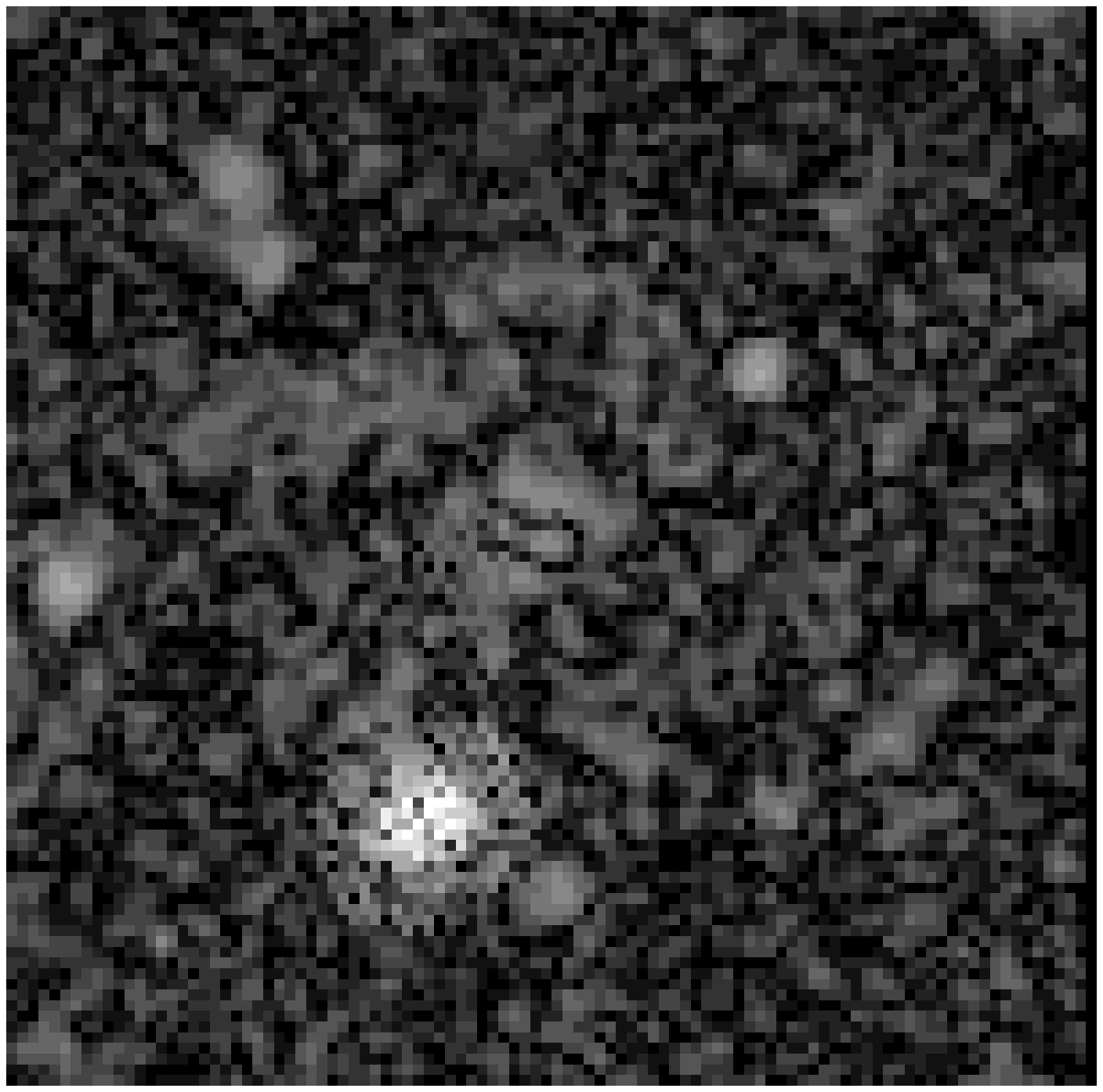}\vspace{0.025cm}
\includegraphics[angle=0,scale=.28]{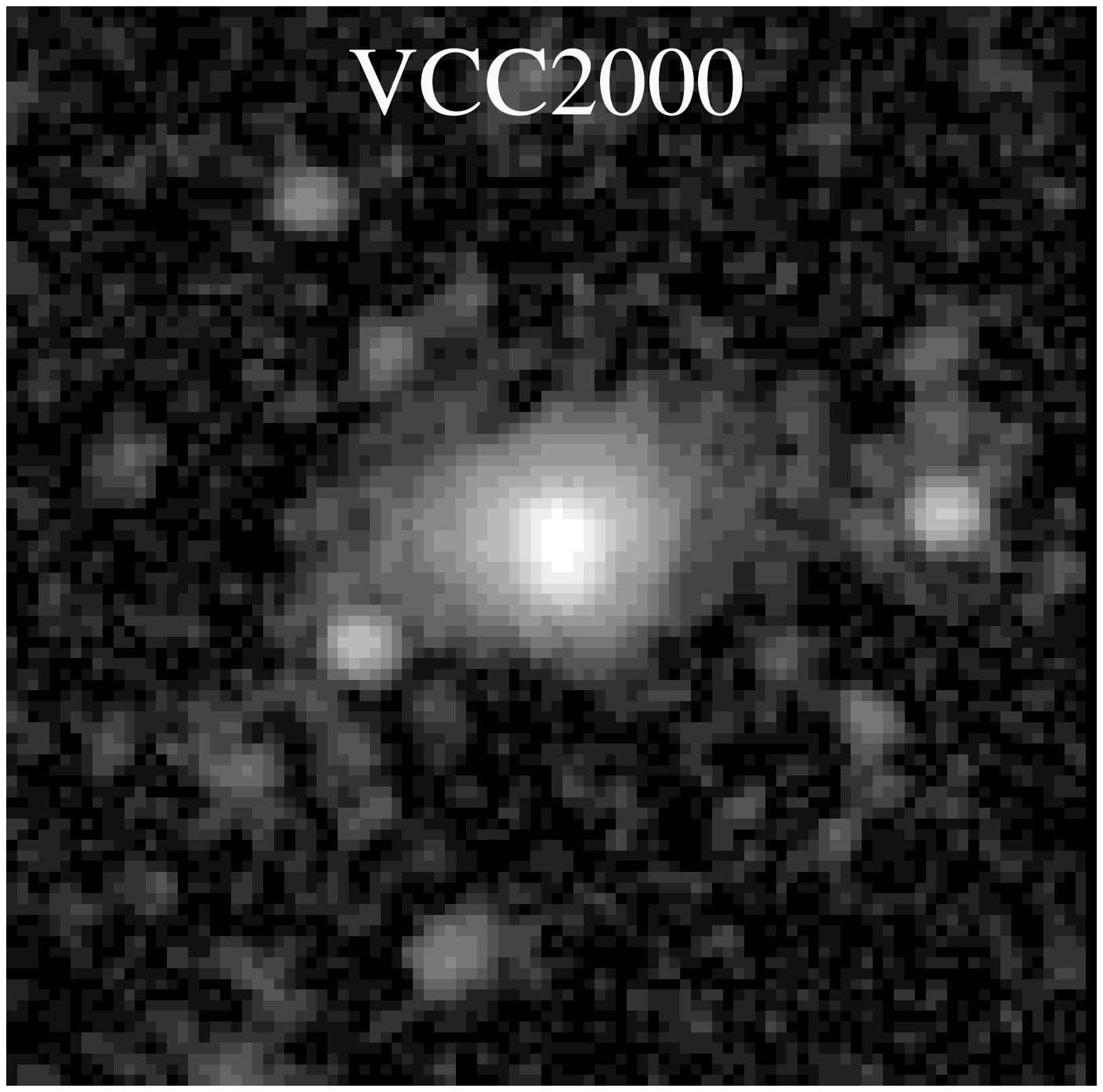}
\includegraphics[angle=0,scale=.28]{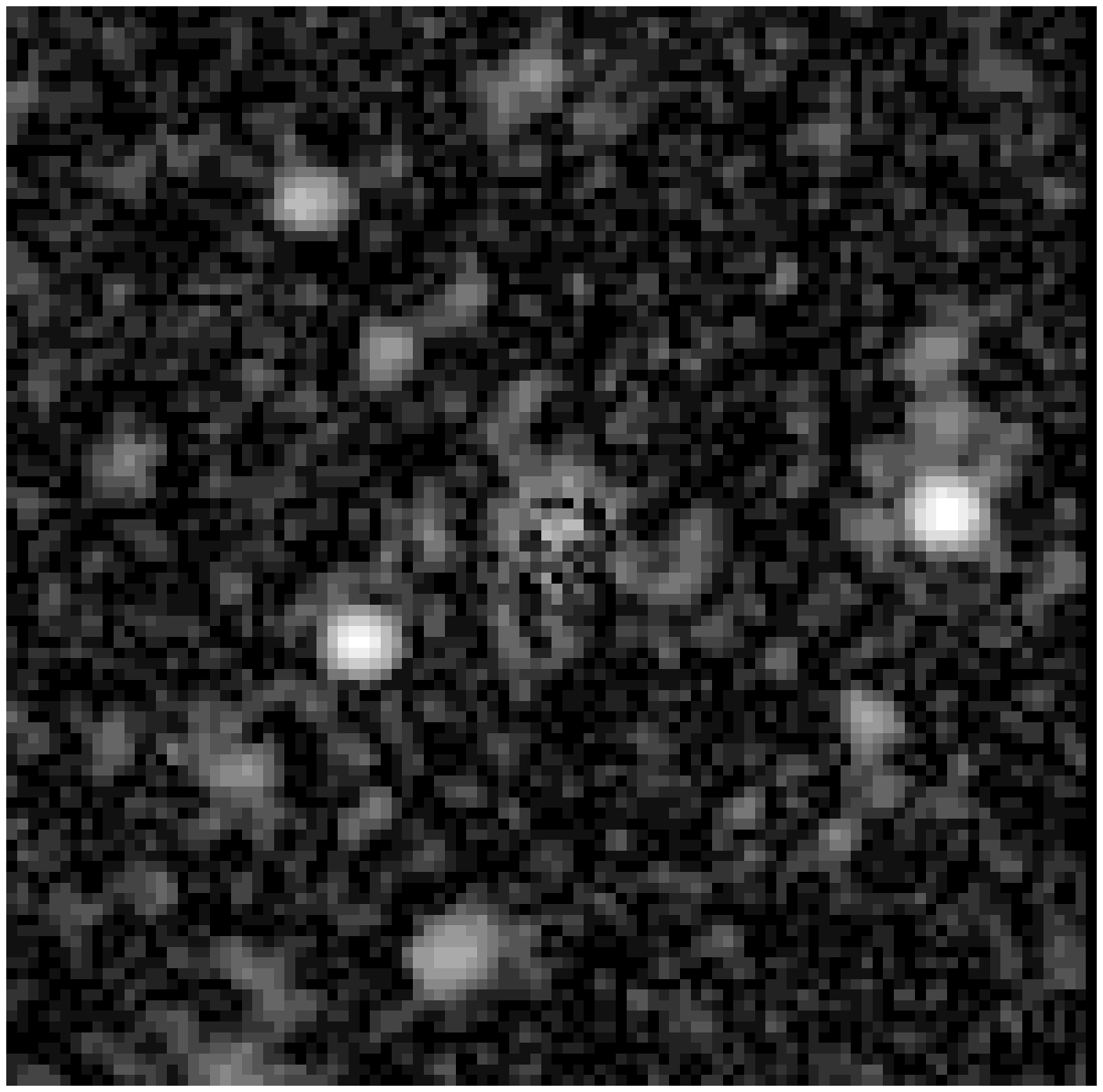}\hspace{0.06cm}
\includegraphics[angle=0,scale=.28]{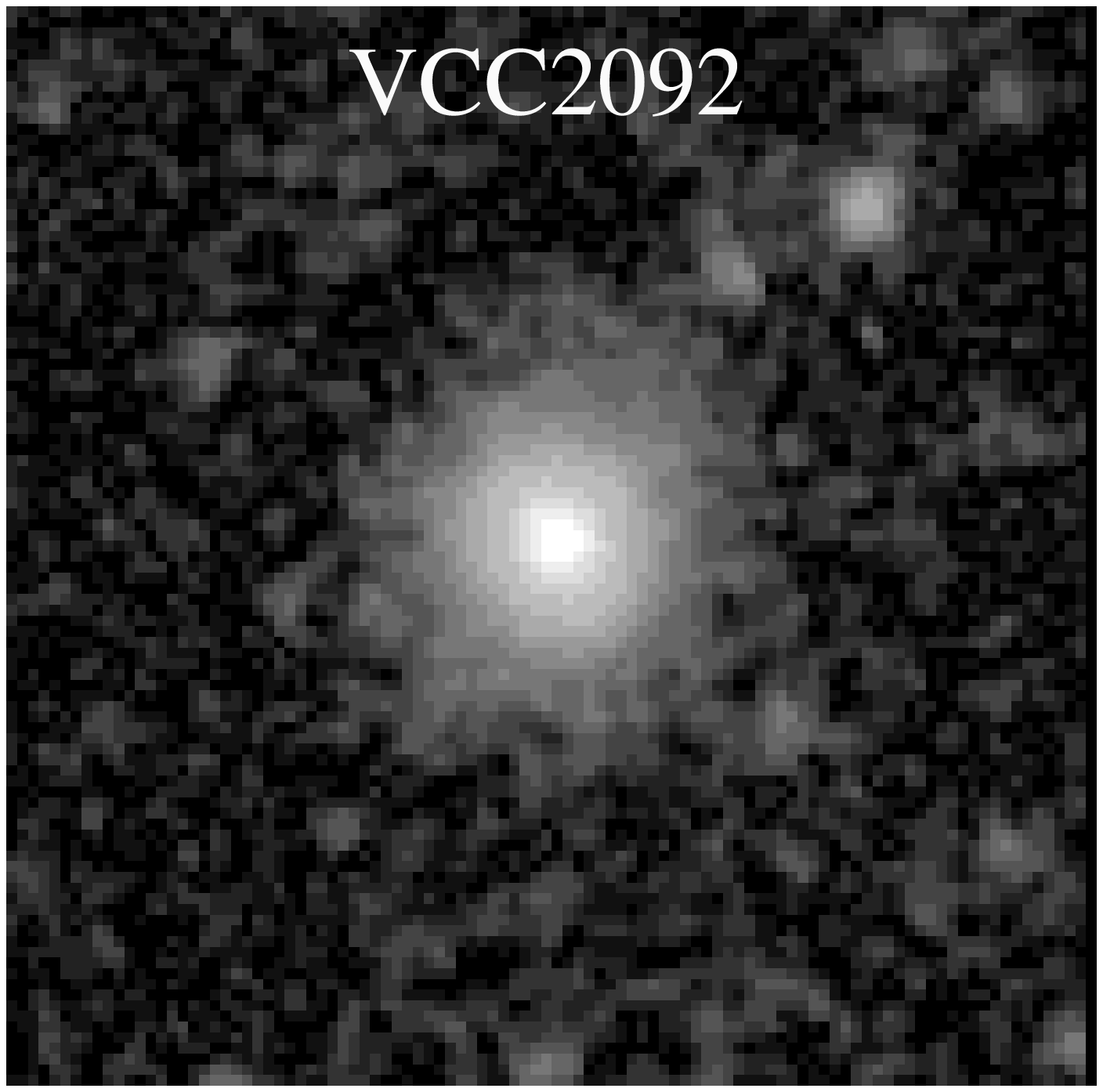}
\includegraphics[angle=0,scale=.28]{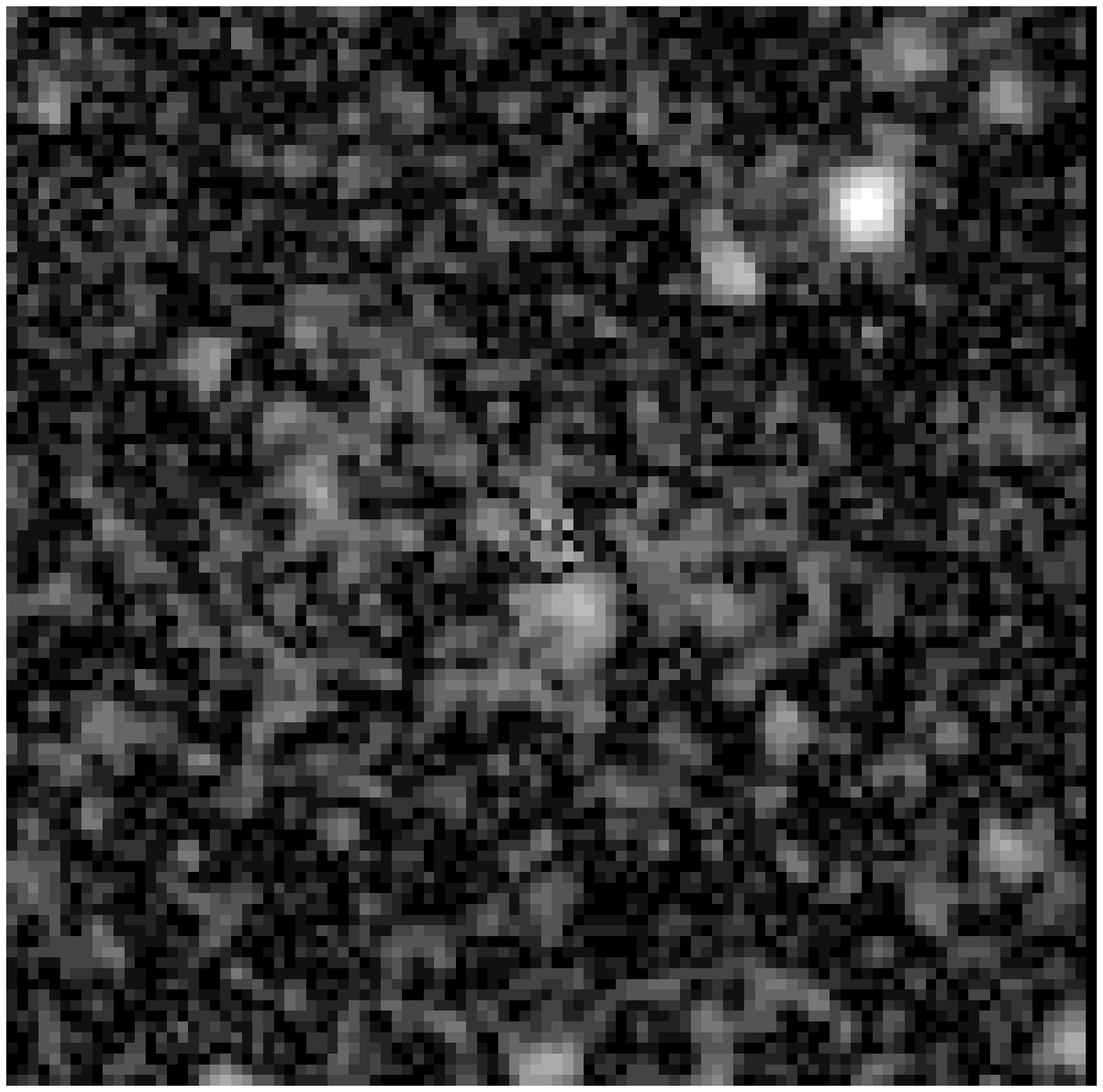}\vspace{0.025cm}
\includegraphics[angle=0,scale=.28]{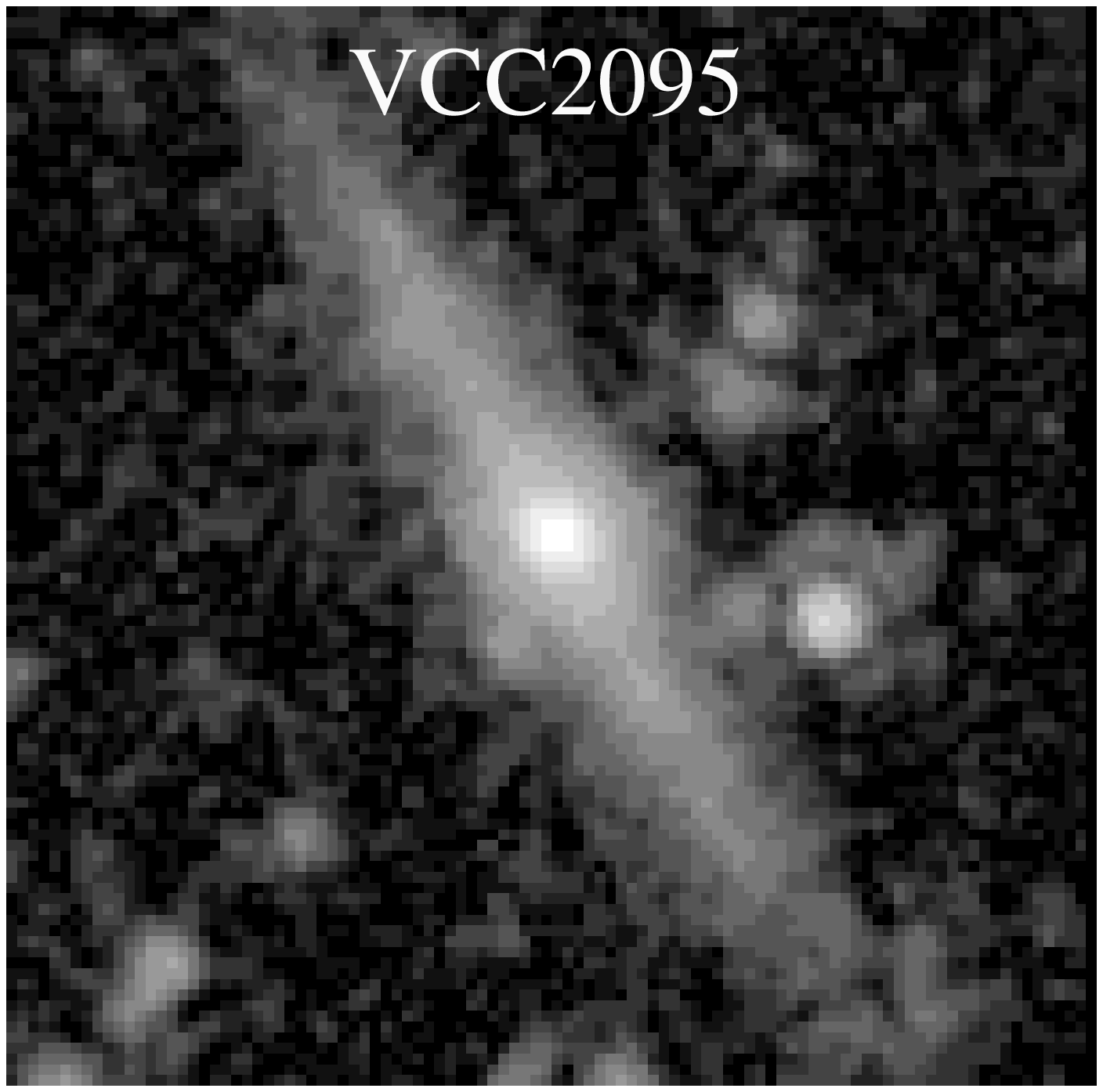}
\includegraphics[angle=0,scale=.28]{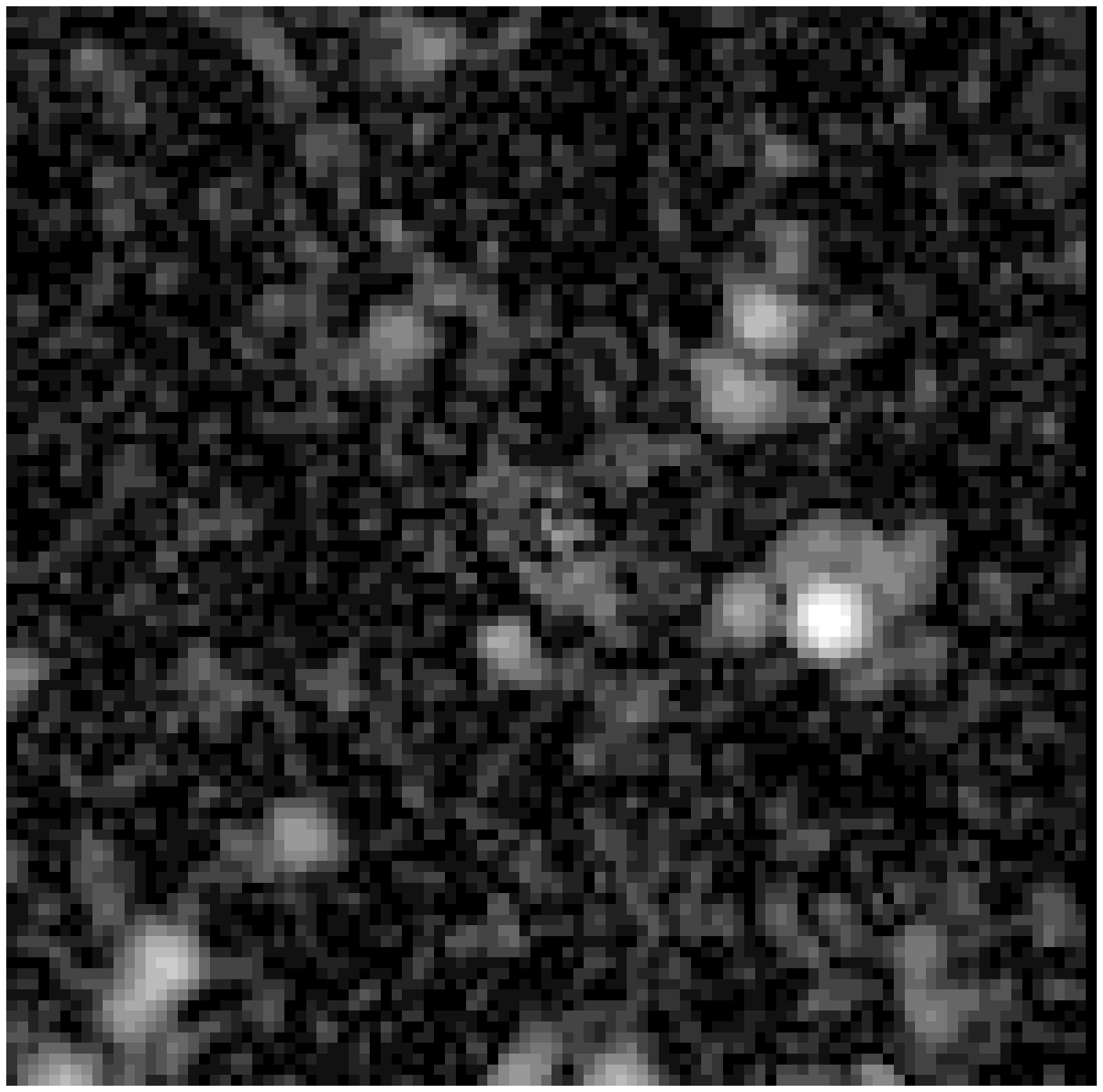}
\caption{{\it continued}}
\end{figure*}

The significance of PSF components in the individual objects was
tested using similar techniques as described earlier for the aperture
photometry: we placed 1000 random apertures in the residual images
(i.e. after subtraction of the best-fit model) and fitted the
resulting distribution of flux values with a Gaussian. The value of
three times the sigma of this Gaussian (including an aperture
correction) was converted into an AB magnitude limit for the PSF
component. For six sources the fitted PSF component was fainter than
the 3$\sigma$ magnitude limit for the respective image and, thus, the
limiting magnitude was taken as the bright limit for the PSF component
in these cases. For the remaining 31 objects the nuclear PSF component
is significant at a $>$\,$3\sigma$ level.

\section{Results and discussion}
\label{sec:results}

\subsection{Origin of the nuclear MIR ecxcess}

The additional nuclear PSF required to fit the 24\,$\mu$m radial
profiles and images could be due to processes like star formation or
low-luminosity AGN activity. However, radial variations in the age
and/or metallicity of the stellar population can also result in a
modified optical/MIR light ratio towards the center of the galaxies
\citep[e.g.][]{bre98,pio03,tem05,bre06}.  Since our method assumes
that the stellar host component has a fixed $z$/24\,$\mu$m ratio
throughout the galaxy, part of the nuclear PSF component could in fact
be due to a variation in the central properties of the stellar
populations. In the following we will discuss such effects.

\subsubsection{Residual star formation}

One possible process to boost the MIR emission in the core would be
residual star formation. In fact, the UV/optical colors of $z<0.1$
early-type galaxies indicate that 10-30\,\% of them have experienced
recent star formation with 1-3\,\% of the stellar mass having ages
$\leq$\,1-2\,Gyr \citep[e.g.][]{tre05,yi05,kav07}. However, if this is
the case for our objects, then the accompanying dust has to be located
on very small scales (or be very diffuse) because it is not detected
in the high resolution ACS optical images.

This aspect can be explored further by studying MIR spectra of sufficient quality 
which are available for 12 out of 37 objects in which no dust is visible from
optical imaging. Genuinely quiescent early-type galaxies usually do
not show aromatic features (like PAHs, strong indicators of recent 
activity) in their MIR spectra and have very low levels of FIR
emission \citep[e.g.][]{bre06,tem09,cle10}. This is consistent with
the findings for the objects in our sample
\citep[e.g.][]{bre06,breg06,tem09}.

\begin{figure}[t!]
\centering
\includegraphics[angle=0,scale=.45]{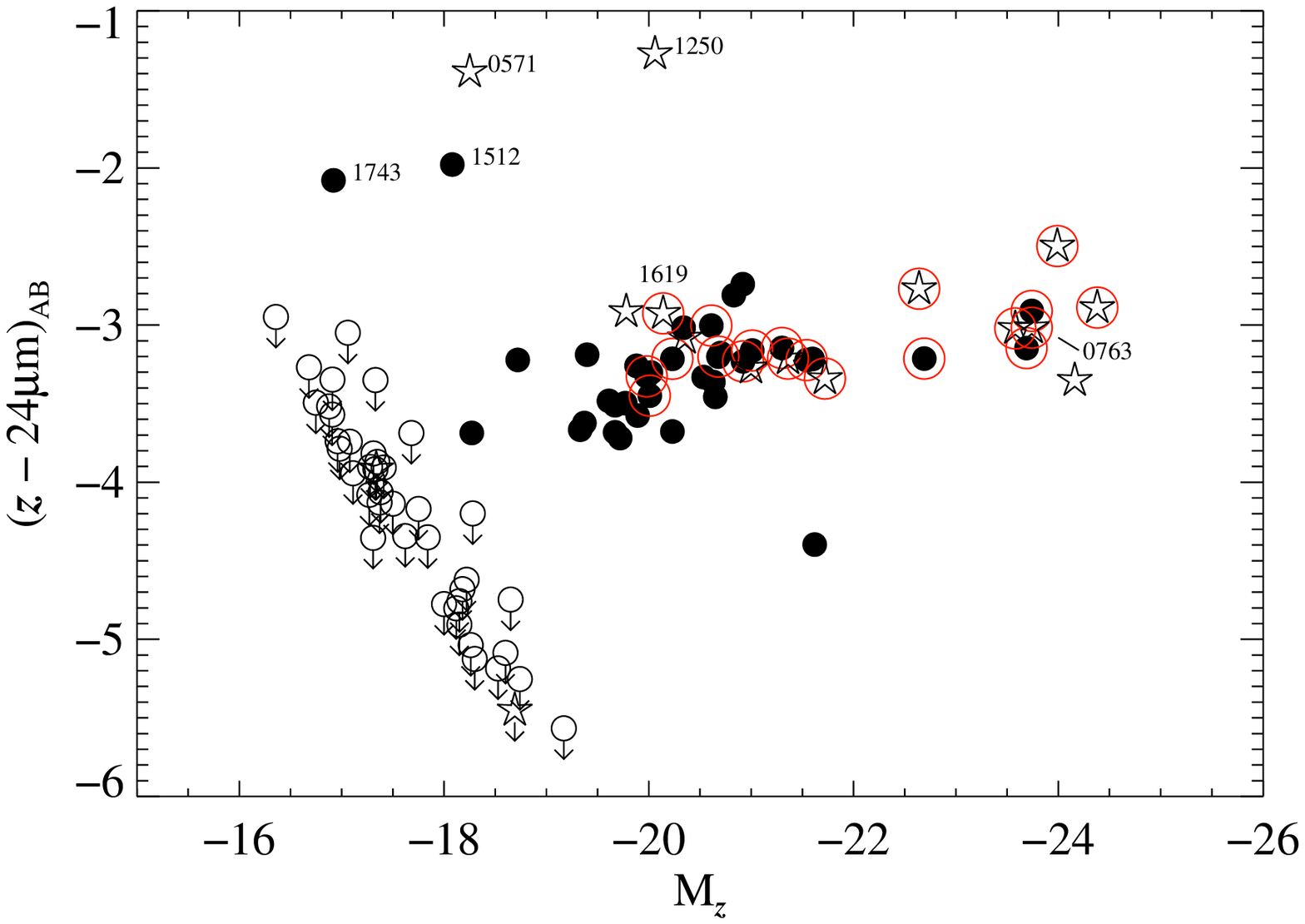}
\caption{Optical/IR color-magnitude relation for the sample objects. 
Solid circles represent objects without dust in the optical and open
stars objects with dust visible in the optical. Open circles are MIR upper
limits. Objects with MIR spectra are marked with red circles. Interestingly, two
objects with MIR spectra show PAH emission (VCC\,0763 and VCC\,1619)
but still fall onto the ``sequence'' defined by the MIR-detected
sources. Note that the upper limits shown here are appropriate for
point sources and not necessarily for resolved objects (see section
3.1).  \label{mir_cmd}}
\end{figure}

\citet[][]{cle09} report a number of ``MIR excess''
sources in their color-magnitude diagram (K$_s$-16$\mu$m vs. M$_{{\rm
K}_s}$, their Fig.\,3) of early-type galaxies which they interpret as objects being younger
(or rejuvenated). We find only a few sources showing significant
offset from most other galaxies in our sample for a comparable diagram
in $z$-24$\mu$m vs. M$_{z}$ (Fig.\,\ref{mir_cmd}). From the four
sources falling noticeably above the rest of the sample, two show dust
in the optical images which possibly boosts the 24\,$\mu$m
emission. 

While residual star formation may contribute some nuclear flux in a
few individual objects, it appears unlikely to be responsible for the
bulk of the nuclear MIR excess observed (to varying degrees) in almost
all of our MIR detected objects.

\subsubsection{Stellar age gradients}

The single stellar population (SSP) models of e.g. \citet[][]{bre98}
and~\citet{pio03}, which include the effects of dust from AGB stars,
predict a strong decline of the MIR emission with the age of the
stellar populations. A younger stellar population in the centers of the 
galaxies would then translate into increased MIR emission. However, radial age gradients in early-type
galaxies are usually small or negligible. This can be explored further
on an individual basis using the results from the optical
spectroscopic study of \citet{spo10} which has eight objects in common
with our sample of MIR-detected sources. For this small number of
objects we do not find any correlation between the strength of the
nuclear MIR component (relative to the host magnitude) and age,
metallicity, and/or the radial gradients thereof. 
If age gradients 
would play a significant role, we would expect that the sources showing 
strong age gradients also show stronger nuclear MIR excess, which is not observed. 
On the contrary, the three objects with the strongest age
gradients (VCC\,0828, VCC\,1025, and VCC\,1630) only have 3$\sigma$
upper limits on the strength of the nuclear MIR component.  
The remaining five sources with measured age gradients show nuclear MIR 
emission at a level $>$3$\sigma$ but the radial age gradients of the 
stellar population are flat within the errors. 
In particluar VCC\,1297 shows a clear nuclear MIR excess which 
accounts for about 45\% of the flux in the central 6\arcsec~but the galaxy has 
a flat radial age gradient.
From these examples it appears that radial variations in
the age of the stellar population are not the main driver of the
nuclear MIR emission in these objects.

A similar result was obtained by \citet{cle11} for some early-type
galaxies in the Virgo cluster in which they also find that the MIR light
profiles are more centrally concentrated than the NIR or optical light
profiles. Based on a near-to-mid-infrared color-magnitude diagram 
(K$_s$-16$\mu$m vs. M$_{{\rm K}_s}$) they demonstrate that age gradients are 
incompatible with the observed trends for lower mass objects to have smaller 
K$_s$-16$\mu$m colors. We also see the same trends in our color-magnitude 
diagram using $z$-24$\mu$m vs. M$_{z}$ (Fig.\,\ref{mir_cmd}).

\begin{figure}[t!]
\centering
\includegraphics[angle=0,scale=.5]{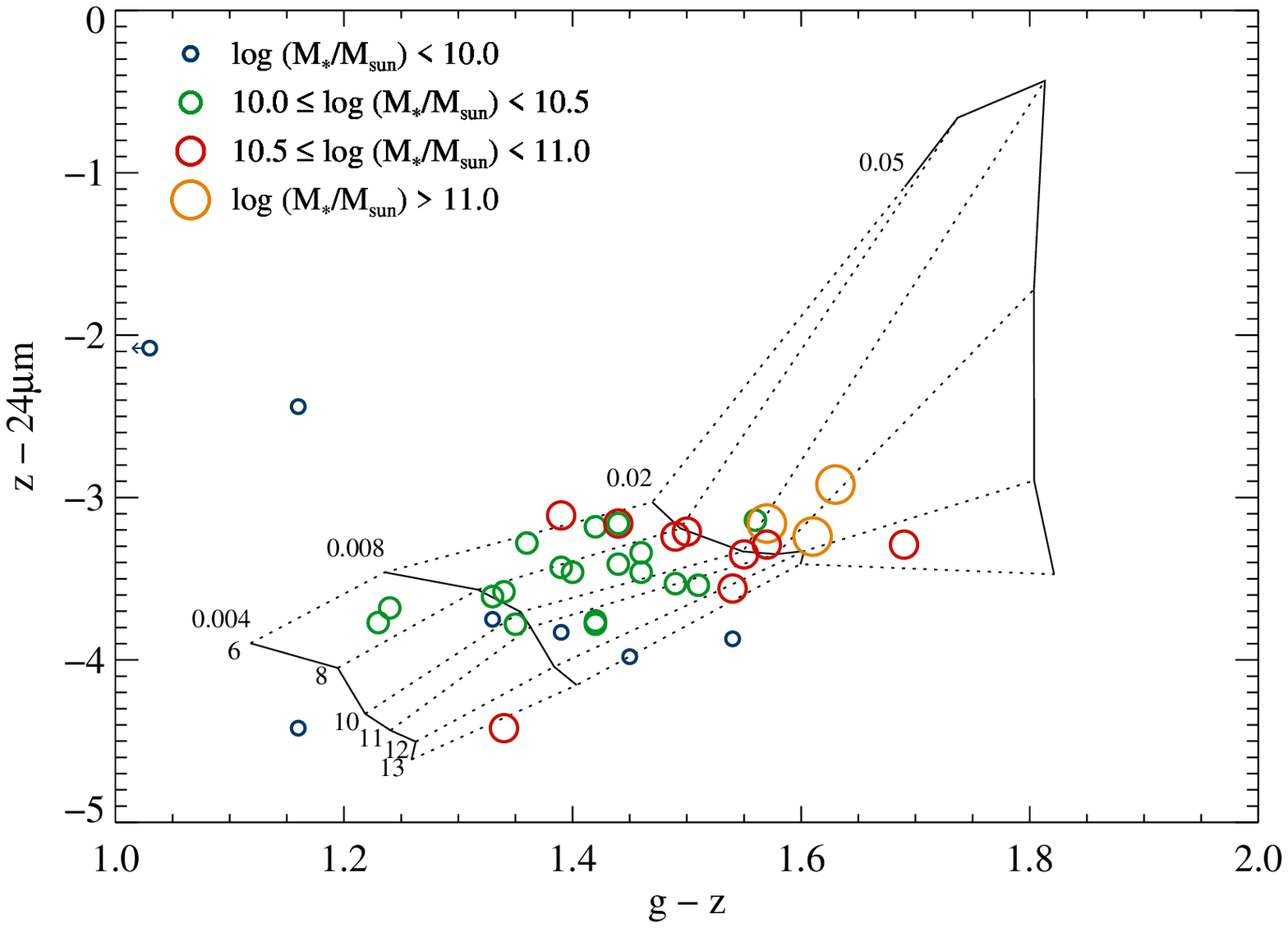}
\caption{Color-color diagram for our sample galaxies. The $g$ magnitudes were 
taken from the same reference as the $z$ magnitudes \citep{fer06}. The
MIR magnitude corresponds to the host (i.e. Sersic) component
only. The dotted (age) and solid (metallicity) lines connect the
colors expected for a representative set of SSP model SEDs taken from
the GRASIL webpage
\citep{sil98}. The symbols for the VCC objects have been color and size 
coded according to their stellar masses. \label{ccmodels}}
\end{figure}

\begin{figure}[t!]
\centering
\includegraphics[angle=0,scale=.45]{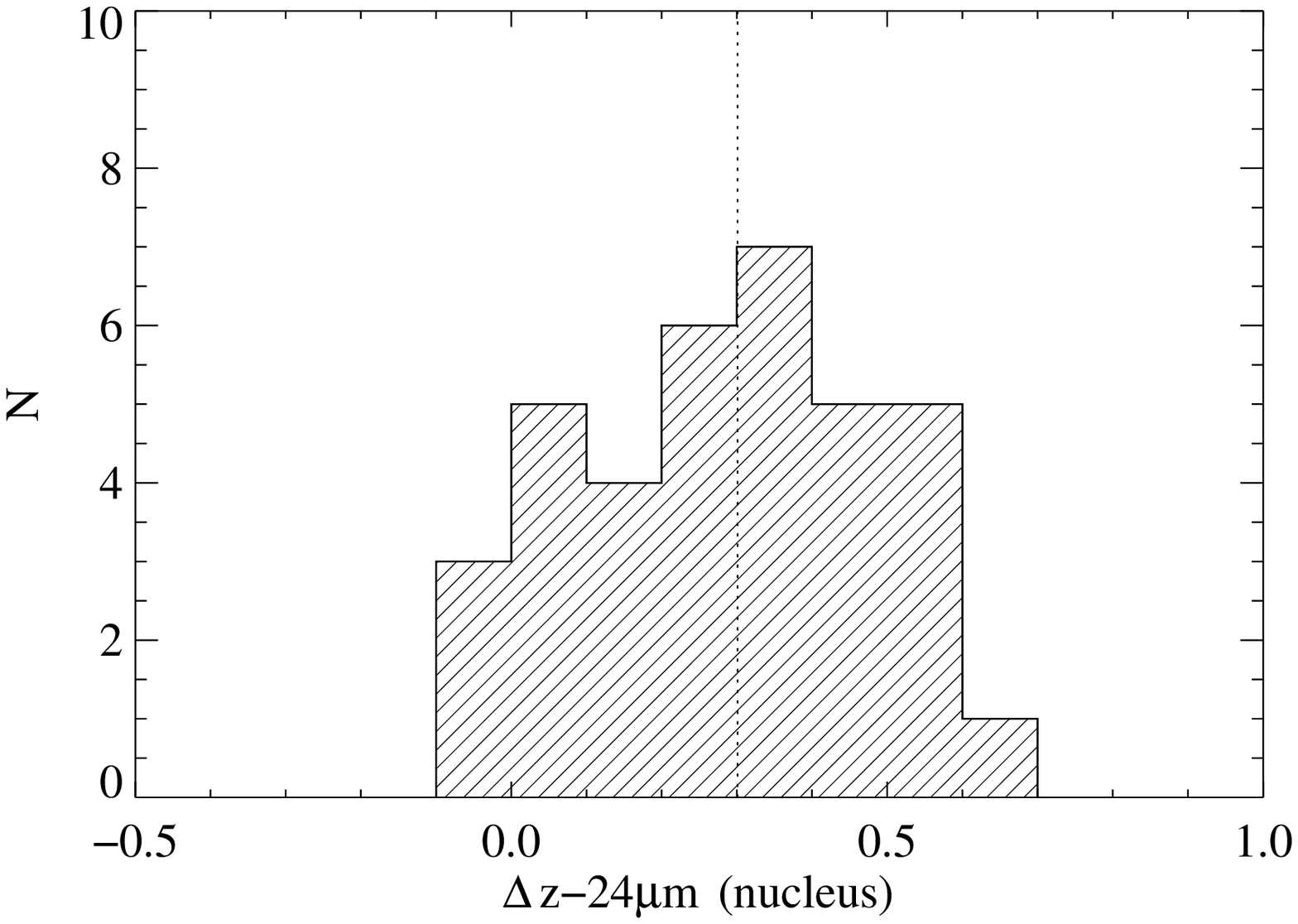}
\caption{Histogram of the nuclear color excess as determined from the radial 
profiles when integrating the inner 3\arcsec~(see
text). The dashed vertical line indicates the median value (0.3\,mag) .
\label{colorhisto}}
\end{figure}

\subsubsection{Stellar metallicity gradients}

Unlike age gradients, negative metallicity gradients are commonly
reported for early-type galaxies
\citep[e.g.][]{meh93,san07,spo10,pip10}, but the SSP models mentioned
above show only a mild dependence of the MIR emission on
metallicity. In Fig.\,\ref{ccmodels} we present an optical/MIR
color-color diagram which shows the colors predicted by the GRASIL SSP
models \citep{sil98,bre98} for our filters in an age-metallicity
grid. For most objects in our sample the host metallicities range from
about solar to half-solar values. In addition, the coding of the
symbols reveals the well known mass-metallicity relation where more
massive galaxies show higher (average) metallicities than less massive
objects \citep[e.g.][]{spo10}

\begin{figure*}[th!]
\centering
\includegraphics[angle=0,scale=.5]{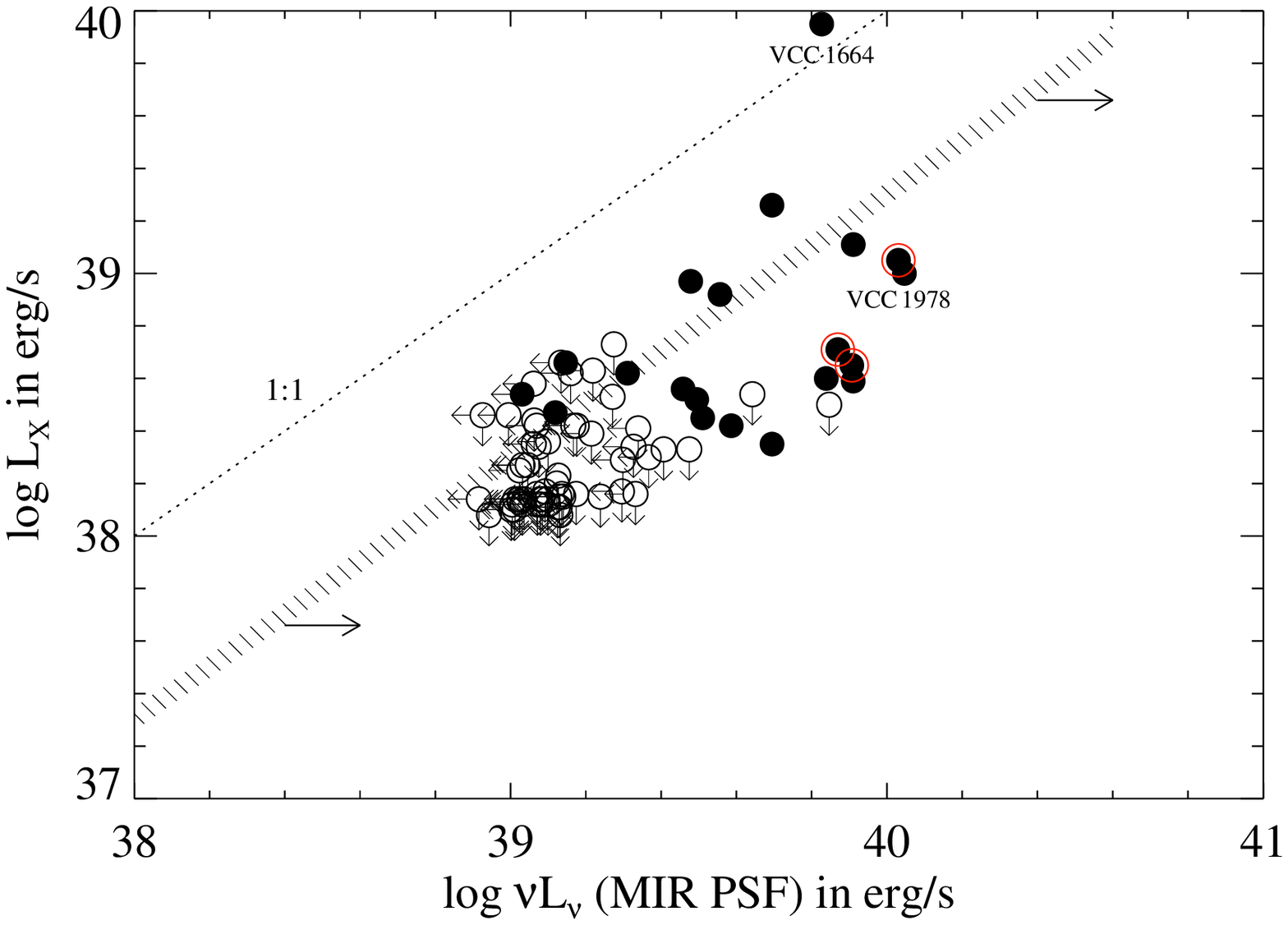}
\includegraphics[angle=0,scale=.5]{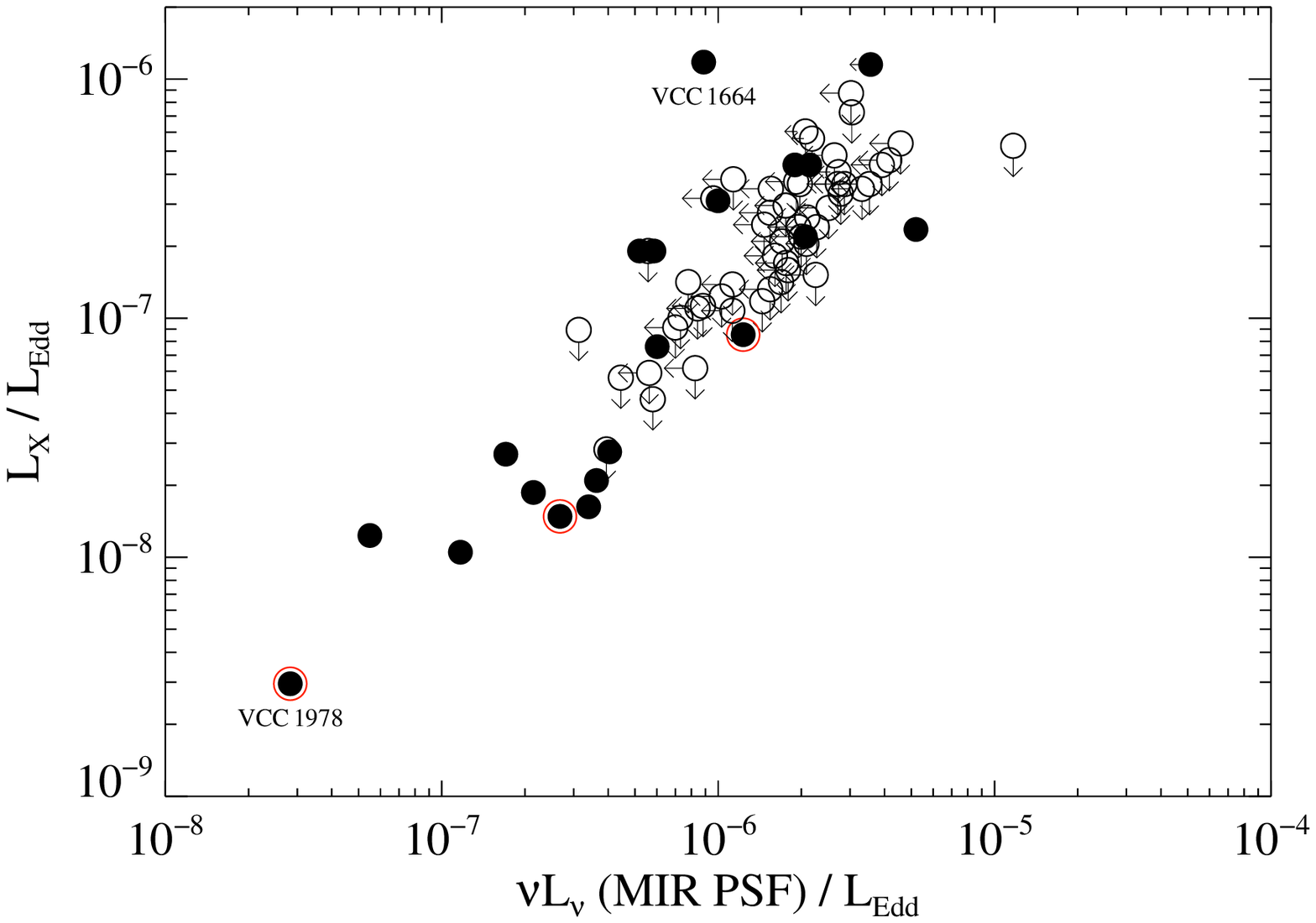}
\caption{Nuclear X-ray luminosity (0.3-10\,keV) plotted versus the luminosity 
of the PSF component at 24\,$\mu$m ({\it left}) and after both
measures have been normalized to the Eddington luminosity of the
source ({\it right}).  Filled symbols represent X-ray detected objects
while open symbols identify the X-ray non-detections. Only the 77
sources with 24\,$\mu$m observations and no optical dust are
shown. The objects circled in red indicate detections of compact radio
emission at 8.4\,GHz \citep{cap09}.  All other objects are either
undetected at cm wavelengths or were not observed. Among the
MIR-detected objects, only VCC\,1512 and VCC\,1743 were not observed
at radio wavelengths. The hashed region in the left panel shows the typical 
MIR/X-ray luminosity ratio for unobscured type-1 AGN \citep{sil04,lam09}, while the arrows 
indicate the direction in which this ratio changes with increasing obscuration. The 
1:1 relation is typical for low-luminosity AGN.\label{lx_over_ledd}}
\end{figure*}

We can use these models to predict the color changes owing to
metallicity gradients and compare them with the observed nuclear
colors.  Using the median metallicity gradient for the high-mass
objects in \citet{spo10} of ${\rm grad}\,[{\rm Z/H}]$\,$=$\,$-0.23$
(and using their definition of the gradient), we estimate an increase
of the metallicity Z by a factor of $\sim$\,1.5 between 15\arcsec~and
2\arcsec~(using the relation between [Z/H] and Z given in
\citealt{cas97}). These radii were chosen for the calculation because
at 15\arcsec~distance from the center we exclude any significant
contributions from a nuclear PSF component and the colors are
dominated by the host. At 2\arcsec~we are able to sample nuclear PSF
emission (if present), while avoiding the very centers of the radial
profiles, which are also not sampled by the optical spectra
\citep{spo10}. For the models shown in Fig.\,\ref{ccmodels} an
increase in metallicity by a factor of 1.5 results in a $z$-$24\,\mu$m
color change of about 0.3\,mag in the metallicity range $\sim$\,0.008
- 0.02.

To estimate the observed nuclear color excess we use our radial
profile fits presented in section \ref{sec:radprof}. The integration
of the scaled $z$-band profile and the 24\,$\mu$m profile 
in the inner 3\arcsec~in radius (which corresponds to the approximate 
size of the MIPS PSF at 24\,$\mu$m) yields a measure of the color excess in the
nucleus over the host galaxy $z$-$24\,\mu$m color. The results are
presented in Fig.\,\ref{colorhisto} in the form of a histogram of the 
nuclear color excess caused by the additional point source. In 
Tab.\,\ref{tab_all} we collect the fractional contribution of this 
additional nuclear point source to the flux within the inner 3\arcsec~radius.
The histogram (Fig.\,\ref{colorhisto}) reflects the
impression from the radial profile plots and demonstrates that the
nuclear MIR excess can change the nuclear colors
significantly. The median of this (fairly broad) distribution
falls close to the value we estimate from the models for a median
metallicity gradient. Thus, the nuclear MIR excess for many of our objects 
with less prominent nuclear MIR PSFs is consistent with the 
metallicity gradient derived from optical observations.
We emphasize that we only used a median value for 
the metallicity gradient to calculate the typical color change of about 0.3\,mag. 
However, the large dispersion in measured metallicity gradients can in principle 
also explain the objects with more prominent nuclear MIR excess emission. At higher 
galaxy masses, metallicity gradients in a range of approximately 0.0 to -0.4 are observed 
\citep[][their Figure\,4]{spo10}. This converts into a nuclear color change between 
zero, and $\sim$\,0.4 or up to a magnitude for lower metallicities ($\sim$\,0.008) 
or higher metallicities ($\sim$\,0.02), respectively, in the models shown 
in Fig.\,\ref{ccmodels}. In addition, we see no dependency of the nuclear color 
excess on the stellar mass of the host galaxy.

In order to explore further the distinction between color gradients
and nuclear point sources, we re-fitted the MIR images treating the
effective radius as a free parameter and not including a point
source. The effective radii from the new fits turn out to be smaller
on average than in the optical, which is not surprising given the
additional MIR light in the center of the objects. The visual
inspection of the residuals from these fits compared with the results
from the fixed $R_{\rm eff}$ (plus an additional nuclear PSF) fitting
does not reveal significant differences in most cases. The model
chi-square values for the fits also remain virtually unchanged.  We
conclude that the resolution of the data is insuffcient to distinguish
a true point source from a color gradient. However, the known stellar
population gradients appear to be a more natural explanation than true
MIR nuclear emission.

As mentioned in the previous section, our findings are 
consistent with those of \citet{cle11} who also argue (based on a 
color-magnitude diagram) that the stronger concentration of the MIR light 
compared to the NIR or optical light is due to metallicity gradients.

\subsection{Comparison with X-ray nuclei}

\citet{gal10} detect nuclear X-ray emission in 32 out of the 100 VCS objects 
(with a limiting 3$\sigma$ flux limit of $3.7 \times
10^{38}$\,erg\,s$^{-1}$ over 0.5-7 keV) and, after accounting for
contamination from low mass X-ray binaries, conclude that for most of
them the X-ray emission originates from a low-luminosity AGN. Spectral
energy distribution studies find that weak-line galaxies and LINERs
typically have (in $\nu L_{\nu}$) MIR/X-ray ratios $\sim1$
\citep{ho99,nem10}. For type-1 AGN this ratio is typically around 5
\citep{elv94,sil04,ric06,lam09}. For type-2 and increasingly obscured sources 
the ratio can go up to $10^{3-4}$ \citep[e.g.,][]{sil04,zak08,lam09}.

In Figure\,\ref{lx_over_ledd} we compare the nuclear X-ray luminosity
of the MIR-detected sources with the luminosity of the 24\,$\mu$m PSF
component. The left panels shows that for almost all the sources 
the MIR component is more luminous than
the X-ray emission by a factor of $\sim$\,$5-10$. Assuming that both originate
from nuclear activity, this ratio would require a spectral energy
distribution typical of largely unabsorbed (or only very mildly absorbed) 
luminous AGN. In contrast, allowing for a (possibly dominating) contribution 
of the stellar population to the nuclear MIR excess (see 4.1.3) would 
move the objects closer to the 1:1 line which is expected for low-luminosity 
AGN. In addition, we cannot identify any correlation between the nuclear color
excess and the X-ray luminosity of the sources. Thus, the observed MIR/X-ray 
ratio adds further circumstancial evidence in favor of metallicity gradients as the 
dominant contribution to the excess of unresolved MIR emission.

More robustly, however, the observed MIR/X-ray ratio points to an
important conclusion. There are no heavily obscured AGN lurking at the
center of these Virgo early-type galaxies which would exhibit large MIR/X-ray ratios. 
The black holes found by our survey in the X-rays have truly low bolometric Eddington ratios
arising from low accretion rates and/or highly radiatively inefficient
accretion.

\section{Summary}

We presented MIR imaging at 24\,$\mu$m of 95 early-type galaxies in
the Virgo Cluster Survey, 53 of which are detected. Of these, 16 
show evidence for dust from optical images. The radial
profiles and two-dimensional images of the remaining 37 detected sources were
modeled as a combination of a convolved optical $z$-band images and an
observed PSF. Our findings can be summarized as follows:

\begin{enumerate}
\item Virtually all the sources show unresolved excess MIR 
emission in their central regions compared to the optical light profiles 
of up to $\sim$\,0.6\,mag at 24\,$\mu$m.
\item Star formation or radial stellar age gradients do not provide a
convincing explanation for the MIR excess in the majority of the sources.
\item Negative radial metallicity gradients seem to be a
likely and natural explanation of the nuclear MIR emission for the
vast majority of the sources. The typical observed metallicity 
gradients of early-type galaxies can account for 
the nuclear color change we observe for our objects.
\item A comparison of the observed MIR excess
with X-ray nuclear emission reveals that the level of excess emission
is higher than expected for low-luminosity AGN, and would require
spectral energy distributions more typical of quasars or moderatly obscured 
but intrinsically luminous AGN. This is no longer the case once the 
stellar populations are taken into account.
\item Highly obscured nuclear activity is largely ruled out.
\item The black holes found by
our survey in the X-rays have truly low bolometric Eddington ratios
arising from low accretion rates and/or highly radiatively inefficient
accretion.
\end{enumerate}

\acknowledgments
Support for this work was provided by NASA through Chandra Award Number 08900784
(C.L., T.T.) and 11112A (B.P.M.), issued by the Chandra X-ray Observatory Center. 
T.T. acknowledges support from the Packard Foundation
through a Packard Research Fellowship. J.H.W. acknowledges support by the Basic Science 
Research Program through the
National Research Foundation of Korea funded by the Ministry of Education, Science
and Technology (2010- 0021558). We thank the referee for a constructive 
report which helped to improve the paper.

{\it Facilities:} \facility{Spitzer}, \facility{HST (ACS)}, \facility{Chandra}.

\appendix

\section{Appendix material}

\begin{center}
\LongTables
\begin{deluxetable}{ccc r@{.}l cc r@{.}l r@{.}l r@{.}l r@{.}l r@{.}l}
\tablecaption{AMUSE-Virgo III.: Source properties.\label{tab_all}}
\tablehead{
\colhead{VCC} &
\colhead{ID} &
\colhead{$d$} &
\multicolumn{2}{c}{$\log M_{\ast}$} &
\colhead{$\log M_{BH}$} &
\colhead{$z$} &
\multicolumn{2}{c}{$\log L_{\rm X}$} &
\multicolumn{2}{c}{m$_{\rm sersic}$} &
\multicolumn{2}{c}{m$_{\rm PSF}$} &
\multicolumn{2}{c}{$f_{\rm PSF}$} &
\multicolumn{2}{c}{$F_{24\,\mu{\rm m}}$} \\
\colhead{} &
\colhead{} &
\colhead{(Mpc)} &
\multicolumn{2}{c}{($M_{\odot}$)} &
\colhead{($M_{\odot}$)} &
\colhead{(mag)} &
\multicolumn{2}{c}{(erg\,s$^{-1}$)} &
\multicolumn{2}{c}{(mag)} &
\multicolumn{2}{c}{(mag)} &
\multicolumn{2}{c}{} &
\multicolumn{2}{c}{(mJy)} \\
\colhead{(1)} &
\colhead{(2)} &
\colhead{(3)} &
\multicolumn{2}{c}{(4)} &
\colhead{(5)} &
\colhead{(6)} &
\multicolumn{2}{c}{(7)} &
\multicolumn{2}{c}{(8)} &
\multicolumn{2}{c}{(9)} &
\multicolumn{2}{c}{(10)} &
\multicolumn{2}{c}{(11)}
}
\startdata
0009 &  53 & 17.14 &  9&7 &  7.20 &   12.43 & $<$\,38&15 & \multicolumn{2}{c}{\nodata} & \multicolumn{2}{r}{\nodata} & \multicolumn{2}{c}{\nodata} & 		    $<$\,0&31 \\
0021 &  73 & 16.50 &  9&0 &  6.70 &   13.74 & $<$\,38&11 & \multicolumn{2}{c}{\nodata} & \multicolumn{2}{r}{\nodata} & \multicolumn{2}{c}{\nodata} & 		    $<$\,0&33 \\
0033 &  70 & 15.07 &  8&9 &  6.60 &   13.93 & $<$\,38&25 & \multicolumn{2}{c}{\nodata} & \multicolumn{2}{r}{\nodata} & \multicolumn{2}{c}{\nodata} & 		    $<$\,0&31 \\
0140 &  58 & 16.37 &  9&4 &  7.00 &   12.86 & $<$\,38&13 & \multicolumn{2}{c}{\nodata} & \multicolumn{2}{r}{\nodata} & \multicolumn{2}{c}{\nodata} & \multicolumn{2}{c}{\nodata} \\
0200 &  71 & 18.20 &  9&2 &  6.80 &   13.55 & $<$\,38&42 & \multicolumn{2}{c}{\nodata} & \multicolumn{2}{r}{\nodata} & \multicolumn{2}{c}{\nodata} & 		    $<$\,0&30 \\
0230 &  85 & 17.78 &  8&9 &  6.50 &   14.34 & $<$\,38&62 & \multicolumn{2}{c}{\nodata} & \multicolumn{2}{r}{\nodata} & \multicolumn{2}{c}{\nodata} & 		    $<$\,0&31 \\
0355 &  28 & 15.42 & 10&3 &  8.00 &   10.59 &      38&77 & \multicolumn{2}{c}{\nodata} & \multicolumn{2}{r}{\nodata} & \multicolumn{2}{c}{\nodata} & 	12&4\tablenotemark{a} \\
0369 &  16 & 15.85 & 10&4 &  7.80 &   10.66 &      39&26 &                       13&80 &                       16&10 &   		      0&28 & 			12&28 \\
0437 &  68 & 17.14 &  9&6 &  7.20 &   12.52 & $<$\,38&17 & \multicolumn{2}{c}{\nodata} & \multicolumn{2}{r}{\nodata} & \multicolumn{2}{c}{\nodata} & 		    $<$\,0&45 \\
0538 &  90 & 22.91 &  8&9 &  6.50 &   14.74 & $<$\,38&41 & \multicolumn{2}{c}{\nodata} & \multicolumn{2}{r}{\nodata} & \multicolumn{2}{c}{\nodata} & 		    $<$\,0&28 \\
0543 &  63 & 15.70 &  9&4 &  4.10 &   12.83 & $<$\,38&27 & \multicolumn{2}{c}{\nodata} & \multicolumn{2}{r}{\nodata} & \multicolumn{2}{c}{\nodata} & 		    $<$\,0&29 \\
0571 &  72 & 23.77 &  9&4 &  7.10 &   13.63 & $<$\,38&46 & \multicolumn{2}{c}{\nodata} & \multicolumn{2}{r}{\nodata} & \multicolumn{2}{c}{\nodata} & 			 3&57 \\
0575 &  54 & 22.08 & 10&8 &  6.60 &   10.10 & $<$\,38&39 &                       14&52 &                  $>$\,18&02 &    0&40\tablenotemark{b} & 	 5&87 \\
0654 &  23 & 16.50 & 10&4 &  7.00 &   10.54 & $<$\,38&46 & \multicolumn{2}{c}{\nodata} & \multicolumn{2}{r}{\nodata} & \multicolumn{2}{c}{\nodata} & \multicolumn{2}{c}{\nodata} \\
0685 &  21 & 16.50 & 10&6 &  8.20 &   10.09 &      39&14 & \multicolumn{2}{c}{\nodata} & \multicolumn{2}{r}{\nodata} & \multicolumn{2}{c}{\nodata} & 	29&8\tablenotemark{a} \\
0698 &  49 & 18.71 & 10&0 &  6.40 &   11.69 & $<$\,38&44 &                       15&47 &                       18&04 &    0&05 & 			 2&58 \\
0731 &   7 & 23.33 & 11&7 &  8.80 &    8.15 &      39&00 &                       11&31 &                       16&06 &    0&14 & 		       110&01 \\
0751 &  88 & 15.78 &  9&4 &  6.70 &   13.12 & $<$\,38&29 & \multicolumn{2}{c}{\nodata} & \multicolumn{2}{r}{\nodata} & \multicolumn{2}{c}{\nodata} & \multicolumn{2}{c}{\nodata} \\
0759 &  17 & 16.98 & 10&8 &  7.30 &    9.79 & $<$\,38&18 & \multicolumn{2}{c}{\nodata} & \multicolumn{2}{r}{\nodata} & \multicolumn{2}{c}{\nodata} & 	13&6\tablenotemark{a} \\
0763 &   6 & 18.45 & 11&7 &  9.10 &    7.59 &      39&73 & \multicolumn{2}{c}{\nodata} & \multicolumn{2}{r}{\nodata} & \multicolumn{2}{c}{\nodata} & 	66&6\tablenotemark{a} \\
0778 &  34 & 17.78 & 10&2 &  7.50 &   11.27 &      38&56 &                       14&73 &                       16&94 &    0&00\tablenotemark{c} & 			 5&26 \\
0784 &  32 & 15.85 & 10&3 &  6.80 &   10.77 &      38&62 &                       14&55 &                       17&06 &    0&23 & 			 6&04 \\
0798 &   5 & 17.86 & 11&6 &  8.30 &    7.68 & $<$\,38&43 & \multicolumn{2}{c}{\nodata} & \multicolumn{2}{r}{\nodata} & \multicolumn{2}{c}{\nodata} & 	52&4\tablenotemark{a} \\
0828 &  36 & 17.95 & 10&2 &  6.70 &   11.25 & $<$\,38&63 &                       14&59 &                  $>$\,17&56 &    0&00\tablenotemark{b} & 	 5&64 \\
0856 &  57 & 16.83 &  9&5 &  4.50 &   12.91 & $<$\,38&16 & \multicolumn{2}{c}{\nodata} & \multicolumn{2}{r}{\nodata} & \multicolumn{2}{c}{\nodata} & 		    $<$\,0&35 \\
0881 &   4 & 16.83 & 11&9 &  8.60 &    6.97 & $<$\,38&64 & \multicolumn{2}{c}{\nodata} & \multicolumn{2}{r}{\nodata} & \multicolumn{2}{c}{\nodata} & 	27&5\tablenotemark{a} \\
0944 &  24 & 16.00 & 10&4 &  7.20 &   10.39 & $<$\,38&53 &                       13&80 &                       17&18 &    0&28 & 			13&85 \\
1025 &  40 & 22.44 & 10&4 &  7.10 &   11.22 &      38&92 &                       14&65 &                       17&20 &    0&28 & 			 5&49 \\
1030 &  19 & 16.75 & 10&8 &  7.90 & \nodata &      38&72 & \multicolumn{2}{c}{\nodata} & \multicolumn{2}{r}{\nodata} & \multicolumn{2}{c}{\nodata} & 	111&\tablenotemark{a} \\
1049 &  56 & 16.00 &  9&0 &  6.70 &   13.71 & $<$\,38&08 & \multicolumn{2}{c}{\nodata} & \multicolumn{2}{r}{\nodata} & \multicolumn{2}{c}{\nodata} & 		    $<$\,0&35 \\
1062 &  14 & 15.28 & 10&7 &  8.20 &    9.62 &      38&47 &                       12&78 &                       17&46 &    0&19 & 			28&43 \\
1075 &  81 & 16.14 &  9&2 &  6.60 &   13.54 & $<$\,38&12 & \multicolumn{2}{c}{\nodata} & \multicolumn{2}{r}{\nodata} & \multicolumn{2}{c}{\nodata} & 		    $<$\,0&31 \\
1087 &  60 & 16.67 &  9&6 &  4.80 &   12.51 & $<$\,38&15 & \multicolumn{2}{c}{\nodata} & \multicolumn{2}{r}{\nodata} & \multicolumn{2}{c}{\nodata} & 		    $<$\,0&33 \\
1125 &  44 & 16.50 &  9&9 &  7.00 &   11.69 & $<$\,38&46 &                       16&11 &                  $>$\,18&11 &    0&29\tablenotemark{b} & 	 4&22 \\
1146 &  39 & 16.37 & 10&0 &  7.00 &   11.46 & $<$\,38&33 &                       15&14 &                       16&89 &    0&24 & 			 3&83 \\
1154 &  13 & 16.07 & 10&9 &  7.90 &    9.31 &      39&03 & \multicolumn{2}{c}{\nodata} & \multicolumn{2}{r}{\nodata} & \multicolumn{2}{c}{\nodata} & 	107&\tablenotemark{a} \\
1178 &  46 & 15.85 &  9&9 &  7.20 &   11.67 &      38&66 &                       15&50 &                       17&47 &    0&41 & 			 2&66 \\
1185 &  96 & 16.90 &  9&1 &  6.50 &   13.81 & $<$\,38&36 & \multicolumn{2}{c}{\nodata} & \multicolumn{2}{r}{\nodata} & \multicolumn{2}{c}{\nodata} & 		    $<$\,0&29 \\
1192 &  79 & 16.50 &  9&5 &  6.40 &   12.98 & $<$\,38&68 & \multicolumn{2}{c}{\nodata} & \multicolumn{2}{r}{\nodata} & \multicolumn{2}{c}{\nodata} & \multicolumn{2}{c}{\nodata} \\
1199 &  93 & 16.50 &  9&0 &  6.30 &   14.18 & $<$\,38&14 & \multicolumn{2}{c}{\nodata} & \multicolumn{2}{r}{\nodata} & \multicolumn{2}{c}{\nodata} & \multicolumn{2}{c}{\nodata} \\
1226 &   1 & 17.14 & 12&0 &  9.10 &    6.79 & $<$\,38&49 & \multicolumn{2}{c}{\nodata} & \multicolumn{2}{r}{\nodata} & \multicolumn{2}{c}{\nodata} & 	747&\tablenotemark{a} \\
1231 &  11 & 15.28 & 10&8 &  8.10 &    9.38 &      38&60 &                       12&67 &                       15&66 &    0&16 & 			33&02 \\
1242 &  31 & 15.56 & 10&3 &  6.60 &   10.73 & $<$\,38&50 &                       14&19 &                       15&68 &    0&14 & 			 9&60 \\
1250 &  37 & 17.62 & 10&2 &  7.80 &   11.17 &      38&73 & \multicolumn{2}{c}{\nodata} & \multicolumn{2}{r}{\nodata} & \multicolumn{2}{c}{\nodata} & 	35&7\tablenotemark{a} \\
1261 &  48 & 18.11 &  9&8 &  5.10 &   12.12 & $<$\,38&42 & \multicolumn{2}{c}{\nodata} & \multicolumn{2}{r}{\nodata} & \multicolumn{2}{c}{\nodata} & 		    $<$\,0&31 \\
1279 &  26 & 16.98 & 10&5 &  7.60 &   10.47 & $<$\,38&73 &                       13&71 &                       17&30 &    0&19 & 			12&35 \\
1283 &  47 & 17.38 & 10&1 &  6.30 &   11.53 &      38&54 &                       15&06 &                  $>$\,17&96 &    0&22\tablenotemark{b} & 	 3&67 \\
1297 &  61 & 16.29 &  9&7 &  7.80 &   12.34 &      38&42 &                       16&21 &                       16&43 &    0&45 & 			 2&16 \\
1303 &  41 & 16.75 & 10&1 &  6.90 &   11.35 & $<$\,38&15 &                       14&96 &                       17&36 &    0&36 & 			 4&18 \\
1316 &   2 & 17.22 & 11&8 &  9.40 &    7.19 &      41&20 & \multicolumn{2}{c}{\nodata} & \multicolumn{2}{r}{\nodata} & \multicolumn{2}{c}{\nodata} & 	154&\tablenotemark{a} \\
1321 &  35 & 15.42 & 10&1 &  7.70 &   11.05 & $<$\,38&33 &                       14&82 &                       16&59 &    0&36 & 			 5&12 \\
1327 &  43 & 18.28 & 10&1 &  7.60 &   11.53 &      38&68 & \multicolumn{2}{c}{\nodata} & \multicolumn{2}{r}{\nodata} & \multicolumn{2}{c}{\nodata} & 			 6&06 \\
1355 &  59 & 16.90 &  9&4 &  6.90 &   13.02 &      38&58 & \multicolumn{2}{c}{\nodata} & \multicolumn{2}{r}{\nodata} & \multicolumn{2}{c}{\nodata} & 		    $<$\,0&27 \\
1407 &  91 & 16.75 &  9&1 &  6.60 &   13.74 & $<$\,38&35 & \multicolumn{2}{c}{\nodata} & \multicolumn{2}{r}{\nodata} & \multicolumn{2}{c}{\nodata} & 		    $<$\,0&27 \\
1422 &  50 & 15.35 &  9&6 &  7.20 &   12.24 & $<$\,38&08 & \multicolumn{2}{c}{\nodata} & \multicolumn{2}{r}{\nodata} & \multicolumn{2}{c}{\nodata} & 		    $<$\,0&30 \\
1431 &  64 & 16.14 &  9&5 &  6.90 &   12.86 & $<$\,38&66 & \multicolumn{2}{c}{\nodata} & \multicolumn{2}{r}{\nodata} & \multicolumn{2}{c}{\nodata} & 		    $<$\,0&35 \\
1440 &  84 & 16.00 &  9&2 &  6.70 &   13.40 & $<$\,38&27 & \multicolumn{2}{c}{\nodata} & \multicolumn{2}{r}{\nodata} & \multicolumn{2}{c}{\nodata} & 		    $<$\,0&29 \\
1475 &  45 & 16.60 &  9&9 &  6.60 &   11.73 & $<$\,38&34 &                       15&48 &                       17&75 &    0&24 & 			 2&62 \\
1488 &  74 & 16.50 &  9&0 &  4.10 &   13.78 & $<$\,38&14 & \multicolumn{2}{c}{\nodata} & \multicolumn{2}{r}{\nodata} & \multicolumn{2}{c}{\nodata} & 		    $<$\,0&20 \\
1489 &  99 & 16.50 &  8&7 &  6.30 &   14.73 & $<$\,38&14 & \multicolumn{2}{c}{\nodata} & \multicolumn{2}{r}{\nodata} & \multicolumn{2}{c}{\nodata} & 		    $<$\,0&31 \\
1499 &  77 & 16.50 &  8&8 &  6.70 &   14.18 &      38&42 & \multicolumn{2}{c}{\nodata} & \multicolumn{2}{r}{\nodata} & \multicolumn{2}{c}{\nodata} & 		    $<$\,0&29 \\
1512 &  98 & 18.37 &  9&2 &  7.30 &   13.24 & $<$\,38&23 &                       15&32 &                       17&84 &    0&42 & 			 2&97 \\
1528 &  65 & 16.29 &  9&3 &  6.90 &   13.06 & $<$\,38&13 & \multicolumn{2}{c}{\nodata} & \multicolumn{2}{r}{\nodata} & \multicolumn{2}{c}{\nodata} & 		    $<$\,0&27 \\
1535 &   8 & 16.50 & 11&0 &  9.20 & \nodata & $<$\,38&21 & \multicolumn{2}{c}{\nodata} & \multicolumn{2}{r}{\nodata} & \multicolumn{2}{c}{\nodata} & 	267&\tablenotemark{a} \\
1537 &  33 & 15.85 & 10&1 &  7.00 &   11.12 &      38&52 &                       14&53 &                       16&60 &    0&12 & 			 6&43 \\
1539 &  95 & 16.90 &  8&9 &  6.60 &   14.06 & $<$\,38&16 & \multicolumn{2}{c}{\nodata} & \multicolumn{2}{r}{\nodata} & \multicolumn{2}{c}{\nodata} & 		    $<$\,0&27 \\
1545 &  78 & 16.83 &  9&2 &  6.80 &   13.45 & $<$\,38&16 & \multicolumn{2}{c}{\nodata} & \multicolumn{2}{r}{\nodata} & \multicolumn{2}{c}{\nodata} & 		    $<$\,0&51 \\
1619 &  29 & 15.49 & 10&2 &  6.60 &   10.81 &      38&68 & \multicolumn{2}{c}{\nodata} & \multicolumn{2}{r}{\nodata} & \multicolumn{2}{c}{\nodata} & 	 9&8\tablenotemark{a} \\
1627 &  83 & 15.63 &  9&1 &  6.60 &   13.64 & $<$\,38&34 & \multicolumn{2}{c}{\nodata} & \multicolumn{2}{r}{\nodata} & \multicolumn{2}{c}{\nodata} & 		    $<$\,0&58 \\
1630 &  38 & 16.14 & 10&2 &  6.80 &   11.03 & $<$\,38&29 &                       14&57 &                  $>$\,17&13 &    0&03\tablenotemark{b} & 	 5&91 \\
1632 &  10 & 15.85 & 11&3 &  8.70 &    8.36 &      39&58 & \multicolumn{2}{c}{\nodata} & \multicolumn{2}{r}{\nodata} & \multicolumn{2}{c}{\nodata} & 	58&5\tablenotemark{a} \\
1661 & 100 & 15.85 &  9&0 &  6.80 &   13.63 & $<$\,38&12 & \multicolumn{2}{c}{\nodata} & \multicolumn{2}{r}{\nodata} & \multicolumn{2}{c}{\nodata} & 		    $<$\,0&29 \\
1664 &  22 & 15.85 & 10&6 &  7.70 &   10.08 &      39&95 &                       13&43 &                       15&77 &    0&36 & 			17&20 \\
1692 &  18 & 17.06 & 10&6 &  8.00 &   10.15 &      38&45 &                       13&36 &                       16&72 &    0&28 & 			17&19 \\
1695 &  66 & 16.52 &  9&5 &  7.00 &   12.83 & $<$\,38&14 & \multicolumn{2}{c}{\nodata} & \multicolumn{2}{r}{\nodata} & \multicolumn{2}{c}{\nodata} & 		    $<$\,0&26 \\
1720 &  27 & 16.29 & 10&4 &  7.70 &   10.41 & $<$\,38&54 &                       13&99 &                       16&29 &    0&34 & 			10&31 \\
1743 &  94 & 17.62 &  8&9 &  6.50 &   14.31 & $<$\,38&20 &                       16&75 &                       17&76 &    0&69\tablenotemark{c} & 	 1&01 \\
1779 &  75 & 16.50 &  9&0 &  6.70 &   13.75 & $<$\,38&14 & \multicolumn{2}{c}{\nodata} & \multicolumn{2}{r}{\nodata} & \multicolumn{2}{c}{\nodata} & 		    $<$\,0&31 \\
1826 &  97 & 16.22 &  8&8 &  6.40 &   14.37 & $<$\,38&12 & \multicolumn{2}{c}{\nodata} & \multicolumn{2}{r}{\nodata} & \multicolumn{2}{c}{\nodata} & 		    $<$\,0&32 \\
1828 &  89 & 16.83 &  9&1 &  6.60 &   13.72 & $<$\,38&16 & \multicolumn{2}{c}{\nodata} & \multicolumn{2}{r}{\nodata} & \multicolumn{2}{c}{\nodata} & 		    $<$\,0&32 \\
1833 &  67 & 16.22 &  9&3 &  6.90 &   13.21 & $<$\,38&11 & \multicolumn{2}{c}{\nodata} & \multicolumn{2}{r}{\nodata} & \multicolumn{2}{c}{\nodata} & 		    $<$\,0&34 \\
1857 &  80 & 16.50 &  9&0 &  6.70 &   13.81 & $<$\,38&14 & \multicolumn{2}{c}{\nodata} & \multicolumn{2}{r}{\nodata} & \multicolumn{2}{c}{\nodata} & 		    $<$\,0&30 \\
1861 &  62 & 16.14 &  9&5 &  6.90 &   12.74 & $<$\,38&12 & \multicolumn{2}{c}{\nodata} & \multicolumn{2}{r}{\nodata} & \multicolumn{2}{c}{\nodata} & 		    $<$\,0&26 \\
1871 &  52 & 15.49 &  9&5 &  6.90 &   12.68 & $<$\,38&08 &                       16&66 &                       17&93 &    0&00 & 			 1&03 \\
1883 &  30 & 16.60 & 10&4 &  6.80 &   10.40 &      38&35 &                       13&68 &                       16&20 &    0&39 & 			13&45 \\
1886 &  92 & 16.50 &  8&8 &  6.50 &   14.21 & $<$\,38&14 & \multicolumn{2}{c}{\nodata} & \multicolumn{2}{r}{\nodata} & \multicolumn{2}{c}{\nodata} & 		    $<$\,0&29 \\
1895 &  76 & 15.85 &  9&0 &  6.60 &   13.73 & $<$\,38&10 & \multicolumn{2}{c}{\nodata} & \multicolumn{2}{r}{\nodata} & \multicolumn{2}{c}{\nodata} & 		    $<$\,0&27 \\
1903 &   9 & 14.93 & 11&3 &  8.50 &    8.18 &      39&11 &                       11&42 &                       15&43 &    0&26 & 		       100&62 \\
1910 &  55 & 16.07 &  9&5 &  7.00 &   12.75 & $<$\,38&30 & \multicolumn{2}{c}{\nodata} & \multicolumn{2}{r}{\nodata} & \multicolumn{2}{c}{\nodata} & 		    $<$\,0&60 \\
1913 &  42 & 17.38 & 10&1 &  6.50 &   11.48 & $<$\,38&46 &                       15&24 &                  $>$\,18&05 &    0&17\tablenotemark{b} & 	 3&13 \\
1938 &  25 & 17.46 & 10&5 &  7.30 &   10.38 &      38&97 &                       13&49 &                       16&85 &    0&00 & 			19&24 \\
1948 &  82 & 16.50 &  8&8 &  6.40 &   14.34 & $<$\,38&14 & \multicolumn{2}{c}{\nodata} & \multicolumn{2}{r}{\nodata} & \multicolumn{2}{c}{\nodata} & 		    $<$\,0&27 \\
1978 &   3 & 17.30 & 11&7 &  9.40 &    7.45 &      39&05 &                       10&37 &                       15&62 &    0&07 & 		       342&10 \\
1993 &  87 & 16.52 &  9&0 &  6.40 &   14.11 & $<$\,38&14 & \multicolumn{2}{c}{\nodata} & \multicolumn{2}{r}{\nodata} & \multicolumn{2}{c}{\nodata} & 		    $<$\,0&25 \\
2000 &  20 & 15.00 & 10&4 &  8.30 &   10.27 &      38&65 &                       13&43 &                       15&45 &    0&04 & 			17&82 \\
2019 &  69 & 17.06 &  9&4 &  6.90 &   13.01 & $<$\,38&17 & \multicolumn{2}{c}{\nodata} & \multicolumn{2}{r}{\nodata} & \multicolumn{2}{c}{\nodata} & 		    $<$\,0&28 \\
2048 &  51 & 16.50 &  9&6 &  6.30 &   12.56 & $<$\,38&12 & \multicolumn{2}{c}{\nodata} & \multicolumn{2}{r}{\nodata} & \multicolumn{2}{c}{\nodata} & 		    $<$\,0&29 \\
2050 &  86 & 15.78 &  9&0 &  6.50 &   13.88 & $<$\,38&10 & \multicolumn{2}{c}{\nodata} & \multicolumn{2}{r}{\nodata} & \multicolumn{2}{c}{\nodata} & 		    $<$\,0&27 \\
2092 &  15 & 16.14 & 10&9 &  8.20 &    9.44 &      38&59 &                       12&73 &                       15&60 &    0&33 & 			31&47 \\
2095 &  12 & 16.50 & 10&6 &  7.60 &   10.17 &      38&71 &                       13&73 &                       15&75 &    0&41 & 			24&89 \\
\enddata
\tablecomments{Col.: (1) VCC source name; (2) ACS/VCS target number; (3) Distance; (4) Stellar mass of the host galaxy, in M$_{\odot}$; (5) Black-hole mass; (6) HST/ACS $z$-band
model AB magnitudes \citep[taken from][]{fer06}; (7) Nuclear luminosity between 0.3-10 keV, corrected for absorption; (8) + (9) Sersic and PSF AB magnitudes from the image
fitting at 24\,$\mu$m; (10) Fractional contribution of the additional PSF component to the nuclear MIR flux within the inner 3\arcsec~in radius as determined from the radial profiles; (11) Total 24\,$\mu$m flux, calculated as 
the sum of all the fitted components. Upper limits were determined as described in the text. Data for (3), (4), (5), and (7) taken from \citet{gal10}.}
\tablenotetext{a}{MIR aperture fluxes for dusty sources taken from \citet{tem09}.}
\tablenotetext{b}{These fractional contributions should be considered upper limits as their significance is low (see Fig.\,\ref{images_radprof}) and the presence of a nuclear
MIR excess is not supported from the two-dimensional images at a 3$\sigma$ level.}
\tablenotetext{c}{Spurious result due to the influence of a strong, nearby point source (see Figs.\,\ref{images_radprof} and \ref{images_galfit}).}
\end{deluxetable}
\end{center}


\end{document}